\newif \ifcomments
\commentsfalse
\newif\ifsubmission
\submissionfalse

\documentclass[12pt]{article}
\usepackage[utf8]{inputenc}
\usepackage{amsfonts,amssymb,amsmath,amsthm}
\usepackage{mathrsfs}
\usepackage{fullpage}
\usepackage[colorlinks=true,linkcolor=blue,citecolor=blue,breaklinks=true]{hyperref}
\usepackage{breakcites}
\usepackage{algorithm}
\usepackage{algorithmicx}
\usepackage[noend]{algpseudocode}
\usepackage{bm} %
\usepackage{dsfont}
\usepackage{graphicx, float, tikz}
\usepackage{colonequals}  %
\usepackage{color}
\usepackage{multicol}
\usepackage{tikz}
\usepackage{quantikz}
\usepackage{fancybox}
\usepackage[breakable, theorems, skins]{tcolorbox}
\usetikzlibrary{matrix,arrows}
\usetikzlibrary{positioning}
\usetikzlibrary{decorations}
\usetikzlibrary{decorations.pathreplacing}
\usetikzlibrary{shapes.geometric, chains, calc, fit, matrix}
\tikzstyle{system}=[rectangle,draw,fill=lightgray,minimum height=0.8cm,minimum
            width=0.8cm,thick]
\tikzstyle{BC}=[system]
\tikzstyle{resource}=[system]
\tikzstyle{RO}=[resource, minimum width=1cm]
\tikzstyle{protocol}=[circle, inner sep=0.7mm, draw]
\tikzstyle{simulator}=[circle, inner sep=0.7mm, draw]
\tikzstyle{memory}=[resource]
\tikzstyle{distinguisher}=[resource,fill=white,minimum width=3.5cm,
minimum height=1.2cm]
\tikzstyle{link}=[]

\usepackage{cleveref}

\usepackage{paralist}
\usepackage{enumerate}
\usepackage{enumitem}
\usepackage[n,
advantage,
operators,
sets,
adversary,
landau ,
probability,
notions,
logic,
ff,
mm,
primitives,
events,
complexity,
asymptotics,
keys]{cryptocode}

\renewcommand{\secpar}{\lambda}
\usepackage{tikz}
\usetikzlibrary{arrows,automata}

\usepackage{braket}

\usepackage[framemethod=tikz]{mdframed}
\usepackage{subcaption}

\usepackage{algpseudocode}

\usepackage{cleveref}

\usepackage[normalem]{ulem}

\newcommand{\bbF}{\mathbb{F}}

\newcommand{\bbN}{\mathbb{N}}

\newcommand{\bbR}{\mathbb{R}}

\newcommand{\calA}{\mathcal{A}}
\newcommand{\calB}{\mathcal{B}}
\newcommand{\calC}{\mathcal{C}}
\newcommand{\calD}{\mathcal{D}}
\newcommand{\calE}{\mathcal{E}}
\newcommand{\calF}{\mathcal{F}}
\newcommand{\calG}{\mathcal{G}}

\newcommand{\calI}{\mathcal{I}}

\newcommand{\calL}{\mathcal{L}}
\newcommand{\calM}{\mathcal{M}}

\newcommand{\calO}{\mathcal{O}}
\newcommand{\calP}{\mathcal{P}}
\newcommand{\calQ}{\mathcal{Q}}
\newcommand{\calR}{\mathcal{R}}
\newcommand{\calS}{\mathcal{S}}
\newcommand{\calT}{\mathcal{T}}
\newcommand{\calU}{\mathcal{U}}
\newcommand{\calV}{\mathcal{V}}
\newcommand{\calW}{\mathcal{W}}
\newcommand{\calX}{\mathcal{X}}
\newcommand{\calY}{\mathcal{Y}}

\newcommand{\Hyb}{\mathsf{Hyb}}

\DeclareMathOperator{\Tr}{\mathsf{Tr}}
\DeclareMathOperator{\Span}{\mathsf{Span}}

\newcommand{\email}[1]{\href{mailto:#1}{#1}}

\newcommand{\Gen}{\mathsf{Gen}}
\newcommand{\QFHE}{\mathsf{QFHE}}
\newcommand{\Auth}{\mathsf{Auth}}
\newcommand{\Hist}{\mathsf{Hist}}

\newcommand{\CX}{\mathsf{CX}}
\newcommand{\CZ}{\mathsf{CZ}}

\newcommand{\Can}{\mathsf{Can}}

\newcommand{\NTCF}{\mathsf{NTCF}}

\newcommand{\Supp}{\mathsf{Supp}}
\newcommand{\Chek}{\mathsf{Check}}
\newcommand{\Synd}{\mathsf{Synd}}
\newcommand{\Good}{\mathsf{Good}}

\newcommand{\PPT}{\mathsf{PPT}}

\newcommand{\A}{\mathsf{A}}
\newcommand{\Q}{\mathsf{Q}}
\newcommand{\C}{\mathsf{C}}

\newcommand{\NP}{\mathsf{NP}}
\newcommand{\secp}{{\lambda}}

\newcommand{\QPT}{\mathsf{QPT}}

\newcommand{\out}{\mathsf{out}}
\newcommand{\PC}{\mathsf{PC}}

\newcommand{\FHE}{\mathsf{FHE}}
\newcommand{\KeyGen}{\mathsf{KeyGen}}
\newcommand{\Eval}{\mathsf{Eval}}
\newcommand{\Enc}{\mathsf{Enc}}
\newcommand{\Dec}{\mathsf{Dec}}
\newcommand{\QEnc}{\mathsf{QEnc}}
\newcommand{\QDec}{\mathsf{QDec}}
\newcommand{\QEval}{\mathsf{QEval}}

\newcommand{\Open}{\mathsf{Open}}

\newcommand{\Real}{\mathsf{Real}}
\newcommand{\Ideal}{\mathsf{Ideal}}

\newcommand{\Recover}{\mathsf{Recover}}
\newcommand{\Exp}{\mathsf{Exp}}
\newcommand{\Ver}{\mathsf{Ver}}

\newcommand{\com}{\mathsf{com}}
\newcommand{\Com}{\mathsf{Com}}
\newcommand{\ct}{\mathsf{ct}}

\newcommand{\QMA}{\mathsf{QMA}}
\newcommand{\lang}{\calL}

\newcommand{\Accept}{\top} %
\newcommand{\Reject}{\bot} %

\newcommand{\Sim}{\mathsf{Sim}}

\newcommand{\Ext}{\mathsf{Ext}}

\newcommand{\Setup}{\mathsf{Setup}}
\newcommand{\pubparams}{\mathsf{pp}}
\newcommand{\td}{\mathsf{td}}

\newcommand{\Invert}{\mathsf{Invert}}

\def\vec0{\mathbf{0}}
\def\vect{\mathbf{t}}

\def\veca{\mathbf{a}}
\def\vece{\mathbf{e}}
\def\vecp{\mathbf{p}}
\def\vecu{\mathbf{u}}
\def\vecv{\mathbf{v}}
\def\vecw{\mathbf{w}}
\def\vecx{\mathbf{x}}
\def\vecy{\mathbf{y}}
\def\vecz{\mathbf{z}}
\def\vecd{\mathbf{d}}

\newcommand{\brho}{\boldsymbol{\rho}}

\newcommand{\inputReg}{\mathcal{I}}

\newcommand{\ketbra}[1]{\mathinner{|{#1}\rangle\langle{#1}|}}

\renewcommand{\vec}[1]{\mathbf{#1}}

\newcommand{\LWE}{\mathsf{LWE}}

\newcommand{\op}{{\mathsf{op}}}
\newcommand{\IF}{\mathsf{IF}}
\newcommand{\twoPC}{\mathsf{2PC}}
\newcommand{\REAL}{\mathsf{REAL}}
\newcommand{\IDEAL}{\mathsf{IDEAL}}

\newcommand{\CNOT}{\mathsf{CNOT}}

\newcommand{\inp}{\mathsf{in}}
\newcommand{\outp}{\mathsf{out}}

\newcommand{\RSP}{\mathsf{RSP}}
\newcommand{\new}{{\color{red}\textbf{(New)} }}
\newcommand{\modified}{{\color{red}\textbf{(Modified)} }}

\newcommand{\KeyEval}{\mathsf{KeyEval}}
\newcommand{\Write}{\mathsf{Write}}
\newcommand{\fDec}{\mathsf{fDec}}
\newcommand{\ZXLin}{\mathsf{ZX\text{-}Lin}}
\newcommand{\ans}{\mathsf{ans}}
\newcommand{\CBOracle}{\mathsf{fMeas}}
\newcommand{\fMeas}{\CBOracle}

\newcommand{\SRExp}{\mathsf{SR\text{-}Exp}_{\secpar, n}}

\newcommand{\tel}{\mathsf{tel}}

\newenvironment{functionalitybox}
    {\begin{tcolorbox}[colback=white,colframe=black,boxrule=0.4pt,arc=0pt,left=4pt,right=4pt,top=4pt,bottom=4pt]}
    {\end{tcolorbox}}

\newcommand{\SampBQP}{\mathsf{SampBQP}}

\newcommand{\CVQS}{\mathsf{CVQS}}

\newcommand{\format}{\mathsf{format}}
\newcommand{\prop}{\mathsf{prop}}

\newcommand{\CollIP}{\mathsf{Coll}\text{-}\mathsf{IP}}

\newcommand{\IndExt}{\mathsf{Ind}\text{-}\mathsf{Ext}}

\newcommand{\vecl}{\boldsymbol{\ell}}

\newcommand{\diffConditionedL}[2]{\Delta_{\vecl}\Big(#1\Big|#2\Big)}

\newcommand{\approxnegl}{\approx_{\negl}}

\newcommand{\vecb}{\vec{b}}
\newcommand{\vecc}{\vec{c}}

\newcommand{\probone}{{\mathsf{Pr}_{1}}}
\hypersetup{
    pdftitle={The Entire Quantum Computer on a Classical Leash},
    pdfauthor={Yael Tauman Kalai, Dakshita Khurana, Justin Raizes}
}

\title{The Entire Quantum Computer on a Classical Leash}

\ifsubmission
\author{}
\else
\author{Yael Tauman Kalai\thanks{MIT. \email{tauman@mit.edu}} \and Dakshita Khurana\thanks{UIUC and NTT Research. \email{dakshita@illinois.edu}} \and Justin Raizes\thanks{NTT Research.\email{justin.raizes@ntt-research.com}}}
\fi
\date{}
\newtheorem{definition}{Definition}
\newtheorem{remark}{Remark}
\newtheorem{lemma}{Lemma}
\newtheorem{theorem}{Theorem}
\newtheorem{claim}{Claim}
\crefname{claim}{Claim}{Claims}

\newtheorem{corollary}{Corollary}

\theoremstyle{definition}
\newtheorem{construction}{Construction}
\crefname{construction}{construction}{constructions}
\Crefname{construction}{Construction}{Constructions}

\newtheorem{protocol}{Protocol}
\begin{document}
\maketitle

\begin{abstract}
    We show that a classical client with no quantum hardware can use a single untrusted and potentially malicious quantum server as if it were its own trusted quantum computer. Through the server, the client can privately choose and run quantum computations, prepare quantum states, and process quantum states supplied by other parties. It can also keep quantum memory on the server, hidden from the server itself, and use that memory in later computations so its quantum state persists across rounds of interaction. Together, these capabilities let a purely classical party act as a quantum party in a variety of interactive protocols.

We expect this virtual quantum computer to be a broadly useful tool. As immediate applications, we resolve two known open problems: (1) classical verification of quantum sampling ($\mathsf{SampBQP}$) with negligible computational soundness error, and (2) remote state preparation ($\mathsf{RSP}$) of arbitrary efficiently preparable quantum states.

Underlying all of these results is a general-purpose, maliciously secure two-party computation protocol between a classical client and a quantum server, which we construct assuming only the post-quantum hardness of learning with errors.

\end{abstract}

\clearpage
\setcounter{tocdepth}{2}
\tableofcontents
\clearpage

\pagenumbering{arabic} 

\section{Introduction}

Can a classical party use an adversary’s quantum computer as its own trusted quantum device?

Mahadev’s breakthrough~\cite{Mahadev18} demonstrated that a classical party can verify quantum decision problems, even when the quantum computation is performed by an untrusted prover. 
A quantum computer, however, provides much more than answers to decision problems. It can sample from distributions, prepare quantum states for other parties, process quantum states received from them, and retain quantum information across successive computations. These capabilities are central to quantum cryptography, where security can depend on information being encoded in a quantum state rather than revealed through a classical description. Participating in such a protocol may require a party to keep private quantum memory, manipulate it in response to incoming messages, and release only selected outputs.
This raises a broader question:
\begin{center}
{\em Can a classical client obtain the capabilities of a trusted quantum computer\\ through classical interaction with a single untrusted quantum server?
}
\end{center}

Even restricted versions of this goal remain unresolved as we discuss below.

\paragraph{$\mathsf{SampBQP}$ Verification.}
A natural first step beyond decision problems is to ask a quantum server to sample the classical output of a prescribed quantum circuit. This task, known as classical verification of $\SampBQP$, requires the client to verify the distribution of the output, rather than a single decision bit. Prior work~\cite{EC:CLLW22} makes progress on this question, but is limited to soundness error that is inversely proportional to the number of repetitions, leaving the following question open. 
\begin{center}
    {\em Can classical verification of $\SampBQP$ be achieved with {\em negligible error}?}
\end{center}

\paragraph{Remote State Preparation.}
Quantum outputs introduce a further challenge: correctness alone may not capture the desired cryptographic guarantee. 
Consider, for example, a classical party that wishes to give a quantum adversary an unclonable quantum ciphertext. The classical party may know how to prepare the ciphertext, but sending these instructions to the quantum device would defeat the purpose: the adversary could generate as many ciphertexts as it likes. Asking the adversary to prove that it prepared the correct state does not resolve this problem either, since correctness alone says nothing about the additional copies or information it may have retained. We would like the adversary to learn only what it could have learned from receiving a single quantum ciphertext, even though it performs every quantum operation involved in preparing that ciphertext! 

The example above is a special case of what is known as (verifiable) remote state preparation (RSP) in the literature~\cite{FOCS:GheVid19}.
RSP allows a classical party to ``send'' a quantum state $\ket{\psi}$ to a quantum party, as long as it knows a classical description of a quantum circuit that efficiently prepares the state. The goal is for the quantum party to receive no more information than if it had directly received $\ket{\psi}$.

There has been meaningful progress on building RSP for restricted state families with inverse-polynomial error~\cite{FOCS:GheVid19,ICALP:GheMetPor23}. For cryptography, such a loss can overwhelm the security guarantee altogether: in the case of the unclonable encryption example discussed earlier, this leaves room for an adversary to succeed with inverse-polynomial probability where the ideal quantum protocol permits only negligible success.
Thus, seven years after its introduction, the main question about RSP has remained open:
\begin{center}
    {\em Can remote state preparation be obtained for {\em arbitrary}\\ efficiently computable states with {\em negligible error}?}
\end{center}

\paragraph{Stateful Multi-Round Protocols, Beyond RSP.}
A key motivation for the introduction of RSP was to allow a classical client to effectively assume the role of a quantum party in an interactive protocol. But  enabling this requires an even stronger form of classical control: we want the classical device to be able to direct the quantum device through multiple rounds of interaction, with assured privacy and a guarantee that any deviations will be detected.

Let us explain what we are looking for in more detail.
Imagine a classical client with a trusted quantum computer of its own. The client could privately choose a computation, apply it to a quantum state received from another party, and return a quantum output. It could then keep part of the state as private memory, and use it in a later computation to return another output, and so on. We would like to faithfully reproduce this entire interaction when {\em all of the quantum hardware is instead controlled by another party, who may be malicious}. The client should communicate only classically, yet retain the privacy and control that its own quantum computer would have provided.

\paragraph{The Obstacle: Testing Destroys the Output.} A central obstacle is that checking a quantum computation can destroy the very state we wanted to use. This is relatively harmless when the goal is to learn a decision: after obtaining enough evidence for the answer, we can discard the computation. It becomes a serious problem when the goal is to obtain a sample or continue computing on a quantum output. Evidence that a state was correct is of little use if producing that evidence consumes the state.

One natural approach is to ask the server for many candidate states, test most of them, and retain an untested state for use. This approach underlies the classical verification of $\SampBQP$~\cite{EC:CLLW22}. However, testing polynomially many candidates leaves an inverse-polynomial opportunity for an incorrect state to escape inspection. For decision computations, independent repetition and majority can reduce a bounded error to negligible. For general sampling tasks and quantum outputs, there is no analogous generic operation that combines several moderately reliable outputs into one negligibly erroneous output. 

There is a further difficulty. Even if a test preserves an honest server’s state, what prevents a malicious server from replacing its state after passing the test? To support a sequence of computations, the client needs a guarantee about the same evolving quantum state throughout the interaction. It must be possible to test that state {\em without destroying it}, continue using the state, and keep the server bound to the result of each operation. Obtaining this guarantee through classical communication is the central technical challenge addressed by our work.

\subsection{Main Result}

\begin{figure}[htbp]
    \centering
    \includegraphics[width=0.9\textwidth]{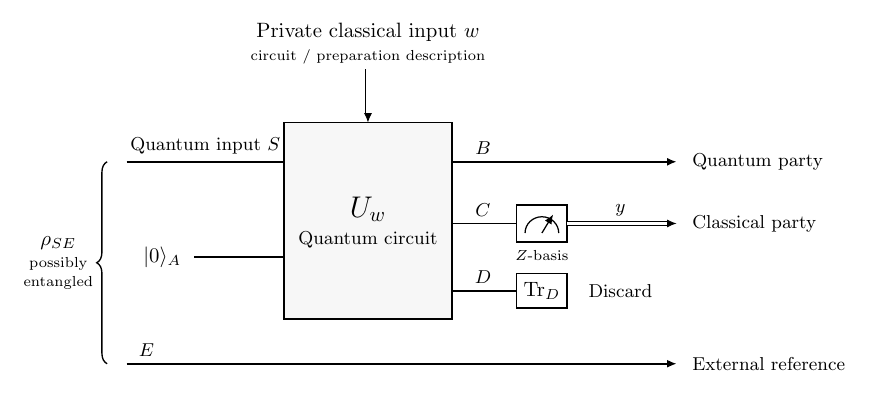}
    \caption{The ideal quantum two-party functionality (abort omitted). The client's private input $w$ specifies $U_w$, which acts on the server's input register $S$ and zero-initialized ancillas $A$. It returns $B$ to the server, measures $C$ in the computational basis to produce the client's output $y$, and discards $D$. The possibly entangled external reference $E$ is untouched. Our protocol realizes this functionality using only classical communication.}
    \label{fig:quantum-2pc-functionality}
\end{figure}

Our main construction is a two-party computation protocol between a classical client and a quantum server for general real quantum circuits.
\begin{itemize}
\item 
The client supplies a private classical input which may specify a {\em classical description of} (1) the desired quantum circuit $C$ and (2) an efficient preparation procedure for an initial state $\ket{\psi_C}$.
\item The server supplies an arbitrary quantum input, possibly entangled with an external reference. 
\item The computation returns designated classical outputs to the client and prescribed quantum outputs to the server. The parties may choose to discard some output wires so their state/value is returned to neither party.
\end{itemize}Figure~\ref{fig:quantum-2pc-functionality} illustrates the ideal functionality. 

\begin{theorem}[Informal]
    Assuming the post-quantum hardness of LWE, there exists an interactive protocol for secure quantum computation of real quantum circuits between a classical client and a quantum server which achieves simulation-based security with abort with an $\epsilon$-simulator.
\end{theorem}

Here, $\epsilon$-simulation means that, for every inverse-polynomial $\epsilon>0$, there is a simulator running in time polynomial in the security parameter, the computation size, and $1/\epsilon$, whose ideal execution is indistinguishable from the real execution up to computational advantage $\epsilon$. Crucially, the protocol itself and the honest parties' running times are independent of $\epsilon$. This guarantee suffices to preserve negligible security loss in any efficiently testable cryptographic game: any inverse-polynomial violation would yield an attack on the ideal functionality by choosing $\epsilon$ sufficiently smaller than the violating advantage. We highlight some example applications below where we get the standard (negligible-error) indistinguishability-based security.
\begin{remark}
At the end of \Cref{sec:overview:2pc}, we isolate a particular primitive whose construction with negligible security would generically lift our results to negligible-loss simulation.\footnote{ChatGPT 6 has claimed a construction of this primitive, thereby lifting the result to negligible simulation loss. We are in the process of evaluating the claim.}
\end{remark}

The restriction to real gates is inherent to any classical-communication protocol and is not overly restrictive.
It is well-established that any quantum circuit with complex gates can be generically converted into one with real gates, with a specific input-encoding~\cite{STOC:BV93}. Moreover, many of the example applications we mention below natively use real quantum circuits.

\paragraph{Private Quantum Memory Across Computations.}
The sequential composition of our protocol also allows the classical client to maintain private quantum memory on the server across a polynomial number of sequential computations. 
At each stage, the client can privately choose the next computation, process the stored memory together with newly supplied quantum inputs, and obtain designated outputs while retaining the remaining state for future use. 
The quantum server is prevented from tampering with the persistent memory by authenticating it between each computation.
The resulting interaction admits $\epsilon$-simulation with abort, including when the memory is entangled with an external system. This gives the client the capabilities of a trusted quantum computer whose state persists across interactions. We describe this extension in Section~\ref{sec:memory-delegation}.

\subsection{Applications.}
We highlight a few consequences and applications of our main result below.

\paragraph{Verification of quantum sampling.}
Chung, Lee, Lin, and Wu~\cite{EC:CLLW22} gave classical verification of $\SampBQP$ with inverse-polynomial computational soundness error and explicitly asked whether negligible soundness error is possible. 
We resolve this question for verification of the output distribution of a prescribed quantum circuit.

\begin{corollary}[Verification of Quantum Sampling, Informal]
    Let $C$ be an efficient quantum circuit with classical output distribution $D$. 
    
    Assuming the post-quantum hardness of LWE, there exists a classical verification of quantum sampling protocol where the classical client's output distribution is computationally indistinguishable from $D$ (conditioned on the quantum server not causing an abort).
\end{corollary}

The negligible error here is the additional error introduced by verification. It is separate from the statistical approximation allowed in the definition of $\SampBQP$: if $C$ samples from $D$ with $\|D-D^*\|_{\mathsf{TV}}\leq\delta$ for a target distribution $D^*$, then the corresponding joint decision-and-output distribution has computational distinguishing error at most $\delta+\negl$ relative to ideal sampling from $D^*$. Our result does not require the quantum sampler's approximation error $\delta$ itself to be negligible. See Section~\ref{sec:CVQS}.

\medskip
\noindent\textbf{Remote state preparation.}
Prior RSP protocols prepare states from restricted families with security error scaling inversely with the protocol length $T$~\cite{FOCS:GheVid19,ICALP:GheMetPor23}. 
When the time $T$ protocol is used to replace a round of quantum communication in a falsifiable game $G_Q$ to obtain $G_C$, the error permits an inverse polynomial increase in the maximum winning probability of any QPT adversary $\mathsf{adv}$:
\[
    \max_{\mathsf{QPT\ adv}} \Pr[\mathsf{win}\ G_C] \leq \max_{\mathsf{QPT\ adv}}  \Pr[\mathsf{win}\ G_Q] + 1/\poly[T]
\]
Gheorghiu, Metger, and Poremba explicitly ask how to obtain negligible security error in the cryptographic applications of RSP. 
Our construction addresses this question through $\epsilon$-simulation and also supports arbitrary (real) efficiently preparable states chosen by the sender.

\begin{corollary}[Remote State Preparation, Informal]
    Let $\ket{\psi}$ be any state which can be efficiently prepared by a real circuit. 
    
    Assuming the post-quantum hardness of LWE, there exists a remote state preparation protocol with $\epsilon$-simulation for $\ket{\psi}$.
    As a result, it has standard (negligible error) falsifiable security:
    \[
        \max_{\mathsf{QPT\ adv}} \Pr[\mathsf{win}\ G_C] \leq \max_{\mathsf{QPT\ adv}}  \Pr[\mathsf{win}\ G_Q] + \negl
    \]

\end{corollary}

Here, arbitrary means any efficiently preparable state chosen through the sender’s private classical description—not an unknown state supplied to the classical sender.

This is an $\epsilon$-simulation guarantee but its cryptographic applications can nevertheless have negligible security loss. For example, the client can remotely prepare an unclonable quantum ciphertext without revealing its preparation procedure. Replacing quantum transmission by this protocol preserves the original unclonability game's security bound up to a negligible additive term: any inverse-polynomial excess winning probability would contradict the original scheme's security by choosing $\epsilon$ sufficiently smaller than that excess. Thus the receiver's participation in preparing the ciphertext does not give it a non-negligible advantage in producing two usable ciphertexts. See Sections~\ref{sec:general-rsp} and~\ref{sec:other-applications}.

\medskip
\noindent\textbf{Classical revocation of leased quantum keys.}
Secure key leasing allows a user to borrow a quantum decryption key, with the guarantee that returning a key that passes verification removes the user's ability to decrypt, as formalized by the scheme's security game~\cite{EC:AKNYY23}. In the quantum-return formulation, revocation requires the lessor to receive and verify a quantum key. A classical revocation protocol removes that requirement, allowing the lessor to verify surrender of the decryption capability through classical interaction alone.

\begin{corollary}[Classical Revocation of Quantum States, Informal]
    Assuming the post-quantum hardness of LWE, every secure key leasing scheme with efficient quantum verification admits an interactive revocation protocol using only classical communication. The compiled scheme preserves security after revocation up to negligible loss.
\end{corollary}

The parties securely evaluate the scheme's verification procedure on the lessee's quantum input and the lessor's private verification key, return the verification decision to the lessor, and discard the submitted key registers. This application illustrates the ability to process and \textit{consume} a quantum input already held by an adversary.
\par
\pagebreak[0]
\begin{samepage}
\paragraph{Other Applications.}
Section~\ref{sec:other-applications} develops the following additional applications:
\begin{itemize}
    \item Unclonable encryption over a classical channel (Section~\ref{sec:classical-unclonable-encryption}).
    \item Public-key quantum money with a classical bank from polynomial hardness (Section~\ref{sec:classical-bank}).
    \item Improved extraction for classical-verifier arguments of quantum knowledge (Section~\ref{sec:cvqc-knowledge}).
    \item Secure key leasing with proofs of no-intrusion (Section~\ref{sec:PoNI-skl}).
    \item Quantum obfuscation with interactive classical communication (Section~\ref{sec:classical-obfuscation}).
\end{itemize}
We provide more details about these applications in \Cref{sec:overview:applications}.
\end{samepage}

\section{Technical Overview}

We begin with an overview of our approach for a slightly less ambitious task: verifying a single general quantum computation where the circuit $U = U_T\circ \dots \circ U_1$ is public. After computing $U$, the classical party's output register $\calC$ is measured in the computational basis and the discarded register $\calD$ is traced out.

\paragraph{History State Verification.}
Feynman's famed clock construction succinctly encodes the entirety of any quantum computation as a single quantum state: the ``history'' state. As computation progresses, the initial input and ancillas $\ket{\psi_0} = \ket{\inp} \otimes \ket{0^n}$ evolves into $\ket{\psi_1} = U_1 \ket{\psi_0}$, then into $\ket{\psi_2} = U_2 \ket{\psi_1}$, and so on. The history state encodes all steps of the evolution in superposition over the timestep $t$:
\[
    \sum_{t=1}^{T} \ket{t} \otimes \ket{\psi_t}
\]
Now measure the clock register, and get a time $t$. If $t$ is maximal at $t=T$, then the residual state is the output of the computation: $\ket{\psi_T}$. 
Moreover, a history state can be tested via a series of projective measurements~\cite{KSV02}.

This suggests a general template for enforcing delegated computation for any publicly known circuit. 
Suppose the client had a trusted measurement device which could hold a state and perform simple measurements on it. Then:
\begin{enumerate}
    \item The server sends a history state of the computation to the trusted measurement device.
    \item The measurement device tests the received state by measuring whether it violates any history state invariants.
    \item The measurement device measures the clock to get a time $t$. 
    \begin{itemize}
        \item If $t < T$: send $\ket{\psi_t}$ back to the server for another try.
        \item If $t = T$, measure register $\calC$ and discard register $\calD$.
    \end{itemize}
\end{enumerate}

\paragraph{Implementing the Trusted Measurement Device.}
Mahadev~\cite{Mahadev18} took the first step towards implementing this template for a classical client. Following her approach, we will replace the trusted measurement device with a commitment which can be opened to measurement results of the committed state:
\begin{enumerate}
    \item The server \textbf{commits} to a history state.
    \item The client picks a measurement test and asks the server to give an \textbf{opening} for that measurement. It repeats this multiple times to amplify soundness.
    \item The client asks the server to \textbf{open} the clock register in the computational basis to get the time $t$. If $t=T$, the server is committed to $\ket{\psi_T}$ and can open to a measurement of it.
\end{enumerate}
The limitation of Mahadev's commitment is that it only supports a \emph{complete} measurement of the committed state (in either the computational or Hadamard basis). This destroys the state completely, preventing amplification and ruining the output step. Still, the collected statistics from one test are sufficient for deciding decision problems like BQP or QMA. 

\cite{EC:CLLW22} opened the approach to verifying quantum sampling tasks (SampBQP) by committing to many states, then testing all-but-one of them. The remaining untested state is likely to be a history state, except with $\approx 1/N$ probability where $N$ is the number of states committed to. So, it can be measured to get the output sample in step 3.

\paragraph{Non-Destructive Testing.}
To transcend the $1/N$ limit, the client must test a state, then \emph{use the same tested state for output}. 
This requires, at a minimum, that the test not destroy the committed history state. 

\cite{KKR26} gave a general approach to making classical verification of QMA non-destructive.
In the context of Mahadev's protocol, an important part of their approach is to avoid directly measuring the commitment openings. Instead, the prover proves the \emph{existence} of openings which the verifier would accept in a special NP argument which avoids collapsing superpositions over NP witnesses. When paired with a high-completeness ZX Hamiltonian~\cite{MN26}, the prover's opening register is entirely supported on openings which the verifier would accept, and so is not disturbed by the witness-preserving argument.

We adopt a similar approach, but with a few important modifications.
\cite{KKR26}'s NP argument requires that the prover knows the statement it is proving -- which requires revealing the decoding key for the commitment. Once the key is revealed, the tested state is \emph{no longer bound} by the original commitment.\footnote{\cite{KKR26} ensures soundness by making a second commitment to the opening before revealing the decoding key. This leaves the opening bound, but not the state.} A malicious server can swap it out before the next test or output step.

\paragraph{Safely Verifying Openings.}
We insulate the client's decoding key by designing a witness-preserving argument for NP where the precise statement being proven is hidden from the prover.
In other words: a secure two-party computation (2PC) for \emph{classical} functionalities which \emph{preserves the quantum party's input}. In its most general form, the goal is to measure some secret classical function $f$ on the quantum party's input, then return the residual state:
\begin{enumerate}
    \item \textbf{Input:} The classical client inputs a classical function $f$. The quantum server inputs a superposition $\ket{\psi} = \sum_{v} \alpha_v\ket{v}$.
    \item \textbf{Output:} Measure $y = f(v)$ on the quantum party's input. Send $y$ to the classical party and return the residual quantum state $\ket{\psi_y} \propto \sum_{v: f(v) = y} \alpha_v \ket{v}$  to the quantum party.
\end{enumerate}
Importantly, the 2PC hides $f$, up to whatever is necessarily revealed by the residual state $\ket{\psi_y}$. Now the client can input its decoding key as part of a function $f$ which checks for accepting openings, and be assured that the decoding key is hidden. 

Constructing such a 2PC for general $f$ is difficult. Instead, we construct a 2PC for constant-degree multilinear polynomials. This suffices for testing committed history states because we use a commitment with linear decoding and the history state Hamiltonian has constant locality, so their composition is a constant degree multilinear polynomial. 
See \Cref{sec:overview:2pc} for more details.

\paragraph{Ensuring Commitment Continuity.}
Simply hiding the decoding key does not ensure binding after a decoding query, though. The adversary can potentially learn enough to break binding through its residual state $\ket{\psi_y}$, or even through the client's decision of whether to continue or abort.

We define and construct a new classical commitment to quantum states with \emph{continual ideal binding}. 
\begin{itemize}
    \item \textbf{Ideal binding} refers to a strong, simulation-based definition wherein there is a well-defined quantum state which the adversarial committer can open to. Notably, the committed state is defined with overwhelming probability whenever the commitment phase passes; prior works allow an adversary to cause a $\sqrt{\delta}$ extraction error by giving invalid openings with probability $\delta$~\cite{GKNV25}.%
    \item \textbf{Continual} refers to the preservation of ideal binding even after making multiple queries to a decoding oracle. Decoding queries allow the adversary to modify the committed state by measuring the decoding results. Continual binding requires that the adversarial committer is still bound to the 
    post-measurement state after each measurement.
\end{itemize}
Together, these two properties allow us to reason about the residual committed state after doing multiple measurement tests. If the adversary passes all of the tests, the residual committed state must be negligibly far from a history state. We also show how the client can help the server to \emph{recover} the committed state after testing it, which allows a coherent output in the overall protocol.

Our construction leverages a close connection between authentication and commitments. Essentially, we build a remote state preparation protocol for \cite{STOC:BarBraVai24}'s coset-state authentication code. The client can help the server to remotely authenticate any state of the server's choice, under the client's key. Moreover, \cite{STOC:BarBraVai24}'s authentication code has linear decoding, so it is compatible with our linear input-preserving 2PC.
See \Cref{sec:overview:commitments} for more details.

\paragraph{Summary of the Approach for Public Computations.} Putting everything together, our approach for verifying public computations is as follows.
\begin{enumerate}
    \item The server commits to a history state.
    \item The client repeatedly runs gentle tests on the committed state by using the state-preserving 2PC for classical (linear) functions. The server inputs an opening, and the state-preserving 2PC measures \emph{only} whether the opening corresponds to a valid history state.
    \item The client asks the server to open the clock register to a time $t$. 
    \begin{itemize}
        \item If $t < T$, the client helps the server recover the whole $\ket{\psi_t}$. Then, the server uncomputes $U_t\circ \dots \circ U_1$ and they restart the protocol.
        \item If $t = T$, the server sends the client's output by opening the classical register $\calC$ in the computational basis. Then, the client helps the server recover its output register $\calB$ from the commitment. Register $\calD$ is \emph{not} recovered.
    \end{itemize}
\end{enumerate}

\paragraph{Wrapping Up: Privacy and Persistent Memory.}
So far, we have only addressed computations where the server knows the description $U = (U_1, \dots, U_T)$ and can choose the input. How can the client hide $U$ from the server? And if it wants to do a follow-up computation $V = (V_1, \dots, V_{T'})$, how can it ensure that $U$'s output $\ket{\psi_T}$ is used as the input to $V$?

The first question can be addressed using quantum fully-homomorphic encryption (QFHE). Instead of directly telling the server the classical description of $U$, the client encrypts it under a QFHE as $\widetilde{U}$ and asks the server to homomorphically evaluate the universal circuit on $\widetilde{U}$ and the input. If the QFHE is circuit-private, the (honestly) evaluated output $\ket{\widetilde{\psi}_T}$ is statistically independent of $\widetilde{U}$ and can be safely decrypted. 
There is a crucial technical subtlety for circuit privacy: the public computation needs to hide the ancillas used for QFHE evaluation. This is implemented in the 2PC by discarding register $\calD$.

Next, we use private computation to implement persistent memory. Suppose that the client wants to perform a computation $U$, save some memory, and later use that memory for another computation $V$. Since $U$ is hidden, the client may safely embed an authentication key $k$ into it. The modified $U'$ performs the original $U$, then authenticates the output $\ket{\psi_T}$ under $k$. This also ensures hiding of $\ket{\psi_T}$ between computations since quantum authentication is necessarily hiding~\cite{FOCS:BCGST02}. Then, the client embeds $k$ into $V$ so that $V'$ first verifies and decodes the authenticated $\ket{\psi_T}$ before using it.

\paragraph{Simulation Error.} Our construction ultimately achieves $\epsilon$-simulation, meaning that for every $\epsilon = 1/\poly[\secpar]$, there is $1/\epsilon$-time simulator that is indistinguishable from the real execution except with probability $\epsilon$. This arises from the input-preserving linear 2PC and propagates through the rest of the primitives we construct. If one could achieve negligible simulation error input-preserving 2PC, then it would immediately improve the main result to also have negligible simulation error. We discuss this point further at the end of \Cref{sec:overview:2pc}.

\subsection{Input-Preserving 2PC for Classical Constant-Degree Polynomials}
\label{sec:overview:2pc}

The first step to our solution is a input-preserving 2PC for classical functions $f$ which are constant degree multilinear polynomials over $\bbF_2$. We also extend the behavior slightly to allow the classical party to abort if $f(\vec{v})$ is not an allowed output. 
This addition is useful for checking the openings of our commitments: decoding applies a linear function, then checks whether a post-fix of the output is $\vec{0}$. In case it is not, the 2PC has measured more than is strictly necessary for decoding and so it might not be safe to return the post-measurement state.
Output-checking can be easily added to our construction since the classical party gets its output before the recovery process begins for the quantum party's output.

\paragraph{Simplifications.}
We make two simplifications for this task.
First, it suffices to construct an input-preserving 2PC only for \emph{linear} $f$. To upgrade to a polynomial with maximum degree $c$, the quantum party may reversibly extend its input to include all degree-$c$ monomials. If $c$ is constant, this is a polynomial blow-up. Then, the classical party uses the linearized version of $f$ whose coefficients are all the monomials of $f$. The goal is thus to realize the following functionality:
\begin{enumerate}
    \item \textbf{Input:} The classical client inputs a classical linear function $L\in \bbF_{2}^{\ell \times m}$. The quantum server inputs a superposition $\ket{\psi} = \sum_{\vecv} \alpha_v\ket{\vecv}$.
    \item \textbf{Output:} Measure $\vecu = L\vecv$ on the quantum party's input. Send $\vecu$ to the classical party. Then, if $\vecu$ is an allowed output, return the residual quantum state $\ket{\psi_{\vecu}} \propto \sum_{v: L\vec{v} = \vecu} \alpha_{\vecv} \ket{\vecv}$ to the quantum party.
\end{enumerate}

Second, we show security against arbitrary malicious quantum parties, but only against honest-but-curious classical parties. An honest-but-curious party follows the protocol exactly, but may nevertheless be a QPT algorithm. Upgrading security for the classical party can be done generically~\cite{STOC:GolMicWig87,C:HalSmiSon11}: the classical party proves in zero-knowledge for NP that each message is correctly computed. The same transformation does not work against the quantum party because the correctness of its messages are not necessarily even QMA statements. Security against a semi-honest classical party is more straightforward, so for the purposes of this overview we just describe security against a malicious quantum party.

\paragraph{Protocol.} For the purposes of the overview, we focus on $L \in \bbF_2^{1 \times m}$ which only contain 1 row.
The main tool we use is a noisy trapdoor claw-free function (NTCF) with the adaptive hardcore bit property~\cite{BCMVV18}. Roughly, this is a $2\rightarrow 1$ function which can be inverted using a secret trapdoor. Moreover, the adaptive hardcore bit property guarantees that it is hard to find inner products with the difference between collisions $(x_0 \oplus x_1)\cdot \vec{d}$, even if $\vec{d}$ is chosen by the adversary.
The protocol can be broadly separated into 4 phases.
\begin{enumerate}
    \item \textbf{Input Commitment.} The classical party samples $m$ NTCF public parameters and trapdoors $(\pubparams_i, \td_i)$. The quantum party now commits to each bit $v_i$ of their input $\vecv$ by coherently evaluating the NTCF $\pubparams_i$ on $v_i$ and measuring the output $\vec{y}_i$. This leaves them with the state
    \[
        \sum_{\vecv} \alpha_\vecv \ket{\vecv} \bigotimes_{i=1}^{m} \ket{\vec{x}_{i, v_{i}}}
    \]
    Crucially, $\vecv$ is not collapsed because the NTCF has two preimages $\vec{x}_{i, 0}$ and $\vec{x}_{i, 1}$ for every $\vec{y}_i$.
    The quantum party sends the measured images $(\vec{y}_i)_{i\in [m]}$ to the classical party.
    
    Then, they perform a \emph{witness-preserving} argument of knowledge for NP~\cite{KKR26} to show knowledge of preimages $(v_i, \vec{x}_{i, v_{i}})$ for each $\vec{y_i}$, \emph{without collapsing their witness superposition}.

    \item \textbf{Function Masking.} The classical party uses $\td_i$ to invert each $\vecy_i$ into its preimages $\vec{x}_{i, 0}$ and $\vec{x}_{i, 1}$, which it uses to compute $\vec{\Delta}_i = (0,\vec{x}_{i, 0}) \oplus (1, \vec{x}_{i, 1})$.
    Then, it samples random $\vecd_i$ and masks each bit of $L$ as
    \[
        p_i = (\vecd_i \cdot \vec{\Delta}_i) \oplus L_i
    \]
    Finally, it sends each masked bit to the classical party:
    \[
        (\vecd_i, p_i)_{i\in [m]}
    \]

    \item \textbf{Evaluation.} The quantum party coherently computes the following into a fresh register, using its superposition over $\ket{v_i, \vec{x}_{i, v_{i}}}$:
    \[
        t = \bigoplus_{i} \big(\vec{d}_i \cdot (v_i,\: \vec{x}_{i, v_{i}})\big) \oplus v_i p_{i})
    \]
    Then, it measures the result and sends $t$ to the classical party.

    The quantum party proves that $t$ is correctly computed using a witness-preserving argument of knowledge for NP, which negligibly disturbs the superposition over $\ket{v_i, \vec{x}_{i, v_{i}}}$.

    Finally, the classical party computes its output $u$ as
    \[
        u = \left(\bigoplus_{i=1}^{m} \vec{d}_i \cdot (0, \vecx_{i,0}) \right) \oplus t
    \]

    \item \textbf{Recovery.} The quantum party measures each of its $\vec{x}_i$ registers in the Hadamard basis to obtain vectors $\vec{e}_i$. It then sends these to the classical party.

    The classical party sends back the following for each $i$:\footnote{There is an additional verification procedure to ensure that $\vec{e}_i$ and $\vec{d}$ are not problematically dependent. We omit its precise description from the overview for simplicity.}
    \[
        \beta_i = \vec{e}_i \cdot (\vec{x}_{i,0} \oplus \vec{x}_{i,1})
    \]

    Finally, the quantum party applies $Z^{\beta_i}$ to the $i$'th index of its input register, for every $i$.
\end{enumerate}
To generalize this to more rows, the classical party samples fresh NTCF instances for each row and the quantum party commits to $\vec{v}$ repeatedly under each of the NTCF rows. Thus, $v_i$ is committed to a total of $\ell$ times. Additionally, only one recovery phase is reported for each $v_i$, which is the sum of the phases it incurs across the $\ell$ rows of $L$.

\paragraph{Correctness and Input Preservation.}
At the end of the \textbf{input commitment} step, the quantum party holds
\[
    \sum_{\vecv} \alpha_\vecv \ket{\vecv} \bigotimes_{i=1}^{m} \ket{\vec{x}_{i, v_{i}}}
\]
The next time it is modified is in the \textbf{evaluation} step, where
\[
    t = \bigoplus_{i} \big(\vec{d}_i \cdot (v_i,\: \vec{x}_{i, v_{i}})\big) \oplus v_i p_{i})
\]
is measured.
The trick is to show that $t$ depends \emph{only} on $(0, \vec{x}_{i,0})$ and $L\vec{v}$. Expanding $p_i = (\vecd_i \cdot \vec{\Delta}_i) \oplus L_i$, we find that
\begin{align*}
    t &= \bigoplus_{i} \left(\big(\vec{d}_i \cdot (v_i,\: \vec{x}_{i, v_{i}})\big) \oplus v_i (\vec{d}_i \cdot \Delta)\right) \oplus \bigoplus_i L_i v_i
    \\
    &= \bigoplus_{i} \big(\vec{d}_i \cdot ((v_{i},\: \vec{x}_{i, v_{i}}) \oplus v_i \Delta\big) \oplus L \vec{v}
    \\
    &= \bigoplus_{i} \big(\vec{d}_i \cdot (0,\: \vec{x}_{i, 0}) )\big) \oplus L \vec{v}
\end{align*}
The last simplification holds because when $v_i=0$, $\Delta_i$ is not added, and when $v_i = 1$, $\Delta_i$ switches $(1, \vec{x}_{i,1})$ to $(0,\: \vec{x}_{i, 0})$. Thus, measuring $t$ only disturbs the quantum party's state by measuring $L\vec{v}$, and the classical party correctly unmasks its output.

The final step to analyze is \textbf{recovery}. When the quantum party measures its $\vec{x}_i$ register in the Hadamard basis, it adds a phase $(-1)^{\vece \cdot \vec{x}_{i, v_i}}$ to the corresponding $v_i$ register. If there were only one $i$, then the residual state would be
\[
    \sum_{v_i} \alpha_{i, v_i} (-1)^{\vece \cdot \vec{x}_{i, v_i}} \ket{v_i}.
\]
Factoring out a global phase $(-1)^{\vece \cdot \vec{x}_{i, 0}}$ removes the phase when $v_i=0$ and changes it to a relative phase $(-1)^{\vec{e}_i \cdot (\vec{x}_{i,0} \oplus \vec{x}_{i,1})}$ when $v_i=1$. Therefore, up to global phase the remaining state is
\[
    \sum_{v_i} \alpha_{i, v_i} ((-1)^{v_i})^{\vec{e}_i \cdot (\vec{x}_{i,0} \oplus \vec{x}_{i,1})} \ket{v_i}.
\]
Therefore applying $Z^{\vec{e}_i \cdot (\vec{x}_{i,0} \oplus \vec{x}_{i,1})}$ removes the relative phase.

\paragraph{Security for 1 Column $L$.}
The protocol above is secure when $L$ only has one column, but has a concrete attack when $L$ has more. 
We first describe the security for one column and why it breaks down for multiple columns.
then describe how to fix the attack by encoding the inputs $L$ and $v$ in a particular way before using them in the above protocol.

Intuitively, security relies on two parts: the extraction of the input via the argument of knowledge and the masking of $L_i$ by a random inner product $\vec{d}_i \cdot \Delta_i$.
For now we assume that the quantum party's input is temporarily exposed during the argument of knowledge and we will revisit extraction at the end of this section.

If the recovery step were skipped, then the quantum party's input looks collapsed at the end of the protocol.
In other words, it is indistinguishable whether $(v_i, \vec{x}_{i,v_i})$ was measured during the argument of knowledge. In this case, $\vec{d}_i \cdot \Delta_i$ is indistinguishable from a uniformly random bit; otherwise quantum Goldreich-Levin~\cite{STACS:AdcCle02} can be used to recover $\Delta$, which can then be combined with the measured $(v_i, \vec{x}_{i,v_i})$ to find the other preimage of $\vecy_i$.

The recovery step complicates the situation by providing an additional bit of leakage about $\Delta$. Security comes from the fact that $L \vec{v}$ is \emph{always} measured. 
If $L = 1$, then $\vec{v}$ itself is measured. In this case, we can show using the adaptive hardcore bit property that the random inner product $\vec{d} \cdot \Delta$ is indistinguishable from a uniform bit, even given the additional recovery leakage. Thus, we may change $p$ to be an encoding of $0$ instead of an encoding of $L=1$. Now $p$ is \emph{always} an encoding of $0$, regardless of $L$.

\paragraph{An Attack on Multiple Columns.}
The issue with the more general multi-column case is that the adversary can isolate individual columns using a slight variation on the honest computation of $t$.\footnote{We thank ChatGPT 6 for pointing out this attack and suggesting the fix via input encoding.}
Suppose that the adversary chose a column $i$ and added the phase
\[
     s_i = \big(\vec{d}_i \cdot (v_i,\: \vec{x}_{i, v_{i}})\big) \oplus v_i p_{i})
\]
to $v_i$. As we computed before, this becomes a global phase $(0,\: \vec{x}_{i,0})$ together with a local phase $L_i v_i$:
\[
    s_i = (0,\: \vec{x}_{i,0}) + L_i v_i
\]
This allows the adversary to distinguish between two possible choices of $L^{(0)}$ or $L^{(1)}$ which have the same row space. Since they have the same row space, measuring $L^{(b)}$ separates the same vectors, and so should not be distinguishable using the post-measurement state. However, the adversary can input a superposition over $\vec{v}$ and $\vec{w}$ such that $L^{(0)}\vecv = L^{(0)}\vecw$ and $L^{(1)}\vecv = L^{(1)}\vecw$, but which differ at an index $i$ where $L^{(0)}_i \neq L^{(1)}_i$. In this case, applying the isolated phase at index $i$ flips $\ket{\vecv} + \ket{\vec{w}}$ to $\ket{\vecv} - \ket{\vec{w}}$ if and only if $L^{(1)}$ is being computed.

\paragraph{Extension to Multi-Column $L$: Input Encoding.}
The solution is to use a redundant randomized encoding of $L$ as in the input. Suppose that we encoded $L$ as
\[
    L' = LD + RP
\]
where $R$ is a random matrix and there is some generator matrix $G$ satisfying $DG = I$ ($D$ is a ``decoding'' matrix) and $PG = 0$ ($P$ is a ``parity-check'' matrix). Then if we encode the quantum input as
\[
    v' = Gv
\]
and compute the protocol using inputs $L'$ and $v'$, the output is correct:
\[
    L'v' = L(DG)v + R(PG) = Lv.
\]

The randomization term $RP$ buys us the ability to randomize individual indices of $L'$. Thus, if the adversary attempts to isolate a specific column $j$ of $L'$, the value $L_{*j}$ will be masked by a random bit.
By choosing an appropriate $G$, we can ensure that every combination of columns from $L'$ will always be masked by a random bit, unless the adversary adds a whole row $L_{i*}$. However, adding a phase from an entire row $L_{i^*}$ is safe. The ideal functionality has already measured $L_{i*}\vec{v}$, which ensures such an attack only introduces a global phase.
See the technical sections for more details on the choice of the encoding matrices.

\paragraph{Indistinguishable Temporary Extraction and $\epsilon$-Simulation.}
The $\epsilon$-error in simulation comes from the extraction of the input in the witness-preserving argument of knowledge.
Intuitively, we need an extraction guarantee similar to \cite{FOCS:LomMaSpo22}'s ``state-preserving extraction'', which allows extracting an NP witness while simultaneously simulating the prover's post-argument state. Using \cite{FOCS:LomMaSpo22}'s extractor (with slight modifications described below) limits us to $\epsilon$-simulation\footnote{Or coherent runtime simulation. We default to $\epsilon$-runtime on the recommendation of \cite{TCC:LaiSpoTro26}.} until a construction with a better extractor is found. The new construction would necessarily need to be a super-constant round protocol to avoid the impossibility results of \cite{FOCS:CCLY21}.

\emph{Witness}-preservation complicates matters further. Since the witness superposition must remain uncollapsed, the extractor cannot simply measure the witness during the extraction process or it has no hope of simulating the post-argument state. Instead, the best it can do is to temporarily expose the witness, then ``replace'' the witness after the simulator is finished. 
We call this \emph{indistinguishable extraction}.
\cite{arxiv:KR26} observes that $\epsilon$-indistinguishable extraction can be accomplished by applying the unitary truncation $U_{\epsilon^2}$ of \cite{FOCS:LomMaSpo22}'s variable-runtime guaranteed extractor to \cite{KKR26}'s argument to expose the witness, then inverting $U_{\epsilon^2}$ to return the witness. Since there is an $\epsilon^2$ error in extraction, temporary extraction in this way causes an $O(\epsilon)$ statistical disturbance of the state.

Thus, obtaining negligible simulation boils down to the following task: construct a witness-preserving argument of knowledge with strict polynomial-time guaranteed extraction.

\subsection{Classical Commitments to Quantum States with Continual Binding}
\label{sec:overview:commitments}

Next, we describe our new commitment scheme. 
To commit to a quantum state, the quantum sender interacts with a classical receiver. At the end of the commitment protocol, the receiver holds a commitment key $\com$ and the sender holds an opening register $\calR_{\op}$. 
The sender can then measure their opening register to obtain openings to a measurement of their committed state in either the Hadamard or computational basis.
Moreover, after giving (partial) openings via the input-preserving 2PC from the previous section, the receiver can help the sender to \emph{recover} what remains of the committed state after these measurements.

\paragraph{Defining Binding.} 
Intuitively, a commitment is binding to a quantum state if there exists a well-defined state $\ket{\psi}$ whose measurement statistics match the adversary's opening distribution.
We formalize this through the real/ideal paradigm. In the ideal world, the adversary must explicitly declare the state $\ket{\psi}$ which they are committing to. Then, they lose access to $\ket{\psi}$ except through a measurement oracle which represents ``opening'' the commitment. 
Ideal binding requires that some QPT simulator forces the adversary to adhere to the ideal behavior: It must extract $\ket{\psi}$ explicitly and generate an indistinguishable view for the adversary, given only access to the measurement oracle.

Continual binding extends the ideal world to permit multiple ``opening''/measurement queries.
This naturally tracks how the committed state $\ket{\psi}$ evolves as the simulated adversary measures it. 
After each measurement partially collapses the committed state, the simulator only has access to measurements on the residual state.

\paragraph{Authentication and Commitment.}
Our construction makes use of a close connection between commitments and authentication. 
Both protect a quantum state against undetected modification, but they differ at the encoding step.
In an authentication scheme, the encoder is trusted. 
In a commitment scheme, the encoder is \emph{adversarial}.
Their goal is to change the message after committing to it and they might cheat during the encoding process to do so. 

Our construction bridges the adversarial encoding gap
by remotely preparing authenticated states underneath a key held by the classical receiver. 
We use the authentication scheme of \cite{STOC:BarBraVai24}, whose continual security was established in \cite{FOCS:HuaTan25}.
To inherit continual binding, we show that a malicious committer’s view can be simulated using only an honestly authenticated state and the oracle access permitted by the authentication guarantee. Such a simulation allows us to transform any attack on the commitment into an attack on the underlying authentication scheme.

Only a fixed resource needs to be prepared: an authenticated \(\lvert+\rangle\) state. We first explain how the scheme’s linear decoding and homomorphic CNOT operations let us use this resource both to authenticate the committer’s chosen input and to recover the state afterward. We then address the main new challenge: preparing $\Auth(\ket{+})$ through classical interaction with the simulation guarantee needed by this reduction.

\paragraph{Committer's Input Choice.}
Suppose for now that the classical receiver can securely prepare an authenticated $\ket{+}$ state on the committer’s hardware. This fixed resource suffices to authenticate an arbitrary input. We first transform it into a half-authenticated EPR pair, then let the committer teleport its chosen input into the authenticated half. The receiver accounts for the teleportation corrections by updating the authentication key. Concretely:

\begin{enumerate}
    \item The classical receiver helps the quantum committer to prepare an authenticated $\ket{+}$ state.
    \item The receiver and committer transform $\Auth_k(\ket{+})$ into a half-authenticated EPR pair $\sum_{b} \ket{b} \otimes \Auth_k(\ket{b})$ using a state-preserving classical 2PC with the following linear functionality: On input $(b_1, \Auth_k(b_2))$ from the quantum party and a key $k$ from the classical party, the 2PC outputs $c = b_1 \oplus b_2$. 
    
    If the committer inputs $\ket{+} \otimes \Auth_k(\ket{+})$, they will hold $\sum_{b} \ket{b\oplus c} \otimes \Auth_k(\ket{b})$ at the end of the 2PC. They can then correct the half-authenticated EPR pair by applying $X^c$ to the plain half.
    \item The committer teleports their chosen input $\ket{\psi}$ through the clear half into the authenticated half. They report the teleportation Pauli corrections to the receiver, who updates the authentication key $k$ accordingly.
\end{enumerate}
The second step uses the continual security of the underlying authentication scheme. The 2PC functionality decodes the authenticated $b_2$ before measuring its parity with $b_1$.
Continual security ensures that the authenticated register remains bound to the resulting logical state, including its entanglement with the unencoded register. Thus, the decoding query used to create the EPR pair does not invalidate the security needed for subsequent operations.

\paragraph{Recovery.}
Recovery uses the same teleportation idea in the opposite direction, but must preserve the security of the preceding interaction. Simply revealing the authentication key could expose information that distinguishes a simulated commitment from a real one. We instead release the state while keeping the key secret.

As \cite{FOCS:HuaTan25} observe, the scheme’s homomorphic CNOT operations allow the authenticated state to be teleported through a second half-authenticated EPR pair, this time into the unencoded half. The committer sends openings to the authenticated teleportation outcomes, which the receiver decodes to determine the corrections. The receiver then communicates these corrections so that the committer can recover the state. This procedure uses only operations and decoding queries covered by the authentication scheme’s continual-security guarantee, allowing recovery to extend the preceding simulation safely.

\paragraph{Reducing Authentication to Coset-State Preparation.}
The remaining step is how to securely prepare $\Auth(\ket{+})$. In \cite{STOC:BarBraVai24}'s scheme, a computational basis state $\ket{b}$ is encoded as a random coset state
\[
    \Auth_k(\ket{b}) \coloneqq 
    X^\vec{x} Z^\vec{z}\ket{S+ b\vec{\Delta}} 
\]
The key $k = (\vec{x}, \vec{z}, S, \vec{\Delta})$ specifies the subspace $S$, masking offsets $(\vec{x}, \vec{z})$, and logical offset $\vec{\Delta}$. 
Multiple qubits are encoded using the same $S$ and $\Delta$, but independent masks $(\vec{x}, \vec{z})$.
Thus, the authenticated $\ket{+}$ state that we need to prepare is simply a random coset state 
\[
    \Auth_k(\ket{+}) = X^\vec{x}Z^\vec{z}\ket{S_\Delta}
\]
of the subspace $S_\Delta = S + \Span(\Delta)$. 

\subsubsection{Remote State Preparation of Coset States.} 
The main technical contribution of our commitment construction is a way to securely and remotely prepare such coset states. The classical sender chooses any subspace $S\subset \bbF_2^{4\secpar}$ of dimension $\approx 2\secpar$ and helps the quantum receiver prepare a random coset state $X^x Z^z \ket{S}$.\footnote{For simplicity during the overview, we omit the $\Delta$ from $S_\Delta$ and ignore constant variations in the dimension.} The classical sender also learns the random masks $(x, z)$. 
In the commitment application, the classical sender is the commitment receiver, and the quantum receiver is the committer.

The goal of this section is to develop an RSP protocol which preserves all of the (falsifiable) security properties of coset states up to a negligible security loss. In particular, we need the binding property from \cite{STOC:BarBraVai24}'s authentication code to hold.
A few prior works have constructed protocols for remotely preparing coset states which preserve specific security properties up to negligible loss~\cite{STOC:Shmueli22,TCC:CheHerVu23,C:ShmZha25,FOCS:BGMS26}.
However, these protocols do not preserve all security properties; indeed, the core technical contributions of \cite{TCC:CheHerVu23} and \cite{FOCS:BGMS26} are to modify \cite{STOC:Shmueli22}'s and \cite{C:ShmZha25}'s protocols to preserve additional specific properties.
We obtain a wholistic solution using a simulation-based notion of security, which we describe below.

\paragraph{Security with Membership Oracles.}
The goal is for the quantum receiver $R_Q^*$ to receive no more information than if it had directly received the intended state, along with some limited leakage oracles. Thus, any adversary breaking some property of coset states prepared remotely could be transformed into an adversary which breaks the same property when the coset states are received directly.

We formalize this notion via simulation. A QPT algorithm $\Sim$ receives $X^xZ^z \ket{S}$ and access to membership oracles for $S + x$ and $S^\perp + z$. Using these, it must generate a view indistinguishable from the receiver’s view in a real execution, if the distinguisher were given the final masks $(x,z)$.

Both the membership oracles and indistinguishability given $(x, z)$ are chosen for compatibility with continual binding. 
Recall that continual binding permits decoding oracle access and binds to the residual state after each decoding query.
Membership queries for $S+x$ and $S^\perp + z$ can be implemented through the computational-basis decoding and Hadamard-basis verification interfaces already covered by that guarantee. 
Furthermore, indistinguishability when also given $(x,z)$ ensures that simulation of the preparation remains indistinguishable even when continual binding exposes additional decoding oracles dependent on the key $k = (S, \Delta, x, z)$.
Consequently, the preparation simulator can be smoothly incorporated into the continual binding reduction.

\paragraph{A Starting Point.} 
We first give some intuition for the construction.
Suppose we instead wanted to construct states of a simpler form 
\[X^x \ket{S}\]
This is a uniform superposition over all elements of the coset $S + x$. $\bbF_2^{4\secpar}$ can be linearly partitioned into such cosets by multiplying with any generator matrix $M_{S^\perp}$ of $S^\perp$.
Thus, preparing a uniform superposition $\sum_{v} \ket{v}$ and measuring $M_{S^\perp} v$ produces a random $X^x \ket{S}$.
Such linear measurements can be carried out underneath our state-preserving linear 2PC.
The classical sender can use $M_{S^\perp} v$ to identify a canonical element $\Can_S(v)$ of $S + v$ to describe the prepared state, so we refer to this measurement as measuring $\Can_S(v)$.

Unfortunately, this basic approach is not simulateable given a pre-determined $X^x \ket{S}$.
There is no way to force the correct $x$ -- the adversary could even force its own $x'$ by just inputting $\ket{x'}$ into the 2PC! 

\paragraph{Hiding $x$ Through $z$.} 
The key to removing adversarial control of $x$ lies in the missing $z$.
We can do the following two-step approach to (still insecurely) prepare $X^x Z^z \ket{S}$:
\begin{enumerate}
    \item Sample a random $3\secpar$-dimensional superspace $T\supset S$ and prepare $X^{z} \ket{T^\perp}$ using the starting approach. Then, apply a Hadamard to produce
    \[
        Z^z \ket{T}
    \]
    \item Use the state-preserving linear 2PC to measure $\Can_S(\bullet)$ on $Z^z \ket{T}$ and obtain $x$.
\end{enumerate}
Now $x$ is drawn randomly from $T$ and its marginal distribution is close to uniform since $T$ is much larger than $S$.
If we had set $T=S$, then we would have been stuck with just $Z^z\ket{S}$!

Setting $T$ to be much larger than $S$ not only enables $x$ to coexist with $z$, it also gives a way to remove the adversary's control of $x$. 
The trick is to view $X^z \ket{T^\perp}$ as authenticated \emph{in the Hadamard basis}. 
Intuitively, for a fixed $S$, the only way to find a vector in $T$ (other than $v\in S$) is to have actually prepared the state $X^z \ket{T^\perp}$ and applied a Hadamard. 
This gives a method to catch cheating provers who attempt to input something other than $Z^z \ket{T}$ in the second step: check that their input is supported on $T\backslash S$ in the computational basis.

\paragraph{Construction.}
We also need to remove adversarial control of $z$, which can be dealt with in a similar manner using a random $R^\perp \supset S^\perp$. This gives the following three-step approach for securely preparing $X^x Z^z \ket{S}$:
\begin{enumerate}
    \item[0.] The sender chooses $S$ and samples $R \subset S \subset T$ where $\dim(R) = \secpar$ and $\dim(T) = 3\secpar$.
    \item The receiver prepares and inputs $\ket{+^{4\secpar}}$ to a state-preserving 2PC which, on computational basis input $\ket{v}$, measures
    \[
        x_R \gets \Can_R(v)
    \]
    This results in the state $X^{x_R}\ket{R}$.
    \item The receiver applies $H$ to their state to get $Z^{x_R}\ket{R^\perp}$ and inputs it to a state-preserving 2PC which measures
    \[
       z
        \gets
        \begin{cases}
            \Can_{T^\perp}(v) &\text{if } v \in R^\perp \backslash S^\perp 
            \\
            \bot &\text{else}
        \end{cases}.
    \]
    This aborts with negligible probability since $S^\perp$ is much smaller than $R^\perp$. If it does not, then the receiver obtains the state $Z^{x_R}X^{z}\ket{T^\perp}$ since $T^\perp \subset R^\perp$.
    
    \item The receiver applies $H$ to their state to get $X^{x_T}Z^{z}\ket{T}$ and inputs it to a state-preserving 2PC which measures
    \[
        x \gets
        \begin{cases}
            \Can_{S}(v) & \text{if } v \in (T + x_{R}) \backslash (S + x_{R})
            \\
            \bot & \text{else}
        \end{cases}.
    \]
    This aborts with negligible probability since $S$ is much smaller than $T$. If it does not, then the receiver obtains the state $X^{x}Z^{z}\ket{S}$ since $S \subset T$.
\end{enumerate}
Finally, the sender outputs the $(x,z)$ pair output by the second two 2PCs.

\paragraph{Simulation Using $X^{x'} Z^{z'} \ket{S}$.}
To show security, $\Sim$ must generate the receiver's view from a prescribed $X^{x'} Z^{z'} \ket{S}$. The main challenge in doing so is that the protocol is classical:
\begin{quote}
\centering
How can $\Sim$ introduce $X^{x'} Z^{z'} \ket{S}$ into the receiver's view when the receiver expects only classical messages?
\end{quote}
Prior works justify the insertion by observable testing and a rigidity argument that isolates a particular state on the adversary's registers, then directly swapping that state. However, the observable tests are destructive, which introduces a $\poly[1/N]$ distinguishing vulnerability scaling with the number of tests $N$.
We design a new method which smoothly inserts $X^{x'} Z^{z'} \ket{S}$ into the receiver's view, without the protocol-dependent error.

$\Sim$ makes two general kinds of modifications to the protocol. First, it replaces each of the 2PCs with a simulated 2PC. This exposes and requires $\Sim$ to internally run the ideal functionalities for each of the 2PCs. 
Second, it changes those ideal functionalities fundamentally.
In the first 2PC, it replaces the $\Can_R(\bullet)$ measurement by a $\CX$ onto the prescribed coset state. Then, in the second 2PC, it replaces $\Can_{T^\perp}(\bullet)$ measurement by a $\CZ$ onto the prescribed state, and in the third, it replaces $\Can_{S}(\bullet)$ by another $\CX$ similarly to the first. The checks $v \in R^\perp \backslash S^\perp$ and $v \in (T + x_{R}) \backslash (S + x_{R})$ are replaced by queries to the $S^\perp + z'$ and $S + x'$ oracles, respectively.

\begin{figure}[H]
    \centering
    \input{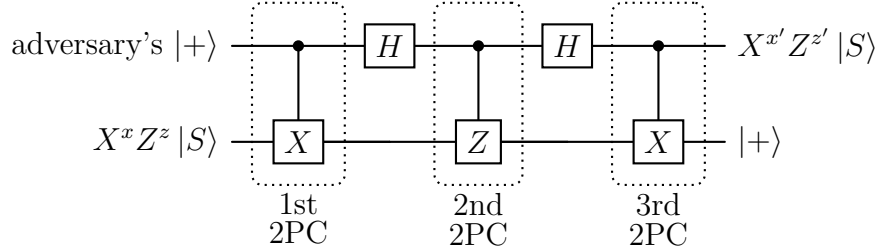}
    \caption{\textbf{Coset RSP Simulator}, when the adversary is honest.}
    \label{fig:coset-rsp-sim}
\end{figure}

To see how $\Sim$ injects $X^{x'} Z^{z'}\ket{S}$ into the receiver's view, consider a receiver which behaves honestly. \Cref{fig:coset-rsp-sim} makes the intuition clear: in this case, $\Sim$ is just a SWAP gate!

\paragraph{Indistinguishability of $\Sim$.}
Of course, $\Sim$ must be indistinguishable even against a dishonest receiver. 
The modified checks guarantee that the sender's output is $(x', z')$, so the state itself matches, but what about the replaced measurements and checks?

As hinted at by our technique used to hide $x$ through $z$, each check goes hand-in-hand with the prior measurement. We focus just on how to replace the first 2PC's $\Can_{R}(\bullet)$ measurement and the second 2PC's $R^\perp \backslash S$ check; the later replacements can be done similarly.

The key to the replacement is a variant of the well-known measurement-phase equivalence: measuring $\Can_{R}(\bullet)$ and discarding the result is equivalent to applying $Z^z$ for a random $z\gets R^\perp$. 
As an intuitive interpretation, measuring $\Can_R(\bullet)$ only reveals a single random vector from $R$. 
Since $R$ is a uniform superspace of $S$ much smaller than the overall space $\bbF_2^{4\secpar}$, it is now hard for the adversary to find any element of $R^\perp$ aside from $S^\perp+z$, even given $S$. Thus, we may indistinguishably shrink the $R^\perp \backslash S^\perp$ check to just $S^\perp + z$.

The phase interpretation also helps with replacing the $\Can_{R}(\bullet)$ measurement by a $\CX$ onto $X^{x'}Z^{z'}\ket{S}$. Since $X^s\ket{S} = \ket{S}$ for every $s \in S$, the next effect of the $\CX$ on any computational basis input $\ket{v}$ is to apply $X^{\Can_S(v)}$ to the prescribed state and phase-kickback $z'$ to the input state. The marginal distribution of a random vector from $R$ is statistically close to the uniform distribution of $z'$, so the phase behavior matches. Moreover, $\Sim$ needs to (temporarily) record $x_S = \Can_S(v)$ anyway for the later $(T + x_S)\backslash (S + x_S)$ check.\footnote{Since $R\subset S \subset T$, these cosets are equivalent to if we had instead used $x_R$.} Therefore the classical-copying of $x_S$ also matches the $\CX$ replacement.

\subsection{Applications}
\label{sec:overview:applications}

We highlight three applications which demonstrate different aspects of our main result.
Two of the three highlighted applications are well-known open problems: negligible security loss in both (1) classical verification of quantum sampling and (2) remote state preparation of arbitrary efficiently preparable states.
The third is a novel application: (3) classical revocation of quantum certificates.
At the end of this section, we also list a number of simpler applications implied by our main results.

\paragraph{Randomized Outputs: Quantum Sampling Verification.}
A natural first step beyond decision problems is to ask for samples from some distribution. 
More concretely, let $D$ be the output distribution for a given a quantum circuit $C$. The task is then to verifiably produce samples from $D$.

Previously, \cite{EC:CLLW22} showed a classical-interaction protocol parameterized by $\epsilon = 1/\poly[\secpar]$ which verifiably samples from a distribution $D'$ that is $\epsilon$-close to $D$. 
This is not a big sacrifice for $\SampBQP$, which already allows $C$'s distribution $D$ to differ from the target distribution $D^*$ by a $1/\poly$ amount. However, for cases where $C$ can sample exactly from the target distribution $D^*$, any non-negligible loss is undesirable.

Our 2PC allows directly implementing the circuit $C$ without dependence on any $\epsilon$, yet with the guarantee that the classical party's output is $\epsilon$-computationally indistinguishable from a true execution of $C$ for \emph{every} $\epsilon = 1/\poly[\secpar]$. 
Therefore the classical party obtains samples from a distribution $D'$ whose computational distance from $D$ is smaller than every $1/\poly[\secpar]$ -- in other words, negligible.

\paragraph{Sending States: Remote State Preparation.}
Remote state preparation (RSP) allows a classical client to send a quantum message, thereby replacing a quantum party in a non-interactive quantum protocol.%
For example, \cite{FOCS:GheVid19} allow a classical verifier to participate in a quantum blind delegation protocol, while \cite{ICALP:GheMetPor23} allow a classical verifier to send an unclonable quantum ciphertext.

So far, existing constructions have had two limitations. 
First, they are limited to preparing specific families of states and the sender cannot the exact state to prepare, e.g. it can decide the basis $X$ or $Z$, but cannot pick between $\ket{+}$ and $\ket{-}$. 
Restrictions on preparable states limit the protocols for which RSP can replace a quantum party.
Second, their protocols take $1/\epsilon$ time to reduce the soundness error to $\epsilon$. 
Many cryptographic protocols require negligible security loss, which a $\epsilon= 1/\poly$ error can overwhelm entirely.

Our 2PC allows directly implementing any efficient preparation procedure, resolving both limitations. Concretely, given any quantum game $G_Q$ where one of the parties sends one quantum message then becomes classical, our 2PC can replace that party with a classical party to produce a new game $G_C$. 
It then guarantees than any $\QPT$ adversary that wins $G_C$ with probability $\delta$ can be turned into an adversary for $G_Q$ with time blow-up $1/\epsilon$ which also wins $G_Q$ with probability $\geq \delta + \epsilon$. Crucially, $\epsilon$ can be tuned independently of $G_C$ itself. Since every $\epsilon = 1/\poly$ still results in a QPT adversary, the increase in winning probability between $G_C$ and $G_Q$ must be negligible:
\[
    \max_{\QPT\ A}\Pr[\mathsf{win}\ G_C(A)] \leq \max_{\QPT\ A}\Pr[\mathsf{win}\ G_Q(A)] + \negl
\]
The replacement can occur regardless of the precise states to be sent. Indeed, we later give several explicit applications where more complicated states must be sent.

\paragraph{Quantum Storage: Classical Revocation of Quantum Certificates.}
Our final highlighted application allows receiving, checking, and confiscating quantum states. For example, consider securely leasing a PKE decryption key as a quantum state~\cite{EC:AKNYY23}. The party borrowing the quantum key can return a certificate (which may be the quantum key itself), and, if the certificate is valid, then the party who borrowed the key becomes unable to decrypt any future ciphertexts. Although there do exist ways to generate a classical certificate of revocation, in many cases it is easier in constructions to directly return the key itself.

Our result gives a way for a classical party to \emph{generically} revoke keys even when the certificate is a quantum state. They can run our 2PC to receive and check the certificate, then discard the remaining state. Our 2PC security guarantees that the discarded state computationally disappears from the server's view, allowing the original secure key leasing security to kick in.%
Since the secure key leasing decryption security game can be $\epsilon$-approximated by the simulator for every $\epsilon = 1/\poly$, any adversarial lessor's advantage in decrypting ciphertexts after revocation must be less than $\epsilon$ -- in other words, negligible.

\paragraph{Other Applications.}
We provide details here about the other previously mentioned example applications.
\begin{itemize}
    \item \textbf{Semi-Quantum Unclonable Encryption.} 
    Unclonable encryption ensures that a ciphertext cannot be split into two parts, each of which can simultaneously be decrypted using the decryption key. 
    Our remote state preparation application allows a classical client to send the quantum ciphertext while incurring only a negligible loss in the cloning security game. 
    This answers an open problem by~\cite{ICALP:GheMetPor23}.

    \item \textbf{Public-Key Semi-Quantum Money from Polynomial Hardness.} \cite{STOC:Shmueli22} gave a public-key quantum money (PKQM) scheme where the bank can mint new money states using only classical interaction with the quantum recipient, assuming iO and \emph{subexponentially} hard LWE. By using our protocol to send banknotes for Zhandry's PKQM scheme~\cite{EC:Zhandry19b}, we obtain the same primitive from weaker assumptions: iO and the \emph{polynomial} hardness of LWE.
    
    \item \textbf{Proofs of No-Intrusion for Decryption Keys.}
    \cite{EC:GoyRai26} gave a classical-communication protocol for non-destructively testing whether a quantum ciphertext has been stolen from an external server. Beyond ciphertexts, one can also check decryption keys. 
    Our result allows generically compiling any secure key leasing scheme with publicly verifiable revocation \emph{in an oracle model} to achieve this. Notably, SKL with public verifiability in an oracle model does not imply public-key quantum money in the plain model, so we can hope to achieve this from weaker assumptions.
    The outline is similar to classical revocation of quantum certificates, except that the certificate is coherently returned to the prover after the 2PC, instead of being confiscated.

    \item \textbf{Improved Extraction in CVQC for QMA.}
    The quality of an extractor can be measured by the difference between the probability that the prover passes the CVQC and the extracted state passes QMA verification (extraction error), the minimum probability with which the prover must pass for this guarantee to hold (knowledge error), and the difference between the number of witness copies required for an honest prover and the number extracted (copy error). 
    Prior works incurred either $1/\poly[T]$ extraction error in a protocol with $T$ rounds and $0$ copy error, or negligible extraction error and $O(\secpar)$ copy error~\cite{VidickZhang21,KKR26}.

    By directly implementing a QMA verifier using our 2PC, we obtain a classical argument of knowledge for QMA with $0$ copy error and whose extraction error depends on the amount of time one is willing to run the extractor for, not on the original protocol. In particular, any $\epsilon=1/\poly[\secpar]$ extraction error can be achieved by running the extractor for $1/\poly[\epsilon]$ time.
    Moreover, given a single copy of any witness which would pass QMA verification with probability $\alpha$, the prover passes CVQC verification with probability $\alpha - \negl$. \cite{KKR26} left this as an open problem for $\alpha$ bounded away from $1$.

    \item \textbf{Quantum Obfuscation with Interactive Classical Communication.}
    A series of works have constructed obfuscation for various types of quantum circuits. For any of these schemes where the circuit $C$ to be obfuscated has a classical description, a classical computer may obfuscate and send $C$ by interacting classically with the quantum recipient using our 2PC. In particular, the obfuscation retains any falsifiable guarantees when compiled in this manner. For example, an indistinguishability obfuscator still satisfies the full notion of indistinguishability obfuscation -- that is, $\iO(C_1)$ is negligibly distinguishable from $\iO(C_2)$ so long as $C_1$ and $C_2$ are functionally equivalent.

    We leave open the question of obtaining a \emph{non-interactive} classical obfuscation procedure.
\end{itemize}

\subsection{AI Disclosure.}
ChatGPT 6.0 was used to proofread and make edits to this paper during writing. It was also used to generate starting text for the applications section, which was subsequently edited by the authors. It found a bug in the original input-preserving 2PC for linear classical functions (see \Cref{sec:overview:2pc}) and suggested a fix for it, together with a revised proof. The authors have checked the proof carefully before incorporating it, and take responsibility for all technical content presented in this paper.

\section{Preliminaries}

\subsection{Notation and Conventions}\label{sec:prelim:notation}

$\secpar$ is the security parameter.

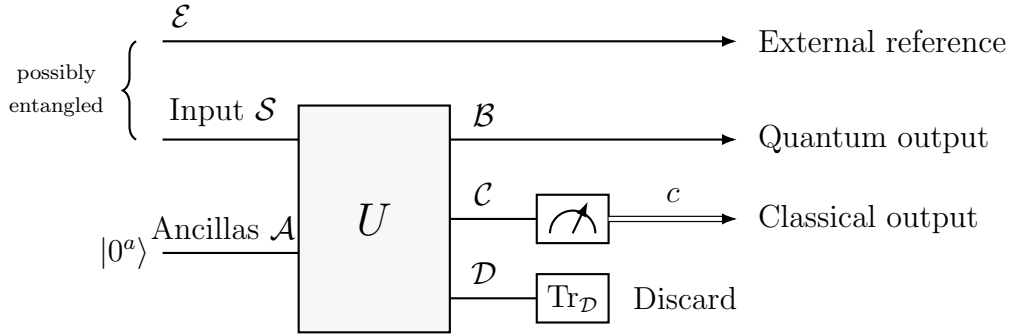
\begin{figure}[H]
    \centering
    \begin{tikzpicture}[
  font=\normalsize,
  >={Latex[length=2mm]},
  wire/.style={line width=.8pt},
  classical/.style={line width=.45pt,double distance=1.5pt},
  box/.style={draw,line width=.8pt,fill=white},
  every node/.style={inner sep=2pt}
]

  \draw[wire,->] (1.2,2.05) -- (8.8,2.05);
  \node[above=3pt] at (1.45,2.05) {$\calE$};
  \node[anchor=west] at (9,2.05) {External reference};

  \draw[wire,decorate,decoration={brace,amplitude=5pt}]
    (.85,.75) -- (.85,2.05);
  \node[anchor=east,align=center] at (.5,1.4)
    {\scriptsize possibly\\[-2pt]\scriptsize entangled};

  \draw[wire] (1.2,.75) -- (3,.75);
  \node[above=3pt] at (2,.75) {Input $\calS$};
  \draw[wire] (1.2,-.75) -- (3,-.75);
  \node[anchor=east] at (1.1,-.75) {$\lvert 0^a\rangle$};
  \node[above=3pt] at (2,-.75) {Ancillas $\calA$};

  \draw[box,fill=black!3] (3,-1.8) rectangle (5,1.2);
  \node at (4,-.3) {\Large $U$};

  \draw[wire,->] (5,.75) -- (8.8,.75);
  \node[above=3pt] at (5.45,.75) {$\calB$};
  \node[anchor=west] at (9,.75) {Quantum output};

  \draw[wire] (5,-.3) -- (6.15,-.3);
  \node[above=3pt] at (5.45,-.3) {$\calC$};
  \draw[box] (6.15,-.62) rectangle (7.1,.02);
  \draw[wire] (6.325,-.49)
    arc[start angle=180,end angle=0,radius=.3];
  \draw[wire,->] (6.625,-.49) -- (6.825,-.07);
  \draw[classical,->] (7.1,-.3) -- (8.8,-.3);
  \node[above=4pt] at (7.95,-.3) {$c$};
  \node[anchor=west] at (9,-.3) {Classical output};

  \draw[wire] (5,-1.35) -- (6.15,-1.35);
  \node[above=3pt] at (5.45,-1.35) {$\calD$};
  \node[box,minimum width=.95cm,minimum height=.64cm]
    at (6.625,-1.35) {$\operatorname{Tr}_\calD$};
  \node[anchor=west] at (7.35,-1.35) {Discard};
\end{tikzpicture}
    \caption{A General Quantum Circuit}
    \label{fig:general-q-circuit}
\end{figure}

\paragraph{Quantum Circuits.}
A unitary circuit $U = U_t \circ \dots U_1$ is composed of unitary gates $(U_1, \dots, U_T)$ applied in sequence to an input register $\calS$ and an ancilla register $\calA$, initialized to $\ket{0^a}$. 
A general quantum circuit additionally specifies the following output registers:  a coherent output register $\calB$, a measured output register $\calC$, and a discarded register $\calD$. $U$ is applied to $\calS$ and $\calA$ to produce $(\calB, \calC, \calD)$, then $\calC$ is measured in the computational basis and $\calD$ is discarded. Any advice is incorporated into the input. See \Cref{fig:general-q-circuit}.

Throughout this work, we will by default consider \emph{real} unitary gates. Any quantum circuit may be generically converted into a real version, with some additional encoding of the input~\cite{STOC:BV93}.

\paragraph{Quantum Turing Machines.}
A quantum Turing machine $M = (U, \Pi_{\mathsf{End}})$ consists of a transition unitary $U$ and an end-state projector $\Pi_{\mathsf{End}}$. It is evaluated on a state $\rho$ in register $\calI$ by initializing a work register $\calW$ to $\ket{0}$, then alternating $U$ and the termination measurement $\{\Pi_{\mathsf{End}}, I - \Pi_{\mathsf{End}}\}$ on $(\calI, \calW)$ until the termination measurement returns $\Pi_{\mathsf{End}}$. The runtime of $M(\rho)$ is the number of applications of $U$ before termination.

\paragraph{Quantum Algorithms.} 
A QPT algorithm is a family of $\poly[n]$-sized unitary circuits $\{U_n\}_{n\in \bbN}$.\footnote{We consider all QPT algoritms to be purified in this work. A strict polynomial time quantum Turing machine may be generically converted to a unitary circuit.}
It is uniform if the family is generated via a deterministic polynomial-time (classical) Turing machine.
An expected QPT algorithm is a quantum Turing machine such that for all $n$-qubit inputs $\rho$, the expected runtime of $M(\rho)$ is $\poly[n]$.

\paragraph{Adversarial Model.} 
We consider QPT adversaries which are non-uniform QPT algorithms with non-uniform quantum advice (auxiliary input) $\{\rho_{\secpar}\}_{\secpar\in \bbN}$. We often omit explicitly mentioning the advice.

For any two distribution ensembles $\calD=\{\calD_\secpar\}_{\secpar\in\mathbb{N}}$ and $\calD'=\{\calD'_\secpar\}_{\secpar\in\mathbb{N}}$, we use $\calD\equiv \calD'$ to denote that these distribution ensembles are statistically close, and use $\calD\approx \calD'$ to denote that they are computationally indistinguishable by any QPT distinguisher with auxiliary input. When an adversary $A_\secpar$ is used to define the distributions, their auxiliary input may be entangled with the distinguisher's.
We use $\overset{\epsilon}{\approx}$ to denote that the distributions may be computationally distinguished with at most probability $\epsilon + \negl$.

When comparing real numbers, $a \approx b$ denotes that $|a - b| = \negl$, i.e. it drops negligible differences. We frequently use $a\approx_{\negl} b$ to be explicit about the dropped difference.

\paragraph{Other Notation.}
We also define some quantum-specific notation.
\begin{itemize}
    
\item
For any $q$-qubit quantum state $\brho$ and string ${\bf b} \in \{0, 1\}^{q}$, we denote by ${\bf m} \gets {\sf Measure}({\bf b}, \brho)$ the result of measuring $\brho$ according to the basis ${\bf b}$, where ${\bf b}_i=0$ corresponds to the $Z$ basis, ${\bf b}_i=1$ corresponds to the $X$ basis.

\item For more general measurements, we use the following notation. Let $U$ be a quantum uniary on $n$ qubits and let $f:\{0,1\}^n\rightarrow \{0,1\}^m$ be a classical function. We denote the projective measurement
\[
    M[f, U] = \left\{ U^\dagger \left(\sum_{x: f(x) = y} \ketbra{x} \right) U \right\}_{y \in \{0,1\}^m}
\]
When $f$ is the identity, we write $M[U] = M[I, U]$ to denote measuring all of the qubits in the basis defined by $U$.

\item We use $\Write(V) = X^V$ to denote coherently writing the value of a random variable $V$ to a register.

\end{itemize}

\subsection{History States}\label{sec:hist}

\paragraph{History States.} Kitaev's proof that the local Hamiltonian problem is $\QMA$-complete makes use of {\em history states}. 
Intuitively, a history state tracks how a quantum computation $U = U_t \dots U_1$ evolves over time by being in superposition over the states of the system $\ket{t} \otimes U_{\leq t}(\ket{\psi}\otimes\ket{0^a})$ at time $t$, where $\ket{0^a}$ are the ancilla registers used in the computation. 
The time $t$ is encoded in unary:
for $T \geq 1$ and $t \in \{0, \ldots, T\}$, define $t_{\mathbf{1},T} \in \{0, 1\}^T$ to be the unary encoding
        \[ t_{\mathbf{1},T} = 1^t 0^{T-t}. \]

\begin{definition}[\cite{KSV02}]\label{def:history}
    Fix any quantum circuit $C$ that takes as input a state of the form $\ket{\psi}\otimes\ket{0^a}$, where the last $a$ qubits are ancillas that are initialized to~$0$,  and consists of $T$ gates $U_1, \ldots, U_T$. Then for any input state $\ket{\psi}\otimes\ket{0^a}$, we define the history state of $C$ on $\ket{\psi}\otimes\ket{0^a}$ to be 
        \[ \Hist_C(\ket{\psi}\otimes\ket{0^a}) = \frac{1}{\sqrt{T+1}} \sum_{t = 0}^T U_{\leq t} (\ket{\psi}\otimes\ket{0^a}) \otimes \ket{t_{\mathbf{1},T}},\]
    where $U_{\leq 0} = I$ and $U_{\leq t} = U_tU_{t-1} \ldots U_1$ for $1 \leq t \leq T$. Define the unitary $U_{\Hist_C}$ (defined on the legal-clock subspace) to be
        \[ U_{\Hist_C} = \sum_{t = 0}^T U_{\leq t} \otimes \ket{t_{\mathbf{1},T}}\bra{t_{\mathbf{1},T}}. \]
    Note that
        \[ U_{\Hist_C} \left((\ket{\psi}\otimes\ket{0^a})\otimes  \frac{1}{\sqrt{T+1}} \sum_{t = 0}^T \ket{t_{\mathbf{1},T}} \right) = {\Hist_C(\ket{\psi}\otimes\ket{0^a})}. \]
\end{definition}

Kitaev showed a local Hamiltonian where history state of a computation $C$ is a low-energy state. The Hamiltonian $H_\Hist$ is comprised of several parts. $H_\inp$ checks that the input ancillas are correctly initialized to $\ket{0^a}$,  $H_\format$ checks that the clock is correctly formatted, and $H_\prop$ checks that each step of the computation is done correctly.

In the context of QMA, there is also a term $H_\outp$ which checks that the output is accepting. 
\textbf{We explicitly omit $H_\outp$ from the Hamiltonian $H$ for this work.} 
As a result, $H_{\Hist}$ always has a non-empty kernel.

\begin{theorem}[Feynman-Kitaev history Hamiltonian and $X/Z$ test]\cite{KSV02,CLN18}
\label{thm:xz-fk}
Let $C=(U_1,\ldots,U_T)$ be a quantum circuit of constant-local
gates that uses $a$ ancilla qubits that are initialized to~$0$.\footnote{We assume without loss of generality that $a\leq T$.} There is a constant-local Hamiltonian
\[
H_{\Hist}(C)
=
H_{\sf format}+H_{\sf prop}+H_{\sf in}
=
\sum_{j=1}^{M}\Pi_{C,j},
\qquad M=O(T),
\]
with $H_{\out}$ omitted, where every $\Pi_{C,j}$ is a projector,
\[
\ker H_{\Hist}(C)
=
\operatorname{span}
\bigl\{
\Hist_C(\ket{\psi}\otimes\ket{0^a}):\ket{\psi}
\bigr\},
\]
and there exists a universal constant $c>0$ such that for $G:=\Pi_{\ker H_{\Hist}(C)}$,
\[
H_{\Hist}(C)
\succeq
\frac{c}{(T+1)^2}
\bigl(I-G\bigr),\footnote{For Hermitian operators $A$ and $B$,  $A\succeq B$ means that
$A-B$ is positive semi-definite. Equivalently,
$\bra{\psi}(A-B)\ket{\psi}\geq 0$ for every state $\ket{\psi}$.}
\]

If, in addition, every $U_t$ is Hermitian and diagonal in either the
$X$ or the $Z$ basis, then for every $j$ there are a product
$X/Z$ basis $\theta_j$ and a constant-size predicate $g_j$ such that
\begin{equation}\label{eqn:Pij}
\Pi_{C,j}
=
H^{\theta_j}
\left(\sum_{z:g_j(z)=1}|z\rangle\langle z|\right)
H^{\theta_j}.
\end{equation}

\end{theorem}

\begin{remark}
    An immediate consequence of \Cref{thm:xz-fk} is the following: Consider the test that samples $j\gets[M]$, performs the projective measurement $\{\Pi_{C,j}, I-\Pi_{C,j}\}$,  and rejects on outcome~$1$. Then for every state $\rho$,
\[
\Pr_{j\leftarrow[M]}[\mathsf{reject}]
=
\frac{1}{M}\Tr(H_{\Hist}(C)\rho)
\ge
\frac{c}{M(T+1)^2}\Tr((I-G)\rho).
\]
Namely, the probability of rejection is at least an inverse-polynomial multiple of the weight of $\rho$ outside the history-state subspace.
In addition, every valid history state is accepted with probability $1$.
\end{remark}

For our work we need each local Hamiltonian term $\Pi_{C,j}$ to be diagonal in a product $X/Z$ basis.  
Fitzsimons, Hajdu\v{s}ek, and Morimae \cite{PRL:FHM18} gave a test for measuring the energy of $XZ$ Hamiltonians using only single-qubit $X/Z$ measurements whose acceptance probability is determined by the energy of an $XZ$ Hamiltonian: the verifier samples a Pauli term from the Hamiltonian and measures that term. This does not directly implement the individual Feynman–Kitaev projectors and does not have perfect completeness.
We instead restrict to circuits over the gate set $\calG_{XZ}$, 
defined by
\[
\calG_{XZ}=\{X,CZ,CCZ,\widehat{G}\}~\mbox{ where }~\widehat{\calG}=(H\otimes H)CZ(H\otimes H).
\]
Building on an observation of Ma and Natarajan \cite{MN26}, the Feynman–Kitaev projectors for such circuits are diagonal in product $X/Z$ bases, yielding the perfect-completeness test below.\footnote{\cite{MN26} uses a different clock construction to ensure that every Hamiltonian term is diagonal entirely in $X$ or entirely in $Z$. For example, this eliminates $X \otimes Z$ checks while preserving $X\otimes X$ or $Z \otimes Z$. The original clock construction together with the new gateset suffices when one does not care about such mixed terms.}

\begin{corollary}[Perfect-completeness $X/Z$ test for history states]
\label{corr:hist}
Let  $C=(U_1,\ldots,U_T)$ be a quantum circuit, where each $U_j\in \mathcal{G}_{XZ}$, and let
$H_\Hist(C)=\sum_{j=1}^M \Pi_{C,j}$.
Then the following holds:
\begin{itemize}
    \item {\bf $X/Z$-Test.}  Every $\Pi_{C,j}$ is 
    diagonal in a constant-size product $X/Z$ basis.

\item {\bf Perfect completeness} For every input state $\ket{\psi}$ and every $j\in [M]$,
\[
\Pr_{j\leftarrow[M]}
\left[
\Pi_{C,j}
\left(
\Hist_C(\ket{\psi}\otimes\ket{0^a})
\right)=0
\right]
=1.
\]

\item {\bf Soundness.} There exists a universal constant $c\in\mathbb{N}$ such that for every state $\rho$,
\[
\Pr_{j\leftarrow[M]}[\Pi_{C,j}(\rho)=1]
=
\frac{1}{M}\Tr\!\left(H_{\Hist}(C)\rho\right)\ge
\frac{c}{M(T+1)^2}\Tr((I-G)\rho).
\]
where $G$ is the projector onto the kernel of $H_\Hist(C)$. %
\end{itemize}

\end{corollary}

\Cref{corr:hist} implies the following theorem.

\begin{theorem}[Repeated history-state test]\label{thm:repeated-hist-test}
Let  $C=(U_1,\ldots,U_T)$ be a quantum circuit, where each $U_j\in \mathcal{G}_{XZ}$, and let
\[
H_{\Hist}(C)=\sum_{j=1}^{M}\Pi_{C,j},
\qquad
G:=\Pi_{\ker H_{\Hist}(C)}.
\]
Consider the test that given a state
\(\ket{\psi}\), independently repeats the following  \(r=M\cdot (T+1)^2\cdot \lambda\) times: sample \(j\gets[M]\),
apply the $X/Z$ projection \(\{I-\Pi_{C,j},\Pi_{C,j}\}\), and abort on the second outcome.
Let \(p_{\sf acc}\) be the probability that all tests accept and let
\(\rho_{\sf acc}\) be the residual state conditioned on acceptance, averaged
over the sampled indices. Then for  \(\gamma=\frac{c}{(T+1)^{2}}\) where $c\in\mathbb{N}$ is the constant from \Cref{thm:xz-fk},

\[
p_{\sf acc}-\|G\ket{\psi}\|^2
\le
\left(1-\frac{\gamma}{M}\right)^r
\le e^{-\gamma r/M}.
\]
Moreover, if \(G\ket{\psi}\neq 0\), then, for
\(\ket{h}:=G\ket{\psi}/\|G\ket{\psi}\|\),
\[
\frac12\bigl\|\rho_{\sf acc}-\ketbra{h}\bigr\|_1
\le
\sqrt{\frac{e^{-\gamma r/M}}{p_{\sf acc}}}.
\]
Consequently,if \(p_{\sf acc}\ge 1/{\sf poly}(\lambda)\), then
\(\rho_{\sf acc}\) is negligibly close to a valid history state.
\end{theorem}

\begin{proof}
Note that 
\[
{\sf image}(G)=\bigcap_{j=1}^M\ker(\Pi_{C,j}),
\]
Let $P_j=I-\Pi_{C,j}$ and $Q:=I-G$.
By Corollary~1, for every $|\phi\rangle\in{\sf image}(Q)$,
\[
\mathbb{E}_{j\leftarrow[M]}\|P_j|\phi\rangle\|^2
= \mathbb{E}_{j\leftarrow[M]}\|(I-\Pi_{C,j})|\phi\rangle\|^2
= 
\|\phi\|^2-\frac{1}{M}
\langle\phi|H_{\Hist}(C)|\phi\rangle
\le
\left(1-\frac{\gamma}{M}\right)\|\phi\|^2.
\]
For for any ${\mathbf j}=(j_1,\ldots,j_r)\in[M]^r$, denote by $
K_{\mathbf j}:=P_{j_r}\cdots P_{j_1}$.
Using this notation, the equation above implies that
\[
\mathbb{E}_{\mathbf j}
\|K_{\mathbf j}Q|\psi\rangle\|^2
\le
\left(1-\frac{\gamma}{M}\right)^r
\|Q|\psi\rangle\|^2.
\]

Moreover,
\[
K_{\mathbf j}|\psi\rangle
=
G|\psi\rangle+K_{\mathbf j}Q|\psi\rangle,
\]
and the two summands are orthogonal. Hence
\[
p_{\sf acc}
=
\|G|\psi\rangle\|^2+
\mathbb{E}_{\mathbf j}
\|K_{\mathbf j}Q|\psi\rangle\|^2,
\]
and therefore
\[
p_{\sf acc}-\|G|\psi\rangle\|^2
\le
\left(1-\frac{\gamma}{M}\right)^r
\le
e^{-\gamma r/M}.
\]
Suppose 
that $G|\psi\rangle\neq0$, and let
\[
|h\rangle:=\frac{G|\psi\rangle}{\|G|\psi\rangle\|}.
\]
The conditioned residual state is
\[
\rho_{\sf acc}
=
\frac{1}{p_{\sf acc}}
\mathbb{E}_{\mathbf j}
\left[
K_{\mathbf j}|\psi\rangle\langle\psi|K_{\mathbf j}^{\dagger}
\right].
\]
The fact that $K_{\mathbf j}^{\dagger}\ket{h}=\ket{h}$ implies that 
\[
\langle h|\rho_{\sf acc}\ket{h}
=
\frac{\|G|\psi\rangle\|^2}{p_{\sf acc}}.
\]
Thus, by the Fuchs--van de Graaf inequality,
\[
\frac12
\left\|
\rho_{\sf acc}-|h\rangle\langle h|
\right\|_1
\le
\sqrt{
1-\frac{\|G|\psi\rangle\|^2}{p_{\sf acc}}
}
\le
\sqrt{\frac{e^{-\gamma r/M}}{p_{\sf acc}}}.
\]

Finally, since $\gamma=\Omega(T^{-2})$ and
$r=\Omega(MT^2\lambda)$,
\[
e^{-\gamma r/M}=e^{-\Omega(\lambda)}.
\]
Therefore, if $p_{\sf acc}\ge 1/{\sf poly}(\lambda)$, then
$\rho_{\sf acc}$ is negligibly close to the valid history state
$|h\rangle$.
\end{proof}

\subsection{Quantum Fully Homomorphic Encryption}\label{sec:QFHE}

\begin{definition}[Quantum Fully-Homomorphic Encryption ($\QFHE$), \cite{C:BroJef15, FOCS:Mahadev18b, C:Brakerski18}]\label{def:QFHE}
    A quantum homomorphic encryption scheme with respect to a message space $\mathcal{M}$ consists of four $\QPT$ quantum algorithms 
    \[ (\Gen, \Enc, \Eval, \Dec) \]
    with the following syntax:
        \begin{itemize}
            \item
                \textbf{Key Generation.} $\Gen$ is a classical ${\sf PPT}$  algorithm that takes as input the security parameter in unary $1^{\lambda}$ and outputs a public and secret key pair $(\pk, \sk)$.\footnote{For simplicity, we think of the evaluation key as being included in $\pk$.}
            \item
                \textbf{Encryption.} $\Enc: \calM \to \calC$ takes as input a public key $\pk$ and a classical or quantum state $\ket{\psi}$ in the message space $\calM$, and outputs a classical or quantum state ciphertext ${\ket{\widehat{\psi}}}$ in the ciphertext space $\calC$. 
                If $\ket{\psi}$ is entirely classical, then so is $\ket{\widehat{\psi}}$.
            \item
                \textbf{Homomorphic Evaluation.} $\Eval$ takes as input a public key $\pk$, a unitary circuit $C$, a ciphertext $\ket{\widehat{\psi}}$, and ancilla registers $\ket{0^a}$ (which we often omit from the notation), and outputs an evaluated quantum ciphertext $\ket{\widehat{\phi}}$. The garbage ancillas are discarded.

            \item
                \textbf{Decryption.} $\Dec$ takes as input a secret key $\sk$ and a quantum ciphertext ${\ket{\widehat{\phi}}}$, and outputs a quantum state $\ket{\phi}$.  
        \end{itemize}
    The algorithms have the following properties:
        \begin{enumerate}
            \item
                \textbf{Correctness.} For every polynomials $\ell$ and $p$, it holds that for every  $\secpar \in \bbN$, every quantum state $\ket{\psi}$ consisting of $\ell(\secpar)$ qubits, and every quantum circuits~$C$ of size $p(\secpar)$, it holds that
                \[ 
                    \Dec(\sk, \Eval(\pk, C, \Enc(\pk, \ket{\psi}))) 
                    \quad \text{and}\quad
                    C(\ket{\psi}) 
                \]
                have negligible trace distance, where $(\pk, \sk) \leftarrow \Gen(1^{\lambda})$.
            \item 
                \textbf{Semantic Security.} For every polynomial $\ell$, every $\lambda \in \bbN$ and every pair of quantum (or classical) $\ell(\secp)$-qubit states  $(\ket{\psi_0}, \ket{\psi_1})$, it holds that 
                \[ \Enc(\pk, \ket{\psi_0}) \approx \Enc(\pk, \ket{\psi_1}), \]
                where $(\pk, \sk) \leftarrow \Gen(1^{\lambda})$. 
            \item 
                \textbf{Compactness.} There exists a polynomial $p$ such that for any family $\{C_{\lambda}\}_{\lambda \in \bbN}$ of polynomial size circuits, where $C_\secpar$ takes as input an $n(\secpar)$-qubit quantum state, and for every $n(\secpar)$-qubit state $\ket{\psi}$, the runtime of $\Dec$ on input  $(\sk,\Eval(\pk, C, \ket{\psi}))$ is at most $p(\lambda)$, where $(\pk,\sk)\gets\Gen(1^\secpar)$. Namely, the decryption runtime is independent of the size of the evaluated circuit~$C$.

        \end{enumerate}
\end{definition}

For the ciphertext rerandomization approach to circuit privacy, see~\cite[Section~3.7 of the full version]{C:BCKM21a}.
\begin{definition}\label{def:QFHE-circuit-private}
    A $\QFHE$ scheme is said to be circuit private if there exists a $\QPT$
simulator $\Sim$ such that for every polynomial-size quantum circuit~$C$,
 that takes as input a quantum register $\calA$, and for every $\QPT$ adversary that given $\pk$ outputs a state $\rho_{\calA \calE}$, where $\calE$ is an arbitrary auxiliary register,\footnote{$\calE$ is an arbitrary auxiliary register that may be entangled with
registers $\calA$. Its inclusion ensures that circuit privacy holds
even in the presence of arbitrary quantum side information.}
\[
\left(
  \pk,\sk,
  \Eval(\pk,C\otimes I_\calE,\widehat{\rho}_{\calA \calE})
  \right)
\approx_c
\left(
  \pk,\sk,\widehat{w},
  (\Sim\otimes I_\calE)
  \bigl(\pk,(C\otimes I_\calE)\rho_{\calA \calE}\bigr)
\right),
\]
where $(\pk,\sk)\gets\Gen(1^\lambda)$,  and $\rho_{\calA \calE}$ is the state that the adversary outputs upon receiving $\pk$.\footnote{Note that $\Eval$ explicitly discards the leftover ancillas from homomorphic evaluation.} Thus, an evaluated ciphertext reveals no information about
$C$ beyond its output, even to a party holding $\sk$.

\end{definition}

\begin{remark}\label{remark:QFHE}  
In \Cref{sec:general}, we use a circuit-private $\QFHE$ with the property that the encryption scheme is of the form:
  \[\QEnc(\pk,\ket{\psi})= X^x Z^z \ket{\psi}, \Enc(\pk,(x, z)).
    \] 
where $\Enc$ is a classical encryption algorithm corresponding to a classical circuit-private $\FHE$ scheme.  We also assume that given any quantum state $\ket{\psi}$ and any $\pk$ generated according to $\Gen(1^\secp)$ one can efficiently generate a ciphertext 
\[\QEval\left(
    \pk,I,
    \QEnc\left(
      \pk,
      \ket{\psi}
    \right)
  \right)
  \]
  together with the corresponding pads $x,z$ in the clear.
    
\end{remark}

\begin{theorem}[\cite{FOCS:Mahadev18b, C:Brakerski18}]\label{thm:qfhe}
    There exists a circuit private $\QFHE$ scheme that satisfies the property in \Cref{remark:QFHE}, assuming the post-quantum hardness of $\LWE$. 
\end{theorem}

\subsection{Functional Authentication via Coset States}

We recall the notion and construction of functional authentication from \cite{STOC:BarBraVai24}, with the strengthened security shown by \cite{FOCS:HuaTan25}.
Although their construction only supports homomorphic evaluations from a limited class of circuits, we state the definition with respect to arbitrary classes of circuits.

\subsubsection{Definition}

\begin{definition}[Functional Authentication]
\label{def:ht-functional-auth}
A functional authentication scheme for a family of unitaries $\calU = \{\calU_n\}_{n\in \bbN}$ consists of the following algorithms.
\begin{itemize}
    \item $\KeyGen(1^\secpar,1^n) \to k$ takes the security parameter and an
    $n$-qubit message length $n=\operatorname{poly}(\secpar)$, and outputs a
    classical authentication key $k$.

    \item $\Enc_k(\rho) \to \sigma$ is an isometry, parameterized by $k$,
    that maps an $n$-qubit state $\rho$ to an authenticated state
    $\sigma=(\sigma_1,\ldots,\sigma_n)$, where each $\sigma_i$ consists of
    $p(\secpar)$ qubits, where $p$ is a polynomial.

    \item $\Eval(U) \to \widetilde{U}$ is a classical algorithm that takes as input a description of a unitary $U\in \calU_n$ on $n$ qubits, then outputs a description of another unitary $\widetilde{U}$ on $np(\secpar)$ qubits. $\widetilde{U}$ acts as the homomorphic implementation of $U$ on authenticated states.

    \item $\KeyEval(k, U) \to k_U$ is a classical algorithm that maps a key $k$ and a unitary $U \in \calU_n$ to an evaluation key $k_U$.

    We emphasize that $k_U$ does not necessarily belong to the same keyspace as $k$.

    \item $\Dec_{k_U}(c)\to m/\bot$ is a classical decoding algorithm that, given a candidate physical measurement outcome $c$, returns either its decoded logical outcome $m$ or $\bot$. It is parameterized by an evaluation key $k_U$.

    We overload $\Dec_{k_U}(c, f)$ to compute $v = \Dec_k(c)$, then output $f(v)$ for some classical function $f$.

    \item $\Ver_{k_U}(c)\to\top/\bot$ is a classical verification
    algorithm that returns $\top$ exactly when
    $\Dec_{k_U}(c)\neq\bot$.
\end{itemize}

Functional authentication must satisfy correctness and functional security.
\end{definition}

Correctness requires that authenticating a state $\ket{\psi}$, then homomorphically measuring the authenticated state  is equivalent to directly measuring $\ket{\psi}$, then authenticating the residual state.

\begin{definition}[Functional Authentication Correctness]  
    A functional authentication scheme for $\calU$ is correct if
    For every function $f:\{0,1\}^n\to\{0,1\}^*$, unitary $U \in \calU$, and key $k\in \Supp(\KeyGen(1^\secpar,1^n))$,
    \begin{equation}
    \label{eq:fa-correctness}
        M\left[f\circ\Dec_{k_U}, \widetilde{U} \right]\circ\Enc_k
        =
        \Enc_k\circ M[f, U],
    \end{equation}
    where $f(\bot)\coloneqq\bot$, and the definition of $M\left[f, U\right]$ can be found in \Cref{sec:prelim:notation}.  Moreover, for every $n$-qubit state $\rho$ and every unitary $U\in\calU_n$,
    \[
        \Pr\left[
            \top\gets
            M\left[\Ver_{k_U},\widetilde{U}\right](\Enc_k(\rho))
        \right]=1.
    \]
    where the probability is over the measurement outcome.
\end{definition}

Functional authentication security considers a bounded-query algorithm $\calA$ which receives an authenticated state $\ket{\widetilde{\psi}}$ of its choice along with quantum oracle access to a pre-defined set of decoding functions. 
Ideally, we want that $\calA$ should behave as if it never received $\ket{\widetilde{\psi}}$, and the oracles were replaced by direct measurements on $\ket{\psi}$. The following definition from \cite{FOCS:HuaTan25} formalizes this intuition.

\begin{definition}[Functional Authentication Security~\cite{FOCS:HuaTan25}]
\label{def:fnl-auth-sec}
    Fix $t,n'=\operatorname{poly}(\secpar)$, functions $f_1,\ldots,f_t:\{0,1\}^n\to\{0,1\}^{n'}$, and unitaries $U_1, \dots, U_t \in \calU$. Let $k_j = \KeyEval(k, U_j)$.

    Define the following coherent oracles $\mathsf{fDec}_k$ and $\mathsf{fVer}_k$. A query contains an index $j\in[t]$, a candidate physical outcome $\widetilde{v}$, and an answer register $\calY$. $\mathsf{fVer}$ additionally takes as input a register $\calV$. 
    \begin{multicols}{2}
    \begin{itemize}
        \item $\mathsf{fDec}_k(j,\widetilde{v},\calY)$:
        \begin{enumerate}
            \item Compute $v\gets\Dec_{k_{j}}(\widetilde{v})$. 
            \item If $v = \bot$, apply $\Write(\bot)$ to $\calY$.
            \item Otherwise, apply $\Write(f_j(v))$ to $\calY$.
        \end{enumerate}

        \item $\mathsf{fVer}_k(j,\widetilde{v},\calY,\calV)$:
        \begin{enumerate}
            \item Compute $b\gets\Ver_{k_{j}}(\widetilde{v})$. 
            \item If $b = \bot$, apply $\Write(\bot)$ to $\calY$.
            \item Otherwise, apply $\Write(M[f, U_j]\calV)$ to  $\calY$.
        \end{enumerate}
    \end{itemize}
    \end{multicols}
    The definitions of $M\left[f, U\right]$ and $\Write$ can be found in the notation subsection \ref{sec:prelim:notation}.

    The tests and writes above are interpreted as controlled operations, and the measurements are implemented coherently.  
    Then, for every polynomial-query quantum adversary $\calA$ and every state $\ket{\psi}_{\calV\calW}$,
    \begin{equation}
    \label{eq:fa-functional-security}
    \begin{aligned}
        &\left\{
        (\widetilde{\calV},\calW):
        \begin{array}{l}
            (\calV,\calW)\gets\ket{\psi}_{\calV\calW},\\
            k\gets\KeyGen(1^\secpar,1^n),\\
            \widetilde{\calV}\gets\Enc_k(\calV),\\
            (\widetilde{\calV},\calW)\gets
            \calA^{\mathsf{fDec}_k}(\widetilde{\calV},\calW)
        \end{array}
        \right\}
        \\
        &\hspace{2cm}\stackrel{2^{-\Omega(\secpar)}}{\equiv}
        \left\{
        (\widetilde{\calV},\calW):
        \begin{array}{l}
            (\calV,\calW)\gets\ket{\psi}_{\calV\calW},\\
            k\gets\KeyGen(1^\secpar,1^n),\\
            \widetilde{\calV}\gets\Enc_k(\ket{0^n}),\\
            (\widetilde{\calV},\calW)\gets
            \calA^{\mathsf{fVer}_k(\cdot,\cdot,\cdot,\calV)}
            (\widetilde{\calV},\calW).
        \end{array}
        \right\}
    \end{aligned}
    \end{equation}
            
\end{definition}

\subsubsection{Coset-state construction}\label{sec:coset-auth-construction}

In this section we present the construction due to Bartusek, Brakerski, and Vaikuntanathan~\cite{STOC:BarBraVai24}.
The construction supports the family
\[
    \calU_n^{\mathsf{ZX\text{-}Lin}}
    \coloneqq
    \left\{
        H^\theta L :
        \theta\in\{0,1\}^n \text{ and }
        L\text{ is a sequence of $\CNOT$ gates on $n$ qubits}
    \right\}.
\]
Let $\calU^{\mathsf{ZX\text{-}Lin}}\coloneqq\{\calU_n^{\mathsf{ZX\text{-}Lin}}\}_{n\in\bbN}$.
A description of $U\in\calU_n^{\mathsf{ZX\text{-}Lin}}$ includes a factorization $U=H^\theta L$.

\paragraph{Notation.}  
A subspace state 
\[
    \ket{S} = \frac{1}{\sqrt{|S|}} \sum_{s\in S} \ket{s}
\]
is a uniform superposition over a subspace $S \subset \bbF^d$ of some ambient vector space $\bbF^d$ which has dimension $d$. Most commonly, we consider binary vector spaces $\bbF_2^d$.
A coset state is a subspace state with $x$ and $z$ offsets: $X^x Z^z \ket{S}$.

$\Can(S)\subset \bbF^d$ is set of canonical elements of each $S+x$, which is isomorphic to the quotient group $\bbF^d/S = \{S + x: x \in \bbF^d\}$. The isomorphism is given by $\Can_S:\bbF^d \rightarrow \bbF^d$. Note that $\Can_S(x) = \Can_S(x')$ if and only if $S + x = S + x'$.
We additionally require that $\Can_S(x) \in S + x$. Thus, we refer to $\Can_S(x)$ as the ``canonical element'' of $S + x$.

Given a $\secpar$-dimensional subspace $S\subseteq\bbF_2^{2\secpar+1}$ and $\Delta\notin S$, define
\[
    S_\Delta\coloneqq S+\Span\{\Delta\}=S\cup(S+\Delta).
\]
Fix a canonical $\Delta_H\in S^\perp\setminus S_\Delta^\perp$. Then
\[
    S^\perp
    =S_\Delta^\perp+\Span\{\Delta_H\}
    =S_\Delta^\perp\cup(S_\Delta^\perp+\Delta_H).
\]

\begin{construction}[Coset functional authentication]
\label{constr:bvv-coset-functional-auth}
The coset authentication scheme of Bartusek, Brakerski, and Vaikuntanathan~\cite{STOC:BarBraVai24} is defined as follows.
\begin{itemize}
    \item \underline{$\KeyGen(1^\secpar,1^n)$:}
    Sample a uniformly random $\secpar$-dimensional subspace $S\subseteq\bbF_2^{2\secpar+1}$, a uniformly random $\Delta\gets\bbF_2^{2\secpar+1}\setminus S$, and uniformly random $x_i,z_i\gets\bbF_2^{2\secpar+1}$ for every $i\in[n]$.\footnote{We remark that $x_i$ and $z_i$ can equivalently be replaced by $\Can_S(x_i) \in \Can(S)$ and $\Can_{S_\Delta^\perp}(z_i) \in \Can(S_\Delta^\perp)$.}
    Output
    \[
        k\coloneqq(S,\Delta,\vec{x},\vec{z}),
        \qquad
        \vec{x}=(x_i)_{i\in[n]},\quad
        \vec{z}=(z_i)_{i\in[n]}.
    \]

    \item \underline{$\Enc_k(\rho)$:}
    Define the isometry $E_{S,\Delta}$ by
    \[
        E_{S,\Delta}\ket b=\ket{S+b\Delta}
        \qquad\text{for }b\in\{0,1\}.
    \]
    Apply $E_{S,\Delta}^{\otimes n}$ to $\rho$, followed by the quantum one-time pad
    \[
        X^{\vec{x}}Z^{\vec{z}}
        \coloneqq\bigotimes_{i\in[n]}X^{x_i}Z^{z_i}.
    \]

    \item \underline{$\Eval(U)$:}
    Parse the description of $U$ as $U=H^\theta L$.
    Define
    \[
        \widetilde\theta
        \coloneqq(\theta_1^{2\secpar+1}\,\|\,\cdots\,\|\,\theta_n^{2\secpar+1}),
    \]
    where $\theta_i^{2\secpar+1}$ denotes $2\secpar+1$ repetitions of $\theta_i$.
    Define $\widetilde L$ by replacing every logical $\CNOT$ in $L$ with the transversal operation $\CNOT^{\otimes(2\secpar+1)}$ on the corresponding authenticated blocks.
    Output
    \[
        \widetilde U\coloneqq H^{\widetilde\theta}\widetilde L.
    \]

    \item \underline{$\KeyEval(k,U)$:}
    Parse $k=(S,\Delta,\vec{x},\vec{z})$ and the description $U=H^\theta L$.
    Process the $\CNOT$ gates of $L$ in order to update the one-time-pad keys.
    For a $\CNOT$ with control $i$ and target $j$, apply the update
    \[
        ((x_i,z_i),(x_j,z_j))
        \longmapsto
        ((x_i,z_i+z_j),(x_i+x_j,z_j)).
    \]
    Denote the resulting offsets by $(\vec{x}_L,\vec{z}_L)$.
    For every $i\in[n]$, define
    \[
        (T_i,\Gamma_i,r_i)
        \coloneqq
        \begin{cases}
            (S,\Delta,x_{L,i}) & \text{if }\theta_i=0,\\
            (S_\Delta^\perp,\Delta_H,z_{L,i}) & \text{if }\theta_i=1.
        \end{cases}
    \]
    Output the evaluation key
    \[
        k_U\coloneqq\bigl((T_i,\Gamma_i,r_i)\bigr)_{i\in[n]}.
    \]

    \item \underline{$\Dec_{k_U}(c)$:}
    Parse $k_U=((T_i,\Gamma_i,r_i))_{i\in[n]}$ and $c=(c_1,\ldots,c_n)$ with $c_i\in\bbF_2^{2\secpar+1}$.
    For every $i\in[n]$, compute
    \[
        m_i\coloneqq
        \begin{cases}
            0 & c_i\in T_i+r_i,\\
            1 & c_i\in T_i+\Gamma_i+r_i,\\
            \bot & \text{otherwise.}
        \end{cases}
    \]
    If some $m_i=\bot$, output $\bot$; otherwise output $m=(m_1,\ldots,m_n)$.

    \item \underline{$\Ver_{k_U}(c)$:}
    Using the same parsing, check whether $c_i\in T_i+\Span\{\Gamma_i\}+r_i$ for every $i\in[n]$.
    Output $\top$ if every check passes, and output $\bot$ otherwise.
\end{itemize}
As a special case, we also define $\Eval$ and $\KeyEval$ on Pauli operators $X^{\vec{x'}}Z^{\vec{z'}}$ where $\vec{x'}, \vec{z'} \in \bbF_2^n$ by
\begin{itemize}
    \item $\Eval(X^{\vec{x'}}Z^{\vec{z'}}) = I$
    \item and $\KeyEval(k, X^{\vec{x'}}Z^{\vec{z'}}) = (S, \Delta, \vec{x} - \vec{x'}\otimes \Delta, \vec{z} - \vec{z'}\otimes \Delta_H)$.
\end{itemize}
In particular, Pauli operators can be applied solely with a key update, and the updated key remains in the original $\KeyGen$ keyspace. Therefore if we let $k' = \KeyEval(k, X^{\vec{x'}}Z^{\vec{z'}})$, then
\begin{equation}\label{eq:BBV-pauli-update}
    \Auth_{k'}(\ket{\psi}) = \Auth_{k}(X^{\vec{x'}}Z^{\vec{z'}}\ket{\psi})
\end{equation}
\end{construction}

Huang and Tang give a stronger, simulation-based definition than Bartusek, Brakerski, and Vaikuntanathan's original definition and show that the same construction satisfies the stronger definition.

\begin{theorem}[\cite{FOCS:HuaTan25}]
\label{thm:bvv-coset-functional-auth}
For $\secpar\geq\Omega(n)$, \Cref{constr:bvv-coset-functional-auth} is a functional authentication scheme for $\calU^{\ZXLin}$ in the sense of \Cref{def:ht-functional-auth}.
\end{theorem}

We mention an additional useful property of BBV's construction: decoding is (almost) linear. 

\begin{lemma}[Affine decoding]
Let $\mathsf{Chk}_{S_\Delta}$ and $\mathsf{Chk}_{S^\perp}$ be parity-check matrices for $S_\Delta$ and $S^\perp$, respectively, and define
\[
    L_{\mathsf C}\coloneqq
    \begin{pmatrix}
        \Delta_H^\top\\
        \mathsf{Chk}_{S_\Delta}
    \end{pmatrix},
    \qquad
    L_{\mathsf H}\coloneqq
    \begin{pmatrix}
        \Delta^\top\\
        \mathsf{Chk}_{S^\perp}
    \end{pmatrix}.
\]
For each $i\in[n]$, let $y_i=L_{\mathsf C}\,x_{L,i}$ if $\theta_i=0$, and let $y_i=L_{\mathsf H}\,z_{L,i}$ if $\theta_i=1$.
Define
\[
    w_i\coloneqq
    \begin{cases}
        L_{\mathsf C}\,c_i-y_i & \text{if }\theta_i=0,\\
        L_{\mathsf H}\,c_i-y_i & \text{if }\theta_i=1,
    \end{cases}
    \qquad
    m_i\coloneqq
    \begin{cases}
        b & \text{if }w_i=b\,\|\,0^{2\secpar}\text{ for some }b\in\{0,1\},\\
        \bot & \text{otherwise}.
    \end{cases}
\]
Then $\Dec_{k_U}(c)$ outputs $\bot$ if some $m_i=\bot$, and otherwise outputs $(m_1,\ldots,m_n)$.
\end{lemma}
\begin{proof}
    The kernel of $\mathsf{Chk}_{S_\Delta}$ and $\mathsf{Chk}_{S^\perp}$ are $S$ and $S^\perp$, respectively, so $w_i$ ends in $0^{2\secpar}$ if and only if $c_i - x_{L,i} \in S_\Delta$ or $c_i - z_{L,i} \in S^\perp$, respectively. Furthermore, the first bit of $w_i$ is $1$ if and only if $c_i - x_{L, i} \in S + \Delta$ or $c_i - z_{L, i} \in S_\Delta^\perp + \Delta_H$, respectively.
\end{proof}

\subsection{Noisy Trapdoor Claw-Free Functions}\label{sec:NTCF}

In what follows we define the notion of a  {\em noisy trapdoor claw-free function family}. This notion is simpler than the notion of a {\em dual-mode noisy trapdoor claw-free function family}.\footnote{Our formulation is from~\cite{C:BKLMMV22,GKNV25} (without the dual-mode requirement).}

\begin{definition}\label{def:NTCF}
    A noisy trapdoor claw-free function ($\NTCF$) family is described by $\PPT$ algorithms: $$(\Setup,\Eval,\Invert, \Chek, \Synd)$$ with the following syntax:
 \begin{description}
      \item [$\Setup(1^\lambda)\rightarrow (\pubparams,\td)$.] This $\PPT$ key generation algorithm takes as input a security parameter $\secp$ (in unary) and outputs  public parameters $\pubparams$ and a secret trapdoor $\td$. 
      
      We denote by $\mathsf{D}_{\pubparams}$ the domain of the (randomized) function defined by $\pubparams$, and assume for simplicity that  $\mathsf{D}_{\pubparams}$ is an efficiently verifiable and samplable subset of $\{0,1\}^{n(\secp)}$, where $n(\secp)=\poly$. We denote by $\mathsf{R}_{\pubparams}$ the range of this (randomized) function.
     \item [$\Eval(\pubparams,b,\vecx)\rightarrow \vecy$.] This $\PPT$ algorithm takes as input a public key $\pubparams$, a bit $b\in\{0,1\}$ and an element $\vecx \in \mathsf{D}_{\pubparams}$, and outputs a string $\vecy$ distributed according to some distribution~$\chi=\chi_{\pubparams,b,\vecx}$.  
      \item [$\Invert(\td,\vecy)\rightarrow ((0,\vecx_0),(1,\vecx_1))$.] This deterministic polynomial time algorithm takes as input a trapdoor $\td$, and an element $\vecy$ in the range $\mathsf{R}_\pubparams$ and outputs two pairs $(0,\vecx_0)$ and $(1,\vecx_1)$ with $\vecx_0,\vecx_1\in \mathsf{D}_\pubparams$, or $\bot$. 
      \item [$\Chek(\pubparams, b, \vecx, \vecy)\rightarrow 0/1$.] This deterministic poly-time algorithm takes as input a public key $\pubparams$, a bit $b\in \{0,1\}$, an element $\vecx \in \mathsf{D}_\pubparams$ and an element $\vecy \in \mathsf{R}_\pubparams$ and outputs a bit.
    
      \item [$\Synd(b,\vecx, \vecd)\rightarrow \vecd'$.] This deterministic poly-time algorithm takes as input a bit $b\in\{0,1\}$, a domain element $\vecx\in \mathsf{D}_\pubparams$, and a string $\vecd \in \{0,1\}^{n + 1}$, and outputs a syndrome $\vecd'\in\{0,1\}^k$.

     \begin{remark}\label{remark:synd}
        $\Synd(b,\vecx,\vecd)$ ignores the first bit of $\vecd$, and thus it is also well defined for $\vecd\in\{0,1\}^n$ (where the first bit is omitted).
        \end{remark}
      We define the set $\Good_{b,\vecx}=\{\vecd:~ \Synd(b,\vecx,\vecd)\neq 0^k\}$.
      The predicate $\Good({\vecx_0,\vecx_1},\vecd)=1$ means that $\vecd\in\Good_{0,\vecx_0}\cap\Good_{1,\vecx_1}$.

\end{description}

\noindent We require that the following properties are satisfied. 
  \begin{enumerate}
      \item \textbf{\em Completeness:} 
      \begin{enumerate}
            \item\label{def:TCF-a} For every  $(\pubparams, \td) \in \Supp(\Setup(1^\lambda))$,  $b\in\{0,1\}$,  $\vecx\in\mathsf{D}_\pubparams$, and $\vecy \in \Supp(\Eval(\pubparams,b,\vecx))$,
      $$ \Invert(\td,\vecy) = ((0,\vecx_0),(1,\vecx_1))$$ such that $\vecx_b=\vecx$ and  $\vecy\in\Supp(\Eval(\pubparams,\beta,\vecx_\beta))$ for every $\beta\in\{0,1\}$. 
            \item\label{def:TCF-b} For all $(\pubparams, \td) \in \Supp(\Setup(1^\lambda))$, there exists a perfect matching $\mathsf{M}_\pubparams \subseteq \mathsf{D}_\pubparams \times \mathsf{D}_\pubparams$ such that for all $(\vecx_0, \vecx_1)$, it holds that $\Eval(\pubparams, 0, \vecx_0)  \equiv \Eval(\pubparams, 1, \vecx_1)$ if and only if $(\vecx_0, \vecx_1) \in \mathsf{M}_\pubparams$.
            \item\label{def:TCF-c} For every  $(\pubparams, \td) \in \Supp(\Setup(1^\lambda))$, $b \in \{0,1\}$ and  $\vecx \in \mathsf{D}_\pubparams$,
                  \begin{equation}
                        \Pr[\Chek(\pubparams, b, \vecx, \vecy) = 1] = 1
                  \end{equation}
            if and only if $\vecy \in \Supp(\Eval(\pubparams, b, \vecx))$.

            \item\label{def:TCF-d} For every $(\pubparams, \td) \in \Supp(\Setup(1^\lambda))$ and every $b\in\{0,1\}$ and $\vecx\in \mathsf{D}_\pubparams$,
            \[
            \Pr_{\vecd\gets\{0,1\}^{n+1}}[\vecd\in\Good_{b,\vecx}]=1 - \negl.
            \]
            
      \end{enumerate}

    \item\label{item:range-spos} \textbf{Efficient Range Superposition:} There is an quantum polynomial-time procedure which samples a state with negligible trace distance from
    \[
        \sum_{\substack{\vec{x}\in \mathsf{D}_{\pubparams}\\ \vec{y}\in \mathsf{R}_{\pubparams}}}
        \sqrt{p_{\pubparams}(b, \vec{x}, \vec{y})}
        \ket{\vec{x}} \ket{\vec{y}}
    \]
    where $p_{\pubparams}(b, \vec{x}, \vec{y})$ is the probability density of $\vec{y}$ in the distribution $\Eval(\pubparams, b, \vec{x})$.
        
    \item\label{item:HCB} \textbf{\em Adaptive Hardcore Bit:}  For every $\QPT$ adversary $\A$ there exists a negligible function~$\mu$ such that for every $\secp\in\mathbb{N}$, 
      \begin{align*}
      & \Pr[\A(\pubparams)=(\vecy,b,\vecx,\vecd,v): \Chek({\pubparams, b, \vecx, \vecy}) = 1  ~\wedge~\Good({\vecx_0, \vecx_1},\vecd)=1~\wedge~ \vecd \cdot (1, \vecx_0\oplus \vecx_1)=v] \\
      &\leq \frac12+\mu(\secp),
      \end{align*}
      where the probability is over $(\pubparams,\td)\leftarrow \Setup(1^\secp)$, and where $((0,\vecx_0),(1,\vecx_1))=\Invert(\td,\vecy)$.
  \end{enumerate}
\end{definition}

\begin{lemma}~\cite{BCMVV18}
There exists an $\NTCF$ family assuming the post-quantum hardness of $\LWE$. 
\end{lemma}

In addition, we will need three non-standard properties of the $\NTCF$. The first is a modified collapsing property in the presence of one bit of leakage. 
It holds generically for any $\NTCF$. 
The second is the hiding of a random inner product predicate when an adaptive hardcore bit is also leaked. This holds generically for $\NTCF$s where $\Synd$ is linear in $\vec{d}$.
The third is a variation of \cite{arxiv:KR26}'s phase injection lemma which holds in the presence of one bit of random inner product leakage. This holds generically for any $\NTCF$ satisfying the second property.
See \Cref{app:NTCF} for the complete proofs.

\begin{lemma}[Collapsing with a leaked inner product]
\label{lem:ntcf-collapsing-inner-product}
Let $(\Setup,\Eval,\Invert,\Chek,\Synd)$ be a $\NTCF$ family.
For a $\QPT$ adversary $\mathcal{A}$ and $c\in\{0,1\}$, define the experiment $\CollIP^{\mathcal{A}}_c(1^\lambda)$ as follows:
\begin{enumerate}
    \item The challenger samples $(\pubparams,\td)\gets\Setup(1^\lambda)$ and sends $\pubparams$ to $\mathcal{A}$.

    \item $\mathcal{A}$ sends a classical image $y$ to the challenger. The challenger computes $((0,x_0),(1,x_1))\gets\Invert(\td,y)$.
    If inversion fails, the experiment outputs $0$ and terminates.

    \item The challenger samples $d\gets\{0,1\}^{n+1}$ and defines $\Delta=(0,x_0)\oplus(1,x_1)$. Then, it sends $(d,d\cdot \Delta)$ to $\mathcal{A}$.

    \item $\mathcal{A}$ sends a $(n+1)$-qubit register $\mathcal{X}$ to the challenger, retaining its private registers. The challenger performs the two-outcome projective measurement $\{\Pi_y,I-\Pi_y\}$ on $\mathcal{X}$, where
    \[
        \Pi_y = \sum_{\substack{b\in\{0,1\},\,x\\ \Chek(\pubparams,b,x,y)=1}} \ket{b,x}\bra{b,x}.
    \]
    If the outcome is $I-\Pi_y$, the experiment outputs $0$ and terminates.

    \item If $c=1$, the challenger measures $\mathcal{X}$ in the computational basis and discards the outcome. If $c=0$, it leaves $\mathcal{X}$ unchanged. It then returns $\mathcal{X}$ to $\mathcal{A}$.

    \item $\mathcal{A}$ outputs a bit, which is the output of the experiment.
\end{enumerate}
Then, for every $\QPT$ adversary $\mathcal{A}$,
\[
    \left| \Pr\!\left[\CollIP^{\mathcal{A}}_0(1^\lambda)=1\right] - \Pr\!\left[\CollIP^{\mathcal{A}}_1(1^\lambda)=1\right] \right| \leq \negl[\secpar].
\]
\end{lemma}

\begin{definition}[$2$-query adaptive hardcore bit]\label{def:two-query-ahc}
    A $\NTCF$ family is said to satisfy the $2$-query adaptive hardcore bit property if every $\QPT$ adversary $\calA$ wins in the following game with probability $\frac12+\negl$.
    \begin{enumerate}
        \item The challenger samples $(\pubparams,\td)\gets\Setup(1^\secpar)$ and sends $\pubparams$ to $\calA$.
        \item $\calA$ sends a classical tuple $(b,\vecx,\vecy)$ to the challenger.  It may retain arbitrary quantum side information.
        \item The challenger checks that $\mathsf{Check}(\pubparams,b,\vecx,\vecy)=1$. If this is not the case then the game is terminated with the default guess $c'=0$.  Otherwise, the challenger does the following:
        \begin{enumerate}
            \item Compute
        $((0,\vecx_0),(1,\vecx_1))=\Invert(\td,\vecy)$.
        \item Sample uniformly at random $c\gets\{0,1\}$ and $\vecd\gets\{0,1\}^{n+1}$.
        \item Compute $p=\vecd\cdot (1,\vecx_0\oplus\vecx_1)\oplus c$.
        \item Send $(\vecd, p)$.
        \end{enumerate}
        \item $\calA$ sends a classical vector $\vece\in\{0,1\}^{n}$.
        \item The challenger checks that $\Synd(b,\vecx_b,\vece)\notin\{\vec0^k,\Synd(b,\vecx_b,\vecd)\}$ for both $b\in\{0,1\}$. If either check fails then the game is terminated with a default guess $c'=0$.  Otherwise, the challenger sends $\beta=\vece\cdot (\vecx_0\oplus \vecx_1)$ to $\calA$.
        \item $\calA$ outputs a bit $c'\in\{0,1\}$.
    \end{enumerate}
    $\calA$ wins if $c'=c$.
\end{definition}

\begin{lemma}\label{lem:2Q-NTCF}
Fix any $\NTCF$ family (satisfying the adaptive hardcore-bit
property) for which $\Synd$ is linear in~$\vecd$; i.e, the map
\[
\vecd\mapsto \Synd(b,\vecx,\vecd)
\]
is linear over $\mathbb{F}_2$. Then this $\NTCF$ family satisfies the two-query adaptive
hardcore-bit property \Cref{def:two-query-ahc}.
\end{lemma}
We note in the $\NTCF$ construction from \cite{BCMVV18}, the function $\Synd$ is indeed linear in $\vecd$.
Thus, by \Cref{lem:2Q-NTCF} this construction also satisfies the 2-query adaptive hardcore bit property, assuming the post-quantum hardness of $\LWE$.

\begin{lemma}[Phase Injection with 1 Bit Leakage]\label{lem:phase-injection-leakage}
    Let $\calF$ be an NTCF with adaptive hardcore bits and linear $\Synd$. Then every  $\QPT$ adversary $\adv$ wins in the following game with probability at most $1/2+\negl$.
    \begin{enumerate}
        \item \textbf{Challenger:} Sample $(\pubparams, \td) \gets \Setup(1^\secpar)$ and sample a random bit $p\gets\{0,1\}$. Send $\pubparams$ to $\adv$.
        \item $\adv$ sends back a classical value $y$ and a register $\calR$ (which should store a preimage of $y$).
        \item \textbf{Challenger:} Perform the projective measurement $\{\Pi_y, I - \Pi_y\}$ on $\calR$ where $\Pi_y$ projects onto $(b, r_b)$ such that $\Chek(\pubparams,b,r_b,y)=1$. If the result is $I - \Pi_y$, then $\calA$ needs to guess $p$ (without being given any information about~$p$), and the game is over.
        \item \textbf{Challenger:} Otherwise, if $p=1$ then apply a $Z$ gate to the first qubit of $\calR$.  
        
        Send $\calR$ back to $\adv$.
        \label{step:phase-inj-phase}
        \item \textbf{Challenger (Leakage):} Compute $((0,r_0),(1,r_1))\gets\Invert(\td,y)$. Set $\Delta = (0, r_0) \oplus (1, r_1)$. Then sample $\vecd\gets\{0,1\}^{n+1}$ and send $(\vecd, \vecd \cdot \Delta)$ to the adversary.
        \item $\adv$ sends a string $\vece$ to the challenger. \label{step:phase-inj-get-d}
        \item \textbf{Challenger:} If $\Synd(b,r_b,\vece)\notin\{\vec0^k,\Synd(b,r_b,\vecd)\}$ for both $b\in\{0,1\}$, send $(\vece \cdot \Delta) \oplus p$ to $\adv$. Otherwise send a random bit $b'$.
    
        \label{step:phase-inj-inner-product}
        \item $\adv$ outputs a bit $p'$. They win if and only if $p'=p$.
    \end{enumerate}
\end{lemma}

\subsection{Witness-Preserving Arguments for NP} \label{sec:witness-pres-np}

\cite{KKR26} showed how a quantum prover holding a superposition $\sum_w \alpha_w \ket{w}$ of NP witnesses for $x\in \lang$ can convince a classical verifier that $x\in \lang$ \emph{without disturbing} their witness superposition. We recall the notion and their result in this section. Then, in the next section we observe that techniques from \cite{FOCS:LomMaSpo22} can make the protocol ``indistinguishably-extractable''; that is, the prover's view can be indistinguishably simulated in such a way that the witness is temporarily extracted, then returned to the prover to obtain the final view.

\begin{definition}\cite{KKR26}\label{def:state-preserving-NP}
    A \emph{witness-preserving argument} for an $\NP$ language $\lang$ is an interactive protocol between a $\QPT$ prover $P$ and a ${\sf PPT}$ verifier $V$ that satisfies the following guarantees.

    \begin{itemize}
        \item \textbf{Completeness and Witness Preservation.} 
        
        For every $x\in \lang$ and every superposition $\ket{\psi} = \sum_{w} \alpha_{w} \ket{w}_{\calW}\otimes \ket{\phi_w}_{\calB}$ over witnesses for $x$ (encoded in the computational basis and potentially entangled with an external state $\ket{\phi_w}$), there exist negligible functions $\mu_1$ and $\mu_2$ such that for every $\secpar\in\mathbb{N}$,
        \[
\ifsubmission
            \Pr\left[\begin{array}{l}
                \left(\frac{1}{2}\big\|\ketbra{\psi} - \rho \big\|_1 \leq \mu_1(\secpar)\right)
                ~~\land~~ \left(b = \Accept\right):\\
                (\rho, b) \gets \langle P(\ket{\psi}), V\rangle(1^\secpar, x)
            \end{array}\right] \geq 1- \mu_2(\secpar),
\else
            \Pr\left[
                \left(\frac{1}{2}\big\|\ketbra{\psi} - \rho \big\|_1 \leq \mu_1(\secpar)\right)
                ~~\land~~ \left(b = \Accept\right):~
                (\rho, b) \gets \langle P(\ket{\psi}), V\rangle(1^\secpar, x)
            \right] \geq 1- \mu_2(\secpar),
\fi
        \]
        where $P$ acts only on register $\calW$ and where the notation \ifsubmission\[(\rho, b) \gets \langle P(\ket{\psi}), V\rangle(1^\secpar, x)\]\else$(\rho, b) \gets \langle P(\ket{\psi}), V\rangle(1^\secpar, x)$\fi{} means that $V$ outputs $b\in\{\Accept, \Reject\}$ and $P$'s residual state is $\rho$.

        When we wish to be more precise about the repair guarantees, we say that the argument is $(1-\mu_1)$-witness-preserving.
       
        \item \textbf{Computational Soundness.} 
        For every polynomial $\ell=\ell(\secpar)$, there exists a negligible function $\mu$ such that for every $\secpar \in \bbN$, every $x\notin \lang$ of size $\leq \ell(\secpar)$ and every $\QPT$ cheating prover $P^*$ with auxiliary input $\ket{\psi}$ consisting of at most ${\sf poly}(\ell(\secpar))$ qubits
        \[
            \Pr[\langle P^*(\ket{\psi}, V\rangle(1^\secpar, x) = \Accept] \leq \mu(\secpar)
        \]

    \end{itemize}
\end{definition}

\begin{theorem}\label{thm:witness-preserving-NP}\cite{KKR26}
    There exists a witness-preserving argument for $\NP$, assuming the existence of dual-mode claw-free trapdoor function family with state recovery (which in turn can be constructed from the post-quantum hardness of {\sf LWE}~\cite{BCMVV18}).
\end{theorem}

\subsection{Quantum Extraction for Arguments of Classical Knowledge}
\label{sec:prelim:extraction}

We next define the notion of indistinguishable-extraction, which is a variation on \cite{FOCS:LomMaSpo22}'s ``state-preserving extraction''.\footnote{We use the ``indistinguishably-extractable'' to differentiate from \cite{KKR26}'s ``witness-preserving'' property which refers to preserving the prover's state in a \emph{real} execution.} State-preserving extraction allows generating an indistinguishable transcript and adversarial state while simultaneously extracting a witness. However, obtaining the extracted witness and transcript simultaneously necessarily means collapsing the prover's witness, which is immediately distinguishable in a witness-preserving argument.

In indistinguishable-extraction, we deliberately avoid collapsing the witness in this way. Instead, we build an interface which temporarily exposes a coherent witness, then ``returns'' the witness to the adversary without collapsing it. The resulting interface is quite similar to \cite{FOCS:LomMaSpo22}'s guaranteed extraction: there is a unitary extractor $\Ext$ which uses a superposition over accepting transcripts to extract a valid witness with overwhelming probability. The difference from guaranteed extraction is that we require $\Ext$ to be \emph{unitary}. Thus, it may be inverted afterwards, returning the witness.

In what follows, we denote by  
\[
{\sf View}_{P^*}\langle {P^*}(\ket{\psi}, V\rangle(1^\secpar, x)
\]
the view of an adversarial prover $P^*$ after interacting in the protocol \ifsubmission\[\langle {P^*}(\ket{\psi}), V\rangle(1^\secpar, x),\]\else$\langle {P^*}(\ket{\psi}), V\rangle(1^\secpar, x)$,\fi{} which includes the transcript in register $\calT$ as well as the residual state of $P^*$ in register $\calA$.

\begin{definition}[Indistinguishably-Extractable Arguments]\label{def:indist-ext}
    An NP argument is said to be \emph{indistinguishably-extractable} if there exist an expected QPT algorithm $\Sim$ and uniform family of $\poly[\secpar]$-sized unitaries $\{\Ext_\secpar\}_{\secpar\in \bbN}$ satisfying the following for every sufficiently large security parameter $\secpar\in \bbN$, every adversarial QPT prover $P^*$ and every $\NP$ statement $x$:
    \begin{itemize}
        \item \textbf{Syntax.} $\Sim(P^*)$ outputs three registers $(\calT, \calA, \calS)$ and a decision bit $b$ denoting whether $\calT$ contains an accepting transcript. $\calT$ contains a superposition over classical transcripts $T$, $\calA$ contains the adversary's state, and $\calS$ is a work register for the simulator.\footnote{$\Sim$ does not receive restricted information like in a zero-knowledge argument. For example, $\Sim$ could simply run the argument coherently and measure whether the verifier would accept. We also allow other strategies for generating transcripts.}

        $\Ext$ operates on registers $(\calT, \calA, \calS)$ and reorganizes them into registers $(\calW, \calE)$. $\calW$ contains the extracted witness and $\calE$ is a work register.
        
        \item \textbf{Simulation.} $\Sim$ produces indistinguishable adversarial views given oracle access to the prover $P^*$:
        \[
            \Tr_{\calS}\left[\ket{\phi}_{\calT,\calA, \calS}\right] \approx {\sf View}_{P^*}\langle {P^*}, V\rangle(1^\secpar, x)
        \]
        where $(\ket{\phi}_{\calT, \calA, \calS}, b)\gets \Sim^{P^*} (1^\secpar,x,\ket{\psi})$. If the indistinguishability is computational (resp. statistical), we say the argument is computationally (resp. statistically) indistinguishably-extractable.
        
        \item \textbf{Extraction.} Let $\Pi_{\Ext, \secpar} = \Ext_\secpar^\dagger \Pi_\Ver \Ext_{\secpar}$ be the projector onto $\Ext_\secpar$ successfully extracting an accepting witness for $x$ to register $\calW$.
        Consider the following experiment $\IndExt_{\secpar}$:
        \begin{enumerate}
            \item Run $(\ket{\phi}_{\calT, \calA, \calS}, b_{\Sim}) \gets \Sim^{P^*}(1^\secpar, x)$.
            \item If $b_\Sim = \Accept$: measure $\ket{\phi}_{\calT, \calA, \calS}$ according to $\{\Pi_{\Ext, \secpar}, I - \Pi_{\Ext, \secpar}\}$. Call the measurement result (representing successful extraction) $b_\Ext$.
        \end{enumerate}
        Then the probability of generating an accepting transcript but failing to extract a valid witness is negligible:
        \[
            \Pr_{\IndExt_{\secpar}}[b_\Sim = \Accept \land b_\Ext = \Reject] = \negl
        \]
    \end{itemize}
\end{definition}

It will also be useful to define an approximate version, where the extractor is allowed to take $\poly[1/\epsilon]$ time to extract a witness with probability $1-\epsilon$, for every $\epsilon$.

\begin{definition}[$\epsilon$-Indistinguishable-Extractability]\label{def:eps-indist-ext}
    We say an argument is $\epsilon$-Indistinguishably-Extractable if 
    there exists an expected QPT $\Sim$ and uniform family of $\poly[\secpar, n]$-sized unitaries $\{\{\Ext_{\lambda, n}\}_{n\in \bbN}\}_{\secpar\in \bbN}$ such that
    \begin{itemize}
        \item \textbf{Simulation.} $\Sim^{P^*}$ produces indistinguishable adversarial views.
        \item \textbf{$\epsilon$-Extraction.} Let $\IndExt_{\secpar, n}$ the be experiment from \Cref{def:indist-ext} using $\Ext_{\secpar, n}$. For all sufficiently large $\secpar \in \bbN$ and all $n = \poly[\secpar]$, the probability of a successful transcript but unsuccessful extraction is at most $1/n$:
        \[
            \Pr_{\IndExt_{\secpar, \epsilon(n)}}[b_\Sim = \Accept \land b_\Ext = \Reject] \leq 1/n
        \]
    \end{itemize}
\end{definition}

The almost-as-good-as-new lemma~\cite{CCC:Aar04} (closely related to the gentle measurement lemma~\cite{TIT:Win99}) limits the damage that indistinguishable extraction can do to a simulated transcript.

\begin{lemma}\label{lem:iext-close}
    For all adversarial provers $P^*$,
    \[
        \frac{1}{2}\left\| \Sim(P^*) - \IndExt_{\secpar, n} \right\|_1 \leq \sqrt{1/n}
    \]
    Combining this with the indistinguishability of simulation,
    \[
        \Tr_{\calS}\left[\IndExt_{\secpar, n^2}\right] 
        \overset{1/n}{\approx}_c 
        {\sf View}_{P^*}\langle {P^*}(\ket{\psi}), V\rangle(1^\secpar, x)
    \]
    Moreover, if $\IndExt_{\secpar, n^2}$ behaves arbitrarily in the case of $b_\Sim = \Accept \land b_\Ext = \Reject$, the distance is at most $1/n + 1/n^2$.
\end{lemma}

Lombardi, Ma, and Spooner show a general class of protocols which have guaranteed extraction (i.e. $\Ext$ is allowed to be an expected QPT algorithm instead of a unitary).
Although the exact conditions are somewhat complex, they specifically list \cite{FOCS:GolMicWig86}'s 3-coloring protocol as satisfying these properties, with a small tweak: commit to the witness first, then run a GMW argument which is amplified in parallel~\cite[Section 5.3 and Corollary 8.1]{FOCS:LomMaSpo22}. 
Moreover, Lombardi, Ma, and Spooner show that there exists series of unitary dilations $\{U_{\epsilon(n)}\}_{n\in \bbN}$ which $\epsilon$-approximate the guaranteed extractor~\cite[Claim 9.4]{FOCS:LomMaSpo22}.\footnote{We briefly mention that Lai, Spooner, and Tromanhauser~\cite{TCC:LaiSpoTro26} recently pointed out a bug in \cite{FOCS:LomMaSpo22}'s definition of coherent-runtime simulation. However, the other results are intact, including guaranteed extraction and the specific unitary dilation used to instantiate the faulty definition.}

\begin{theorem}
    For every (computationally or statistically) collapsing commitment $\Com$, the following protocol is statistically $\epsilon$-indistinguishably-extractable.
    \begin{enumerate}
        \item Commit to the witness $w$ using $\Com$.
        \item Prove knowledge of an opening to a witness $w$ using \cite{FOCS:GolMicWig86}'s 3-coloring protocol, instantiated with $\Com$.
    \end{enumerate}
\end{theorem}

Finally, we observe that this theorem can be combined with \cite{KKR26}'s construction of \emph{witness}-preserving arguments. Their construction is a parallel repetition of \cite{FOCS:GolMicWig86}'s 3-coloring protocol using a specially chosen commitment which admits future recovery of the committed superposition, followed by a recovery procedure after the verifier makes its decision. An additional commitment and recovery step can be added at the start and end of their protocol. 
The following corollary is obtained by applying the above theorem to the portion of the protocol before recovery and observing that the recovery step afterwards cannot increase the \emph{statistical} distance.\footnote{If the simulation indistinguishability were computational, it is possible that revealing additional information during recovery allows breaking the indistinguishability. However, the recovery information is information-theoretically determined prior to the recovery step and so revealing it does not increase the statistical distance.}

\begin{corollary}\label{coro:wp-np-eps-iext}
    There exists a  -witness-preserving argument for $\NP$ with approximate indistinguishable-extraction, assuming the existence of dual-mode claw-free trapdoor function family with state recovery (which in turn can be constructed from the post-quantum hardness of {\sf LWE}~\cite{BCMVV18}).
\end{corollary}

\subsection{Semi-Quantum Secure 2-Party Computation}
\label{sec:prelim:2pc}
\label{sec:def-2PC}

In this section, we define 2PC for general quantum circuits $G(\ket{\psi}, w)$ between a classical party $P_C$ and a quantum party $P_Q$. $P_Q$ provides the quantum input $\ket{\psi}$ and $P_C$ provides the classical input $w$. 
Without loss of generality, $P_Q$ outputs the coherent output register $\calB$ and $P_C$ outputs the measurement result of $\calC$.
Recall that $G$ also discards the subregister $\calD$ resulting from its evaluation.

\begin{definition}\label{def:HBC-2PC}
    Let $G$ be a general quantum circuit.
    We say that a 2-party protocol $(P_\Q,P_\C)$, where $P_\Q$ is a $\QPT$ party and $P_\C$ is a $\PPT$ (classical) party, securely computes $G$ if the following two conditions hold: 
    \begin{itemize}
        \item {\bf Correctness:}  For every quantum state $\ket{\psi}$ and every classical string $w$, the output distribution of $(P_\Q(\ket{\psi}),P_\C(w))$ has negligible trace distance to $G(\ket{\psi}, w)$.

        \item {\bf Security against $P_Q$:}  
        For every (possibly malicious) $\QPT$ party $P^*_\Q$  there exists a $\QPT$ simulator $S_\Q$ such that for every $(\ket{\psi},w)$, 
        \[
        \Ideal_{S_\Q}(g, \ket{\psi},w)\approx \Real_{P^*_\Q}(g, \ket{\psi},w)\]
        Here, $\Real_{P^*_\Q}(g, \ket{\psi},w)$ is the following distribution:
        \begin{enumerate}
            \item The outputs of $P^*_\Q$ and $P_\C$ after interacting in the $2\PC$ protocol, where $P_\C$ simply outputs its output string $y$, or $\bot$ if the protocol was prematurely terminated, and $P^*_\Q$ outputs its entire view.
        \end{enumerate}
        $\Ideal_{S_\Q}(g, \ket{\psi},w)$ is the following distribution:
        \begin{enumerate}
            \item The classical party $P_\C$ sends its input $w$ to the ideal functionality, and $S_\Q(\ket{\psi})$ sends an arbitrary register $\calS$.
            \item The ideal functionality applies the isometry $G_w=G(\cdot, w)$, obtaining a coherent output register $\calB$ and measurement result $y$. Then, it sends $\calB$ to the simulator.
            \item The simulator decides whether to abort the functionality or not.
            \item If not, the ideal functionality sends $y$ to $P_C$.
        \end{enumerate}

    \item \textbf{Security Against $P_C$:}
    For every (possibly malicious) $\QPT$ party $P^*_\C$ there exists a $\QPT$ simulator $S_\C$ such that for every input pair $(\ket{\psi},w)$,
        \[
        \Ideal_{S_\C}(g,\ket{\psi},w)\approx \Real_{P^*_\C}(g,\ket{\psi},w).\]
        Here $\Real_{P^*_\C}(g, \ket{\psi},w)$ is the following distribution:
        \begin{enumerate}
            \item The outputs of $P^*_\C$ and $P_\Q$ after interacting in the $2\PC$ protocol, where $P_\Q$ simply outputs its residual state $\ket{\psi_y}$, or $\bot$ if the protocol was prematurely terminated, and $P^*_\C$ outputs its entire view.
        \end{enumerate}
        $\Ideal_{S_\C}(g, \ket{\psi},w)$ is the following distribution:
        \begin{enumerate}
            \item $P_\Q$ sends its input state  $\ket{\psi}$ to the ideal functionality, and $S_\C(w)$ sends an arbitrary input $w'$.
            \item The ideal functionality computes $G(\ket{\psi}, w)$, obtaining a coherent output register $\calB$ and measurement result $y$. Then, it sends $y$ to the simulator.
            \item The simulator decides whether to abort the functionality or not.
            \item If not, the ideal functionality sends register $\calB$ to $P_Q$.
        \end{enumerate}
\end{itemize}
\end{definition}

\begin{definition}
    We say a 2-party protocol $\epsilon$-securely computes $g$ if correctness holds and if there exist uniform $\QPT$ oracle circuits $\{S_{Q, \secpar, n}\}_{\secpar, n\in \bbN}$ and $\{S_{C, \secpar, n}\}_{\secpar, n\in \bbN}$ such that for all $\QPT$ adversaries $P_\C^*$ or $P_\Q^*$, all sufficiently large $\secpar$ and all $n = \poly[\secpar]$,
    \begin{align*}
        \Ideal_{S_\C}(g,\ket{\psi},w, n) &\overset{1/n}{\approx} \Real_{P^*_\C}(g,\ket{\psi},w)
        \\
        \Ideal_{S_\Q}(g, \ket{\psi},w, n) &\overset{1/n}\approx \Real_{P^*_\Q}(g, \ket{\psi},w)
    \end{align*}
    where the ideal experiments are defined as above using $S^{P_\Q^*}_{Q, \secpar, n}$ or $S^{P_\C^*}_{C, \secpar, n}$ instead of $S_Q$ or $S_C$, respectively.
\end{definition}

\paragraph{Honest-But-Curious $P_C$ Adversaries.} 
By default, we consider any adversary who controls the quantum party $P_Q$ to be ``malicious'', meaning it may deviate from the protocol arbitrarily. 
For the classical party $P_C$, however, we consider adversaries controlling them to be ``honest-but-curious''. These adversaries behave honestly during the protocol, but attempt to learn additional information from the transcript. They output their entire view at the end of the protocol instead of just the prescribed output. 

It is well-known that honest-but-curious security can be generically upgraded to malicious security \emph{for parties whose role during the protocol consists only of classical computation}. In particular, this consists of applying the GMW compilation to the classical party, using quantum-secure classical subprotocols.~\cite{STOC:GolMicWig87,C:HalSmiSon11}.
This is done by committing to $P_C$'s input at the start of the protocol, and generating committed randomness for the rest of the protocol by coin-tossing-in-the-well. Then the party proves honest computation of each message with respect to the committed input and randomness, in zero-knowledge. Since $P_C$ only does honest computation during the protocol, the statements to be proved are in NP, so standard zero-knowledge for NP suffices.

\part{Input-Preserving 2PC for Classical Functions}
\section{Input-Preserving 2-$\PC$ for Classical Constant-Degree Functions with Post-Processed Abort}
\label{sec:classical-2PC}
In this section we construct our input-preserving 2-$\PC$ for classical constant-degree functions, where the classical party can decide if the output is delivered adaptively, based on the function output. We start by focusing on linear functionalities.  The extension to constant-degree functionalities is pretty straightforward, and is described in \Cref{sec:const-deg}.

 Throughout this section all vector spaces, inner products, and sums are over
$\mathbb F_2$.  
We consider the functionality where the quantum party $P_{\Q}$ takes as input an arbitrary quantum state $\sum_{\vecv\in\{0,1\}^m} \alpha_\vecv \ket{\vecv}$ and the classical party $P_{\C}$ takes as input a linear function $L\in \{0,1\}^{\ell\times m}$ and a poly-size circuit $C:\{0,1\}^\ell\rightarrow\{0,1\}$ which determines if the output should be delivered or not.  Their goal is to securely implement the following quantum circuit $G$:

\paragraph{Ideal Linear Function with Post-Processed Abort.} 
$G(\ket{\psi}, (L, C))$ uses the following inputs. 
$P_Q$ provides a quantum state $\sum_{\vecv\in\{0,1\}^m} \alpha_\vecv \ket{\vecv}$.
$P_C$ provides a linear function $L\in\{0,1\}^{\ell\times m}$ and a poly-size classical circuit $C:\{0,1\}^\ell\rightarrow\{0,1\}$.
Then, $G$ works as follows:
\begin{enumerate}
    \item Compute 
    \[
        \sum_{\vecv\in\{0,1\}^m} \alpha_\vecv \ket{\vecv}\ket{L\vecv}
    \]
    \item Measure the last register to obtain $\vecu\in\{0,1\}^\ell$. Keep the residual state $\ket{\varphi}$, where 
    \[
        \ket{\varphi}\propto \sum_{\vecv\in\{0,1\}^m:~L\vecv=\vecu} \alpha_\vecv \ket{\vecv}
    \]
    \item If $C(\vec{u}) = 0$ then place $\ket{\varphi}$ in register $\calB$. Otherwise discard $\ket{\varphi}$.
    \item Output register $\calB$ (which is sent to $P_Q$) and $\vec{u}$ (which is sent to $P_C$).

\end{enumerate}
Recall that the ideal functionality for $G$ also permits the adversary to abort the computation after seeing its output. We slightly modify the ideal functionality so that $P_C$ always gets its input \emph{before} $P_Q$ does. In particular, an adversarial $P_Q$ does not get to see its own output before deciding whether to abort.

\begin{theorem}\label{thm:classical2PC}
    There exist a $\epsilon$-simulateable input-preserving $2\PC$ for classical linear functions with post-processed abort under the post-quantum hardness of $\LWE$. 

    If there additionally exists an indistinguishable extractable witness-preserving argument for NP, then the $\twoPC$ has negligible-loss security instead.
\end{theorem}

\subsection{Construction}\label{sec:classical-protocol}

\subsubsection{Setup}

\paragraph{Cryptographic primitives}  Our protocol makes use of the following primitives:
\begin{enumerate}
    \item Noisy trapdoor claw-free function family with the adaptive 2-query hardcore bit property, as defined in \Cref{sec:NTCF}. We denote this family by 
    \[
    (\Setup,\Eval,\Invert,\Chek, \Synd),
    \]
    \item Witness-preserving indistinguishably-extractable argument for $\NP$, as defined in \Cref{sec:witness-pres-np}. 

\end{enumerate}
We note that both these primitives can be instantiated from the $\LWE$ assumption.

\paragraph{Encoding Matrices.}
We will use the following matrices to encode the inputs of each party. 
 Fix a logical input length $m\geq1$ and let $N:=m+\binom{m}{2}$ be the encoded input length. Let $K_m$ be the $\binom{m}{2}\times m$ edge--vertex incidence matrix of the complete graph on $m$ vertices. 
 Its rows are all possible combinations of 2 standard basis vectors $e_j + e_k$.

Define the generator matrix $G$, decoder matrix $D$, and parity-check matrix $P$ by
\begin{equation}
\label{eq:encoding-matrices}
\begin{aligned}
G&:=\begin{pmatrix}I_m\\ K_m\end{pmatrix}
&&\in\mathbb F_2^{N\times m},\\
D&:=\begin{pmatrix}I_m & 0_{m\times(N-m)}\end{pmatrix}
&&\in\mathbb F_2^{m\times N},\\
P&:=\begin{pmatrix}K_m & I_{N-m}\end{pmatrix}
&&\in\mathbb F_2^{(N-m)\times N}.
\end{aligned}
\end{equation}
These matrices satisfy
\[
DG=I_m,\qquad PG=0,\qquad \operatorname{rank}(P)=N-m.
\]
We will encode the inputs of each party as
\begin{equation}
\begin{aligned}
    L' &= LD + RP
    \\
    \vec{v'} &= G\vecv
\end{aligned}
\end{equation}
where $R$ is a random $\ell\times (N-m)$ matrix.
For a column vector $\vec{v}\in\mathbb F_2^m$, the encoding $\vec{v}'=G\vec{v}$ consists of the $m$ original coordinates followed by all $\binom{m}{2}$ pairwise parities $v_j+v_k$.
Therefore
\[
    L'\vec{v'} = L(DG)\vecv + R(PG) = L\vecv
\]

\subsubsection{Protocol}

\begin{protocol}[Input-Preserving 2-$\PC$ for the linear functionality]
\label{prot:classical-linear}
The quantum party $P_\Q$ takes as input a register $V$ containing an arbitrary state
\[
    \ket{\psi}
    =
    \sum_{\vecv\in\{0,1\}^m}
    \alpha_\vecv\ket{\vecv}_V
\]
The classical party $P_\C$ takes as input a linear function
$L\in\{0,1\}^{\ell\times m}$ and a poly-size circuit $C:\{0,1\}^\ell\rightarrow\{0,1\}$.
For every $(i,j)\in[\ell]\times[m]$ let $L_{i,j}\in\{0,1\}$ denote the $(i,j)$ entry of $L$ and let $v_j$ denote the $j$th entry of $\vecv$.
\paragraph{Phase $0$: Input Encoding}  Denote $\vecv'=G\vecv$. The quantum party $P_\Q$ encodes its state to be the state
\[
    \ket{\psi'}=\sum_{\vecv\in\{0,1\}^m}\alpha_\vecv\ket{\vecv, G\vecv}.
\]
The classical party $P_\C$ samples a random matrix $R \gets \bbF_2^{\ell \times (N - m)}$ and sets
\[
    L' = LD + RP.
\]
\paragraph{Phase 1: entry-wise $\NTCF$ commitments.}
\begin{enumerate}
    \item $P_\C$ samples for every $(i,j)\in[\ell]\times[N]$, 
\[
    (\pubparams_{i,j},\td_{i,j})
    \leftarrow
    \NTCF.\Setup(1^\lambda),
\]
and sends to $P_\Q$ the public parameters
\[
    \pubparams:=\{\pubparams_{i,j}\}_{i\in[\ell],\,j\in[N]}
\]
\item
$P_\Q$ parses $\vecv' = (v'_1, \dots, v'_N) \in \bbF_2^N$ and does the following:
\begin{enumerate}
    \item Using the efficient range superposition preparation property of the NTCF, add randomness registers for every $(i,j)$ and coherently evaluate $\Eval(   \pubparams_{i,j},v'_j,\vecx_{i,j})$ using the same input bit $v_j'$ for every $i\in[\ell]$, to obtain the state
    \[
        \sum_{\vecv\in\{0,1\}^m}
        \alpha_\vecv\ket{\vecv,\vecv'}_{\calW}
        \bigotimes_{i=1}^\ell\bigotimes_{j=1}^N
        \sum_{\vecx_{i,j}\in\mathbb{F}_2^n}
        \sqrt{p_{\pubparams}(v'_{j}, \vec{x}_{i,j}, \vec{y}_{i,j})} \ket{\vecx_{i,j}}_{\calR_{i,j}}\ket{\vec{y}_{i,j}}.
    \]
    where $p_{\pubparams}(v'_{j}, \vec{x}_{i,j}, \vec{y}_{i,j})$ is the probability density of $\vec{y}$ in the distribution $\Eval(\pubparams_{i,j}, v'_{j}, \vec{x}_{i,j})$.
\item Measure the image registers to obtain $\vecy=(\vecy_{i,j})_{(i,j)\in[\ell]\times[N]}$, and send $\vecy$
to $P_\C$.
\end{enumerate}
The honest residual state is of the form
\[
    \sum_{\vecv\in\{0,1\}^m}
    \alpha_\vecv\ket{\vecv,\vecv'}_{\calW}
    \bigotimes_{i=1}^\ell\bigotimes_{j=1}^N
    \ket{\vecx_{i,j,v'_j}}_{R_{i,j}},
\]
where
\[
    \mathsf{Check}(\pubparams_{i,j},v'_j,\vecx_{i,j,v'_j},\vecy_{i,j})=1.
\]
\item \label{item:sim-ext1} 
The parties then run a witness-preserving indistinguishably-extractable
argument for the relation
\[
\mathcal R_{\mathsf{com}}
=
\left\{
\left(
(\pubparams,\vecy),
(\vecv, \vecv',\{\vecx_{i,j}\}_{(i,j)\in[\ell]\times[N]})
\right):~
\begin{array}{c}
    \vecv' = G\vecv \\
    \land\:\forall i,j,\; \Chek(\pubparams_{i,j}, v'_{j}, \vecx_{i,j}, \vecy_{i,j}) = 1
    \end{array}
\right\}.
\]
If the verifier rejects, $P_\C$ aborts.
\end{enumerate}
 
\paragraph{Phase 2: Masking $L'$.}
$P_\C$ does the following for every $(i,j)\in[\ell]\times [N]$: 
\begin{enumerate}
    \item  Compute $\Invert(\td_{i,j},\vecy_{i,j})$, if the output is $\bot$ then abort.  Otherwise the output is of the form  $((0,\vecx_{i,j,0}),(1,\vecx_{i,j,1}))$, in which case define
\[
    \Delta_{i,j}:=(1,\vecx_{i,j,0}\oplus \vecx_{i,j,1}).
\]
\item 
Sample
$\vecd_{i,j}\leftarrow \{0,1\}^{n+1}$
and compute
\[
    p_{i,j}:=
    (\vecd_{i,j}\cdot\Delta_{i,j}) \oplus L'_{i,j}.
\]
\end{enumerate}
$P_\C$ sends $(\vecd,\vecp)$ to $P_\Q$ where
\[
    \vecd=(\vecd_{i,j})_{(i,j)\in[\ell]\times[N]}~~\mbox{ and }~~\vecp=(p_{i,j})_{(i,j)\in[\ell]\times[N]}
\]

\paragraph{Phase 3: measuring the hidden linear function.}

\begin{enumerate}
    \item 
$P_\Q$ coherently computes into a fresh register for every $i\in[\ell]$ 
\[
    t_i
    =
    \bigoplus_{j=1}^N
    \left(
        (\vecd_{i,j}\cdot (v'_j,\vecx_{i,j,v'_j}))
        \oplus
        p_{i,j} v'_j
    \right)
\]
It measures the resulting register
$\vect=(t_1,\ldots,t_\ell)\in\{0,1\}^\ell$,
and sends $\vect$ to $P_\C$.

Note that 
\begin{align*}
  \vect &=\left(\bigoplus_{j=1}^N
    \left(
        (\vecd_{i,j}\cdot (v'_j,\vecx_{i,j,v'_j}))
        \oplus
        p_{i,j} v'_j
    \right)\right)_{i\in[\ell]}\\
    &= \left(\bigoplus_{j=1}^N
        \vecd_{i,j}\cdot (v'_j,\vecx_{i,j,v'_j})\right)_{i\in[\ell]}\oplus \left(\bigoplus_{j=1}^N
        v'_j\cdot  ((\vecd_{i,j}\cdot\Delta_{i,j})\oplus L'_{i,j})\right)_{i\in[\ell]}\\
        &= \left(\bigoplus_{j=1}^N
        \vecd_{i,j}\cdot (v'_j,\vecx_{i,j,v'_j})\right)_{i\in[\ell]} \oplus \left(\bigoplus_{j=1}^N v'_j(\vecd_{i,j}\cdot\Delta_{i,j})\right)_{i\in[\ell]} \oplus \left(\bigoplus_{j=1}^N
        v'_j\cdot  L'_{i,j}\right)_{i\in[\ell]}\\
        &= \left(\bigoplus_{j=1}^N
        \vecd_{i,j}\cdot ((v'_j,\vecx_{i,j,v'_j})\oplus v'_j\Delta_{i,j})\right)_{i\in[\ell]}\oplus L'\vecv'\\
        &= \left(\bigoplus_{j=1}^N
        \vecd_{i,j}\cdot (0,\vecx_{i,j,0})\right)_{i\in[\ell]}\oplus L\vecv.
\end{align*}
Note that the first summation, denoted by 
\begin{equation}\label{eqn:a}
    \veca:=\left(\bigoplus_{j=1}^N
        \vecd_{i,j}\cdot (0,\vecx_{i,j,0})\right)_{i\in[\ell]}
\end{equation}
is fixed. This implies that this measurement collapses the state to $\vecv$ such that $L\vecv=\vect\oplus \veca$.

\item \label{item:sim-ext2} 
The parties run a second witness-preserving 
indistinguishably-extractable argument for the relation
\[
    \mathcal R_{\mathsf{eval}}
    =
    \left\{
        \left(
            (\pubparams,\vecy,\vecd,\vecp,\vect),
            (\vecv, \vecv',\{\vecx_{i,j}\}_{i,j})
        \right)
        :
        \begin{array}{l}
            \vecv' = G\vecv \\
            \land\:\forall i,j,\; \Chek(\pubparams_{i,j}, v'_{j}, \vecx_{i,j}, \vecy_{i,j}) = 1
            \\[1mm]
            \land \forall i,\;
            t_i = \displaystyle\bigoplus_{j=1}^N 
            \left(
                \vecd_{i,j}\cdot (v'_j,\vecx_{i,j})
                \oplus 
                p_{i,j} v'_j
            \right)
        \end{array}
    \right\}.
\]
If the verifier rejects, $P_\C$ aborts.

\item $P_\C$ computes $\veca=(a_1,\ldots,a_\ell)\in\{0,1\}^\ell$ as in \Cref{eqn:a}, and stores the output
\[
    \vecu=\vect\oplus \veca.
\]
\item If $C(\vecu)=1$ then send $\bot$ to $P_\Q$, abort the protocol, and output $\vecu$.
\item Otherwise, continue to the recovery phase.

\end{enumerate}

\textbf{Phase 4: recovery.}
This phase is executed only if $C(\vecu)=0$.
Party $P_\C$ has already stored $\vecu$ as its output.

\begin{enumerate}
    \item $P_\Q$ applies $H^{\otimes n}$ to every preimage register
    $R_{i,j}$, for $(i,j)\in[\ell]\times[N]$, and measures these
    registers in the computational basis. Denote the outcomes by
    \[
        \mathbf e_{i,j}\in\{0,1\}^n.
    \]
    It sends
    $(\mathbf e_{i,j})_{i\in[\ell],\,j\in[N]}$ to $P_\C$.

    \item $P_\C$ checks that the message has the prescribed format
    and that, for every $(i,j)\in[\ell]\times[N]$ and every
    $b\in\{0,1\}$,
    \[
        \operatorname{Synd}
          (b,\mathbf x_{i,j,b},\mathbf e_{i,j})  
        \notin
       \left\{
            \mathbf0,\,
            \operatorname{Synd}
              (b,\mathbf x_{i,j,b},{\mathbf d}_{i,j})
        \right\}.
    \]
    If any check fails, $P_\C$ sends $\bot$ and terminates,
    retaining its stored output $\mathbf u$.
    No recovery bits are released in this case, and no further
    recovery query is answered.

    Otherwise, $P_\C$ computes
    \[
        \beta_{i,j}
        :=
        \mathbf e_{i,j}\cdot
        (\mathbf x_{i,j,0}\oplus\mathbf x_{i,j,1}),
        \qquad
        \zeta_j
        :=
        \bigoplus_{i=1}^{\ell}\beta_{i,j},
        \quad j\in[N].
    \]
    It sends
    \[
        \boldsymbol\zeta
        :=
        (\zeta_1,\ldots,\zeta_N)\in\{0,1\}^N
    \]
    to $P_Q$, and outputs the stored value $\mathbf u$.

    \item If $P_Q$ receives $\bot$, it discards its quantum
    registers and outputs $\bot$.

    Otherwise, $P_Q$ applies $Z^{\boldsymbol\zeta}$ to the
    encoded register $W$. It then applies the inverse of the
    encoding circuit from Phase~0, retains the original $m$
    input qubits, discards the resulting zero ancillas, and
    outputs the retained register.
\end{enumerate}

\end{protocol}

We next prove \Cref{thm:classical2PC} by arguing that the protocol above satisfies the required correctness and security guarantees.

\subsection{Correctness of Protocol~\ref{prot:classical-linear}}

\begin{lemma}[Correctness]
\label{lem:linear-2pc-correctness}
Assume $P_C(L, C)$ and $P_Q(\rho_{\calA, \calB})$ honestly execute Protocol~\ref{prot:classical-linear} using inputs $(L, C)$ and register $\calA$, respectively.
Then for all $(L, C)$ and $\rho_{\calA, \calB}$, the joint output distribution of $P_C$ and $P_Q$ together with register $\calB$ has $\negl$ trace distance to
\[
    \sum_{\vec{u}} \ketbra{\vec{u}} \otimes \bigg(\left(\Pi_{L,\vec{u}} \otimes I_\calB\right) \rho_{\calA, \calB} \left(\Pi_{L,\vec{u}} \otimes I_\calB\right)\bigg)
\]
where $\Pi_{L,\vec{u}} = \sum_{\vec{v}: L\vec{v} = \vec{u}} \ketbra{\vec{v}}$ projects onto $\ket{\vec{v}}$ such that $L\vec{v} = \vec{u}$ and where the first register is $P_C$'s output and the second is $P_Q$'s output together with register $\calB$.
The only exception occurs when $C(\vec{u}) = 1$, in which case the relation holds when register $\calA$ is traced out.

\end{lemma}

\begin{proof}
The witness-preserving property of the two NP arguments ensures that the quantum party's witness register is negligibly disturbed by the arguments. Thus, we omit them in the following analysis at the cost of a negligible error.

After Phase 1, the perfect matching and  efficient range superposition properties of the NTCF imply that the honest residual state is statistically close to
\[
    |\Phi\rangle
    =
    \sum_{\vecv\in\mathbb \{0,1\}^m}
    \alpha_\vecv \ket{\vecv, G\vecv}_V
    \bigotimes_{i=1}^{\ell}\bigotimes_{j=1}^N
    \ket{\vecx_{i,j,v'_j}}_{R_{i,j}},
\]
where
\[
    \mathsf{Check}(\pubparams_{i,j},v'_j,\vecx_{i,j,v'_j},\vecy_{i,j})=1.
\]
For every $(i,j)$ the two valid preimages satisfy
\[
    (v'_j,\vecx_{i,j,v'_j})=(0,\vecx_{i,j,0})\oplus v'_j\Delta_{i,j},
\]
where
\[
    \Delta_{i,j}:=(0,\vecx_{i,j,0})\oplus (1,\vecx_{i,j,1}).
\]

We first analyze the measurement in Phase 3. For each row $i\in[\ell]$, the
honest quantum party computes
\[
    t_i
    =
    \bigoplus_{j=1}^N
    \left(
        \vecd_{i,j}\cdot (v'_j,\vecx_{i,j,v'_j})
        \oplus
        p_{i,j}v'_j
    \right).
\]
Using the fact that 
\[
    (v'_j,\vecx_{i,j,v'_j})=(0,\vecx_{i,j,0})\oplus v'_j\Delta_{i,j},
\]
we have
\[
    \vecd_{i,j}\cdot (v'_j,\vecx_{i,j,v'_j})
    =
    \vecd_{i,j}\cdot (0,\vecx_{i,j,0})
    \oplus
    v'_j(\vecd_{i,j}\cdot\Delta_{i,j}).
\]
Therefore
\[
\begin{aligned}
    \vecd_{i,j}\cdot (v'_j,\vecx_{i,j,v'_j})
    \oplus
    p_{i,j}v'_j
    &=
    \vecd_{i,j}\cdot (0,\vecx_{i,j,0})
    \oplus
    v'_j(\vecd_{i,j}\cdot\Delta_{i,j}\oplus p_{i,j})  \\
    &=
    \vecd_{i,j}\cdot (0,\vecx_{i,j,0})
    \oplus
    L'_{i,j} v'_j ,
\end{aligned}
\]
where the last equality follows from the fact that 
\[
    p_{i,j} = \vecd_{i,j}\cdot\Delta_{i,j} \oplus L'_{i,j}.
\]
Hence
\[
   t_i
    =
    \left(\bigoplus_{j=1}^N \vecd_{i,j}\cdot (0,\vecx_{i,j,0})\right)
    \oplus
    \left(\bigoplus_{j=1}^N L'_{i,j}v'_j\right)       =
    a_i\oplus (L'\vecv')_i,
\]
where
\[
    a_i:=\bigoplus_{j=1}^N \vecd_{i,j}\cdot (0,\vecx_{i,j,0}).
\]
Denoting by $\veca=(a_1,\ldots,a_\ell)$, we conclude that
\[
    \vect=\veca\oplus L'\vecv' = \veca\oplus L\vecv.
\]
Thus measuring $\vect$ is exactly the same projective measurement as measuring $L\vecv$ (with the outcome one-time padded by $\veca$), and hence the residual state becomes
\[
    \ket{\Phi_\vecu}
    =
    \frac{1}{\sqrt{p_\vecu}}
    \sum_{\vecv:L\vecv=\vecu}
    \alpha_\vecv \ket{\vecv, G\vecv}_\calV
    \bigotimes_{i=1}^{\ell}\bigotimes_{j=1}^N
    \ket{\vecx_{i,j,v'_j}}_{R_{i,j}}.
\]
If $C(\vect\oplus\veca)=C(L\vecv)=1$ then $P_\C$ sends $\bot$ to $P_\Q$, in which case $P_\Q$ outputs $\bot$ and $P_\C$ outputs $L\vecv$, as desired.  Otherwise, if $C(L\vecv)=0$, then the parties continue to the state recovery phase.

In the state-recovery phase, $P_\Q$ Hadamard-measures every
$R_{i,j}$ register. Conditioned on the outcomes $\vece_{i,j}\in\mathbb \{0,1\}^{n}$,
the residual state becomes
\[\frac{1}{\sqrt{p_\vecu}}
    \sum_{\vecv:L\vecv=\vecu}
    \alpha_\vecv (-1)^{
    \sum_{i=1}^{\ell}\sum_{j=1}^N
    \vece_{i,j}\cdot \vecx_{i,j,v'_j}
    }\ket{\vecv, G\vecv}_V
\]
Again using the fact that
\[
    \vecx_{i,j,v'_j}=\vecx_{i,j,0}\oplus v'_j(\vecx_{i,j,0}\oplus \vecx_{i,j,1}),
\]
we get
\[
\begin{aligned}
    &=
    \sum_{i,j} \vece_{i,j}\cdot \vecx_{i,j,0}
    \oplus
    \sum_{i,j} v'_j(\vece_{i,j}\cdot (\vecx_{i,j,0}\oplus \vecx_{i,j,1}))     \\
    &=
    \sum_{i,j} \vece_{i,j}\cdot \vecx_{i,j,0}
    \oplus
    \sum_{i,j} v'_j\beta_{i,j},
\end{aligned}
\]
where
\[
    \beta_{i,j}:=\vece_{i,j}\cdot(\vecx_{i,j,0}\oplus \vecx_{i,j,1}).
\]
The first term is global. The second term is exactly cancelled by $P_\Q$'s
phase correction
\[
    \ket{\vecv, G\vecv}
    \longmapsto
    (-1)^{\sum_{j=1}^N v'_j\zeta_j}
    \ket{\vecv, G\vecv}.
\]
Thus the output of $P_\Q$ after inverting the encoding $\ket{\vec{v}, \vec{0}} \mapsto \ket{\vecv, G\vecv}$ is the residual state 
\[
    \frac{1}{\sqrt{p_\vecu}}
    \sum_{\vecv:L\vecv=\vecu}
    \alpha_\vecv \ket{\vecv}_V,
\]
and the output of $P_\C$ is $L\vecv$, unless there exists $(i,j)\in[\ell]\times[N]$ and $b\in\{0,1\}$ such that
\[
    \Synd(b,\vecx_{i,j,b},\vece_{i,j})\in\{\vec0,\Synd(b,\vecx_{i,j,b},\vecd_{i,j})\}
\]
which will trigger an abort.
We next argue that this occurs only with negligible probability.
To this end, note that for each $(i,j)$, the honest Hadamard outcome $\vece_{i,j}$ is uniform in $\{0,1\}^n$.
Hence, for each $b\in\{0,1\}$, $\Synd(b,\vecx_{i,j,b},\vece_{i,j})$ is uniform in $\{0,1\}^k$.
Therefore,
\[
    \Pr\!\left[\Synd(b,\vecx_{i,j,b},\vece_{i,j})\in\{\vec0,\Synd(b,\vecx_{i,j,b},\vecd_{i,j})\}\right]\leq 2\cdot 2^{-k}.
\]
By a union bound over all $2\ell N$ choices of $(i,j,b)$,
\[
    \Pr[\text{honest recovery abort}]
    \leq 4\ell N\cdot 2^{-k}=\negl,
\]
as desired.
\end{proof}

\subsection{Security Against Malicious $P_\Q$}

\begin{theorem}[$\epsilon$-Security against malicious $P_\Q$]
\label{thm:linear-2pc-malicious-pq}
Assume that the $\NTCF$ family satisfies the two-query adaptive hardcore bit property, and assume witness-preserving indistinguishably-extractable arguments for $\NP$. 
Then Protocol~\ref{prot:classical-linear} $\epsilon$-securely realizes the linear functionality with post-processed abort against malicious $\QPT$ $P^*_\Q$.
\end{theorem}

\begin{proof}
Let $P_\Q^*$ be an arbitrary malicious $\QPT$ quantum party. We construct an
ideal-world simulator $\mathsf{Sim}^{P_\Q^*}$ for the ideal functionality.

\paragraph{Simulator.}
Let $(S_{\secpar}, \{\Ext_{\secpar, \eta}\}_{n\in \bbN})$ be the simulator and extractor for indistinguishable-extraction of the $\NP$ argument. Set $\eta = (n/3)^2$.
The simulator $\Sim_{\secpar, n}$ is given the state of the adversary $P^*_\Q$ and does the following:
\begin{enumerate}
    \item 
For every $(i,j)\in[\ell]\times[N]$ generate
$(\pubparams_{i,j},\td_{i,j})\gets\Setup(1^\secpar)$.
\item Send 
$\pubparams=\{\pubparams_{i,j}\}_{i\in[\ell],j\in[N]}$
to $P_\Q^*$ and receive image values
$\vecy=\{\vecy_{i,j}\}_{i\in[\ell],j\in[N]}$.

\item Run the $S_{\secpar}$ for the first argument to obtain a state $\ket{\phi}_{\calT,\calA,\calS}$. 

\item Measure whether $\calT$ contains an accepting transcript. If it is rejecting then set $b_\Q=0$, indicating abort, so $P_\C$ receives $\perp$, and output the view of the simulated $P^*_\Q$. 

Otherwise, measure $\{\Pi_{\Ext,\secpar,\eta}, I - \Pi_{\Ext, \secpar, \eta}\}$ which determines if the extractor successfully extracts a witness. If not, immediately output $\bot$. If extraction succeeds, discard register $\calS$ and continue.

\item Otherwise, for every $(i,j)\in[\ell]\times[N]$ compute $\Invert(\td_{i,j},\vecy_{i,j})$. If any of these output $\bot$ then abort.\footnote{This happens with only negligible probability.}
Otherwise, denote the $(i,j)$ output by $((0,\vecx_{i,j,0}),(1,\vecx_{i,j,1}))$. and let 
    \[\Delta_{i,j}:=(0,\vecx_{i,j,0})\oplus (1,\vecx_{i,j,1}).
\]

\item Sample $R\gets\bbF_2^{\ell\times(N-m)}$ and set $L'=RP$.
For each $(i,j)$:
\begin{enumerate}
    \item Sample $\vecd_{i,j}\leftarrow \mathbb \{0,1\}^{n+1}$.
    \item Compute $p^0_{i,j}:=\vecd_{i,j}\cdot\Delta_{i,j}\oplus L'_{i,j}$.\footnote{Recall that in the real execution, the classical party computes $p_{i,j}:=\vecd_{i,j}\cdot\Delta_{i,j}+L'_{i,j}$.  Since the simulator does not know the linear function $L$, it uses the encoding $L'=RP$ of the zero function.}
\end{enumerate}
Set $\vecd=\{\vecd_{i,j}\}$ and $\vecp^0=\{p^0_{i,j}\}$ and send $(\vecd,\vecp^0)$ to 
$P_\Q^*$.

\item Upon receiving $\vect\in\mathbb \{0,1\}^\ell$ from $P_\Q^*$, run $S_\secpar$ for the second argument, to obtain a state $\ket{\phi}_{\calT,\calA,\calS}$. 
\item Measure only the bit indicating if the transcript in register $\calT$ is accepting or not.  If it is rejecting, then set $b_\Q=0$ indicating abort, and output the view of the simulated $P^*_\Q$. 
\item Otherwise, set $b_\Q=1$ and measure $\ket{\phi}_{\calT,\calA,\calS}$ using $\{\Pi_{\Ext,\secpar,\eta}, I - \Pi_{\Ext, \secpar, \eta}\}$ to see if the second extraction succeeds. If not, immediately output $\bot$. Otherwise, continue.

\item Compute the witness extraction unitary $\Ext_{\secpar, \eta}$ on the state $\ket{\phi}_{\calT,\calA,\calS}$. Conditioned on not aborting in the prior step, the extracted witness register $\calW$ contains a state wholly supported on values $(\vecv,\vecv',\{\vecx_{i,j}\}_{(i,j)\in[\ell]\times[N]})$ satisfying $\vecv'=G\vecv$ and
\[
    \Chek(\pubparams_{i,j},v'_j,\vecx_{i,j},\vecy_{i,j})=1
\]
for all $(i,j)$, and
\[
    t_i
    =
    \bigoplus_{j=1}^N
    \left(
        (\vecd_{i,j}\cdot (v'_j,\vecx_{i,j}))
        \oplus
        p^0_{i,j}v'_j
    \right)
\]
for all $i\in[\ell]$.

\item Initialize a register $\calV$ to $\ket{0}$ and classical-copy $\vec{v}$ from register $\calW$ to register $\calV$ using a $\CNOT$. Send $\calV$ to the ideal functionality.

\begin{itemize}
    \item \textbf{Ideal Functionality.} The ideal functionality coherently computes and measures $\vec{u} = L\vec{v}$. If $C(\vec{u}) = 1$, it returns $\bot$ to the simulator. Otherwise, it returns the post-measurement state in register $\calV$ to the simulator.
\end{itemize}

If the ideal functionality returns register $\calV$, erase it by applying a second $\CNOT$ from $\calW$ to $\calV$.

\item Compute $\Ext_{\secpar, \eta}^\dagger$ and trace out register $\calS$.

\item If the ideal functionality sent $\bot$, then output the view of the simulated $P^*_\Q$ so far. 
\item Otherwise, continue the interaction
with $P_\Q^*$.

When $P_\Q^*$ sends recovery vectors $\vece_{i,j}\in\{0,1\}^n$, check that the message has the prescribed format and that for every $(i,j)\in[\ell]\times[N]$ and every $b\in\{0,1\}$,
\[
    \Synd(b,\vecx_{i,j,b},\vece_{i,j})\notin\{\vec0,\Synd(b,\vecx_{i,j,b},\vecd_{i,j})\}.
\]
If any check fails, simulate $P_\C$ sending $\bot$ and output the view of the simulated $P^*_\Q$.
\item Otherwise, compute
\[
    \beta_{i,j}:=\vece_{i,j}\cdot (\vecx_{i,j,0}\oplus\vecx_{i,j,1}),
    \qquad
    \zeta_j:=\bigoplus_{i=1}^{\ell}\beta_{i,j},
    \quad j\in[N],
\]
and send $\boldsymbol\zeta=(\zeta_1,\ldots,\zeta_N)$ to $P_\Q^*$.
\end{enumerate}

We prove that the ideal execution with this simulator is computationally
indistinguishable from the real execution.

\paragraph{Hybrid $H_0$: real execution.}
Hybrid $H_0$ is the real execution of Protocol~\ref{prot:classical-linear} between
malicious $P_\Q^*$ and honest $P_\C$. In this hybrid, for every $(i,j)$,
$P_\C$ sends
\[
    p_{i,j}
    =
    L'_{i,j}\oplus (\vecd_{i,j}\cdot\Delta_{i,j}).
\]

\paragraph{Hybrid $H_1$: indistinguishable extraction.}
Hybrid $H_1$ is the same as $H_0$, except we replace the two witness-preserving arguments. Instead, run the indistinguishable-extraction simulator $S_\secpar$ and measure if the extractor $\Ext_{\secpar, \eta}$ succeeds. If the extractor fails, immediately output $\bot$.

By \Cref{lem:iext-close}, each replacement changes the trace distance by at most $\sqrt{1/\eta} + 1/\eta$ since the extractor succeeds with probability $\geq 1/\eta$. Since we set $\eta = (3n)^2$, the total change in trace distance from the two replacements is less than $1/n$ for all $n \in \bbN$.

\paragraph{Hybrid $H_2$: explicit ideal measurement.}
Hybrid $H_2$ is the same as $H_1$, except for an additional measurement right after measuring whether the extractor for the second argument succeeds:
\begin{enumerate}
    \item Proceed as in $H_1$ until the second witness-preserving argument.
    \item Run $S_\secpar$ and measure whether the extractor succeeds, as in $H_1$.
    \item Run $\Ext$ to extract a valid witness. 
    \item Measure the register containing the extracted $\vec{v}$ using the projective measurement $\{\Pi_{\vec{u}}^{L} = \sum_{\vec{v}:L\vec{v} = \vec{u}} \ketbra{\vec{v}}\}_{\vec{u}}$ to obtain a vector $\vec{u} = L\vec{v}$.
    \item Run $\Ext^\dagger$.
    \item Continue as in $H_1$.
\end{enumerate}

We show that measuring $L\vecv$ does not disturb the state, and that the output of $P_\C$ in hybrid $H_1$ (if it doesn't abort) is distributed identically to the extracted vector $\vecu=L\vecv$. In $H_1$, for a
valid extracted witness,
\[
    (v'_j,\vecx_{i,j,v'_j})=
    (0,\vecx_{i,j,0})\oplus v'_j\Delta_{i,j}.
\]
Using the fact that
\[
    p_{i,j}= (\vecd_{i,j}\cdot\Delta_{i,j}) \oplus L'_{i,j},
\]
we observe that
\[
\begin{aligned}
    (\vecd_{i,j}\cdot (v'_j,\vecx_{i,j,v'_j}))
    \oplus
    p_{i,j}v'_j
    &=
    (\vecd_{i,j}\cdot
    ((0,\vecx_{i,j,0})\oplus v'_j\Delta_{i,j}))
    \oplus
    v'_j(L'_{i,j}\oplus \vecd_{i,j}\cdot\Delta_{i,j})        \\
    &=
    (\vecd_{i,j}\cdot (0,\vecx_{i,j,0}))
    \oplus
    L'_{i,j}v'_j.
\end{aligned}
\]
Therefore
\[
\begin{aligned}
    t_i
    &=
    \bigoplus_{j=1}^N
    \left(
        (\vecd_{i,j}\cdot (v'_j,\vecx_{i,j,v'_j}))
        \oplus
        p_{i,j}v'_j
    \right)        \\
    &=
    \left(\bigoplus_{j=1}^N \vecd_{i,j}\cdot (0,\vecx_{i,j,0})\right)
    \oplus
    \bigoplus_{j=1}^N L'_{i,j}v'_j        \\
    &=
    a_i\oplus (L'\vec{v'})_i.
\end{aligned}
\]
Thus $\vect=\veca\oplus L'\vec{v'} = \veca\oplus L\vec{v}$.
Hence, conditioned on the observed value $\vect$, the extracted $V$ register is already
supported only on $\vecv$'s such that 
\[
    L\vecv=\vect\oplus \veca.
\]
Therefore, applying the projective measurement $\{\Pi_\vecu^L\}_\vecu$ does not further
disturb the state.
Therefore
\[
    H_1 = H_2.
\]

\paragraph{Hybrid $H_3$: Moving the Ideal Measurement.} 
$H_3$ adds a measurement of $L\vec{v}$ in between extraction for the first argument. The same measurement is also done in the second argument.

\begin{claim}
    $ H_2\approx_c H_3$
\end{claim}
\begin{proof}
    Measuring $L\vec{v}$ is equivalent to measuring $L'\vec{v'}$. For any row $L'_{i*}$ of $L'$, the measurement of $L'_{i*}\cdot \vec{v'}$ is equivalent to applying $Z^{L'_{i*}}$ with probability $1/2$ during the extraction for either measurement. We show that it is indistinguishable to add these operations during the first extraction.
    
    By the phase injection lemma with 1 bit of inner product leakage (\Cref{lem:phase-injection-leakage}), applying $Z^{L'_{ij}}$ to the extracted $v'_j$ for any $(i, j)$ during the first extraction is indistinguishable so long as the correction $\zeta_j$ is reported as $\zeta_j \oplus L'_{ij}$. Applying the phase injection lemma a second time during the second extraction reverts the correction to $\zeta_j \oplus L'_{ij} \oplus L'_{ij} = \zeta_j$. Moreover, the combined $Z^{L'_{i*}}$ operation during the second extraction only induces a global phase because $L'_{i*}\cdot \vec{v'}$ is measured anyway during the second extraction.
\end{proof}

\paragraph{Hybrid $H_4$: explicit ideal abort.}
$H_4$ is the same as $H_3$ except that after measuring $\vec{u} = L\vec{v}$ in between $\Ext$ and $\Ext^\dagger$ for the second argument, if $C(\vec{u}) = 1$ then it also measures the extracted $\vec{v}$ in the computational basis before continuing.

\begin{claim}
    $H_3 \approx_c H_4$.
\end{claim}
\begin{proof}
    Let $H_{3,i}$ be the hybrid experiment where in the case of an abort, every $v_k$ is measured for $k\leq i$. Then $H_{3,0} = H_3$ and $H_{3,m} = H_4$.
    We will show that for every $i\in\{0,\ldots,m-1\}$, $H_{3,i} \approx_{c} H_{3,i+1}$ by reducing to the collapsing property of the NTCF with 1 bit of inner product leakage (\Cref{lem:ntcf-collapsing-inner-product}). Repeating the reduction for every $i$ shows that $H_3 \approx H_4$.

    Fix any $j\in [\ell]$. Since $v_{i+1}=v'_{i+1}$ and $\vec{x}_{j,i+1}$ uniquely determine each other given $y_{j,i+1}$, is is equivalent to argue collapsing of $(v'_{i+1},\vec{x}_{j,i+1})$.
    Moreover, the adversary has $0$ advantage in distinguishing $H_{3,i}$ from $H_{3,i+1}$ conditioned on the non-abort case $C(\vec{u}) = 0$ because the two hybrids are identical in this case.

    The reduction works as follows. It plays $H_{3,i}$ using the $\pubparams$ chosen by the collapsing challenger for index $(j,i+1)$ and the others chosen internally. It forwards $y_{j,i+1}$ to the collapsing challenger, then receives $(d, d\cdot \Delta)$ from the collapsing challenger and uses this to mask $L'_{j,i+1}$.

    At this point, $H_{3,i}$ runs $\Sim$ and $\Ext$ to extract $\ket{v, \vec{x}}$ from the second witness-preserving argument of knowledge. If $C(Lv) = 0$, it immediately guesses $0$ in the collapsing experiment. Otherwise, it sends the extracted register containing $(v'_{i+1},\vec{x}_{j,i+1})$ to the collapsing challenger.
    The challenger either measures this register or does not, at random, before returning it. After the register is returned, the reduction completes $H_{3,i}$ and outputs the distinguisher's output.

    Observe that if the challenger measures, this is distributed as $H_{3,i+1}$ and otherwise it is distributed as $H_{3,i}$. Since the distinguisher's advantage is entirely concentrated on the abort case, the reduction has the same advantage in the collapsing game with 1 bit of inner product leakage, which must be negligible.
\end{proof}

\paragraph{Hybrid $H_5$: Changing $L'$ to Encode $0$.}
Hybrid $H_5$ is the same as $H_4$, except that the simulator computes $L'$ as an encoding of $0$ instead of $L$:
\[
    L' = RP
\]
for random matrix $R$. It still masks $L'_{i,j}$ as $p_{i,j}=(\vecd_{i,j}\cdot\Delta_{i,j}) \oplus L'_{i,j}$ like in the previous hybrid $H_4$.

The following lemma gives a route to replacing each row $L'_{i*}$ by an encoding of $0$ instead of an encoding of $L_{i*}$.
\begin{lemma}
    \label{lem:2pc-encoding-replacement}
    Let $G$, $D$, and $P$ be the encoding matrices defined at the start of this section.
    For all $\vecl \in \bbF_2^{m}$ and any $\QPT$ adversary,
    \[
        \Exp_{\vecl}(\vecl) \approx_c \Exp_{\vecl}(\vec{0})
    \]
    where $\Exp_{\vecl}(\vec{a})$ is the following game.
    \begin{enumerate}
        \item The challenger samples $N$ public parameters $(\pubparams_{i},\td_{i}) \leftarrow \NTCF.\Setup(1^\lambda)$ and send $(\pubparams_{i})_{i\in [N]}$ to the adversary.
        \item The adversary sends images $y_1, \dots, y_N$ and joint state on registers $(\calV, \calV', \calX)$ to the challenger.\footnote{These registers may be arbitrarily entangled with the adversary's internal state.}
        
        \item \label{step:2pc-encoding-replacement:test}
        The challenger measures whether $(\calV,\calV',\calX)$ contain a state entirely supported on computational basis states $\ket{\vec{v}, \vec{v}', (\vec{x}_1, \dots, \vec{x}_N)}$ such that
        \[
            \vec{v'} = G\vec{v} \quad\text{and}\quad \Chek(\pubparams_i,v'_i,\vecx_i,\vecy_i)=1 \:\forall i\in [N]
        \]
        If the result rejects, the experiment immediately outputs $0$.
        Otherwise, measure $\vecl\cdot \vec{v}$ and return registers $(\calV, \calV', \calX)$ to the adversary.
        
        \item The challenger does the following:
        \begin{enumerate}
            \item Compute $((0,\vecx_i^{(0)}),(1,\vecx_i^{(1)}))=\Invert(\td_i,\vecy_i)$ for each $i\in [N]$.\footnote{Note that the face that $(\pubparams,\vecy)\in\calR$ implies that with overwhelming probability $\Invert(\td,\vecy)\neq\bot$.} 
            \\Then, set $\Delta_i :=(0,\vecx_i^{(0)})\oplus (1,\vecx_i^{(1)})$.
            \item Sample $\vec{d}_i \gets \bbF_2^{n +1}$ for each $i\in [N]$
            \item\label{step:2pc-encoding-replacement:encoding} Sample a uniform vector $\vec{r} \gets \bbF_2^{1\times (N - m)}$ and encode $\vec{a}' = \vec{a}D + \vec{r}P$.
            \item Send $(\vecd_i, (\vecd_i \cdot \Delta_i) \oplus\vec{a}'_i)_{i\in [N]}$ to the adversary.
        \end{enumerate}
    
        \item $\calA$ sends vectors $\vece_i\in\mathbb \{0,1\}^{n}$ for $i \in [N]$.
        \item If the recovery message is malformed or $\Synd(b,\vecx_i^{(b)},\vece_i)\in\{\vec0,\Synd(b,\vecx_i^{(b)},\vecd_i)\}$ for any $i\in[N]$ and $b\in\{0,1\}$, then the challenger sends $\perp$ to $\calA$.
        Otherwise, the challenger sends $\beta_i:=\vece_i\cdot (\vecx^{(0)}_i\oplus \vecx^{(1)}_i)$ to $\calA$ for each $i\in[N]$.
        \item $\calA$ chooses a bit $u'$ to the be experiment's output.
    \end{enumerate}
\end{lemma}
We prove the lemma in \Cref{subsec:c2PC-encoding-replacement}. For now, we describe how it shows that $H_4 \approx_c H_5$. It suffices to show that we can indistinguishably replace any row $L_{i*}$ by $\vec{0}$ in the encoding. Iterating this for each of the $\ell$ rows shows that $H_4 \approx_c H_5$. Let $H_{4,i}$ be the hybrid where all rows before $i$ have been replaced, so that $H_{4} = H_{4,1}$ and $H_5 = H_{4,\ell+1}$.

Fix any row $i$ and consider the experiments $\Exp_{L_{i*}}(L_{i*})$ versus $\Exp_{L_{i*}}(\vec{0})$. 
The reduction runs $H_{4,i}$ using the $(\pubparams_{j})_{j\in [N]}$ for that row of the encoded matrix $L'_{i*}$. It samples the others internally. Upon the first extraction, it sends the registers containing the extracted $\vec{v}$, $\vec{v'}$, and $(\vecx_{i,j})_{j\in [N]}$ to the challenger, who measures $L_{i*}\cdot \vec{v}$ and returns the register. It then completes $H_4$ by forwarding the masked values (corresponding to $\vec{d}_{j}$) for this row and aggregating its corrections (corresponding to $\vec{e}_j$) with the internally computed corrections for all other rows into $(\zeta_j)_{j\in[N]}$ before sending them to the adversary. The reduction running on $\Exp_{L_{i*}}(L_{i*})$ is distributed according to $H_{4,i}$ and it running on $\Exp_{L_{i*}}(\vec{0})$ is distributed according to $H_{4,i+1}$. Therefore any QPT adversary's advantage in distinguishing the two must be negligible by \Cref{lem:2pc-encoding-replacement}.

\begin{remark}
    Note that in $H_5$, 
    $t_i$
    and hence 
    has no dependence on $L$. Thus,  the only
    $L$-dependent operation remaining in $H_5$ is the explicit ideal measurement inserted during the first and second extractions.
\end{remark}

\paragraph{Hybrid $H_6$: Removing the First $L\vec{v}$ Measurement.} $H_6$ removes the measurement of $L\vecv$ during the first extraction. The same argument as $H_3$ shows that 
\[
    H_5 \approx_c H_6
\]

\paragraph{Hybrid $H_7$: the ideal simulation.}   Hybrid $H_7$ is the ideal execution with the simulator
$\mathsf{Sim}^{P_\Q^*}$ described above.  It is easy to see that this hybrid is statistically close to $H_6$.

Combining the hybrids,
\[
    H_0
    \approx_c
    H_1 
    =
    H_2
    \approx_c
    H_3
    \approx_c
    H_4
    \approx_c
    H_5
    \approx_c
    H_6
    \equiv
    H_7
   .
\]
Since $H_0$ is the real execution and $H_7$ is the ideal execution with
$\mathsf{Sim}^{P_\Q^*}$, we conclude that
\[
    \REAL_{P_\Q^*}(\ket{\psi},L)
    \approx_c
    \IDEAL_{\mathsf{Sim}^{P_\Q^*}}(\ket{\psi},L).
\]
This proves security against malicious $P^*_\Q$.
\end{proof}

\subsubsection{Replacement of the Encoded $L$}\label{subsec:c2PC-encoding-replacement}

We recall the statement of \Cref{lem:2pc-encoding-replacement} here.
Informally, it considers an experiment where an adversary commits to $\vec{v'} = G\vec{v}$ using many NTCF instances, then some $\vec{a} \cdot \vec{v'}$ is measured. Afterwards, the adversary is given an encoding of $\vec{a}$ which is masked by a random inner product with the difference $\Delta$ between the collisions of the NTCF. Finally, the adversary may query for an adaptive hardcore bit. 
The lemma states that since $\vec{a} \cdot \vec{v'}$ is measured, we may indistinguishably replace the encoding of $\vec{a}$ by an encoding of $\vec{0}$.

\begin{lemma}
    Let $G$, $P$, and $D$ be the generator, parity-check, and decoding matrix for a minimal binary code. 
    For all $\vecl \in \bbF_2^{m}$ and any $\QPT$ adversary,
    \[
        \Exp_{\vecl}(\vecl) \approx_c \Exp_{\vecl}(\vec{0})
    \]
    where $\Exp_{\vecl}(\vec{a})$ is the following game.
    \begin{enumerate}
        \item The challenger samples $N$ public parameters $(\pubparams_{i},\td_{i}) \leftarrow \NTCF.\Setup(1^\lambda)$ and send $(\pubparams_{i})_{i\in [N]}$ to the adversary.
        \item The adversary sends images $y_1, \dots, y_N$ and joint state on registers $(\calV, \calV', \calX)$ to the challenger.\footnote{These registers may be arbitrarily entangled with the adversary's internal state.}
        
        \item \label{step:2pc-encoding-replacement:test:restated}
        The challenger measures whether $(\calV,\calV',\calX)$ contain a state entirely supported on computational basis states $\ket{\vec{v}, \vec{v}', (\vec{x}_1, \dots, \vec{x}_N)}$ such that
        \[
            \vec{v'} = G\vec{v} \quad\text{and}\quad \Chek(\pubparams_i,v'_i,\vecx_i,\vecy_i)=1 \:\forall i\in [N]
        \]
        If the result rejects, the experiment immediately outputs $0$.
        Otherwise, measure $\vecl\cdot \vec{v}$ and return registers $(\calV, \calV', \calX)$ to the adversary.
        
        \item The challenger does the following:
        \begin{enumerate}
            \item Compute $((0,\vecx_i^{(0)}),(1,\vecx_i^{(1)}))=\Invert(\td_i,\vecy_i)$ for each $i\in [N]$.\footnote{Note that the face that $(\pubparams,\vecy)\in\calR$ implies that with overwhelming probability $\Invert(\td,\vecy)\neq\bot$.} 
            \\Then, set $\Delta_i :=(0,\vecx_i^{(0)})\oplus (1,\vecx_i^{(1)})$.
            \item Sample $\vec{d}_i \gets \bbF_2^{n +1}$ for each $i\in [N]$
            \item Sample a uniform vector $\vec{r} \gets \bbF_2^{1\times (N - m)}$ and encode $\vec{a}' = \vec{a}D + \vec{r}P$.
            \item Send $(\vecd_i, (\vecd_i \cdot \Delta_i) \oplus\vec{a}'_i)_{i\in [N]}$ to the adversary.
        \end{enumerate}
    
        \item $\calA$ sends vectors $\vece_i\in\mathbb \{0,1\}^{n}$ for $i \in [N]$.
        \item If the recovery message is malformed or $\Synd(b,\vecx_i^{(b)},\vece_i)\in\{\vec0,\Synd(b,\vecx_i^{(b)},\vecd_i)\}$ for any $i\in[N]$ and $b\in\{0,1\}$, then the challenger sends $\perp$ to $\calA$.
        Otherwise, the challenger sends $\beta_i:=\vece_i\cdot (\vecx^{(0)}_i\oplus \vecx^{(1)}_i)$ to $\calA$ for each $i\in[N]$.
        \item $\calA$ chooses a bit $u'$ to the be experiment's output.
    \end{enumerate}
    
\end{lemma}

\paragraph{Notation.}
For the proof of this lemma, it will be useful to define a slightly modified experiment $\Exp_{\vecl}(\vec{a}, \vec{z})$ for a vector $\vec{z} \in \bbF_2^{m}$.
This experiment is identical to $\Exp_{\vecl}(\vec{a})$, except for a modification in step \ref{step:2pc-encoding-replacement:test:restated}. If the test measurement rejects, apply $Z^{\vec{z}}$ to register $\calV$ in addition to measuring $\vecl\cdot\vec{v}$.
As a consequence,
\[
    \Exp_{\vecl}(\vec{a}) = \Exp_{\vecl}(\vec{a}, \vec{0})
\]
It will be useful to have shorthand for the probability that any experiment outputs $1$:
\[
    \probone_{,\vecl}(\vec{a}, \vec{z})
    \coloneqq \Pr[1\gets \Exp_{\vecl}(\vec{a}, \vec{z})]
\]

We also consider experiments where the phase being inserted is not fixed. Given a vector $\vec{z'}$, define
\[
    \Exp_{\vecl}(\vec{a}, \vec{z} + *\vec{z}') 
    \coloneqq
    \frac{1}{2} \Exp_{\vecl}(\vec{a}, \vec{z}) + \frac{1}{2} \Exp_{\vecl}(\vec{a}, \vec{z} + \vec{z}') 
\]
In other words, insert phase $\vec{z}$ with probability $1/2$ or phase $\vec{z} + \vec{z}'$ with probability $1/2$ during the check step \ref{step:2pc-encoding-replacement:test:restated}.
Using the equivalence of random phases and measuring, $\Exp_{\vecl}(\vec{a}, \vec{z} + *\vec{z}')$ corresponds to also measuring $\vecv \cdot \vec{z}'$ during this step.

The final piece of notation we will need is to quantify the signed advantage with which an adversary distinguishes adding a vector $\vec{b}$ to the encoding, conditioned on a variety of baseline scenarios $(\vec{a}, \vec{z})$.
We define the signed distinguishing gap between $\Exp_{\vecl}(\vec{a}, \vec{z})$ and $\Exp_{\vecl}(\vec{a} + \vec{b}, \vec{z})$ to be
\[
    \diffConditionedL{\vec{b}}{(\vec{a}, \vec{z})}
    \coloneqq 
    \probone_{,\vecl}(\vec{a}, \vec{z}) - \probone_{,\vecl}(\vec{a} + \vec{b}, \vec{z})
\]
The $*$ notation extends to the output probabilities $\probone_{,\vecl}$ and the gaps $\Delta_{\vecl}$ by using the corresponding experiment, e.g. $\Exp_{\vecl}(\vec{a}, \vec{z} + *\vec{z'})$.

\begin{proof}[Proof of \Cref{lem:2pc-encoding-replacement}]
    We show that 
    \[
        \Exp_{\vecl}(\vecl, \vec{0}) 
        = \Exp_{\vecl}(\vecl, *\vecl)
        \approx_c \Exp_{\vecl}(\vec{0}, *\vecl)
        = \Exp_{\vecl}(\vec{0}, \vec{0})
    \]
    For the first and last transitions, we claim that $\Exp_{\vecl}(\vec{a}, 0) = \Exp_{\vecl}(\vec{a}, *\vecl)$ for all $\vec{a}$. This holds because $ \Exp_{\vecl}(\vec{a}, *\vecl)$ is the experiment which measures $\vecl\cdot \vec{v}$, which is already done in $\Exp_{\vecl}(\vec{a})$.

    The middle transition remains to be shown. Consider any QPT adversary's distinguishing advantage
    \[
        |\probone_{,\vecl}(\vecl, *\vecl) -
        \probone_{,\vecl}(\vec{0}, *\vecl)|
    \]
    Expanding the $*\vecl$ notation and grouping the terms with the same inserted phase, their advantage is
    \[
        \frac{1}{2}\left|
        \diffConditionedL{\vecl}{(\vec{0}, \vec{0})} 
        + 
        \diffConditionedL{\vecl}{(\vec{0}, \vecl)} 
        \right|
    \]
    \Cref{lem:2pc-encoding-replacement:gap-transfer-general} shows that 
    \[
        \diffConditionedL{\vecl}{(\vec{0}, \vecl)} 
        =
        \diffConditionedL{\vecl}{(\vecl, \vec{0})} \pm \negl
    \]
    against any QPT adversary.
    Expanding the notation, we have
    \begin{align*}
        \diffConditionedL{\vecl}{(\vecl, \vec{0})}
        &=
        \probone_{,\vecl}(\vecl, \vec{0}) - \probone_{,\vecl}(\vec{0}, \vec{0})
        \\
        &= -\diffConditionedL{\vecl}{(\vec{0}, \vec{0})}
    \end{align*}
    The $\diffConditionedL{\vecl}{(\vec{0}, \vec{0})}$ gap terms cancel and the adversary's advantage is $\negl$.
\end{proof}

\subsubsection{Distinguishing Gap Transport Lemma}

The following lemma guarantees that the signed distinguishing gap changes negligibly between two baseline scenarios $(\vec{a},\vec{z})$ and $(\vec{a}+\vec{b},\vec{z}+\vec{b})$ where both the encoding and phase differ by the same arbitrary vector $\vec{b}$.
Since $\vec{b}+\vec{b} = \vec{0}$ over $\bbF_2$, this can be interpreted as ``moving'' $\vec{b}$ from the encoding to the phase in the baseline scenario.

\begin{lemma}[Distinguishing Gap Transport, General]
\label{lem:2pc-encoding-replacement:gap-transfer-general}
    For all $\vec{a}, \vec{b},\vec{c}, \vec{z} \in \bbF_2^{m}$ and all $\QPT$ adversaries,
    \[
        \diffConditionedL{\vec{b}}{(\vec{a}, \vec{z})} 
        \approx_{\negl} 
        \diffConditionedL{\vec{b}}{(\vec{a}+\vec{c}, \vec{z}+\vec{c})} 
    \]
\end{lemma}

We start by proving the same relation when $\vec{b}$ is a row of $G$ (\Cref{claim:2pc-encoding-replacement:gap-transfer-limited}), then prove the general version at the end of this subsection.
In preparation for the limited version, we show that when $\vec{g} \cdot \vec{v}$ is measured and $\vec{g}$ is a row of $G$, no QPT adversary can distinguish adding $\vec{g}$ to the value to be encoded.

\begin{claim}\label{claim:2pc-encoding-replacement-measured}
    Let $\vec{g}$ be the $i$'th row of $G$ for any $i$. Then for all $\vec{a}$ and $\vec{z}$,
    \[
        \diffConditionedL{\vec{g}}{(\vec{a}, \vec{z} + *\vec{g})} = \pm \negl
    \]
\end{claim}
\begin{proof}
    If the test passes in the check step \ref{step:2pc-encoding-replacement:test:restated}, then $\vec{v}'_i = \vec{g}\cdot \vec{v}$. Applying $Z^{\vec{g}}$ to $\calV$ with probability $1/2$ is equivalent to measuring $\vec{v}'_i$ in the computational basis. Together with $\vec{y}_i$, this uniquely determines $\vec{x_{i}}$, so $\vec{x_{i}}$ can also be measured without disturbing the state.
    
    Now that $(\vecv_i', \vec{x_{i}})$ is measured, we can invoke the 2-query adaptive hardcore bit property on index $i$ (\Cref{def:two-query-ahc,lem:2Q-NTCF}). This undetectably replaces $(\vecd_i, (\vecd_i \cdot \Delta_i) \oplus \vec{a}'_i)$ by $(\vecd_i, (\vecd_i \cdot \Delta_i) \oplus \vec{a}'_i \oplus 1)$ in $\Exp_{\vecl}(\vec{a}, \vec{z} + *\vec{g})$.
    Finally, we show that this replacement yields the other experiment $\Exp_{\vecl}(\vec{a} + \vec{g}, \vec{z} + *\vec{g})$.
    We have replaced the encoded $\vec{a'}$ by 
    \[
        \vec{a'} + \vec{e_i} = (\vec{a} D + \vec{r}P) + \vec{e_i}
    \]
    where $\vec{e_i}$ is the $i$'th standard basis vector. 
    Observe that $\vec{g}DG = \vec{e_i}G$, so $(\vec{g}D - \vec{e_i}) = (\vec{g}D + \vec{e_i})$ is in the kernel of $G$. Since the row span of $P$ is the kernel of $G$, there exists a vector $\vec{s}$ such that $\vec{s}P = \vec{g}D + \vec{e_i}$.
    Therefore
    \[
        \vec{a'} + \vec{e_i} = (\vec{a} + \vec{g}) D + (\vec{r} + \vec{s})P
    \]
    Finally, the uniform distribution $\vec{r}$ is invariant under linear shifts, so this is distributed precisely as $(\vec{a} + \vec{g}) D + \vec{r}'P$ for uniform $\vec{r}'$, which yields $\Exp_{\vecl}(\vec{a} + \vec{g}, \vec{z} + *\vec{g})$.
\end{proof}

\begin{claim}[Distinguishing Gap Transport, Limited]
\label{claim:2pc-encoding-replacement:gap-transfer-limited}
    Let $\vec{b}$ be any standard basis vector. Then for all $\vec{a}, \vec{b},\vec{c}, \vec{z} \in \bbF_2^{m}$ and all $\QPT$ adversaries,
    \begin{equation}\label{eq:gap-transport-conclusion}
        \diffConditionedL{\vec{b}}{(\vec{a}, \vec{z})} 
        \approxnegl
        \diffConditionedL{\vec{b}}{(\vec{a}+\vec{c}, \vec{z}+\vec{c})}
    \end{equation}
\end{claim}
\begin{proof}
    First, suppose that the claim holds only when $\vec{c}$ is also a standard basis vector. Any general $\vec{c}$ can be decomposed into a linear combination of standard basis vectors.
    Since $\vec{c}$ has polynomial length, applying the assumed claim one at a time for each row shows the more general result.

    We now show that the claim holds when $\vec{c}$ is a standard basis vector.
    We show that the following two relations hold for such $\vec{c}$:
    \begin{gather}
        \label{eq:gap-transport-phase-average}
        \diffConditionedL{\vec{b}}{(\vec{a},\vec{z}+*\vec{c})} 
        \approxnegl
        \diffConditionedL{\vec{b}}{(\vec{a}+\vec{c},\vec{z}+*\vec{c})}
        \\
        \label{eq:gap-transport-coefficient-average}
        \diffConditionedL{\vec{b}}{(\vec{a}+*\vec{c},\vec{z})}
        \approxnegl
        \diffConditionedL{\vec{b}}{(\vec{a}+*\vec{c},\vec{z}+\vec{c})}
    \end{gather}
    The first shows that when the phase is randomized, the baseline encoding can be switched from $\vec{a}$ to $\vec{a}+ \vec{c}$ without changing the gap noticeably.
    The second shows that when the baseline encoding is randomized, the phase can be switched from $\vec{z}$ to $\vec{z} + \vec{c}$ without changing the gap noticeably.

    Combining these two yields the claim for $\vec{c}$ using the following expansion:
    \begin{equation}
    \label{eq:gap-transport-square}
        \begin{aligned}
        &\diffConditionedL{\vec{b}}{(\vec{a},\vec{z})}-\diffConditionedL{\vec{b}}{(\vec{a}+\vec{c},\vec{z}+\vec{c})}\\
        &\quad=\left[\diffConditionedL{\vec{b}}{(\vec{a},\vec{z}+*\vec{c})}-\diffConditionedL{\vec{b}}{(\vec{a}+\vec{c},\vec{z}+*\vec{c})}\right]\\
        &\qquad+\left[\diffConditionedL{\vec{b}}{(\vec{a}+*\vec{c},\vec{z})}-\diffConditionedL{\vec{b}}{(\vec{a}+*\vec{c},\vec{z}+\vec{c})}\right].
        \end{aligned}
    \end{equation}
    To verify this identity, expand the two starred shifts. The terms at $(\vec{a},\vec{z}+\vec{c})$ and $(\vec{a}+\vec{c},\vec{z})$ cancel, while the terms at $(\vec{a},\vec{z})$ and $(\vec{a}+\vec{c},\vec{z}+\vec{c})$ each occur twice with coefficient $1/2$, with positive and negative signs, respectively. \eqref{eq:gap-transport-phase-average} implies the first grouping is negligible and ~\eqref{eq:gap-transport-coefficient-average} implies the second grouping is negligible.

    \medskip
    \noindent\emph{Changing Encoding with Randomized Phase \eqref{eq:gap-transport-phase-average}.}
    Expanding the $\Delta_{\vecl}$ notation and recombining the first terms of each against the last terms of each
    yields the identity
    \begin{align*}
        &\diffConditionedL{\vec{b}}{(\vec{a},\vec{z}+*\vec{c})}-\diffConditionedL{\vec{b}}{(\vec{a}+\vec{c},\vec{z}+*\vec{c})}\\
        &\qquad=\diffConditionedL{\vec{c}}{(\vec{a},\vec{z}+*\vec{c})}-\diffConditionedL{\vec{c}}{(\vec{a}+\vec{c},\vec{z}+*\vec{c})}.
    \end{align*}
    Both terms on the lower line are negligible by \Cref{claim:2pc-encoding-replacement-measured} since $\vec{c}$ is a standard basis vector. Therefore
    \begin{equation*}
        \left| 
            \diffConditionedL{\vec{b}}{(\vec{a},\vec{z}+*\vec{c})} 
            - \diffConditionedL{\vec{b}}{(\vec{a}+\vec{c},\vec{z}+*\vec{c})} 
        \right|
        =\negl
    \end{equation*}

    \medskip
    \noindent\emph{Changing the Phase with a Randomized Baseline Encoding \eqref{eq:gap-transport-coefficient-average}.}
    Since the baseline $(\vec{a} + *\vec{c}, \vec{z})$ adds $\vec{c}$ to the encoding with probability $1/2$, it is invariant under additions of $\vec{c}$. Therefore we can shift the distinguishing term by $\vec{c}$:
    \begin{equation*}
        \label{eq:gap-transport-randomized-change}
        \diffConditionedL{\vec{b}}{(\vec{a}+*\vec{c},\vec{z})}=\diffConditionedL{\vec{b}+\vec{c}}{(\vec{a}+*\vec{c},\vec{z})}.
    \end{equation*}
    Next, expanding the starred notation in \Cref{claim:2pc-encoding-replacement-measured} gives a method to shift the phase by $\vecb$ which is a row of $G$:
    \[
        \frac{1}{2}\left(\diffConditionedL{\vecb}{(\veca, \vecz)} + \diffConditionedL{\vecb}{(\veca, \vecz+\vecb)}\right)
        = 
        \diffConditionedL{\vecb}{(\veca, \vecz+*\vecb)}
        \approxnegl 0
    \]
    Combining the phase shift with the distinguishing term shift, we can transform
    \begin{alignat*}{2}
        \diffConditionedL{\vecb}{(\veca+*\vecc, \vec{z})}
        &\approxnegl & -\diffConditionedL{\vecb}{(\veca+*\vecc, \vec{z} + \vecb)}
        \\
        &= & -\diffConditionedL{\vecb + \vecc}{(\veca+*\vecc, \vec{z} + \vecb)}
        \\
        &\approxnegl &\diffConditionedL{\vecb+\vecc}{(\veca+*\vecc, \vec{z} + \vecc)}
        \\
        &= &\diffConditionedL{\vecb}{(\veca+*\vecc, \vec{z} + \vecc)}
    \end{alignat*}
    The first phase shift is by by $\vecb$, while the second is by $\vecb+\vecc$. Both are rows of $G$ since $\vecb$ and $\vecc$ are both standard basis vectors.
\end{proof}

So far, we have proved the property holds when the $\vec{b}$ is a row of $G$ and $\vec{c}$ is arbitrary. We now generalize it to general $\vec{b}$.

\begin{proof}[Proof of the General Version (\Cref{lem:2pc-encoding-replacement:gap-transfer-general})]

    First, we decompose $\vec{b}$ in terms of rows of $G$:
    \[
        \vec{b} = \sum_{i=1}^{m} \beta_i \vec{g_i}
    \]
    where $\vec{g_i}$ is the $i$'th row of $G$ and $\beta_i \in \bbF_2$. One such decomposition is given by $(\beta_1, \dots, \beta_N) = \vec{b}D$, since $\vec{b}DG = \vec{b}$.
    Next, we use these to define a path from $\vec{0}$ to $\vec{b}$. The $k$'th step along the path is
    \[
        \vec{s_k} \coloneqq  \sum_{i=1}^{k} \beta_i \vec{g_i}
        = \vec{s_{k-1}} + \beta_i \vec{g_i}
    \]
    We have $\vec{s_0} = \vec{0}$ and $\vec{s_{m}} = \vec{b}$.
    Now we expand  $\diffConditionedL{\vec{b}}{(\vec{a}, \vec{z})}$ by both adding and subtracting every intermediate step, to obtain
    \begin{align*}
         \diffConditionedL{\vec{b}}{(\vec{a}, \vec{z})} 
         &= \probone_{,\vecl}(\vec{a}, \vec{z}) - \probone_{,\vecl}(\vec{a} + \vec{b}, \vec{z})
         \\
         &= \sum_{k=0}^{m-1}\probone_{,\vecl}(\vec{a} + \vec{s_k}, \vec{z}) - \sum_{k=1}^{m}\probone_{,\vecl}(\vec{a} + \vec{s_{k}}, \vec{z})
         \\
         &= \sum_{k=0}^{m-1} \diffConditionedL{\beta_{k}\vec{g_k}}{(\vec{a} + \vec{s_k}, \vec{z})} 
    \end{align*}
    The last line follows by grouping the positive $k$'th and negative $(k+1)$'th terms.
    \Cref{claim:2pc-encoding-replacement:gap-transfer-limited} shows that translating the baseline in every gap $k$ by $(\vec{c}, \vec{c})$ only incurs negligible error $\mu_{k}(\secpar)$. We may group the $m$ negligible terms into one because $m = \poly[\secpar]$. 
    Then, expanding the $\Delta_{\vecl}$ terms and canceling the intermediate steps yields
    \begin{align*}
         \diffConditionedL{\vec{b}}{(\vec{a}, \vec{z})} 
         &= \sum_{k=0}^{m-1} \left(\diffConditionedL{\beta_{k}\vec{g_k}}{((\vec{a} + \vec{c})+ \vec{s_k}, \vec{z} + \vec{c})} \pm \mu_k(\secpar)\right)
         \\
         &= \left(\sum_{k=0}^{m-1}\probone_{,\vecl}((\vec{a} + \vec{c}) + \vec{s_k}, \vec{z} + \vec{c}) - \sum_{k=1}^{m}\probone_{,\vecl}((\vec{a} + \vec{c}) + \vec{s_{k}}, \vec{z} + \vec{c})\right) \pm \negl
         \\
         &= \left(\probone_{,\vecl}(\vec{a} + \vec{c}, \vec{z} + \vec{c}) - \sum_{k=1}^{m}\probone_{,\vecl}((\vec{a} + \vec{c}) + \vec{b}, \vec{z} + \vec{c})\right) \pm \negl
         \\
         &= \diffConditionedL{\vec{b}}{(\vec{a}+\vec{c}, \vec{z}+\vec{c})} \pm \negl
    \end{align*}
\end{proof}

\subsection{Security Against Honest-But-Curious $P_C$}

In this section we show that the functionality specification itself implies security against honest-but-curious classical parties $P_C$. Intuitively, the state-preservation property ensures honest-but-curious \emph{statistical} indistinguishability between any two inputs which have the same output distribution; otherwise, the classical party's registers would collapse a superposition between those two inputs. Then, the linearity allows the simulator to find some arbitrary input which matches the output.

We remark that it is also possible to show security by relying on the semi-honest zero-knowledge of \cite{KKR26}'s witness-preserving argument (inherited from GMW's 3-coloring ZK).

\begin{theorem}[Statistical Privacy against honest-but-curious $P_\C$]
\label{lem:linear-2pc-hbc-pc}
Protocol~\ref{prot:classical-linear} is statistically secure against any honest-but-curious $P^*_\C$. 
That is, there exists a QPT algorithm $\Sim$ with access to $P^*_\C$'s code and inputs $(L, C)$ which uses the output $\vec{u}$ prescribed by the ideal functionality to produce a view statistically indistinguishable from a real execution:
\[
    \{\calV^* :(\calV^*, \rho_u) \gets \twoPC\langle P_C^*(L, C), P_Q(\rho)\rangle\}
    \equiv
    \left\{\calV^* : 
    \begin{array}{c}
        \vec{u} \gets M[L](\rho)
        \\
         \calV^*\gets \Sim(P^*_\C, (L, C), \vec{u}) 
    \end{array}
    \right\}
\]
where 

\end{theorem}
\begin{proof}
We describe a simulator for the honest-but-curious view of $P^*_\C$.
The simulator begins with the input of $P^*_\C$, which consists of a linear function $L\in\{0,1\}^{\ell\times m}$ and a poly-size circuit $C:\{0,1\}^\ell\rightarrow\{0,1\}$.
It sends $(L,C)$ to the ideal functionality to get $\vecu$, then computes any $\vec{v}$ such that $L\vec{v}' = \vec{u}$. Finally, it interacts with $P^*_C$ in an honest execution of $\twoPC$ using $\ket{\vec{v}'}$ as the input of $P_Q$.

It suffices to show statistical indistinguishability of $P_C^*$'s view between $P_Q$'s use of any two inputs
\[
    \ket{\psi} = \sum_{u} \Pi_{\vec{u}} \ket{\psi}
    \quad \text{and}\quad
    \ket{\phi} = \sum_{u} \Pi_{\vec{u}} \ket{\phi}
\]
where for all $u$,
\[
    \Tr[\Pi_{\vec{u}}\ket{\psi}] = \Tr[\Pi_{\vec{u}}\ket{\phi}].
\]
It suffices to prove input-indistinguishability for $C$ which never aborts. For general $C$, the classical view is obtained from the always-deliver execution by recording the actual input $C$ and, whenever $C(\vec{u})=1$, truncating at that decision and appending the prescribed abort message.

Let $\rho,\sigma$ be two input states such that
\[
    \Tr(\Pi_{L,\vec{u}}\rho)=\Tr(\Pi_{L,\vec{u}}\sigma)=p_{\vec{u}}
    \qquad\text{for every }\vec{u}.
\]
Choose purifications $\ket{\psi}_{\calA\calB}$ and $\ket{\phi}_{\calA\calB}$ of $\rho$ and $\sigma$, respectively, and consider an honest execution on register $\calA$ of
\[
    \ket{\Omega}_{F\calA\calB}
    =\frac{\ket{0}_F\ket{\psi}_{\calA\calB}+\ket{1}_F\ket{\phi}_{\calA\calB}}{\sqrt{2}},
\]
where $F\calB$ is an untouched reference register. For every $\vec{u}$ with $p_{\vec{u}}>0$, the ideal normalized residual state is
\[
    \ket{\Omega_{\vec{u}}}
    =\frac{(I_F\otimes\Pi_{L,\vec{u}}\otimes I_\calB)\ket{\Omega}}{\sqrt{p_{\vec{u}}}}.
\]
By \Cref{lem:linear-2pc-correctness}, the test that, controlled by the classical output $\vec{u}$, projects $F\calA\calB$ onto $\ket{\Omega_{\vec{u}}}$ succeeds with probability at least $1-\varepsilon$, where $\varepsilon=\negl$; the test rejects aborts and outcomes with $p_{\vec{u}}=0$. By gentle measurement, conditioning on success changes the full joint state, including $P_\C^*$'s view, by trace distance at most $\sqrt{\varepsilon}$.

After this projection, conditioned on $\vec{u}$, register $F\calA\calB$ is pure and hence independent of $P_\C^*$'s view. Moreover, the equality of the two output distributions implies that measuring $F$ in $\ket{\Omega_{\vec{u}}}$ yields a uniform bit for every $\vec{u}$. Thus, after the projection, this bit is independent of the classical view even without conditioning on $\vec{u}$. Before the projection, measuring $F$ gives a uniform bit whose two conditional views are exactly the real views on inputs $\rho$ and $\sigma$. Contractivity of trace distance and the triangle inequality therefore bound the distance between these two views by $2\sqrt{\varepsilon}=\negl$.

Finally, for each $\vec{u}\in\operatorname{im}(L)$, fix a vector $\vec{v}'_{\vec{u}}$ satisfying $L\vec{v}'_{\vec{u}}=\vec{u}$, computable by Gaussian elimination. Apply the preceding argument to the actual input $\rho$ and
\[
    \sigma=\sum_{\vec{u}\in\operatorname{im}(L)}\Tr(\Pi_{L,\vec{u}}\rho)\ketbra{\vec{v}'_{\vec{u}}}.
\]
These states induce the same ideal output distribution, and the real view on $\sigma$ is exactly the simulator's view averaged over the ideal output. This proves the claimed statistical indistinguishability.

\end{proof}

\subsection{Input-Preserving Classical $2\PC$ for Constant Degree Functions}\label{sec:const-deg}
We note that our input-preserving classical $2\PC$ for linear functions can be easily extended to a $2\PC$ for  {\bf constant degree} functions.  Specifically, suppose the classical party $P_\C$ takes as input an $m$-variate polynomial $f$ of total degree $d$, where $d$ is a constant, and the quantum party takes as input a state $\sum_{\vecv\in\{0,1\}^m} \alpha_\vecv\ket{\vecv}$.  
The state preserving $2\PC$ protocol for computing $f(v)$ is identical to the linear $2\PC$, but where the quantum party, rather than using its input $\sum_{\vecv\in\{0,1\}^m} \alpha_\vecv\ket{\vecv}$, it uses the extended input $\sum_{\vecv\in\{0,1\}^m} \alpha_\vecv\ket{{\sf mon}_d(\vecv)}$ where ${\sf mon}_d(\vecv)$ consists of all the monomials of $\vecv$ of degree $\leq d$.  Namely, 
\[
{\sf mon}_d(\vecv)=(v_{i_1}\cdot\ldots \cdot v_{i_j})_{j\leq d, (i_1,\ldots,i_j)\in[m]^j}
\]
Similarly, the classical party, rather than using its polynomial $f$ as input, it uses its linearized version which is the linear function whose coefficients are all the monomials of $f$.  In addition, in both the indistinguishably-extractable argumentss the quantum party also proves that the preimages are of the form ${\sf mon}_d(\vecv)$ for some $\vecv$.  Therefore we obtain the following theorem
\begin{theorem}
    There exists a input-preserving classical $2\PC$ for all constant degree functions, assuming the existence of a noisy trapdoor claw-free function family with the adaptive 2-query hardcore predicate property and assuming a witness-preserving indistinguishably-extractable argument for $\NP$, both which can be instantiated under the $\LWE$ assumption.
\end{theorem}

\part{Classical Commitments to Quantum States}
\section{Remote State Preparation of Coset States}

In this section, we construct a remote-state preparation protocol for random coset states 
\[X^{x}Z^{z}\ket{S}.\] 
Our protocol allows the classical sender to choose $S$, while the $x$ and $z$ offsets are randomized.
Intuitively, the security guarantees that any information learned from the protocol could also be learned from \emph{directly} receiving the coset state, along with oracle access to $S + x$ and $S^\perp + z$.

\paragraph{Notation.} We briefly recall some notation for coset states from \Cref{sec:coset-auth-construction}. $\Can(S)\subset \bbF_2^d$ is a set of canonical elements of each $S+x$ and is isomorphic to the quotient group $\bbF_2^d/S = \{S + x: x\in \bbF_2^d\}$. We denote one such isomorphism by $\Can_S:\bbF_2^d\rightarrow \bbF_2^d$. Thus, $\Can_S(x)$ identifies a unique canonical element of $S + x$.

\subsection{Definition}
A remote state preparation protocol for coset states $\RSP_d$ is an interactive protocol between a classical sender $C$ and a quantum sender $Q$. It is parameterized by a polynomial $d = d(\secpar)$ representing the dimension of the ambient space $\bbF_2^d$.

Both parties input the security parameter $1^\secpar$ in unary.
The classical party additionally inputs a description of a subspace $S\in \bbF_2^{d}$.
The quantum party $Q$ outputs a register $\calR$ containing a coset state $X^x Z^z \ket{S}$ and the classical party $C$ outputs the description $(S, x, z)$ of the coset state.

We denote an execution of $\RSP_d$ with quantum output register $\calR$ and classical output $(x, z)$ by
\[
    (\calR, (x, z)) \gets \RSP_d\langle Q, C(S)\rangle(1^\secpar)
\]

A remote state preparation protocol for coset states must satisfy \emph{correctness} and \emph{security}. 

\begin{definition}[Coset RSP Correctness]
    A remote state preparation protocol $\RSP_d$ for coset states is correct with respect to a family of subspace sets $\{\calS_{\secpar}\}_{\secpar\in \bbN}$ if for all $\secpar \in \bbN$ and all $S \in \calS_{\secpar}$,
    \[
        \Pr\left[\calR \textnormal{ contains } X^x Z^z \ket{S}: (\calR, (S, x, z)) \gets \RSP\langle Q, C(S)\rangle(1^\secpar)\right] = 1-\negl
    \]
\end{definition}

\noindent For any coset $S + x$, we denote the oracle which decides membership in $S + x$ as $\calO_{S + x}$.

\begin{definition}[Coset RSP Security]
    A remote state preparation protocol $\RSP_d$ for a family of subspace sets $\{\calS_{\secpar}\}_{\secpar\in \bbN}$ is \emph{secure up to membership oracles}
    if there exists a $\QPT$ algorithm $\Sim$ such that for all adversarial quantum receivers $Q^*$, all sufficiently large $\secpar \in \bbN$ and all $S \in \calS_{\secpar}$,
    \[
        \left\{
            (\calR^*, Y) \gets \RSP_d\langle Q^*, C(S)\rangle\left(1^\secpar\right)
        \right\}
        \approx_c
        \left\{
            \left(\calR^*, 
            Y \right) :
            \begin{array}{c}
                x\gets \Can(S), z\gets \Can(S^\perp)
                \\
                (\calR^*, a) \gets \Sim^{\calO_{S + x}, \calO_{S^\perp + z}}\big(Q^*, X^{x}Z^{z} \ket{S} \big)
            \end{array}
        \right\}
    \]
    where the classical sender's output is 
    \[
        Y = \begin{cases}
                (x, z) & \textnormal{if } C \textnormal{ accepts (respectively, } a = 0\textnormal{)}
                \\
                \bot &\textnormal{if } C \textnormal{ rejects (respectively, } a = 1\textnormal{)}
            \end{cases}
    \]
    
    \noindent We say it is $\epsilon$-secure up to membership oracles if there exists a uniform family of $\poly[\secpar, n]$-sized QPT algorithms $\{\{\Sim_{\secpar, n}\}_{n\in \bbN}\}_{\secpar\in \bbN}$ such that the two distributions above (where $\Sim$ is replaced by $\Sim_{\secpar, n}$) are $1/n$-indistinguishable for every $\secpar \in \bbN$ and $n = \poly[\secpar]$.
    
\end{definition}

\noindent Note that we require the \emph{joint} distribution of the receiver's view and the sender's output is indistinguishable.

\subsection{Construction}

Before describing our protocol, we first give some intuition about the construction. Suppose that we instead wanted to construct states of the simpler form 
\[
    X^x \ket{S}.
\]
Such a state can be constructed by using a state-preserving 2PC for linear functions to partition a $\ket{+^\secpar}$ state into cosets $S+x$ of $S$, then measuring which coset the state is supported on. In other words, measure $\Can_S(\bullet)$ on a $\ket{+^\secpar}$ state.

Although this approach provides a good starting point, a single repetition does not suffice for our purposes. 
In fact, an adversarial receiver can even \emph{control} the $x$ offset which the sender outputs by simply inputting $\ket{x}$ instead of $\ket{+^\secpar}$ to the 2PC.
The solution to hiding $x$ lies in the missing $z$ offset. To add a (mostly) random $z$ offset, we can add an extra measurement round, implemented via 2PC, and slightly adjust the first round as follows:
\begin{enumerate}
    \item Measure the coset of a random $R\subset S$ to get $x_R$ and residual state $\ket{R+x_R}$.
    \item Hadamard $\ket{R + x_R}$ to get $Z^{x_R}\ket{R}$, then measure the coset of $S^\perp$ to get $z$.
\end{enumerate}
Now the $z$ offset is randomized within $R^\perp \supset S^\perp$. Moreover, if the adversary is forced to input $Z^{x_R}\ket{R^\perp}$, then they have \emph{no control} over $z$. We force this by checking that the input to the second 2PC is supported on $R^\perp$.
Intuitively, the adversary should not be able to find a vector in $R^\perp$, \emph{except} by inputting $H \ket{R + x_R}$.

Now we have forced a random $z$ which is hidden within $R^\perp$. The adversary also cannot directly pick $x$ by inputting $\ket{x}$ to the first measurement because that would leave them unable to find a valid input in $R^\perp$ for the second measurement. This does not rule out indirect influence over $x$, however.

For achieving simulation security, we also add a third round to be able to treat $x$ in the same way as $z$. Concretely, the three measurements become:
\begin{enumerate}
    \item Measure a coset of a random $R \subset S$ to get $x_R$.
    \item Check membership in $R^\perp$, then measure a coset of a random $T^\perp \subset S^\perp$ to get $z$.
    \item Check membership in $T + x_R$, then measure a coset of $S$ to get $x$.
\end{enumerate}
In the final version, $R^\perp$ randomizes $z$, while $T$ randomizes $x$. Moreover, $R^\perp$ and $T$ are already quite randomized, resulting in nearly uniform $x$ and $z$. We describe the protocol formally below.

\begin{protocol}[Remote State Preparation of Coset States]
\label{protocol:rsp-auth}
    The parties start with the security parameter $1^\secpar$ and the parameterized polynomial $d=d(\secpar)$. Additionally, the classical sender $C$ starts with a subspace $S \subset \bbF_2^{d}$ such that $\dim(S) \in [2\secpar, d - 2\secpar + O(1)]$.
    The additive constant in the upper bound is fixed independently of $\secpar$ and $d$; the construction is defined for all sufficiently large $\secpar$.
    The parties interact as follows.
    \begin{enumerate}
        \item \textbf{Classical Sender:} 
        Sample subspaces $R$ and $T$ of dimension $\secpar$ and $d - \secpar$, respectively, such that $R \subset S \subset T$.\footnote{For example, we may set $d = 4\secpar$, $\dim(T) = 3\secpar$, $\dim(S) = 2\secpar$, and $\dim(R) = \secpar$.}

        \item \textbf{Quantum Receiver:} 
        Prepare the uniform superposition $\frac{1}{\sqrt{2^{d}}}\sum_{v \in \bbF_2^{d}} \ket{v}$ in register $\calV$.

        \item \label{step:com-first-2pc} \textbf{First 2PC:} 
        Do a state-preserving 2PC with the following functionality and inputs.
        \begin{functionalitybox}
            \textbf{Quantum Party Input:} Register $\calV$.

            \textbf{Classical Party Input:} The subspace $R$.

            \textbf{Functionality:} On computational-basis input $v$, output
            \[
                x_{R} = \Can_{R}(v).
            \]
        \end{functionalitybox}
        If the 2PC output is $x_R$, then the quantum receiver's resulting state in register $\calV$ is
        \begin{align*}
            \ket{R + x_{R}}
            = X^{x_{R}}\ket{R}
        \end{align*}

        \item \textbf{Quantum Receiver:} 
        Hadamard register $\calV$. At this point, an honest quantum receiver's state is
        \[
            Z^{x_{R}} \ket{R^\perp}
        \]

        \item \label{step:com-second-2pc} \textbf{Second 2PC:} 
        Do a state-preserving 2PC with the following functionality.
        \begin{functionalitybox}
            \textbf{Quantum Party Input:} The register $\calV$.

            \textbf{Classical Party Input:} The subspaces $R$ and $T^\perp$.

            \textbf{Functionality:} On computational-basis input $v$, output
            \[
                z_{T^\perp}
                =
                \begin{cases}
                    \Can_{T^\perp}(v) &\text{if } v \in R^\perp \backslash S^\perp 
                    \\
                    \bot &\text{else}
                \end{cases}.
            \]
        \end{functionalitybox}
        If any output is $\bot$, abort. Otherwise, the classical sender's output is $z$ where $z_{T^\perp} \in R^\perp$ and the quantum receiver's state is
        \[
            Z^{x_{R}} \ket{T^\perp + z_{T^\perp}}
            =
            Z^{x_{R}} X^{z_{T^\perp}} \ket{T^\perp}
        \]

        \item \textbf{Quantum Receiver:} Hadamard register $\calV$.
        At this point, an honest quantum receiver's state in each register $\calR_{\outp, i}$ is
        \[
             X^{x_R} Z^{z_{T^\perp}} \ket{T}
        \]

        \item \label{step:com-third-2pc} \textbf{Third 2PC:} 
        Do a state-preserving 2PC with the following functionality.
        \begin{functionalitybox}
            \textbf{Quantum Party Input:} The register $\calV$.

            \textbf{Classical Party Input:} The subspace $S$ and the coset $T + x_R$.

            \textbf{Functionality:} On computational-basis input $v$, output
            \[
                x =
                \begin{cases}
                    \Can_{S}(v) & \text{if } v \in (T + x_{R}) \backslash (S + x_{R})
                    \\
                    \bot & \text{else}
                \end{cases}.
            \]
        \end{functionalitybox}
        If any output is $\bot$, abort. Otherwise, the classical sender's output is $x$ where $x \in T + x_{R}$ and the quantum receiver's state is
        \[
            X^{x} Z^{z_{T^\perp}} \ket{S}
        \]
        
        We can equivalently interpret $Z^{z_{T^\perp}}$ as $Z^{z}$ for any $z \in S^\perp + z_{T^\perp}$. 

        \item \textbf{Classical Sender:} Compute $z = \Can_{S^\perp}(z_{T^\perp})$, then output $(S, x, z)$.

        \item \textbf{Quantum Receiver:} Output register $\calV$.
    \end{enumerate}
\end{protocol}

\begin{theorem}
    If $\twoPC$ is an $\epsilon$-secure (respectively, secure) state-preserving 2PC for linear functions with post-processed abort, then \Cref{protocol:rsp-auth} is an $\epsilon$-secure (respectively, secure) remote state preparation protocol for coset states for all polynomials $d$ and all $S$ such that $\dim(S) \in [2\secpar, d - 2\secpar + O(1)]$.
\end{theorem}

\begin{proof}
    We begin with correctness.
    The evolution of the receiver's state when no abort occurs is documented as part of the protocol description. By definition, $z = s + z_{T^\perp}$ for some $s \in S^\perp$, so $X^{x} Z^{z_{T^\perp}} \ket{S} = X^{x} Z^{z} \ket{S}$.
    
    An abort occurs only if the quantum receiver inputs $v \in S^\perp$ in the \textbf{second 2PC} or if they input $v \in S+ x_{R}$ in the \textbf{third 2PC}. For $\dim(S)\leq d-2\secpar$, these events happen with negligible probability because $v$ is a uniform element from $R^\perp$ and $T + x_{R}$, respectively, and the two failure cases form a negligible portion of these spaces: $|S^\perp|/|R^\perp| \leq 2^{-\secpar}$ and $|S + x_{R}|/|T + x_{R}| \leq 2^{-\secpar}$.
    Increasing $\dim(S)$ by a constant above $d-2\secpar$ multiplies the latter bound by at most $2^{O(1)}$, so it remains negligible.

    We show security in \Cref{sec:coset-rsp-security}.
\end{proof}

\subsection{Security}\label{sec:coset-rsp-security}
Before proving security, we first prove an equivalence between measurement coset representatives and applying a $Z^{z}$ operation for a random $z\gets S^\perp$. This generalizes the well-known equivalence between measuring in the computational basis and randomly applying a $Z$ gate.
\begin{lemma}\label{lem:measure-canonical-equiv-phase}
    For any subspace $S \subset \bbF_2^n$ and any mixed state $\rho$, the following two mixed states are equivalent:
    \begin{itemize}
        \item Measure $\Can_{S}(\rho)$, discard the measurement result, and output the residual state.
        \item Sample $s \gets S^\perp$ and output $Z^s \rho Z^s$.
    \end{itemize}
\end{lemma}
\begin{proof}
    By linearity, it suffices to consider the effect on any $\ket{x}\!\bra{y}$ term. The measurement channel preserves the term if $S + x = S + y$, i.e. if $x + y \in S$, and otherwise erases the term. The phase channel results in
    \[
        \ket{x}\!\bra{y} \frac{1}{|S^\perp|}\sum_{s\in S^\perp} (-1)^{s \cdot x + s \cdot y}
    \]
    Similarly, the summation becomes $1$ if $x + y \in S$ and $0$ otherwise.
\end{proof}
The equivalence of measurement and random $Z$ operations will play a crucial role in the proof of security for remote state preparation, which we give next.

\subsubsection{The Simulator}
    The simulator $\Sim(\adv)$ works as follows.
    \begin{enumerate}
        \item Take as input the state
        \[
            X^{x'}Z^{z'} \ket{S}
        \]
        in register $\calC$.
        
        \item Run the adversary until the \textbf{first 2PC} (step \ref{step:com-first-2pc}).

        \item \textbf{First 2PC.} Run the 2PC simulator for the following functionality.
        \begin{functionalitybox}
            \textbf{Quantum Party Input:} Register $\calV_1$.

            \textbf{Functionality:} Apply $\CX$ from register $\calV_1$ to register $\calC$.
        \end{functionalitybox}

        \item Run the adversary until the \textbf{second 2PC} (step \ref{step:com-second-2pc}).

        \item \textbf{Second 2PC.} Run the 2PC simulator for the following functionality.
        \begin{functionalitybox}
            \textbf{Quantum Party Input:} Register $\calV_2$.

            \textbf{Functionality:} 
            \begin{enumerate}
                \item Query $\calO_{S^\perp + z}$ on $\calV_2$. If it rejects, immediately return $\calV_2$ to $Q^*$ and output $Q^*$'s internal state.
                \item Otherwise, apply $\CZ$ from register $\calV_2$ to register $\calC$.
            \end{enumerate}
        \end{functionalitybox}

        \item Run the adversary until the \textbf{third 2PC} (step \cref{step:com-third-2pc}).

        \item \textbf{Third 2PC.} Run the 2PC simulator for the following functionality.
        \begin{functionalitybox}
            \textbf{Quantum Party Input:} Register $\calV_3$.

            \textbf{Functionality:} 
            \begin{enumerate}
                \item Query $\calO_{S + x}$ on $\calV_3$. If it rejects, immediately return $\calV_3$ to $Q^*$ and output $Q^*$'s internal state.
                \item Otherwise, apply $\CX$ from register $\calV_3$ to register $\calC$.
            \end{enumerate}
        \end{functionalitybox}

        \item \textbf{Output.} Output the adversary's internal register $\calR^*$.
    \end{enumerate}

\subsubsection{Indistinguishability of the Simulator}

 We now show indistinguishability of $\Ideal(\Sim(\adv), 1^\secpar)$ from the real experiment via a series of hybrid experiments. 
The ideal functionality prescribes the output $(x'_\Can, z'_\Can) \in \Can(S) \times \Can(S^\perp)$ to the classical party and sends $X^{x'_{\Can}}Z^{z'_\Can}\ket{S}$ to $\Sim$. 
It is equivalent, up to a global phase, to assume that $\Sim$ instead receives $X^{x'}Z^{z'}\ket{S}$ where $x' \gets S + x'_\Can$ and $z' \gets S^\perp + z'_\Can$.

We divide the hybrids into 4 high-level phases: 1) setup, 2) force the $Z$ key to be $z'_\Can$, 3) force the $X$ key to be $x_\Can'$, and 4) replace the 2PC ideal functionalities by $\CX$ and $\CZ$ operations with the prescribed state $X^{x'}Z^{z'}\ket{S}$.
\begin{itemize}
    \item \underline{$\Hyb_{0}$} is the real execution of the remote state preparation protocol. Its output is $(\calR^*, Y)$ for $Y \in \{(x, z), \bot\}$, the joint distribution of the adversary's state along with the classical sender's output $Y$.
    
    \item \underline{$\Hyb_1$} changes $\Hyb_{0}$ by running the 2PC simulator for each of the three 2PC executions and internally computing the ideal functionalities for them. We set the $\epsilon$-security for each simulator to be $1/3n$. 

    \item \underline{$\Hyb_2$} changes $\Hyb_{1}$ by modifying the outputs of the \textbf{first} and \textbf{second 2PCs}. 
    Specifically, compute the ideal functionality for both 2PCs as before, then in the output step:
    \begin{itemize}
        \item \textbf{First 2PC:} Instead of outputting $x_{R} = \Can_{R}(v)$, output $x_{S} \coloneqq \Can_S(x_{R})$.
        \item \textbf{Second 2PC:} Instead of outputting $z_{T^\perp} = \Can_{T^\perp}(v)$, output $z \coloneqq \Can_{S^\perp}(z_{T^\perp})$.
    \end{itemize}
\end{itemize}

\paragraph{Forcing $\vec{z'}$.} 
The next set of hybrids modifies the ideal functionalities of the \textbf{first} and \textbf{second 2PCs} to force the classical sender's output to be $(x, z'_\Can)$ where $z'_\Can$ is sampled uniformly from $\Can(S^\perp)$ instead of being chosen by the adversary in the second 2PC.
\begin{itemize}
    \item $\underline{\Hyb_3}$ modifies the ideal functionality of the \textbf{first 2PC}. 
    On computational basis input $v_1$:
    \begin{enumerate}
        \item \new Measure $x_{S} \gets \Can_{S}(v_1)$.
        \item \modified Sample $z' \gets R^\perp$. 
        Apply $Z^{z'}$ to register $\calV$. 
        This replaces the measurement with respect to $R$.
        \item Output $x_{S}$.
    \end{enumerate}

    \item \underline{$\Hyb_4$} changes the distribution of $z'$. In $\Hyb_3$, one may equivalently first sample $z'$ from its marginal distribution, then sample $R^\perp \supset S^\perp$ subject to $z' \in R^\perp$.

    In $\Hyb_4$, instead of sampling $z'$ from its marginal distribution, we sample it uniformly from $\bbF_2^{d}$.
    Equivalently, sample $z'_\Can\gets \Can(S^\perp)$ and $z' \gets S + z'_\Can$.
    
    \item $\underline{\Hyb_5}$ changes ideal functionality of the \textbf{second 2PC}.
    On computational basis input $v_2$:
    \begin{enumerate}
        \item \modified Check that $v_2 \in S^\perp + z'$. If not, output $\bot$.\footnote{Outputting $\bot$ here causes an abort in the overall protocol.}
        \item Measure $z_{T^\perp} \gets \Can_{T^\perp}(v_2)$. Discard the measurement result.
        \item \modified Output $z'_{\Can}$.
    \end{enumerate}

    Note that the check forces the output to be $z'_{\Can} = \Can_S(z')$, which is eventually output as part of the classical party's output $(S, x, z'_{\Can})$.
\end{itemize}

\paragraph{Forcing $x'$.}
The next set of hybrids modifies the ideal functionalities of the \textbf{second} and \textbf{third 2PCs} to force the classical sender's output to be $(x'_{\Can}, z'_\Can)$ where $x'_{\Can}$ is sampled uniformly from $\Can(S)$ instead of being chosen by the adversary in the third 2PC.
\begin{itemize}
    \item $\underline{\Hyb_6}$ modifies the ideal functionality of the \textbf{second 2PC}. 
    On computational basis input $v_2$:
    \begin{enumerate}
        \item Check that $v_2 \in S^\perp + z'$. If not, output $\bot$.
        \item \modified Sample $x'' \gets T^\perp$. 
        Apply $Z^{x''}$ to register $\calV$. 
        This replaces the measurement with respect to $T^\perp$.
        \item Output $z'_{\Can}$.
    \end{enumerate}

    \item \underline{$\Hyb_7$} changes the distribution of $x''$. In $\Hyb_6$, one may equivalently first sample $x''$ from its marginal distribution, then sample $T \supset S$ subject to $x'' \in T$. 

    In $\Hyb_7$, instead of sampling $x''$ from its marginal distribution, we sample it uniformly from $\bbF_2^{d}$.
    
    \item $\underline{\Hyb_8}$ changes ideal functionality of the \textbf{third 2PC}.
    On computational basis input $v_3$:
    \begin{enumerate}
        \item \modified Check that $v_3 \in S + x_S + x''$. If not, output $\bot$.
        \item Measure $x \gets \Can_{S}(v_3)$. Discard the measurement result.
        \item \modified Output $\Can_S(x_S + x'')$.
    \end{enumerate}

    Note that the check forces the output to be $x'' + x_S = $, which is eventually output as part of the classical party's output $(S, x_S + x'', z')$.

    \item \underline{$\Hyb_9$} changes the distribution of $x''$ again. Now sample $x' \gets \bbF_2^{d}$ and set $x'' = x' - x_S$.

    Note that this modifies the \textbf{third 2PC} to check membership in $S + x'$ and to always output either $\bot$ or $x'_\Can = \Can_S(x')$.
\end{itemize}

\paragraph{Switching to Controlled $X$ and $Z$.}
Finally, we switch the remaining measurements and $Z$ operations to use the prescribed $X^{x'} Z^{z'} \ket{S}$ state as a black box.
\begin{itemize}
    \item \underline{$\Hyb_{10}$} changes the ideal functionality of the \textbf{first 2PC}. On input a register $\calV_1$:
    \begin{enumerate}
        \item Measure $x_S \gets \Can_S(\cdot)$ from register $\calV_1$.
        \item \modified Apply $\CX$ from register $\calV_1$ to register $\calC$.
        \item Output $x_S$.
    \end{enumerate}

    \item \underline{$\Hyb_{11}$} changes the ideal functionality of the \textbf{second 2PC}. On input a register $\calV_2$:
    \begin{enumerate}
        \item Check that $\calV_2$ is supported entirely on $S^\perp + z'$ using $\calO_{S^\perp + z'}$. If not, abort the protocol.
        \item \modified Apply $\CZ$ from register $\calV_2$ to register $\calC$.
        \item Output $z'_{\Can}$.
    \end{enumerate}

    \item \underline{$\Hyb_{12}$} changes the ideal functionality of the \textbf{third 2PC}. On input a register $\calV_3$:
    \begin{enumerate}
        \item Check that $\calV_3$ is supported entirely on $S + x'$ using $\calO_{S + x'}$. If not, abort the protocol.
        \item \modified Apply $\CX$ from register $\calV_3$ to register $\calC$.
        \item Output $x'_{\Can}$.
    \end{enumerate}

    \item \underline{$\Ideal(\Sim, 1^\secpar)$} changes $\Hyb_{12}$ as follows.
    \begin{itemize}
        \item Remove all output steps from the three 2PC functionalities (not including abort checks). Directly output $(x'_{\Can}, z'_{\Can})$ as the classical sender's output.
        \item Do not measure $x_S \gets \Can_S(\cdot)$ on register $\calV_1$ during the \textbf{first 2PC}.
    \end{itemize}
\end{itemize}

$\Hyb_0$ is $1/n$-indistinguishable from $\Hyb_1$ by the security of the 2PC, invoked 3 times sequentially. We show indistinguishability of the remaining hybrids in \Cref{claim:hyb1-2,claim:hyb2-3,claim:hyb3-4,claim:hyb4-5,claim:hyb5-6,claim:hyb6-7,claim:hyb7-8,claim:hyb8-9,claim:hyb9-10,claim:hyb10-11,claim:hyb11-12,claim:hyb12-ideal}. By triangle inequality, $\Hyb_0$ can be distinguished from $\Ideal(\Sim, 1^\secpar)$ with at most $(1/n + \negl)$ probability for any QPT receiver.

\begin{claim}\label{claim:hyb1-2}
    $\Hyb_1$ and $\Hyb_2$ are identically distributed.
\end{claim}
\begin{proof}
    $z_{T^\perp}$ is only used to compute the output $z = \Can_{S^\perp}(z_{T^\perp})$ even in the real protocol. $x_R$ is only used to check membership in $T+x_R$. Since $S \subset T$, the cosets $T+x_R = T + x_S$ are equal and so it is equivalent to use $x_S$ instead.
\end{proof}

\begin{claim}\label{claim:hyb2-3}
    $\Hyb_2$ and $\Hyb_3$ are identically distributed.
\end{claim}
\begin{proof}
    First, observe that $x_S$ may be directly measured from $v_1$ without modifying the experiment. Then, measuring and discarding $x_{R} \gets \Can_R(v_1)$ is equivalent to inserting a random $Z^{z'}$ where $z' \gets R^\perp$ by \Cref{lem:measure-canonical-equiv-phase}.
\end{proof}

\begin{claim}\label{claim:hyb3-4}
    $\Hyb_3$ and $\Hyb_4$ are statistically close.
\end{claim}
\begin{proof}
    The marginal distribution of $z'$ is 
    \[
        \begin{cases}
            z' \gets S^\perp &\textnormal{with probability } \frac{|S^\perp|}{|R^\perp|} \leq 2^{-\secpar}
            \\
            z' \gets \bbF_2^{d}\backslash S^\perp &\textnormal{with probability } 1 - \frac{|S^\perp|}{|R^\perp|}
        \end{cases}
    \]
    Since $\dim(S^\perp) \leq d - 2\secpar$, the total variation distance from uniform is
    \begin{align*}
        \frac{1}{2}|S^\perp| \left|\frac{\frac{|S^\perp|}{|R^\perp|}}{|S^\perp|}- \frac{1}{|\bbF_2^{d}|} \right| + \frac{1}{2}|\bbF_2^{d}\backslash S^\perp|\left|\frac{1 - \frac{|S^\perp|}{|R^\perp|}}{|\bbF_2^{d}\backslash S^\perp|} - \frac{1}{|\bbF_2^{d}|}\right|
        &= |S^\perp|\frac{2^{\secpar} - 1}{2^{d}}
        \\
        &\leq
        2^{-\secpar}
    \end{align*}
\end{proof}

\begin{claim}\label{claim:hyb4-5}
    $\Hyb_4$ and $\Hyb_5$ are statistically close.
\end{claim}
\begin{proof}
    The primary difference between the two experiments is the check during the \textbf{second 2PC}.
    \begin{itemize}
        \item $\Hyb_4$ checks $v_2 \in R^\perp \backslash S^\perp$.
        \item $\Hyb_5$ checks $v_2 \in S^\perp + z'$.
    \end{itemize}
    Observe that the experiments are identical whenever $v_2 \notin R^\perp \backslash S^\perp$, in which case they both output $\bot$, or whenever $v_2 \in S^\perp + z'$, in which case 
    they both output $\Can_{S^\perp}(v_3) = z'_{\Can}$.
    Thus, it suffices to show that $v_2 \in R^\perp \backslash (S^\perp + \Span(z'))$ with negligible probability.

    Observe that before the \textbf{second 2PC}, only $S$ and $z'$ are used, as part of the \textbf{first 2PC}. 
    Therefore we may consider sampling $R^\perp \supset (S^\perp + \Span(z'))$ \emph{after} the adversary chooses its inputs for the \textbf{second 2PC}.

    Thus, the probability of a divergence in the two hybrid experiments is bounded by
    \[ 
        \max_{v_2} \Pr_{R^\perp \supset (S + \Span(z'))}\left[v_2 \in R^\perp \backslash (S^\perp + \Span(z'))\bigg| S, z'\right]
    \]
    The probability is $0$ for all $v_2 \in (S^\perp + \Span(z'))$. 
    Otherwise, the probability is identical for all $v_2 \notin (S^\perp + \Span(z'))$, using the observation that the distribution of $R^\perp$ is invariant under all linear transformations which leave $S^\perp + \Span(z')$ invariant, and in particular linear transformations swapping any pair $v \leftrightarrow v'$ not in $S^\perp + \Span(z')$. 
    Therefore this probability is $\frac{|R^\perp\backslash (S^\perp + \Span(z'))|}{|\bbF_2^{d}\backslash(S^\perp + \Span(z'))|} < 2^{-\secpar+1}$.
\end{proof}

\begin{claim}\label{claim:hyb5-6}
    $\Hyb_5$ and $\Hyb_6$ are identically distributed.
\end{claim}
\begin{proof}
    This is immediate from the equivalence of measuring then discarding $x_{R} \gets \Can_R(v_1)$ and inserting a random $Z^{z'}$ where $z' \gets R^\perp$ (\Cref{lem:measure-canonical-equiv-phase}).
\end{proof}

\begin{claim}\label{claim:hyb6-7}
    $\Hyb_6$ and $\Hyb_7$ are statistically close.
\end{claim}
\begin{proof}
    First assume $\dim(S)\leq d-2\secpar$.
    The marginal distribution of $x''$ is
    \[
        \begin{cases}
            x'' \gets S &\textnormal{with probability } \frac{|S|}{|T|} \leq 2^{-\secpar}
            \\
            x'' \gets \bbF_2^{d}\backslash S &\textnormal{with probability } 1 - \frac{|S|}{|T|}
        \end{cases}
    \]
    Since $\dim(S) \leq d - 2\secpar$, the total variation distance from uniform is
    \begin{align*}
        \frac{1}{2}|S| \left|\frac{\frac{|S|}{|T|}}{|S|}- \frac{1}{|\bbF_2^{d}|} \right| + \frac{1}{2}|\bbF_2^{d}\backslash S|\left|\frac{1 - \frac{|S|}{|T|}}{|\bbF_2^{d}\backslash S|} - \frac{1}{|\bbF_2^{d}|}\right|
        &= |S|\frac{2^{\secpar} - 1}{2^d}
        \\
        &\leq 2^{-\secpar}
    \end{align*}
    Allowing $\dim(S)\leq d-2\secpar+O(1)$ multiplies the final bound by at most $2^{O(1)}$, which preserves negligibility.
\end{proof}

\begin{claim}\label{claim:hyb7-8}
    $\Hyb_7$ and $\Hyb_8$ are statistically close.
\end{claim}
\begin{proof}
    The primary difference between the two experiments is the check during the \textbf{third 2PC}.
    \begin{itemize}
        \item $\Hyb_7$ checks $v_3 \in (T + x_S) \backslash (S + x_S)$.
        \item $\Hyb_8$ checks $v_3 \in S + x'' + x_S$.
    \end{itemize}
    Observe that the experiments are identical whenever $v_3 \notin (T + x_S) \backslash (S + x_S)$, in which case the \textbf{third 2PC} outputs $\bot$, or whenever $v_3 \in S + x''+ x_S $, in which case the the \textbf{third 2PC} outputs $\Can_S(v_3) = \Can_{S^\perp}(x_S + x'')$.
    Thus, it suffices to show that $v_3 \in (T + x_S) \backslash (S + \Span(x'') + x_S)$ with negligible probability.

    Observe that before the \textbf{third 2PC} only the following information is used:
    \begin{itemize}
        \item \textbf{First 2PC}: $S$ and $z' \gets \bbF_2^{d}$.
        \item \textbf{Second 2PC:} $S^\perp$ and $x''$ such that $x'' \in T$.
    \end{itemize}
    Therefore we may consider sampling $T \supset (S + \Span(x''))$ \emph{after} the adversary chooses its inputs for the \textbf{third 2PC}.

    Thus, the probability of a divergence in the two hybrid experiments is bounded by
    \[ 
        \max_{v_3} \Pr_{T \supset (S + \Span(x''))}\left[(v_3 - x_S) \in T \backslash S + \Span(x'')| S, x'', x_S\right]
    \]
    The probability is $0$ for all $v_3 \in (S + \Span(x''))$. 
    Otherwise, the probability is identical for all $v_3 \notin (S + \Span(x''))$, using a similar argument to the proof of \Cref{claim:hyb4-5}. Therefore this probability is 
    $\frac{|T \backslash (S + \Span(x''))|}{|\bbF_2^{d}\backslash(S + \Span(x''))|} < 2^{-\secpar+1}$.
\end{proof}

\begin{claim}\label{claim:hyb8-9}
    $\Hyb_8$ and $\Hyb_9$ are identically distributed.
\end{claim}
\begin{proof}
    The uniform distribution is invariant under linear shifts.
\end{proof}

\begin{claim}\label{claim:hyb9-10}
    $\Hyb_{9}$ and $\Hyb_{10}$ are identically distributed.
\end{claim}
\begin{proof}
    Consider the hybrid experiment $\Hyb_9'$ where at the end of the \textbf{first 2PC}, the simulator applies $Z^{z'}X^{x_S}Z^{z'}$ to register $\calC$. This is equivalent since register $\calC$ is disjoint from the adversary's view. 

    The two hybrids differ only in the \textbf{first 2PC}. It suffices to show that the following two operations are equivalent on the residual state.
    \begin{itemize}
        \item $\Hyb_{9}$: Apply $Z^{z'}_{\calV_1} \otimes (Z^{z'}X^{x_S}Z^{z'})_{\calC}$ to registers $\calV_1 \otimes \calC$.
        \item $\Hyb_{10}$: Apply $\CX_{\calV_1 \rightarrow \calC}$ to registers $\calV_1 \otimes \calC$.
    \end{itemize}
    After the measurement, the state on registers $\calV_1 \otimes \calC$ is supported on
    \[
        \ket{s + x_S}_{\calV_1} \otimes X^{x'}Z^{z'}\ket{S}_{\calC}
    \]
    for $s \in S$. By linearity it suffices to show equivalence on such states.
    We may write $\CX$ as
    \[
        \CX = \sum_{w} Z^{z'} \ketbra{w} \otimes Z^{z'} X^{w} Z^{z'}
    \]
    for any $z'$. Incorporating the existing $X^{x'}Z^{z'}$ and observing that $X^{s}$ commutes with $Z^{z'}X^{x'}Z^{z'}$, the state in $\Hyb_{10}$ becomes
    \[
        \left(Z^{z'} \otimes Z^{z'} X^{x_S} Z^{z'}\right)\left(\ket{s + x_S}_{\calV_1} \otimes X^{x'}Z^{z'}\left(X^{s}\ket{S}_{\calC}\right)\right)
    \]
    Finally, $X^{s}\ket{S} = \ket{S}$ since $s \in S$. Therefore the net effect of $\Hyb_{10}$ is to apply $Z^{z'} \otimes Z^{z'} X^{x_S} Z^{z'}$, the same as $\Hyb_9'$.
\end{proof}

\begin{claim}\label{claim:hyb10-11}
    $\Hyb_{10}$ and $\Hyb_{11}$ are identically distributed.
\end{claim}
\begin{proof}
    Consider the hybrid experiment $\Hyb_{10}'$ where at the end of the \textbf{second 2PC}, the simulator applies $X^{x' + x_S}Z^{z'}X^{x' + x_S}$ to register $\calC$. This is equivalent since register $\calC$ is disjoint from the adversary's view after the \textbf{first 2PC}.

    The two hybrids differ only in the \textbf{second 2PC}. If the $v_2 \in S^\perp + z'$ check rejects, then the two hybrids are identical. Otherwise, it suffices to show that the following two operations are equivalent on the residual state.
    \begin{itemize}
        \item $\Hyb_{10}$: Apply $Z^{x'+x_S}_{\calV_2} \otimes (X^{x' + x_S}Z^{z'}X^{x' + x_S})_{\calC}$ to registers $\calV_2 \otimes \calC$.
        \item $\Hyb_{11}$: Apply $\CZ_{\calV_2 \rightarrow \calC}$ to registers $\calV_2 \otimes \calC$.
    \end{itemize}
    If the check passes, then the state on registers $\calV_2 \otimes \calC$ is supported on
    \[
        \ket{s + z'}_{\calV_2} \otimes X^{x'+x_S}Z^{z'}\ket{S}_{\calC}
    \]
    for $s \in S^\perp$. By linearity it suffices to show equivalence on such states.
    We may write $\CZ$ as
    \[
        \CZ = \sum_{w} Z^{x''} \ketbra{w} \otimes X^{x''} Z^{w} X^{x''}
    \]
    for any $x''$, and in particular $x'' = x' + x_S$. Incorporating the existing $X^{x' + x_S}Z^{z'}$ and observing that $Z^{s}$ commutes with $X^{x'+x_S}Z^{z'}X^{x'+x_S}$, the state in $\Hyb_{10}$ becomes
    \[
        \left(Z^{x'+ x_S} \otimes X^{x'+x_S}Z^{z'}X^{x'+x_S}\right)\left(\ket{s + z'}_{\calV_2} \otimes X^{x'+ x_S}Z^{z'}\left(Z^{s}\ket{S}_{\calC}\right)\right)
    \]
    Finally, $Z^{s}\ket{S} = \ket{S}$ since $s \in S^\perp$. Therefore the net effect of $\Hyb_{11}$ is to apply $Z^{x'+ x_S} \otimes X^{x'+x_S}Z^{z'}X^{x'+x_S}$, the same as $\Hyb_{10}'$.
\end{proof}

\begin{claim}\label{claim:hyb11-12}
    $\Hyb_{11}$ and $\Hyb_{12}$ are identically distributed.
\end{claim}
\begin{proof}
     Consider the hybrid experiment $\Hyb_{11}'$ where at the end of the \textbf{second 2PC}, the simulator applies $X^{x'}$ to register $\calC$. This is equivalent since register $\calC$ is disjoint from the adversary's view after the \textbf{second 2PC}.

    The two hybrids differ only in the \textbf{third 2PC}. If the $v_3 \in S + x'$ check rejects, then the two hybrids are identical. Otherwise, it suffices to show that the following two operations are equivalent on the residual state.
    \begin{itemize}
        \item $\Hyb_{11}$: Apply $I_{\calV_3} \otimes X^{x'}_{\calC}$ to registers $\calV_3 \otimes \calC$.
        \item $\Hyb_{12}$: Apply $\CX_{\calV_3 \rightarrow \calC}$ to registers $\calV_3 \otimes \calC$.
    \end{itemize}
    If the check passes, then $\calV_3$ is supported on $\ket{s + x'}$ where $s \in S$. Thus, the effect of the $\CX_{\calV_3 \rightarrow \calC}$ in $\Hyb_{12}$ is to apply $X^{s + x'}$ to register $\calC$. Note that at the start of the \textbf{third 2PC}, register $\calC$ contains $X^{x_S + x'}\ket{S}$. Commuting $X^{s}$ through and observing that $X^{s} \ket{S} = \ket{S}$ for all $s \in S$, the net effect of $\Hyb_{12}$ is to apply $X^{x'}$ to register $\calC$, the same as $\Hyb_{11}$.
\end{proof}

\begin{claim}\label{claim:hyb12-ideal}
    $\Hyb_{12}$ and $\Ideal(\Sim, 1^\secpar)$ are identically distributed.
\end{claim}
\begin{proof}
    First, observe that the outputs of the three 2PCs are never used. The classical party's output is guaranteed to be $(x'_{\Can}, z'_{\Can})$ or an abort, and $x_S$ is unused. Therefore the output steps can be omitted.

    Second, we show that the measurement of $x_S$ can be commuted to \emph{after} the end of the \textbf{third 2PC}, at which point $\calC$ is discarded and therefore it does not matter whether $x_S$ was measured from it.

    To commute past the \textbf{first 2PC}: the $x_S$ measurement commutes with the $\CX_{\calV_1 \rightarrow \calC}$ operation since both are diagonal in the computational basis on register $\calV_1$. After the $\CX$ operation, registers $\calV_1$ and $\calC$ are supported on $\ket{s + x_S}_{\calV} \otimes \ket{S + x' + x_S}_{\calC}$ for $s \in S$, so it is equivalent to measure $x_S \gets \Can_{S}(\calC - x')$ on register $\calC$ instead of measuring it from register $\calV_1$.

    To commute past the \textbf{second 2PC}: the $x_S \gets \Can_{S}(\calC - x')$ measurement commutes with the adversary's operations in between the first and second 2PC ideal functionalities since it is on a disjoint register. Furthermore, it commutes with the $\CZ_{\calV_2 \rightarrow \calC}$ operation since both are diagonal in the computational basis.

    To commute past the \textbf{third 2PC}: the $x_S \gets \Can_{S}(\calC - x')$ measurement commutes with the adversary's operations in between the first and second 2PC ideal functionalities since it is on a disjoint register.
    Furthermore, the check forces $\calV_3$ to be entirely supported on $S + x'$. Therefore the state before the $\CX_{\calV_3\rightarrow \calC}$ is supported on $\ket{s + x'}_{\calV} \otimes \ket{S + x' + x_S}_{\calC}$. For such states, afterwards register $\calC$ is supported on $\ket{S+x_S}$ and so $x_S \gets \Can_{S}(\calC)$ may be directly measured.

\end{proof}

\section{Classical Commitments to Quantum States with Ideal Binding}
\label{sec:fnal-commitments}

In this section, we show how to classically commit to a quantum state with a new, strong notion of binding. We call this binding ``ideal binding'' because it is formalized via the real/ideal paradigm.
In the ``real'' binding game, an adversarial committer commits, then opens their commitment in ``one shot''.
Ideally, the initial commitment defines a single quantum state $\rho$ and the adversary should only be able to open to measurements of $\rho$.
One-shot ideal binding formalizes this by requiring that the real binding game should be indistinguishable from the ideal scenario (see \Cref{sec:com-ideal-binding}).

We also formulate and achieve a stronger notion of ideal binding which we call \emph{continual} ideal binding. Whereas one-shot binding considers an adversary who commits then gives a single opening, continual binding allows the adversary to see the results of opening their state many times. This necessarily allows them to \emph{change} the committed state by measuring it. 
The ideal world for continual binding requires that after each time the adversary does so, the adversary remains bound to the post-measurement state (see \Cref{sec:com-continual-binding}).

We achieve a single protocol which is $\epsilon$-indistinguishable from the ideal continual binding world for every $\epsilon = 1/\poly[\secpar]$.
Since the protocol description is independent of $\epsilon$, this is sufficient for our later applications.
\footnote{Previous works~\cite{GKNV25} achieve a weaker notion of binding where it is possible to extract a state which is consistent with the opening distribution. This state is only required to be $\sqrt{\epsilon}$-indistinguishable when the adversarial committer gives a valid opening with probability $1-\epsilon$. Our construction's $\epsilon$-indistinguishability is guaranteed regardless of the adversary's successful opening probability.}

\begin{theorem}\label{thm:linear-functional-com}
    Assuming the existence of $\epsilon$-secure (respectively, secure) state-preserving $2\PC$ for classical linear functions with post-processed abort (\Cref{sec:classical-2PC}) there exists a classical commitment scheme for quantum states with state recovery (\Cref{def:com:recover}) and $\epsilon$-secure (respectively, secure) continual ideal binding (\Cref{def:fnal-com-continual-binding}) for the set of ZX-and-linear measurements (\Cref{sec:coset-auth-construction}). Moreover, its state recovery is compatible with continual binding (\Cref{def:cont-bind-recovery}).
\end{theorem}

This theorem, together with \Cref{thm:classical2PC}, immediately yields the following corollary.
\begin{corollary}
    There exists a classical commitment scheme for quantum states with binding-compatible state recovery and $\epsilon$-secure continual ideal binding for the set of ZX-and-linear measurements $\{M[H^\theta L]\}_{\theta\in \{0,1\}^n, L}$ assuming the post-quantum hardness of $\LWE$. 
\end{corollary}

\paragraph{Authentication versus Commitment.}
A similar notion of binding was already shown by \cite{FOCS:HuaTan25} to hold for \cite{STOC:BarBraVai24}'s coset-based quantum authentication scheme.
Conceptually, commitment is quite similar to authentication -- the key difference is that in a commitment, the \emph{adversary} gets to ``authenticate'' their own state instead of a trusted party. Showing binding for commitments requires additional work since the adversary may deviate from the honest authentication procedure.

Our construction shows how to use classical interaction to force a dishonest committer to correctly authenticate any state of the committer's choice according to \cite{STOC:BarBraVai24}'s authentication scheme, without learning the authentication key.
Thus, we inherit essentially all properties of their scheme, including continual binding and the ability to open to any ``ZX-and-linear'' measurement. Such measurements are defined by a change of basis $H^\theta L$, where $L$ is a series of $\CNOT$ gates, followed by a computational basis measurement (see \Cref{sec:coset-auth-construction}).

Moreover, we show the classical receiver can later help the quantum committer to \emph{recover} the committed state. This property generalizes \cite{KKR26}'s notion of state recovery beyond commitments to classical messages. The recovery protocol composes safely with continual binding, so that the sender can recover a partially measured state.

\subsection{Definition} \label{sec:def:com}

\paragraph{Syntax.}
A classical commitment scheme to quantum states with classical communication, with respect to a family of projective measurements $\calM[\calU] = \{M[U]\}_{U \in \calU}$, consists of the following algorithms. Here, $M[U]$ denotes the logical measurement implemented by applying $U$ and then measuring in the computational basis.
\begin{itemize}
    \item $\Com\langle S(\calS_{\inp}), R \rangle(1^\secpar, 1^n) \rightarrow (\calR_{\op}, \com)$ is an interactive protocol between a $\QPT$ sender $S$ and a classical $\PPT$ receiver $R$. $S$ inputs a state on register $\calS_{\inp}$, and both parties input the security parameter $1^\secpar$ along with the number $n$ of qubits to be committed to.

    At the end of the protocol, $S$ outputs an opening register $\calR_{\op}$ and $R$ outputs a classical commitment key $\com$.

    \item $\Eval(U) \to \widetilde{U}$ is a classical algorithm that takes as input a description of a unitary $U \in \calU$, then outputs a description of an efficient unitary $\widetilde{U}$ which operates on register $\calR_{\op}$. $\widetilde{U}$ acts as the homomorphic implementation of $U$ on committed states.

    A classical opening string $\op_U$ for the measurement $M[U]$ may be obtained by applying $\widetilde{U}$ to $\calR_{\op}$, then measuring the state in the computational basis.

    \item $\KeyEval(\com, U) \to \com_U$ is a classical algorithm that maps a commitment key $\com$ and a unitary $U \in \calU$ to an evaluation key $\com_U$.

    We emphasize that $\com_U$ does not necessarily belong to the same keyspace as $\com$.

    \item $\Dec_{\com_U}(\op) \rightarrow y / \bot$ is a $\PPT$ decoding algorithm that, given a classical opening string $\op$, returns either its decoded outcome $y$ for the measurement $M[U]$ or $\bot$. It is parameterized by an evaluation key $\com_U$.

    \item $\Ver_{\com_U}(\op) \rightarrow \top / \bot$ is a $\PPT$ verification algorithm that returns $\top$ exactly when $\Dec_{\com_U}(\op) \neq \bot$.
\end{itemize}
In addition, a classical commitment to a quantum state may be equipped with a $\Recover$ algorithm that allows it to recover the committed state~\cite{KKR26}.
\begin{itemize}
    \item $\Recover\langle S(\calR_{\op}), R(\com)\rangle \rightarrow \calS_{\outp}$ is an interactive protocol between a $\QPT$ committer $S$ and a classical $\PPT$ receiver $R$. $S$ inputs its opening register $\calR_{\op}$ and $R$ inputs the commitment key $\com$. At the end of the protocol, $S$ outputs a register $\calS_{\outp}$ containing the recovered state.
\end{itemize}

A functional commitment must satisfy correctness (\Cref{def:fnal-com-correctness}).
Additionally, it must satisfy either one-shot ideal binding (\Cref{def:fnal-com-security}) or continual ideal binding (\Cref{def:fnal-com-continual-binding}).
If equipped with a $\Recover$ protocol, it must also satisfy recovery correctness.

\begin{definition}[Commitment Correctness]
\label{def:fnal-com-correctness}
    For any $U \in \calU$ and $\com$ in the support of $\Com$, let $\widetilde{U} \gets \Eval(U)$ and $\com_U \gets \KeyEval(\com, U)$. 

    A commitment scheme is correct if for every $U \in \calU$, every $n, \secpar \in \bbN$, and every $n$-qubit state in register $\calS_{\inp}$,
    \[
        \left\{
        \begin{array}{c}
             (\calR_\op, \com) \gets \Com\langle S(\calS_{\inp}), R\rangle (1^\secpar, 1^n)
            \\
            (y, \calC) \gets M[\Dec_{\com_U}, \widetilde{U}](\calR_{\op})
            \\
            \\
            \textnormal{Output } (y, \calC, \com)
        \end{array}
        \right\}
        =
        \left\{
        \begin{array}{c}
             (y, \calC) \gets M[U](\calS_{\inp})
            \\
             (\calR_\op, \com) \gets \Com\langle S(\calC), R\rangle (1^\secpar, 1^n)
             \\
             \\
             \textnormal{Output } (y, \calR_\op, \com)
        \end{array}
        \right\}
    \]
    In other words, committing then homomorphically measuring is equivalent to first measuring and then committing.
\end{definition}

\begin{definition}[Recovery Correctness]\label{def:com:recover}
    An interactive commitment to quantum states \[(\Com, \Eval, \KeyEval, \Dec, \Ver, \Recover)\] has \emph{state recovery} if the following holds.

    Consider the following quantum channel $\mathsf{SR\text{-}Exp}_{\secpar, n}$, parameterized by the security parameter $\secpar$ and a number of qubits $n$.
    \begin{enumerate}
        \item \textbf{Input:} An $n$-qubit state in register $\calA$.
        \item Compute $(\calR_{\op}, \com) \gets \Com\langle S(\calA), R\rangle(1^\secpar, n)$, a commitment to the contents of register $\calA$.
        \item Compute $\calA' \gets \Recover\langle S(\calR_{\op}), R(\com)\rangle(1^\secpar, n)$, the recovered state.
        \item \textbf{Output:} The contents of register $\calA'$.
    \end{enumerate}

    $\Recover$ is correct if for all $\secpar, n \in \bbN$ and all states $\ket{\psi}_{\calA, \calB}$ contained in an $n$-qubit register $\calA$ and arbitrary register $\calB$, 
    \[
        \frac{1}{2}\left\| 
            \left(I_{\calA} \otimes I_{\calB}\right)(\ket{\psi}_{\calA, \calB}) 
            -
            \left(\SRExp\otimes I_{\calB}\right)(\ket{\psi}_{\calA, \calB})
        \right\|_1 \leq \negl
    \]
    In other words, the diamond distance between the identity channel $I_{\calA} \otimes I_{\calB}$ and the commit-and-recover channel $\SRExp \otimes I_{\calB}$ is negligible. 
    As a more granular definition, we say the commitment has $(1-\delta)$-state recovery if the distance is at most $\delta$.
\end{definition}

It is also useful to define \emph{partial} recovery where the input register $\calA$ is divided into $\calA = (\calA_1, \calA_2)$ and only $\calA_1$ is recovered. In this case, we require the same distance bound when the identity channel first traces out $\calA_2$. We denote the experiment where $\Recover$ only recovers $\calA_1$ as $\SRExp^{\calA_1}$.

\begin{definition}[Partial Recovery]
    An interactive commitment to quantum states satisfies partial recovery of register $\calA_1$ if for all states $\ket{\psi}_{\calA, \calB}$ where $\calA = (\calA_1, \calA_2)$,
    \[
        \frac{1}{2}\left\| 
            \left(I_{\calA} \otimes I_{\calB}\right)\Tr_{\calA_2}[\ket{\psi}_{\calA, \calB}]
            -
            \left(\SRExp^{\calA_1}\otimes I_{\calB}\right)(\ket{\psi}_{\calA, \calB})
        \right\|_1 \leq \negl
    \]
\end{definition}

    Typically a commitment necessarily must satisfy hiding as well. We omit this property since 1) we do not need it for our applications and 2) it can be generically added by encrypting the state with a one-time pad $X^x Z^z$ before committing, then separately committing to the classical keys $(x, z)$.

\paragraph{Binding.}  We define two notions of binding.
The first covers the simplest (real world) setting of a commitment: the adversarial receiver commits, then opens their commitment in ``one-shot.'' 
After the adversary commits, there is a well-defined (if potentially entangled) state $\ket{\psi}$ which is committed to. Furthermore, any valid opening $(U, \op)$ which the adversary gives in the real world should match the result of directly measuring $\ket{\psi}$ using $M[U]$ in the ideal world.
We formalize this notion in \Cref{sec:com-ideal-binding}.

Second, we define a notion of binding for a more strenuous setting wherein an adversarial committer gains query access to a set of evaluation keys $\{\com_{U_i}\}$. Using these keys, it may decode certain measurements on the committed state. In fact, correctness necessarily permits the adversary to \emph{modify} the committed state by measuring it.
Despite this, there is a well-defined evolution of the committed state from the original state $\ket{\psi_0}$ to some modified state $\ket{\psi_t}$ which is the result of making $t$ sequential measurements on the state.
Thus, we may ask that the next valid opening $(U, \op)$ which the adversary gives in the real world matches the result of directly measuring $\ket{\psi_t}$.
We formalize this notion in \Cref{sec:com-continual-binding}.

\subsubsection{One-Shot Ideal Binding}\label{sec:com-ideal-binding}

We formalize binding via the real/ideal paradigm. 
Intuitively, the definition compares the value that an adversarial committer is able to open to after playing a real execution of the commitment protocol to an idealized model where the adversary is forced to explicitly declare their state $\ket{\psi}$ and a measurement is performed on $\ket{\psi}$ by a trusted party.
For binding to hold, the adversary's real opening must match the result of directly measuring their declared state $\rho$.

\paragraph{Real Experiment.} 
In the real experiment $\Real(\adv, 1^n, 1^\secpar, U)$, an adversarial $\QPT$ committer $\adv = (\adv_1, \adv_2)$ interacts with an honest classical receiver as follows. 
\begin{enumerate}
    \item Run the commitment protocol $(\calR_{\adv}, \com) \gets \Com\langle \adv_1, R\rangle(1^n, 1^\secpar)$.
    \item $(\op, \calR_{\adv}') \gets \adv_2(\calR_{\adv})$ outputs an opening string $\op$ and an arbitrary state in register $\calR_{\adv}'$.
    \item Compute $\com_U \gets \KeyEval(\com, U)$ and $y \gets \Dec_{\com_U}(\op)$.
    \item The experiment outputs $(y, \calR_{\adv}')$.
\end{enumerate}

\paragraph{Ideal Experiment.}
In the ideal experiment, the adversarial sender and classical receiver directly interact with a trusted party implementing the ideal functionality of a commitment. The ideal functionality $\IF$ works as follows:
\begin{enumerate}
    \item \textbf{Commitment.} $\IF$ receives from the sender a register $\calV$ containing the state to be committed to.
    \item \textbf{Opening.} $\IF$ receives a bit $b$ denoting whether the sender wishes to open its commitment and a choice of unitary $U \in \calU_n$. If $b = 0$, it outputs $\bot$. 
    Otherwise, it measures register $\calV$ using $M[U]$ and outputs the result $y$ to the receiver.
\end{enumerate}
\noindent The ideal experiment $\Ideal(\calB, 1^n, 1^\secpar, U)$ is played by an idealized adversary $\calB = (\calB_1, \calB_2)$ as follows.
\begin{enumerate}
    \item \textbf{Commitment.} $(\calV, \calR_{\adv}) \gets \calB_1$ outputs a register $\calV$ containing the state they wish to commit to and an internal register $\calR_\adv$. It sends $\calV$ to the commitment ideal functionality $\IF$.
    \item \textbf{Optional Abort.} 
    $(b, \calR_{\adv}') \gets \calB_2(\calR_\adv)$ chooses a bit $b$ denoting whether it wishes to open the commitment or not, along with an arbitrary state in register $\calR_{\adv}'$.
    \item \textbf{Opening.} $y\gets \IF(b, U)$ is computed by the commitment ideal functionality.
    \item The experiment outputs $(y, \calR_{\adv}')$.
\end{enumerate}

\begin{definition}[One-Shot Ideal Binding]\label{def:fnal-com-security}
    A commitment scheme $(\Com,\allowbreak \Eval,\allowbreak \KeyEval,\allowbreak \Dec,\allowbreak \Ver)$ satisfies one-shot ideal binding for the family of measurements $\calM$ if there exists a $\QPT$ algorithm $\Sim = (\Sim_1, \Sim_2)$ such that for every $n, \secpar \in \bbN$, every unitary $U \in \calU_n$, and every $\QPT$ adversary $\adv = (\adv_1, \adv_2)$, the real and ideal experiments are computationally indistinguishable:
    \[
        \{(y, \calR_{\adv}') \gets \Real(\adv, 1^n, 1^\secpar, U)\}
        \approx_c
        \{(y, \calR_{\adv}') \gets \Ideal(\Sim(\adv), 1^n, 1^\secpar, U)\}
    \]
   \end{definition}

We omit the definition of $\epsilon$-security for this notion because we do not directly use this notion for later applications.

\subsubsection{Continual Ideal Binding}\label{sec:com-continual-binding}

Next, we consider a setting where the adversary receives oracle access to the family of decoding algorithms $\{\Dec_{\com_U}\}_{U \in \calU_n}$ after committing. This grants a huge amount of power to the adversary, even allowing them to modify the committed state by making targeted measurements.

In this case, the idealized model still requires the adversary to explicitly declare their state $\ket{\psi_0}$, but now allows them to modify it into some well-defined state $\ket{\psi_t}$ by making $t$ decoding queries. Then, rather than requiring binding to the originally committed state $\ket{\psi_0}$, we require binding to the most recent state -- that is, the adversary's real opening must match the result of directly measuring $\ket{\psi_t}$.

\paragraph{Real Experiment.} 
The experiments are parameterized by $\ell = \poly$ many unitaries $U_1, \dots, U_\ell \in \calU_n$ and classical functions $f_1, \dots, f_\ell:\{0,1\}^{n}\rightarrow \{0,1\}^{n'}$.\footnote{The classical functions act as a syntactic tool for easier analysis of protocols using continual binding. If an adversary queries for a particular $f$, the analysis is guaranteed that there is no other disturbance to the committed state.}

In the real continual binding experiment $\Real_c(\adv, 1^n, 1^\secpar, (U_i, f_i)_{i\in [\ell]})$, an adversarial $\QPT$ committer $\adv$ interacts with the honest classical receiver, as follows.
\begin{enumerate}
    \item \textbf{Commitment.} Run the commitment protocol $(\calR_{\adv}, \com) \gets \Com\langle\adv, R\rangle(1^\secpar, 1^n)$.
    \item Define the oracle $\fDec$ which works as follows.
    \begin{enumerate}[align=left]
        \item[\textbf{Input.}] On input $(i, \op, \calR_{\ans})$:
        \item Compute $v \gets \Dec_{\com_i}(\op)$, where $\com_i = \KeyEval(\com, U_i)$.
        \item Apply $\Write(f_i(v))$ to $\calR_{\ans}$, where we define $f_i(\bot) = \bot$.
        
    \end{enumerate}
   
    \item \textbf{Opening.} Output $\calR_{\adv}$ and coherent quantum oracle access to $\fDec$.
\end{enumerate}

\paragraph{Ideal Functionality.} In the ideal experiment, an idealized adversary $\calB$ interacts directly with the ideal functionality $\IF$ defined below.
\begin{itemize}[align=left]
    \item \textbf{Initialization.} $\IF_{\calV}$ is initialized with a register $\calV$.
    \item \textbf{Query $\IF_\calV(i, \calR_{\ans})$.} Coherently measure and apply $\Write(M[f_i, U_i] \calV)$ to register $\calR_{\ans}$.
\end{itemize}

\paragraph{Ideal Experiment.}
The ideal continual binding experiment $\Ideal_c(\calB, 1^\secpar, 1^n, (U_i, f_i)_{i\in [n\ell]})$ is played by an idealized adversary $\calB = (\calB_1, \calB_2)$ as follows.
\begin{enumerate}
    \item \textbf{Commitment.} 
    $(\calV, \calR_\outp, \calR_{\calB}) \gets \calB_1$ outputs a register $\calV$ containing the state it wants to commit to along with two additional registers.
    \item Initialize $\calB_2$ with $\calR_{\calB}$, then define the oracle $\CBOracle[\calB_2, \IF_\calV]$ which works as follows.
    \begin{enumerate}[align=left]
        \item[\textbf{Input.}] On input $(i, \op, \calR_{\ans})$:
        \item Compute $b \gets \calB_2(i, \op)$.
        \item If $b = 0$, apply $\Write(\bot)$ to register $\calR_{\ans}$.
        Otherwise, run $\IF_{\calV}(i, \calR_{\ans})$.
    \end{enumerate}

    \item \textbf{Opening.} Output register $\calR_{\adv}$ and coherent quantum oracle access to $\CBOracle[\calB_2, \IF_\calV]$.
\end{enumerate}

\begin{definition}[Continual Ideal  Binding]\label{def:fnal-com-continual-binding}
    A commitment scheme $(\Com,\allowbreak \Eval,\allowbreak \KeyEval,\allowbreak \Dec,\allowbreak \Ver)$ satisfies \emph{continual} ideal binding for the family of measurements $\calM[\calU]$ if there exists a $\QPT$ algorithm $\Sim = (\Sim_1, \Sim_2)$ such that for every $n, \secpar \in \bbN$, every $\ell = \poly$, every set of $\ell$ unitaries $U_1, \dots, U_\ell \in \calU_n$, every set of ($\PPT$-computable) classical functions $f_1, \dots f_\ell:\{0,1\}^n\rightarrow \{0,1\}^{n'}$,
    \begin{align*}
        &\{(\calR_{\adv}', \fDec) \gets \Real_c(S^*, 1^n, 1^\secpar, (U_i, f_i)_{i\in[\ell]})\}
        \\
        \approx_c
        &\{(\calR_{\adv}', \CBOracle[\Sim_2, \IF_\calV]) \gets \Ideal_c(\Sim(S^*), 1^n, 1^\secpar, (U_i, f_i)_{i\in[\ell]})\}
    \end{align*}
\end{definition}

\begin{definition}[$\epsilon$-secure Continual Ideal Binding]\label{def:fnal-com-continual-binding-eps}
    We say the scheme satisfies $\epsilon$-secure continual ideal binding if $\Sim_1$ and $\Sim_2$ are uniform families of $\poly[\secpar, 1/\epsilon]$-sized circuits such that the above distributions are $\epsilon$-indistinguishable for every $\epsilon = 1/\poly[\secpar]$ where $\Sim_{1, \secpar, \epsilon}$ replaces $\Sim_1$ and $\Sim_{2, \secpar, \epsilon}$ replaces $\Sim_2$.
\end{definition}

We mention a few useful properties of this definition. 
First, the real-world queries are ``one-to-one'' with the ideal world queries, in the sense that a query to $i$ can only become either a query to $i$ or a skipped query.
Second, if an adversarial opener at any point produces an opening $(i, \op)$, the result of $f(\Dec_{\com_i}(\op))$ is either $\bot$ or distributed according to measuring $M[f_i, U_i]$ on the committed state in register $\calV$.

\subsubsection{Continual Binding with Recovery}\label{sec:cont-binding-recoverty}

Sometimes it is useful to recover the committed state \emph{after} doing a sequence of measurements on it, as is allowed in continual binding. However, one must be careful that $\Recover$ does not allow distinguishing the ideal binding world from the real after the fact. For example, what if $\Recover$ involves giving $\com$ to the sender?
We augment the definition of continual binding to ensure safe recovery after several openings.

The real and ideal experiments from continual binding are augmented with an additional recovery step. 
After preparing the oracles $\fDec$ (in the real experiment) or $\CBOracle$ (in the ideal experiment), the adversary gets an opportunity to query them. Then, the new step is to run $\Recover$ in the real world. This is matched in the ideal world by a QPT algorithm $\Sim$ which directly receives a register $\calV$ containing the recovered state and must produce an indistinguishable view.

\begin{definition}[Continual Binding with Recovery]
    \label{def:cont-bind-recovery}
    Let $\Real_c$ and $\Ideal_c$ be the continual binding experiments from \Cref{def:fnal-com-continual-binding}.
    
    A commitment scheme with state recovery satisfies \emph{continual binding with recovery} for the family of measurements $\calM[\calU]$ if 
    there exists a $\QPT$ algorithm $\Sim = (\Sim_{1,2}, \Sim_3)$ 
    such that for every $\secpar \in \bbN$, every $n = \poly$ and $\ell = \poly$, every set of $\ell$ unitaries $U_1, \dots, U_\ell \in \calU_n$, every set of ($\PPT$-computable) classical functions $f_1, \dots, f_\ell: \{0,1\}^n \rightarrow \{0,1\}^{n'}$, and every $\QPT$ adversarial committer $\adv = (\adv_{1,2}, \adv_3)$, the real and ideal experiments with appended recovery step are computationally indistinguishable:
    \begin{align*}
        &\left\{\calR_{\adv}'' :
        \begin{array}{l}
             (\calR_{\adv}, \fDec) \gets \Real_c(\adv_{1,2}, 1^n, 1^\secpar, (U_i, f_i)_{i\in[\ell]}) \\
             \calR_{\adv}' \gets \adv_2^{\fDec}(\calR_{\adv}) \\
             \calR_{\adv}'' \gets \Recover\langle \adv_{3}(\calR_{\adv}'), R(\com)\rangle
        \end{array}
        \right\}
        \\
        \approx_c
        &\left\{\calR_{\adv}'':
        \begin{array}{l}
             (\calR_{\adv}, \CBOracle_\calV) \gets \Ideal_c(\Sim(S^*), 1^n, 1^\secpar, (U_i, f_i)_{i\in[\ell]}) \\
             \calR_\adv' \gets \adv_2^{\CBOracle_{\calV}}(\calR_{\adv})\\
             \calR_{\adv}'' \gets \Sim_{3}(\adv_3(\calR_{\adv}), \calV)
        \end{array}
        \right\}
    \end{align*}
    Here $\com$ is the commitment key from $\Real_c$ and $\Sim_3$ receives state from $\Sim_2$ after $\adv_2$ finishes using $\CBOracle$.
\end{definition}

\begin{definition}[$\epsilon$-secure Continual Binding with Recovery]\label{def:cont-bind-recovery-eps}
    We say the scheme satisfies $\epsilon$-secure continual binding with recovery if $\Sim_1$, $\Sim_2$, and $\Sim_3$ are uniform families of $\poly[\secpar, 1/\epsilon]$-sized circuits such that the above distributions are $\epsilon$-indistinguishable for every $\epsilon = 1/\poly[\secpar]$ where $\Sim_{i, \secpar, \epsilon}$ replaces $\Sim_i$ for $i\in \{1,2,3\}$.
\end{definition}

Both definitions above can also be easily adapted for \emph{partial} recovery. In this case, we divide the register which contains the committed state into $\calV = (\calV_1, \calV_2)$ and $\Sim_3$ only receives register $\calV_1$.

\subsubsection{Continual Binding for Authentication}

Continual binding can also be defined for authentication. The only difference is in the real world's initial commitment step, which is replaced by
\begin{enumerate}
    \item Receive a register $\calV$ from the adversary.
    \item Sample $k \gets \Auth.\KeyGen(1^\secpar, 1^n)$, then send $\widetilde{\calV}\gets \Auth.\Enc_k(\calV)$ back to the adversary.
\end{enumerate}

\cite{FOCS:HuaTan25}'s definition of functional authentication security can be seen as continual binding with a specific simulator.

\begin{lemma}
    Functional authentication security (\Cref{def:fnl-auth-sec}) implies continual binding for quantum authentication.
\end{lemma}
\begin{proof}
    Fix any sequence of unitaries $U_1, \dots, U_t \in \calU_n$. 
    The simulator $\Sim_1$ works as follows:
    \begin{enumerate}
        \item Receive $\calV$ from the adversary and forward it to the continual binding ideal functionality $\IF$.
        \item Sample $k \gets \Auth.\KeyGen(1^\secpar, 1^n)$ then send $\widetilde{\calV}\gets \Auth.\Enc_k(\ket{0^n})$ to the adversary. 
        Let $\com_j = \Auth.\KeyEval(\com, U_j)$ and send it to $\Sim_2$.
    \end{enumerate}
    The oracle simulator $\Sim_2$ acts as follows on queries $(i, \op)$:
    \begin{enumerate}
        \item Output $b\gets \Auth.\Ver_{\com_i}(\op)$, where $\com_i = \Auth.\KeyEval(\com, U_i)$.
    \end{enumerate}
    Observe that $\CBOracle[\Sim_2, \IF_\calV]$ is precisely the oracle $\mathsf{fVer}$ from \Cref{def:fnl-auth-sec}, and therefore oracle access to it along with $\widetilde{\calV} = \Auth.\Enc_k(\ket{0^n})$ is indistinguishable from $\Auth.\Enc(\calV)$ and oracle access to $\fDec$.
\end{proof}

\subsection{Construction}

In the following, we set $\Auth$ to be the coset-based authentication scheme from \cite{STOC:BarBraVai24}.

\begin{construction}[$\Com\langle S(\calS_\inp), R\rangle(1^\secpar, 1^n)$]
\label{constr:com-main}
    To commit to an $n$-qubit state contained in register $\calS_\inp$, the quantum sender $S(\calS_\inp)$ and classical receiver $R$ interact as follows.
    \begin{enumerate}
        \item \textbf{Classical Receiver:} 
        Sample a $2\secpar$-dimensional subspace $S\subset \bbF_2^{4\secpar}$ and a vector $\Delta \in \bbF_{2}^{4\secpar}\backslash S$. Let $S_\Delta = S + \Span(\Delta)$.
        
        \item \textbf{Both:}
        Set $d(\secpar) = 4\secpar n$ and remotely prepare a random coset state of $S_{\Delta}^n \subset \left(\bbF_2^{4\secpar}\right)^{n}$
        \[
            \left(\bigotimes_{i=1}^{n} X^{x_i} Z^{z_i} \ket{S_{\Delta}}, (\vec{x}, \vec{z})\right)
            \gets \RSP_{d}\left\langle Q, C\left(S_\Delta^{n}\right)\right\rangle\left(1^\secpar\right)
        \]
        where the classical receiver $R$ acts as the classical party $C$ and the quantum sender acts as the quantum party $Q$.\footnote{For every $n\geq1$, $2\secpar\leq\dim(S_\Delta^n)=n(2\secpar+1)\leq4\secpar n-2\secpar+1$, so the call lies in the RSP domain, including when $n=1$.}
        Let $Q$'s output coset state be stored in register $\widetilde{\calR}_{\outp} = (\widetilde{\calR}_{\outp, 1}, \dots, \widetilde{\calR}_{\outp, n})$.

        \item \textbf{Classical Receiver:} Set $k = (S, \Delta, \vec{x}, \vec{z})$.
        Note that the overall prepared coset state in registers $(\widetilde{\calR}_{\outp, i})_{i\in [n]}$ can be interpreted as the authentication of $\ket{+^n}$ using key $k$:
        \[
            \bigotimes_{i=1}^{n} X^{x_i}Z^{z_i}\ket{S_\Delta}
            =
            \Auth.\Enc_{k}(\ket{+^n})
        \]
        
        \item \textbf{Quantum Sender:} Initialize an $n$-qubit register $\calR_{\inp} = (\calR_{\inp, 1}, \dots, \calR_{\inp, n})$ to $\ket{+^n}$.
        \item \textbf{Both:} Execute a state-preserving $2\PC$ for the following classical linear-or-abort functionality.
        \begin{functionalitybox}
            \textbf{Quantum Party Input:} Registers $(\calR_{\inp, 1}, \widetilde{\calR}_{\outp, 1}) \dots, (\calR_{\inp, n}, \widetilde{\calR}_{\outp, n})$.
            
            \textbf{Classical Party Input:} The key $(S, \Delta, \vec{x}, \vec{z})$.

            \textbf{Functionality:} On input $\vec{b} \in \bbF_2^{n}$ and $\vec{v}\in \bbF_2^{4\secpar n}$:
            \begin{enumerate}
                \item Let $k_I = \Auth.\KeyEval(k, I)$ be the homorphic key for the identity unitary.
                \item Compute $\vec{b'} = \Auth.\Dec_{k_I}(\vec{v})$.
                \item If $\vec{b'} = \bot$, output $\bot$. Otherwise output $\vec{b} - \vec{b'}$.
            \end{enumerate}
        \end{functionalitybox}
        If the classical party receives output $\vec{b''}$, then the quantum sender's state in registers $(\calR_{\inp, i}, \widetilde{\calR}_{\outp, i})_{i\in [n]}$ is now
        \[
            \frac{1}{\sqrt{2^n}} \sum_{\vec{b}\in \bbF_2^{n}} \ket{\vec{b}}_{\calR_{\inp}} \otimes \Auth.\Enc_k\left(X^{\vec{b''}} \ket{\vec{b}}\right)_{\widetilde{\calR}_{\outp}}
            \
        \]
        by the correctness of authentication.

        \item \textbf{Quantum Sender:} 
        Perform the sender half of quantum teleportation of $\calS_{\inp, i}$ through register $\calR_{\inp, i}$ into register $\widetilde{\calR}_{\outp, i}$ for each $i\in [n]$.
        This yields a list of teleportation corrections $(x_{\tel, i}, z_{\tel, i})_{i\in [n]}$. Let $\vec{x_{\tel}}$ and $\vec{z_{\tel}}$ be the concatenations of the $X$ and $Z$ corrections, respectively. 
        
        Note that if the state being teleported was $\ket{\psi_i}$, the resulting state in register $\widetilde{\calR}_{\outp}$ is
        \[
            \Auth.\Enc_k\left(X^{\vec{b''} + \vec{x_{\tel}}} Z^{\vec{z_{\tel}}} \ket{\psi_i}\right)
        \]

        Send $(\vec{b''} + \vec{x_{\tel}}, \vec{z_{\tel}})$ to the classical receiver and output register 
        \[
            \calR_{\op} \coloneqq \widetilde{\calR}_{\outp}. 
        \]

        \item \textbf{Classical Receiver:} Perform the key update
        \[
            \com \coloneqq \Auth.\KeyEval\left(k, X^{\vec{b''} + \vec{x_{\tel}}} Z^{\vec{z_{\tel}}}\right)
        \]
        Output $\com$.
    \end{enumerate}
\end{construction}

\begin{construction}[Other Commitment Algorithms]\label{constr:com-other}
    The remaining commitment algorithms are identical to \cite{STOC:BarBraVai24}'s authentication code, i.e.
    \begin{align*}
        &\Eval = \Auth.\Eval
        \\
        &\KeyEval = \Auth.\KeyEval
        \\
        &\Dec_{\com_U} = \Auth.\Dec_{\com_U}
        \\
        &\Ver_{\com_U} = \Auth.\Ver_{\com_U}
    \end{align*}
    where $\com$ is interpreted as $k$.
\end{construction}

\begin{theorem}\label{thm:com}
    \Cref{constr:com-main,constr:com-other} together are a classical commitment to quantum states with $\epsilon$-secure continual ideal binding (\Cref{def:fnal-com-continual-binding}) for the family of ZX-and-Linear measurements $\calU^{\ZXLin}$. Furthermore, $\Dec$ is a linear function with post-processed abort.
\end{theorem}
\begin{proof}
    The protocol description already shows that the sender prepares $\Auth.\Enc_{\com}(\calV)$, using the correctness of $\RSP_d$, the correctness of $\Auth$, and the correctness of teleportation. Moreover, $\com$ is in the keyspace of $\Auth.\KeyGen$ for \cite{STOC:BarBraVai24}'s construction. Therefore correctness follows from the correctness of $\Auth$.

    Next, we show security. Fix any sequence of unitaries $U_1, \dots, U_t \in \calU^{\ZXLin}$ and classical functions $f_1, \dots, f_t$. The commitment-phase simulator $\Sim_1$ works as follows.
    \begin{enumerate}
        \item Prepare $\ket{+^n}$ in register $\calV$.
        \item Sample $k \gets \Auth.\KeyGen(1^\secpar, 1^n)$ and prepare
        \[
            \Auth.\Enc_k(\ket{0}^n) 
            =
            \bigotimes_{i=1}^{n} X^{x_i} Z^{z_i} \ket{S} 
        \]
        \item Simulate the RSP protocol using $\bigotimes_{i=1}^{n} X^{x_i} Z^{z_i} \ket{S}$ and the membership oracles for $\bigtimes_{i=1}^{n} (S_\Delta + x_i)$ and $\bigtimes_{i=1}^{n} (S^\perp + z_i)$.\footnote{Note that these are the verification spaces for the authentication scheme. The Hadamard space in particular is changed from the original $\RSP$.}
        \item Simulate the state-preserving 2PC protocol. During its ideal functionality, replace the decoding step by checking $\Auth.\Ver_k(\vec{v})$ and, if it is not $\bot$, computing the output $\vec{b} - \vec{b'}$ by the XOR of the input register and register $\calV$.
        \item Receive the teleportation corrections $(\vec{x_\tel}, \vec{z_\tel})$ from the adversary.
        \item Apply $X^{\vec{x_\tel}}Z^{\vec{z_\tel}}$ to register $\calV$. Compute $k' = \Auth.\KeyEval(X^{\vec{x_\tel}}Z^{\vec{z_\tel}})$.
        \item Send register $\calV$ to the commitment ideal functionality and send $k'$ to $\Sim_2$.
    \end{enumerate}
    On any query, $\Sim_2$ simply checks $\Auth.\Ver_{k'}(\bullet)$ and allows the query to go through if and only if the output is accepting.

    We now show indistinguishability from the real experiment. Fix any $\epsilon = 1/\poly[\secpar]$.
    \begin{itemize}
        \item $\Hyb_0$ is the real experiment.
        \item $\Hyb_1$ runs the $\RSP$ simulator instead of the real $\RSP$ protocol, using the state
        \[
            \Auth.\Enc_k(\ket{+}^n) 
            =
            \bigotimes_{i=1}^{n} X^{x_i} Z^{z_i} \ket{S_\Delta} 
        \]
        Set the simulation error to $\epsilon/2$. This is $\epsilon/2$-computationally indistinguishable by the $\epsilon$-security of $\RSP$.
        \item $\Hyb_2$ simulates the state-preserving 2PC with simulation error $\epsilon/2$. This is $\epsilon/2$-computationally indistinguishable from $\Hyb_1$ by the $\epsilon$-security of the 2PC.
        \item $\Hyb_3$ replaces the $\Auth.\Enc_k(\ket{+}^n)$ by
        \[
            \Auth.\Enc_k(\ket{0}^n) 
            =
            \bigotimes_{i=1}^{n} X^{x_i} Z^{z_i} \ket{S}.
        \]
        Additionally, it changes all usage of the decoding oracles $\Auth.\Dec$ to verification queries $\Auth.\Ver$ in combination with operations directly on register $\calV$. This includes the $\RSP$ preparation.
        This is statistically indistinguishable by the functional authentication security of $\Auth$ (\Cref{def:ht-functional-auth}).
    \end{itemize}
    $\Hyb_3$ is precisely the simulator listed above with a slight syntactic reinterpretation: since the key update only affects $\Auth.\Ver$, the decoding is updated to match by applying $X^{\vec{x_\tel}}Z^{\vec{z_\tel}}$ to register $\calV$ (\Cref{eq:BBV-pauli-update}).

\end{proof}

\subsection{Construction with Recovery}

Next, we generically add a $\Recover$ protocol to the construction from the prior section by teleporting the committed state into an uncommitted register.

\begin{construction}[Commitment with Recovery] 
    \label{constr:com-recovery-core}
    Here we compile any commitment \[(\Com, \Eval, \KeyEval, \Dec, \Ver)\] which supports ZX-and-linear measurements and satisfies continual ideal binding into a commitment \[(\Com', \Eval', \KeyEval', \Dec', \Ver', \Recover)\] which supports the same measurements and additionally has continual binding with recovery.
    The core algorithms, except for recovery, are as follows.
    \begin{itemize}
        \item $\Com'\langle S(\calS_{\inp}), R\rangle(1^n, 1^\secpar)$: The quantum sender $S$ prepares $n$ EPR pairs in registers $(\calE_{\inp}, \calE_{\outp})$, where the $i$'th EPR pair is in register $(\calE_{\inp,i}, \calE_{\outp, i})$. Then, it commits to the $2n$-qubit state in registers $(\calS_{\inp}, \calE_{\inp})$ using the commitment from \Cref{constr:com-main}:
        \[
            (\calR_{\op}, \com) \gets \Com\langle S(\calS_{\inp}, \calE_{\inp}), R\rangle(1^{2n}, 1^\secpar)
        \]
        The sender outputs $(\calR_{\op},\calE_{\outp})$, which together form the opening register of the compiled commitment, and the receiver outputs $\com$.
        \item $\Eval'$, $\KeyEval'$, $\Dec'$, $\Ver'$: On input $U\in \calU_n$, set $U' = U \otimes I_n$. The algorithms $\Eval'$ and $\KeyEval'$ run the corresponding underlying algorithms on $U'$, and $\Eval'$ acts on $\calR_{\op}$ while leaving $\calE_{\outp}$ unchanged. The algorithm $\Dec'$ runs the underlying decoder and returns $\bot$ if the underlying decoder returns $\bot$; otherwise, it returns only the first $n$ decoded bits. In the coherent implementation of $\Dec'$, all workspace, including the last $n$ decoded bits, is uncomputed. The algorithm $\Ver'$ runs the underlying verification algorithm.
    \end{itemize}

\end{construction}

\noindent Next, we describe the recovery protocol. It uses the Bell-measurement unitary
\[
    U_{\tel}
    \coloneqq
    \left(H^{\otimes n}\otimes I_n\right)
    \left(\prod_{i=1}^{n}\CNOT_{\calS_{\inp,i}\rightarrow\calE_{\inp,i}}\right).
\]
$U_\tel$ is a ZX-and-linear unitary: the CNOT layer is linear, and it is followed by Hadamards on the first $n$ logical qubits.

\begin{construction}[Recovery]
    \label{constr:com-recovery-recovery}
    To recover the state, $S$ and $R$ interact in $\Recover\langle S(\calR_{\op},\calE_{\outp}), R(\com)\rangle$ as follows.
    \begin{enumerate}
        \item \textbf{Homomorphic teleportation.} Let the first $n$ logical qubits of the underlying commitment correspond to $\calS_{\inp}$ and the last $n$ logical qubits correspond to $\calE_{\inp}$. 
        The sender measures the opening
        \[
            \op_{\tel} \gets M\left[\widetilde{U}_{\tel}\right](\calR_{\op})
        \]
        for the homomorphic teleportation measurement
        \[
            \widetilde{U}_{\tel}
            \gets
            \Eval(U_{\tel}).
        \]
        It sends $\op_{\tel}$ to the receiver.

        \item \textbf{Decode the teleportation corrections.} The receiver computes
        \[
            \com_{\tel}
            \gets
            \KeyEval(\com,U_{\tel}).
        \]
        It then computes
        \[
            (a,b)
            \gets
            \Dec_{\com_{\tel}}(\op_{\tel}),
        \]
        where $a,b\in\{0,1\}^{n}$ are respectively the decoded outcomes of the first and last $n$ logical qubits. If decoding returns $\bot$, the receiver sends $\bot$ to the sender and both parties output $\bot$. Otherwise, the receiver sends $(a,b)$ to the sender.

        \item \textbf{Apply the teleportation corrections.} The sender applies $Z^aX^b$ to $\calE_{\outp}$ and outputs the resulting register as the recovered state.
    \end{enumerate}
\end{construction}

\begin{theorem}
    \Cref{constr:com-recovery-core} and \Cref{constr:com-recovery-recovery} together are a classical commitment to quantum states satisfying $\epsilon$-continual ideal binding with recovery (\Cref{def:cont-bind-recovery}) for the family of ZX-and-Linear measurements $\calU^{\ZXLin}$. 
    Furthermore, $\Dec'$ is a linear function with post-processed abort.
\end{theorem}

\begin{remark}
    We note that the same construction can be adapted to partial recovery of any sub-register $\calV_1$ simply by having the receiver only reveal the teleportation corrections corresponding to $\calV_1$. Essentially the same security proof applies.
\end{remark}

\begin{proof}
    Correctness and the additional property of $\Dec'$ follow from the underlying commitment. 
    Recovery correctness follows from the correctness of the baseline commitment, which implies that it is equivalent to run $U_\tel$ before committing, and the correctness of teleportation.

    For continual binding with recovery, fix $(U_i,f_i)_{i\in[\ell]}$ and define
    \[
        U_i'\coloneqq U_i\otimes I_n
        \qquad\text{and}\qquad
        f_i'(v,w)\coloneqq f_i(v)
    \]
    for $i\in[\ell]$, together with the recovery pair
    \[
        U_{\ell+1}'\coloneqq U_{\tel}
        \qquad\text{and}\qquad
        f_{\ell+1}'(v,w)\coloneqq(v,w).
    \]
    We implicitly pad their outputs to a common length. All of these unitaries are ZX-and-linear. For any adversary $\adv=(\adv_{1,2},\adv_3)$, let $(\Sim_1,\Sim_2)$ be the underlying continual-binding simulator for this family and the adversarial $2n$-qubit commitment induced by $\adv_{1,2}$, with $\calE_{\outp}$ included in its auxiliary state. For any desired $\epsilon^* = 1/\poly$, we set their $\epsilon$-security identical to $\epsilon^*$ and omit the subscripts in the following.

    The simulator $\Sim_{1,2}'$ runs $\Sim_1$, writes its $2n$-qubit ideal register as $(\calV_1,\calV_2)$, sends $\calV_1$ to the compiled ideal functionality, and retains $\calV_2$ and the private simulator state for $\Sim'_2$. To queries $(i,\op)$ made to $\fMeas$ by $\adv_2$, $\Sim'_{2}$ runs $\Sim_2$ on the corresponding query $(i,\op)$ for $(U_i',f_i')$. Then, it returns $\bot$ if $\Sim_2$ rejects and otherwise forwards $i$ to the compiled ideal functionality. This exactly emulates the underlying ideal oracle since
    \[
        M[f_i',U_i']_{\calV_1\calV_2}
        =
        M[f_i,U_i]_{\calV_1}\otimes I_{\calV_2}.
    \]
    For recovery, $\Sim_3'$ receives the current $\calV_1$, then runs $\adv_3$. Upon receiving $\op_{\tel}$, it runs $\Sim_2$ on $(\ell+1,\op_{\tel})$. On rejection it returns $\bot$. Otherwise it measures $(\calV_1,\calV_2)$ using $M[f_{\ell+1}',U_{\tel}]$, sends the resulting $(a,b)$ to $\adv_3$, and outputs its final state.

    $\epsilon^*$-indistinguishability of the above simulator follows from the $\epsilon^*$-indistinguishability of the underlying simulator $(\Sim_{1},\Sim_{2})$ from the real experiment and the equivalence of $(\Sim'_{1},\Sim'_{2},\Sim'_{3})$ to $(\Sim_{1},\Sim_{2}, \Sim_{3})$ with their respective ideal functionalities.

\end{proof}

\part{State-Preserving 2PC for General Quantum Functions}
In this section, we combine the tools we have developed to obtain our general 2PC for quantum functionalities. We abstract out a way to verifiably commit to a history state in \Cref{sec:com-hist}. Once we have this primitive, it is possible to verify any public computation. We show how to do so while also combining it with quantum fully homomorphic encryption for privacy in \Cref{sec:general}.

\section{Certified Commitment to History States} \label{sec:com-hist}

In this section we show how to commit to a history state (as defined in \Cref{sec:hist}), prove that indeed the state committed to is a history state, in a state-preserving manner, and later open the commitment. 

\paragraph{Admissible opening requests.}
Recall that the continual-binding guarantee (\Cref{def:fnal-com-continual-binding}) is selective; namely, it is parameterized by a polynomial-size family $S=\bigl\{(U_i,f_i)\bigr\}_{i\in[q]}$, 
fixed before the commitment phase, where each $U_i$ belongs to the family of unitaries supported by the commitment scheme, and each $f_i$ is a constant-degree function.  

During the opening phase of our commitment, $P_\C$ may adaptively invoke
any $(U_i,f_i)\in S$, with repetitions.

\paragraph{Ideal Experiment.}  In the ideal experiment the adversary takes as input $(C,\ket{\psi})$, where $C$ is a classical description of a quantum circuit and $\ket{\psi}$ is an input to $C$. The ideal functionality, denoted by $\IF$, is given only $C$.  
\begin{enumerate}
\item {\bf Commitment.} The adversary sends a register $\calV$ to $\IF$.  
Denote by $H_\Hist$ the Hamiltonian defined in \Cref{thm:xz-fk}. 
Let $G$ be the projector onto the kernel of $H_\Hist(C)$, namely
    \[
    G=\Pi_{\ker H_{\Hist}(C)}.
    \]
    \item {\bf Certify.} 
    $\IF$ checks that $\calV$ is a history state corresponding to the circuit~$C$ by projecting $\calV$ onto $\{G,I-G\}$ (see \Cref{corr:hist}).  If it projects  onto $G$ then set $b_\Hist=1$ and otherwise set $b_\Hist=0$.  In either case, the
functionality stores the post-measurement register $\calV$ and the bit
$b_\Hist$.
    \item {\bf Adversarial abort.} The adversary sends a bit $a\in\{0,1\}$ to $\IF$, where $a=0$ indicates abort and $a=1$ is an indication to continue.  
If $a=0$, then $\IF$ sends $0$ to both parties and terminates.
Otherwise, $\IF$ sends $b_{\Hist}$ to both parties.  If
$b_{\Hist}=0$, it terminates; if $b_{\Hist}=1$, it proceeds to
the opening phase.

   \item {\bf Opening phase (repeated polynomially many times).} $P_\C$ sends an opening request of the form $(U_i,f_{i})\in S$, where $(U_i,f_i)$ are public and are sent to $P_\Q$.  $P_\Q$ sends a bit indicating an abort or a continue request, and if it does not abort then $\IF$ applies ${\sf M}[f_{i},U_i]$ on register $\calV$, sends the outcome to $P_\C$, and retains the post-measurement residual state as the new~$\calV$.

    \item {\bf Recover.}
On a recovery request from $P_\C$, the functionality requests a bit
$a_{\sf rec}\in\{0,1\}$ from $P_\Q$.  If
$a_{\sf rec}=0$, it sends $\perp$ to $P_\C$ and terminates.
Otherwise, it sends the current residual register $\calV$ to $P_\Q$ and
terminates.\footnote{It is also useful to consider recovery of a sub-register $\calV_1 \subset \calV$.}
\end{enumerate} 

We denote this ideal experiment, with an an honest classical party with input $\{(U_i, f_{i})\}$ and adversarial quantum party $\Sim$, with input $(C,\ket{\psi})$, by
\[
\Ideal_{\Sim}(C,\ket{\psi}, \{(U_i, f_{i})\}).
\]

We realize this ideal functionality via  a 3-phase protocol 
\[
\Pi_{\sf HistCom}=({\sf CertCom}, \Open, \Recover)
\]
where
\begin{itemize}
    \item ${\sf CertCom}$ implements Steps 1-3,
    \item $\Open(U,f)$ implements Step 4 and may be invoked polynomially many times,
    \item $\Recover$ is invoked at most once and implements Step 5.
\end{itemize}
These protocols make use of the following
ingredients:
\begin{itemize}
   
    \item A classical commitment scheme for quantum states with continual binding with recovery, as defined in \Cref{def:cont-bind-recovery}, denoted by
    \[
    (\Com, \Eval, \KeyEval, \Dec, \Ver, \mathsf{Recover})
    \] 
    with the following properties:
    \begin{itemize}
        \item $\Dec$ can be implemented via a constant-degree computation with post-processed abort.
        \item The continual-binding simulator is query-preserving. 
        
    \end{itemize}
    We note that the commitment scheme from \Cref{constr:com-main} satisfies these properties (see proof of \Cref{thm:com}). 
    \item State-preserving   2-$\PC$ for constant-degree classical functions with classical post-processed abort, as defined and constructed in \Cref{sec:classical-2PC}.   
\end{itemize}

Our $\mathsf{Recover}$ protocol is the same $\mathsf{Recover}$ protocol from the underlying commitment scheme, which can also be adapted to partial recovery.  We define the protocols $\mathsf{CertCom}$ and $\Open$ in \Cref{sec:certcom,sec:hist-open}, respectively. We provide the security analysis in \Cref{sec:hist-analysis}

\subsection{Protocol ${\sf CertCom}$}\label{sec:certcom}

In what follows, we construct the protocol ${\sf CertCom}$ that implements Steps 1-3 in the ideal experiment above.  
\begin{protocol}\label{protocol:hist}
  The quantum party $P_\Q$ takes as input $(C,\ket{\psi},1^\secpar)$, where $C$ is a classical description of a $\poly$-size quantum circuit over gates in $\calG_{XZ}$, and $\secpar$ is the security parameter. The classical party $P_\C$ takes as input only $(C, 1^\secpar, 1^n)$, where  $n$ is the number of qubits in the history state $\Hist_C(\ket{\psi}\otimes\ket{0^a})$. 

\begin{enumerate}
\item \label{item:hist-preparation} $P_\Q$ computes the history state corresponding to running $C$ on $\ket{\psi}\otimes\ket{0^a}$.  Denote this state by $\Hist_C(\ket{\psi}\otimes\ket{0^a})$.
\item  $P_\Q$ and $P_\C$ execute protocol $\Com$ on input $(1^\secpar,1^n)$, and $P_\Q$ takes as additional input the history state $\Hist_C(\ket{\psi}\otimes\ket{0^a})$.  Denote the output of $P_\C$ by $\com$ and the output of $P_\Q$ by $\calR_\op$. 

\item \label{item:cert} $P_\Q$ proves to $P_\C$ that it committed to a history state, by running the repeated history state test from \Cref{thm:repeated-hist-test} (\Cref{sec:hist}). This is done by repeating the following sub-protocol $r=M\cdot (T+1)^2\cdot\secpar$ times, where $M$ is the number of local Hamiltonians in the history state Hamiltonian $H_\Hist(C)=\sum_{j=1}^M \Pi_{C,j}$ (\Cref{thm:xz-fk}), and $T$ is the number of gates in the circuit~$C$.

\begin{enumerate}
    \item $P_\C$ samples $j\gets[M]$ uniformly at random, and sends $j$ to $P_\Q$. 

    Let $U_j=H^{\theta_j}$ as defined in \Cref{thm:xz-fk}.\footnote{Note that $U_j=H^{\theta_j}$ is the rotation needed to project register $\calR_\op$ onto $\{\Pi_{C,j},I-\Pi_{C,j}\}$.}
    \item $P_\Q$ computes $\widetilde{U}_j\gets \Eval(U_j)$ and  applies $\tilde{U}_j$ to its $\calR_\op$ register. 
    \item $P_\C$ computes $\com_{U_j}\gets\KeyEval(\com, U_j)$.  
   
    \item $P_\C$ and $P_\Q$ run a state-preserving classical $2\PC$, where $P_\Q$ takes as input register $\calR_\op$, $P_\C$ takes as input $\com_{U_j}$, and $P_\C$ obtains the output of the computation
    \[
    F_j(\com_{U_j},\op):=(\Ver_{\com_{U_j}}(\op),(g_j\circ \Dec_{\com_{U_j}})(\op))
    \]
    where $\op$ is the value in register $\calR_\op$, and  $g_j$ is defined in \Cref{eqn:Pij} in \Cref{thm:xz-fk} (\Cref{sec:hist}).  Note that $g_j\circ \Dec_{\com_{U_j}}$ is constant degree because $g_j$ is constant degree and $\Dec_{\com_{U_j}}$ is linear.
    $P_\C$ outputs~$b=1$ if $\Ver_{\com_{U_j}}(\op)=\bot$ or if $(g_j\circ \Dec_{\com_{U_j}})(\op)=1$.  Namely, $b$ is the rejection bit and that the residual state is released only when $b=0$.
   
    \item If $b=1$ then $P_\C$ aborts and both parties output ${\sf reject}$. 
    \item $P_\Q$ applies $\widetilde{U}_j^\dagger$ to $\calR_\op$.
    
\end{enumerate}

\item $P_\C$ outputs~$\com$ and $P_\Q$ outputs $\calR_\op$.

\end{enumerate}

\subsection{Protocol $\Open(U,f)$}\label{sec:hist-open}

 We next describe the protocol  $\Open$.  In this protocol, $P_\Q$ takes as input a state $\calR_{\op}$ and $P_\C$ takes as input 
$\com$, corresponding to an output of the commitment protocol $\Com$. They both take as an additional input ae  pair $(U,f)$, where $U$ is a unitary that acts on the committed register and $f$ is a constant degree polynomial. %

\begin{protocol}\label{protocol:open}
 The protocol proceeds as follows.   

\begin{enumerate}
  \item $P_\Q$ computes
 $\widetilde U \leftarrow \Eval(U)$,
  and applies $\widetilde U$ to $\calR_\op$.

  \item $P_\C$ computes
  $\com_U
      \leftarrow
      \mathsf{KeyEval}(\com,U)$.

  \item \label{item:open:2PC} The parties execute the state-preserving classical $2\PC$ from \Cref{sec:classical-2PC} for the following functionality:
  \[
  F_{f,U}(\mathsf{com}_U,\mathsf{op})
  :=
  \left(
     \mathsf{Ver}_{\mathsf{com}_U}(\mathsf{op}),
     (f(
       \mathsf{Dec}_{\mathsf{com}_U}(\mathsf{op}))
  \right),
  \]
  where $P_\Q$ supplies the contents $\op$ of
  $\calR_{\op}$, and $P_\C$ supplies $\com_U$.\footnote{Recall that $\Dec_{\com_U}$ and $\Ver_{\com_U}$ are both linear functions, and $f$ is a constant-degree polynomial. Hence we can use the  state-preserving classical $2\PC$ from \Cref{sec:classical-2PC}.}

  The post-processed abort predicate releases the residual quantum
  state to $P_\Q$ if and only if
  $\Ver_{\com_U}(\op)=\top$.

  \item If the $2\PC$ aborts or $\Ver_{\com_U}(\op)=\bot$, then
  $P_\C$ outputs $\bot$ and the protocol terminates.

  \item Otherwise, let
  $y=f(
      \Dec_{\com_U}(\op))$
  be the value obtained by $P_\C$.  Party $P_\Q$ applies
  $\widetilde U^\dagger$ to the residual register
  $\calR_{\op}$.

  Party $P_\C$ outputs $y$, while $P_\Q$ retains the resulting state
  $\calR_\op$.  
\end{enumerate}
\end{protocol}

\subsection{Analysis}\label{sec:hist-analysis}

In this section we prove the security of the protocol
\[\Pi_{\mathsf{HistCom}}=(\mathsf{CertCom},\mathsf{Open},
\mathsf{Recover}).
\]
We focus on the case of full recovery, but the same proof also shows partial recovery when $\Recover$ is replaced by its partial recovery counterpart from the underlying commitment scheme.

\begin{theorem}\label{thm:quantum}
There exists a uniform sequence of QPT oracle circuits $\{\Sim_{\secpar, n}\}_{\secpar, n \in \bbN}$ such that for every $\QPT$ malicious quantum party $P_Q^*$ in protocol $\Pi_\mathsf{HistCom}$, every input $(C,\ket{\psi},(U_i, f_{i})_{i\in [\ell]})$, every sufficiently large $\secpar \in \bbN$ and every $n = \poly[\secpar]$,
\[
\Real^{\Pi_\mathsf{HistCom}}_{P^*_\Q}(C,\ket{\psi},(U_i, f_{i})_{i\in [\ell]},1^\secp)
\overset{1/n}{\approx} 
\Ideal_{\Sim_{\mathsf{HistCom}}}(C,\ket{\psi},(U_i, f_{i})_{i\in [\ell]},1^\secp).
\]
\end{theorem}
We prove \Cref{thm:quantum} by defining for any fixed $P^*_\Q$, $\secpar$, and $n$, a simulator $\Sim_{\mathsf{HistCom}}$ composed of three simulators:
\[
    \Sim_{\mathsf{HistCom}}
      =
    \bigl(
      \Sim_{\mathsf{Cert}},
      \Sim_{\mathsf{Open}},
      \Sim_{\mathsf{Recover}}
    \bigr)
\]
with the following interfaces. We suppress the dependence on $\secpar$ and $n$ to ease notational density and explicitly mention them where they are directly relevant.

\begin{itemize}
    \item
    $\Sim_{\mathsf{Cert}}(C,\ket{\psi})$ emulates the communication in 
    $\mathsf{CertCom}$ and outputs an ideal register $\calV$, a private simulator state $\tau$,
   and the certification outcome
    $b_{\mathsf{Hist}}$.

    \item
    $\Sim_{\mathsf{Open}}
       (U,f; \calV,\tau)$
    emulates the communication in one invocation of $\mathsf{Open}(U,f)$, given oracle access to the ideal functionality that takes as input $(U,f)$ from $P_\C$ and from a register $\calV$ from $\Sim_{\mathsf{Open}}$. It also receives a bit $a\in\{0,1\}$ from $\Sim_{\mathsf{Open}}$, where $a=1$ indicates abort. If $a=1$
    the ideal functionality outputs $\perp$.
    Otherwise, it performs the ideal measurement
    \[
        (y,\calV')
          \leftarrow
        M[f,U](\calV),
    \]
    and sends $y$ to $P_\C$, and
    sends the updated post-measurement ideal register $\calV'$ to  $\Sim_{\mathsf{Open}}$, who outputs $\calV'$ together with a private auxiliary state $\tau'$.

    \item
    $\Sim_{\mathsf{Recover}}(\calV',\tau')$
    emulates $\mathsf{Recover}$ using the current ideal register $\calV'$ and the current simulator state $\tau'$, and terminates the
    certified-history execution.
\end{itemize}

We prove that for every polynomial-size admissible opening family and every
polynomial-length adaptively chosen sequence of opening requests,
followed by at most one recovery request, the execution obtained by
invoking these simulators sequentially, always passing the updated
pair $(\calV,\tau)$ from one invocation to the next, is computationally
indistinguishable from the corresponding real execution.
More formally, the proof of \Cref{thm:quantum} relies on  \Cref{thm:certcom-security} and \Cref{thm:open-security}, stated and proved in \Cref{sec:CertCom:sec} and \Cref{sec:Open:sec}, respectively.

\subsubsection{Security of $\mathsf{CertCom}$}\label{sec:CertCom:sec}

  \begin{theorem}\label{thm:certcom-security}
      For every $\QPT$ adversary $P_\Q^*$ in \Cref{protocol:hist} there exists a uniform family of $\QPT$ simulators $\{\Sim_{\mathsf{Cert}, n}\}_{n\in \bbN}$\footnote{We suppress the dependence on $\secpar$ to ease notational density.} in the ideal experiment, such that 
      \[
        \Real^{\mathsf{CertCom}}_{P_{\Q^*}}(C,\ket{\psi},1^\secpar)
        \overset{1/n}{\approx} 
        \Ideal^{\mathsf{CertCom}}_{\Sim_\mathsf{Cert}}(C,\ket{\psi}).
      \]
      where $\Ideal^{\mathsf{CertCom}}_{\Sim_\mathsf{Cert}}(C,\ket{\psi})$ is the output of the simulator $\Sim_\mathsf{Cert}$  that only executes Steps 1-3 of the ideal experiment.

  \end{theorem}

 \begin{proof}
Fix a $\QPT$ adversary $P_\Q^*$, and as above, let 
\[
    H_{\Hist}(C)=\sum_{j=1}^M \Pi_{C,j},
    \qquad
    G:=\Pi_{\ker H_{\Hist}(C)}.
\]
By \Cref{thm:xz-fk},
\[
    H_{\Hist}(C)\succeq \gamma (I-G),
    \qquad
    \gamma:=\frac{c}{(T+1)^2}.
\]
Set
\[
    \varepsilon
      :=
      \left(1-\frac{\gamma}{M}\right)^r
      \leq e^{-\gamma r/M}.
\]
The fact that $r=M(T+1)^2\lambda$ implies that $\varepsilon\leq e^{-c\secpar}=\negl$. Fix an $n =\poly[\secpar]$ and set $\eta = 2rn$.

The proof proceeds through the following hybrids.

\paragraph{Hybrid $\mathsf{H}_0$.}
This is the real execution of \Cref{protocol:hist} with $P_\Q^*$.

\paragraph{Hybrid $\mathsf{H}_1$: idealizing the $2\PC$ executions.}
Replace each of the $r$ state-preserving $2\PC$ executions by $1/\eta$-simulated transcripts using calls to the ideal
functionality.
For the $j$'th test, the functionality measures the 
\[
    F_j(\com_{U_j},\op)
    =
    \left(\Ver_{\com_{U_j}},\left(g_j\circ \Dec_{\com_{U_j}}\right)(\op)\right),
\] 
and aborts if $\Ver_{\com_{U_j}}(\op)=\bot$ or if $(g_j\circ \Dec_{\com_{U_j}})(\op)=1$.
By the $1/\eta$-security of the state-preserving $2\PC$ against a malicious quantum party, and a standard sequential hybrid over the $r=\poly$ executions where $\eta = 2rn$,
\[
    \mathsf{H}_0 \overset{1/(2n)}{\approx_c} \mathsf{H}_1.
\]

\paragraph{Hybrid $\mathsf{H}_2$: idealizing the commitment.}

Note that Hybrid $\mathsf{H}_1$ can be thought of as a real continual-binding-with-recovery experiment (\Cref{def:cont-bind-recovery}), with an adversary that is defined by~$P_\Q^*$, where the permitted logical measurements are the fixed family
\[
    \mathcal{M}_{\mathsf{Hist}}=\bigl\{{\sf M}[g_j,U_j]\bigr\}_{j\in[M]},
    \qquad U_j=H^{\theta_j}.
\]
Let
\begin{gather*}
  \mathcal{M}_{\mathsf{Open}}
  :=
  \left\{
    \mathsf{M}[f_{i},U_i]
  \right\}_{i\in[q]},
  \\
  \mathcal{M}_{\mathsf{all}}
  :=
  \mathcal{M}_{\mathsf{Hist}}
  \cup
  \mathcal{M}_{\mathsf{Open}}.
\end{gather*}

$H_2$ instantiates the continual-binding-with-recovery simulator (\Cref{def:cont-bind-recovery})
w.r.t.\ the
entire fixed family $\mathcal{M}_{\mathsf{all}}$ and with indistinguishability parameter $2n$.  
During the present $\mathsf{CertCom}$ phase, only measurements from $\mathcal{M}_{\mathsf{Hist}}$ are invoked.

By \Cref{eqn:Pij},
\[
  U_j^\dagger
  \left(
    \sum_{z:g_j(z)=1}
    \ket{z}\!\bra{z}
  \right)
  U_j
  =
  \Pi_{C,j}.
\]
By the query-preserving continual ideal binding property (\Cref{def:fnal-com-continual-binding}), the commitment can therefore
be replaced by a single ideal register $\calV$, possibly entangled with the
adversary's remaining registers, denoted by $\calE$, such that every valid
$j$-th query performs the measurement
\[
  \{I-\Pi_{C,j},\Pi_{C,j}\}
\]
on $\calV$, while an invalid query causes an abort and induces no operation on
$\calV$.  Hence
\[
  H_1 \overset{1/(2n)}{\approx_c} H_2.
\]

\paragraph{Comparison with the history-state ideal functionality.}

We next construct the simulator $\Sim$ that can subsequently
be continued through the protocols $\Open$ and
$\Recover$.  $\Sim$ proceeds as follows:
\begin{enumerate}
    \item Emulate
the entire ${\sf CertCom}$ protocol using the internally run continual-binding
functionality.  For each sampled index~$j$, the continual-binding functionality
performs the measurement
\[
    \{I-\Pi_{C,j},\Pi_{C,j}\}
\]
on the current ideal register $\calV$.  
Thus, the simulator reproduces 
the sequence of sampled indices and the first rejecting test, if one
occurs.
\item Transfer the
current private register $\calV$ from the continual-binding ideal functionality
to the history-state ideal functionality, which projects $\calV$ onto $\{G,I-G\}$.  

If the emulated ${\sf CertCom}$ protocol aborts then send $a=0$. Otherwise, send $a=1$.

\end{enumerate}

It remains to show that inserting the measurement $\{G,I-G\}$ changes the experiment only negligibly.  
We argue this separately for the aborting branch  and the non-aborting branch.

On the aborting branch (i.e., the branch that eventually rejects), the simulator sends an abort message $a=0$ to the ideal functionality.  As a result, $\calV$ is not included in the simulated adversary's view, and the outcome of the projection $\{G,I-G\}$ is ignored on the aborting branch. Hence the measurement $\{G,I-G\}$ does not change the output of the experiment.

We next focus on the non-aborting branch.  Let $\rho_{\calV \calE}$ be the state immediately
before the first local test, where $\calV$ is the ideal committed register and
$\calE$ contains all other registers.  Let $\sigma_{\mathrm{acc}}$ denote the
resulting subnormalized state on the non-aborting branch,\footnote{Note that $\sigma_{\mathrm{acc}}$ accepts the $r$ tests of the protocol, but is not necessarily in ${\sf im}(G)$.} and
set
\[
\delta
:=
\Tr\!\left(
  ((I-G)\otimes I_\calE)\sigma_{\mathrm{acc}}
\right).
\]

For a sequence
$\mathbf{j}=(j_1,\ldots,j_r)\in[M]^r$, denote by $P_{j_i}:=I-\Pi_{C,j_i}$, and let 
\[
K_{\mathbf{j}}
:=
P_{j_r}\cdots P_{j_1}.
\]

Note that all operations other than the
projections $P_{j_i}$ act only on register $\calE$.  This follows from the fact that in the continual-binding ideal experiment, after register $\calV$ is sent to the ideal functionality, the functionality keeps $\calV$, and the adversary retains only register $\calE$. 
Moreover, the simulator constructed in the proof of \Cref{thm:com}
is query-preserving, in the sense that after the commitment phase, every valid protocol query with index~$j$ induces exactly one query with the same index to the
continual-binding ideal functionality, whereas an invalid query induces
no operation on the ideal register $\calV$.  Consequently, on the
non-aborting branch, the only operations on $\calV$ are
$P_{j_1},\ldots,P_{j_r}$.  Conditioned on this branch, all remaining
operations form a positive trace-nonincreasing map acting
only on $\calE$. Therefore,

\begin{align*}
\delta
&\le
\mathbb{E}_{\mathbf{j}\leftarrow[M]^r}
\Tr\!\left[
  \bigl(K_{\mathbf{j}}(I-G)\otimes I_\calE\bigr)
  \rho_{\calV \calE}
  \bigl((I-G)K_{\mathbf{j}}^\dagger\otimes I_\calE\bigr)
\right]
\\
&=
\mathbb{E}_{\mathbf{j}\leftarrow[M]^r}
\Tr\!\left[
  \bigl(K_{\mathbf{j}}^\dagger(I-G)K_{\mathbf{j}}
        \otimes I_\calE\bigr)\rho_{\calV \calE}
\right].
\end{align*}
Note that by the definition of $P_j$,
\begin{align*}
P_j(I-G)P_j
&=
(I-\Pi_{C,j})(I-G)(I-\Pi_{C,j})
\\
&=
(I-G)
-\Pi_{C,j}(I-G)
-(I-G)\Pi_{C,j}
+\Pi_{C,j}(I-G)\Pi_{C,j}
\\
&=
(I-G)-\Pi_{C,j}-\Pi_{C,j}+\Pi_{C,j}^{2}\\
&=
(I-G)-\Pi_{C,j}.
\end{align*}
This, together with \Cref{thm:xz-fk} (\Cref{sec:hist}), implies that 
\begin{align*}
\mathbb{E}_{j\leftarrow[M]}
\left[P_j(I-G)P_j\right]
&=
(I-G)-\frac{1}{M}H_{\Hist}(C)
\\
&\preceq
\left(1-\frac{\gamma}{M}\right)(I-G).
\end{align*}
where $\gamma:=\frac{c}{(T+1)^2}$.
Iterating this operator inequality gives
\[
\mathbb{E}_{\mathbf{j}\leftarrow[M]^r}
\left[
  K_{\mathbf{j}}^\dagger(I-G)K_{\mathbf{j}}
\right]
\preceq
\left(1-\frac{\gamma}{M}\right)^r(I-G).
\]
Therefore,
\begin{align*}
\delta
&\le
\left(1-\frac{\gamma}{M}\right)^r
\Tr\!\left(
  ((I-G)\otimes I_\calE)\rho_{\calV \calE}
\right)
\le
\left(1-\frac{\gamma}{M}\right)^r
.
\end{align*}

In Hybrid $H_2$, all of $\sigma_{\mathrm{acc}}$ is labeled with verifier
output $1$.  After inserting the global measurement, the part
$G\sigma_{\mathrm{acc}}G$ is labeled with verifier output $1$, while the
part
\[
(I-G)\sigma_{\mathrm{acc}}(I-G)
\]
is labeled with verifier output $0$.  In what follows, we suppress the identity
operator on $\calE$.  By the subnormalized gentle-measurement bound,
\[
\left\|
  \sigma_{\mathrm{acc}}
  -
  G\sigma_{\mathrm{acc}}G
\right\|_1
\le
2\sqrt{
  \Tr(\sigma_{\mathrm{acc}})
  \Tr((I-G)\sigma_{\mathrm{acc}})
}
\leq
2\sqrt{
   \Tr((I-G)\sigma_{\mathrm{acc}})
}
=
2\sqrt{\delta}.
\]
Note that the acceptance branch in Hybrid $H_2$ is \[\rho_{H_2}:=\ket{1}\!\bra{1}\otimes\sigma_{\mathrm{acc}}.
\]The accepted branch in $ \Ideal_{\Sim}(C,\ket{\psi})$ is
\[
\rho_{\sf Ideal}:=\ket{1}\!\bra{1}\otimes G\sigma_{\mathrm{acc}}G
  +
  \ket{0}\!\bra{0}\otimes
  (I-G)\sigma_{\mathrm{acc}}(I-G).
\]
Therefore, the trace distance is
\begin{align*}
&
\frac{1}{2}\left\|\rho_{H_2}-\rho_{\sf Ideal}\right\|_1=\\
&\frac12\left\|\ket{1}\!\bra{1}\otimes\sigma_{\mathrm{acc}}
  -
  \ket{1}\!\bra{1}\otimes G\sigma_{\mathrm{acc}}G
  -
  \ket{0}\!\bra{0}\otimes
  (I-G)\sigma_{\mathrm{acc}}(I-G)
\right\|_1
\\
&\qquad\le
\frac12\left\|
  \sigma_{\mathrm{acc}}
  -
  G\sigma_{\mathrm{acc}}G
\right\|_1
+
\frac12\Tr((I-G)\sigma_{\mathrm{acc}})
\\
&\qquad\le
\sqrt{\delta}+\frac{\delta}{2},
\end{align*}
which is negligible.
Thus
\[
    \mathsf{H}_2
      \equiv_s
      \Ideal^{1\mbox{-}3}_{\Sim}(C,\ket{\psi})
\]
as desired.  Combining the hybrids gives the claimed result
\[
    \Real_{P_\Q^*}(C,|\psi\rangle,1^\lambda)
      \overset{1/n}{\approx_c}
    \Ideal^{1\mbox{-}3}_{\Sim}(C,|\psi\rangle,1^\lambda).
\]
for every $n = \poly[\secpar]$.
\end{proof}

\end{protocol}
\subsubsection{Security of $\Open$ }\label{sec:Open:sec}

\paragraph{The hybrid opening protocol.}
Let $\mathsf{Open}^{\mathsf{hyb}}(U,f)$ denote the protocol obtained
from $\mathsf{Open}(U,f)$ by replacing the state-preserving classical
$2\PC$ in Step~\ref{item:open:2PC} with its ideal functionality.  All other steps of the
protocol remain unchanged.

\begin{theorem}[Security of $\mathsf{Open}$]
\label{thm:open-security}
For every $n=\poly[\secpar]$ and every $\QPT$ malicious quantum party $P_\Q^*$, there exists a  $\QPT$ party $P^{\mathsf{Hyb}}_\Q$ such that for every admissible opening request $(U,f)$, every classical input string $\mathsf{com}$, and every quantum input $\rho$, containing register $\calR_{\mathsf{op}}$ and auxiliary register $\calE$, and thus denoted by $\rho_{\calR_{\mathsf{op}}\calE}$, 
\[
    \mathsf{Open}_{P_\Q^*}
      (U,f;\mathsf{com},\rho_{\calR_{\mathsf{op}}\calE})
    \overset{1/n}{\approx_c}
    \mathsf{Open}^{\mathsf{Hyb}}_{P^{\mathsf{Hyb}}_\Q}
      (U,f;\mathsf{com},\rho_{\calR_{\mathsf{op}}\calE}).
\]
In both experiments, $P_\C$ starts with $\mathsf{com}$ and
$P_Q^*$ starts with the state $\rho_{\calR_{\mathsf{op}}\calE}$, and both are given $(U,f)$.
The experiments output the output of $P_\C$ and the entire final
view of $P_\Q^*$.

The experiment $\mathsf{Open}$ executes \Cref{protocol:open}, whereas
$\mathsf{Open}^\mathsf{Hyb}$ is identical except that the state-preserving classical
$2\PC$ in Step~\ref{item:open:2PC} is replaced by its ideal functionality.
\end{theorem}

\begin{proof}
The two experiments start with the same inputs and differ only in
Step~\ref{item:open:2PC} of \Cref{protocol:open}. In $\mathsf{Open}$, the parties execute the real
state-preserving classical $2\PC$, whereas in $\mathsf{Open}^\mathsf{Hyb}$ they invoke
its ideal functionality with $1/n$ simulation. All operations before and after Step~\ref{item:open:2PC}
are identical.

The claim therefore follows directly from the security of the
state-preserving classical $2\PC$ against a malicious quantum party,
with $\calE$ included as part of the auxiliary input and residual state
of $P_\Q^*$.
\end{proof}
  
\subsubsection{Security of the Composed Protocol}

We next show how to use \Cref{thm:certcom-security} and \Cref{thm:open-security} to prove \Cref{thm:quantum}.

\paragraph{Proof of \Cref{thm:quantum}}

Fix a $\QPT$ malicious quantum party $P_\Q^*$, and let
$L=\poly$ be an upper bound on the number of invocations of
$\mathsf{Open}$. Set $\eta = Ln$.

For every $j\in\{0,\ldots,L\}$, let $H_j$ denote the execution in which the
first $j$ invocations of $\mathsf{Open}$ are replaced by
$\mathsf{Open}^{\mathsf{hyb}}$, while all subsequent invocations use
the real protocol $\mathsf{Open}$. Thus, $H_0$ is the real execution
of $\Pi_{\mathsf{HistCom}}$.  In what follows we prove that 
\begin{equation}\label{eqn:H0L}
     H_0\overset{1/n}{\approx_c} H_L.
\end{equation}
By a standard hybrid argument, it suffices to prove that for every $j\in[L]$, 
\[
    H_{j-1} \overset{1/\eta}{\approx_c} H_j.
\]
This follows immediately by \Cref{thm:open-security} with indistinguishability parameter $\eta$ applied to
the actual inputs of the two parties immediately before the $j$'th invocation. The fact that \Cref{thm:open-security} includes the entire residual state of
$P_\Q^*$ in its output, the remainder of the execution is $\QPT$
post-processing. Hence $H_{j-1}$ is computationally $1/\eta$-indistinguishable from $H_j$ as desired.

It remains to prove that
\[
    H_L
      \approx_c
    \Ideal_{\Sim_{\mathsf{HistCom}}}.
\]
Recall that the continual-binding-with-recovery simulator was
instantiated with respect to
\[
    \mathcal M_{\mathsf{all}}
      =
    \mathcal M_{\mathsf{Hist}}
      \cup
    \mathcal M_{\mathsf{Open}}.
\]
Consider an invocation
$\mathsf{Open}^{\mathsf{hyb}}(U_i,f_{i})$.
The ideal $2\PC$ invocation evaluates the corresponding decoding query
on the current opening register.  The simulator
$\Sim_{\mathsf{Open}}$ runs the query simulator from the
continual-binding with recovery experiment on this query (see \Cref{def:cont-bind-recovery}).  If the query simulator
rejects, $\Sim_{\mathsf{Open}}$ requests abort, outputs $\perp$, and
induces no operation on the ideal register $\calV$.  Otherwise, by query
preservation, it forwards the same request
$(U_i,f_{i})$ to the history-state ideal functionality.  The
functionality applies
\[
    M[f_{i},U_i]
\]
to its current register $\calV$, sends the outcome to $P_\C$, and retains
the post-measurement residual state as the new $\calV$.

Thus, every invocation of $\Open^{\mathsf{hyb}}$ is simulated by
$\Sim_{\Open}$, with all invocations acting on the same evolving ideal
register $\calV$. Upon a recovery request, the recovery component of the
continual-binding-with-recovery simulator receives the final residual
register $\calV$ and emulates $\Recover$. The $\mathsf{CerCom}$ protocol  is
simulated by $\Sim_{\mathsf{Cert}}$, as established in \Cref{thm:certcom-security}.

By the continual-binding-with-recovery guarantee (\Cref{def:cont-bind-recovery}),
replacing the underlying commitment together with all subsequent
opening and recovery queries by this single stateful simulation changes
the execution only negligibly. 
Moreover, inserting the projection $\{G,I-G\}$ can only change the experiment by a negligible statistical
error, as argued in the proof of
\Cref{thm:certcom-security}. Therefore,
\[
    H_L
      \approx_c
    \Ideal_{\Sim_{\mathsf{HistCom}}}.
\]
Combining this with \Cref{eqn:H0L} proves the theorem.

\qed

\section{State-Preserving $2\PC$ for General Quantum Functions}\label{sec:general}

Next, we give a secure two-party computation protocol which computes \emph{any} polynomial-time quantum functionality which can be implemented using only real gates. That, is, it computes a general quantum circuit $G$ which takes as input $\ket{\psi}$ from the quantum party and input $x$ from the classical party, applies a unitary, discards part of the output, measures the classical output, and sends the coherent and classical outputs to the quantum and classical parties, respectively.
The 2PC guarantees privacy for both parties: neither party learns any more information other than their part of the output $G(\ket{\psi}, x)$. Moreover, the parties receive the correct outputs (up to computational indistinguishability) or $\bot$, regardless of the other party's behavior.

This allows a classical client to implement any ``single-shot'' quantum computation on untrusted quantum hardware, including those with randomized or quantum outputs. We show in \Cref{sec:persistent-2pc} how to extend this to maintain persistent quantum memory across executions.

\begin{theorem}
    There exists an $\epsilon$-secure (respectively, secure) semi-quantum 2PC for general real quantum circuits, assuming the following primitives:
    \begin{itemize}
        \item A classical commitment to quantum states with $\epsilon$-secure (respectively, secure) continual binding with partial recovery.
        \item An $\epsilon$-secure (respectively, secure) state-preserving 2PC for constant-degree classical functions with classical post-processed abort.
        \item A circuit-private quantum fully-homomorphic encryption scheme.
    \end{itemize}
    All of these primitives exist with $\epsilon$-security assuming the post-quantum hardness of LWE.
\end{theorem}

\subsection{Construction}

In what follows we give a protocol that is secure against a $\QPT$ malicious quantum party $P^*_\Q$ and an honest-but-curious classical party $P_\C$.  Using standard techniques one can upgrade this protocol to make it secure against a malicious poly-size classical party as well. See \Cref{sec:prelim:2pc} for more details.

Our protocol makes use of the following ingredients:
\begin{itemize}
    \item A circuit-private quantum fully homomorphic encryption scheme ($\QFHE$) \[(\Gen_\QFHE,\QEnc,\QEval,\QDec),
    \]as defined in \Cref{sec:QFHE}.

    We assume that this $\QFHE$ uses as a building block a circuit-private classical $\FHE$ scheme
    \[
    (\Gen_\FHE, \Enc, \Eval, \Dec)
    \]
    where $\Gen_\FHE=\Gen_\QFHE$ and $\Enc$ is a bit-by-bit encryption,
    and we assume that $\QEnc$ is  of the form
    \[\QEnc(\pk,\ket{\psi})= X^x Z^z \ket{\psi}, \Enc(\pk,(x, z)).
    \] 
Moreover, we assume that for every quantum circuit $C$ that takes as input a state $\ket{\psi}$ and classical input $w$, the state obtained by running $\QEval(C, ({\ket{\widehat{\psi}}},\widehat{w}))$, where $\ket{\widehat{\psi}}\gets \QEnc(\pk,\ket{\psi})$ and  $\widehat{w}\gets\QEnc(\pk,w)$, is of the form 
\[X^x Z^z C(\ket{\psi},w),  \widehat{x},\widehat{z},
    \]
    where $x=\Dec(\sk,\widehat{x})$ and  $z=\Dec(\sk,\widehat{z})$.
    
    \item Protocols $({\sf CertCom}, \Open, \Recover)$ from \Cref{sec:com-hist}, where $\Recover$ allows partial recovery.
    These protocols in turn rely on a  classical commitment scheme for quantum states with continual binding with recovery, as defined in \Cref{sec:fnal-commitments} (specifically, \Cref{sec:cont-binding-recoverty}), and on a state-preserving   $2\PC$ for constant-degree classical functions with classical post-processed abort (see \Cref{sec:classical-2PC}).
       
\end{itemize}

\begin{protocol}\label{protocol:quantum}
  The quantum party $P_\Q$ takes as input a quantum state $\ket{\psi}$,  the classical party $P_\C$ takes as input a classical input $w\in\{0,1\}^n$,  and their goal is to securely compute $G(\ket{\psi}, w)$, where $G$ is a general quantum circuit. 
  
  Let $U_G$ be the unitary circuit component of $G$ and recall that its output may be divided into a coherent output register $\calB$, a measured output register $\calC$, and a discarded output register $\calD$.  

    \begin{enumerate}
    \item \label{item:enc}{\bf Encrypting $w$.} $P_\C$ does the following:
    \begin{enumerate}
        \item Sample $(\pk,\sk)\gets\Gen_\QFHE(1^\secp)$.
        \item Compute $\widehat{w}\gets\QEnc(\pk,w)$.
        \item Choose random pads $x_0,z_0\gets\{0,1\}^k$, where $k$ is the number of output qubits of $U_G$, and compute
        $\widehat{x}_0,\widehat{z}_0\gets\QEnc(\pk,(x_0,z_0))$.
        
         \item Send $(\pk,\widehat{w}, \widehat{x}_0, \widehat{z}_0)$ to $P_\Q$.
         \end{enumerate} 
        In what follows we let $C=C_{\pk,\widehat{w},\widehat{x}_0,\widehat{z}_0}$ be the unitary circuit that takes as input a quantum state $\ket{\psi}$ along with ancilla registers $\ket{0^a}$, and uses the ancilla registers to compute $\ket{\widehat{\psi}}\gets\QEnc(\pk,\ket{\psi})$  and compute $\QEval(\widetilde{U}_G,\ket{\widehat{\psi}},\widehat{w}, \widehat{x}_0,\widehat{z}_0)$,
            where 
            \[
                \widetilde{U}_G(\ket{\psi},w,x_0,z_0)=X^{x_0} Z^{z_0} U_G(\ket{\psi},w).
            \]
           
            If one were to compute $C$ on $(\ket{\psi}, \hat{w})$ and discard the garbage ancillas, then the resulting state would be of the form    
            \[
                X^{x\oplus x_0} Z^{z\oplus z_0} U_G(\ket{\psi},w),\ \widehat{x}, \widehat{z},
            \]
            where $\Dec(\sk,\widehat{x})=x$ and $\Dec(\sk,\widehat{z})=z$. We denote the register containing the output ciphertext (and excluding the ancillas) as $\calR_{\outp}$. 
            By partitioning the padded state, $\calR_{\outp}$ may be further subdivided into $\calR_{\outp} = (\widetilde{\calB}, \widetilde{\calC}, \widetilde{\calD}, \widetilde{\calP})$, the padded registers for coherent output ($\widetilde{\calB}$), measured output ($\widetilde{\calC}$), discards ($\widetilde{\calD}$), and the encrypted pads ($\widetilde{\calP}$).

   We assume without loss of generality that all the gates of $C$ are in the set $\calG_{X,Z}$ and denote the number of such gates by~$T$.  This assumption is without loss of generality since $\calG_{X,Z}$ is universal for real-valued quantum computation.
    We denote $C=U_T\cdot\ldots\cdot U_1$.

    \item \label{item:hist} {\bf Certified commitment to the history state.} $P_\Q$ and $P_\C$ run protocol ${\sf CertCom}$ (\Cref{protocol:hist}), where party $P_\Q$ takes as input $(C,\ket{\psi},1^\secpar)$ and $P_\C$ takes as input $(C,1^\secpar,1^N)$, where $N$ is the number of qubits in the history state $\Hist_C(\ket{\psi}\otimes\ket{0^a})$.

    If at the end of ${\sf CertCom}$  $P_\C$ or $P_\Q$ output $\mathsf{reject}$  , then this protocol is terminated and the parties output $\bot$. Otherwise, at the end of ${\sf CertCom}$, $P_\Q$ has a register $\calR_\op$, and $P_\C$ has a classical string $\com$ that is used to decode the openings.
    
    \item {\bf Clock measurement.} \label{item:clock} $P_\Q$ and $P_\C$ run the protocol $\Open(I, f_{\sf clock})$ (\Cref{protocol:open}, \Cref{sec:hist-open}) where $I$ is the identity unitary and $f_{\sf clock}$ is the classical function that projects onto the clock register, to obtain an output~$t$. 
    
    Note that $f_{\sf clock}$  is a linear function. 
    
    \item\label{item:smallt} If $t<T$ then the parties do the following:
    \begin{enumerate}
        \item Run the $\Recover$ protocol (of the underlying commitment scheme) where $P_\Q$ takes as input $\calR_\op$ and $P_\C$ takes as input $\com$, and they both take as input $(1^\secpar,1^N)$.  At the end of this protocol $P_\Q$ obtains a state $\ket{\varphi}$. 
          \item $P_\Q$ computes  $\ket{\psi}=U^\dagger_1\cdot \ldots\cdot U^\dagger_t\ket{\varphi}$.
         \item The parties restart the protocol from \Cref{item:hist}.
    \end{enumerate} 

    This step is repeated at most $K:=\secpar(T+1)$ times, and if $t<T$ for $K$ times then the parties output $\bot$.

    \item If $t=T$ then $P_\Q$ has a commitment to the state 
    \[
        X^{x\oplus x_0} Z^{z\oplus z_0} U_G(\ket{\psi},w)\otimes \ket{\widehat{x}}\otimes\ket{\widehat{z}}\otimes\ket{T}
    \]
    together with the ancilla registers from homomorphic evaluation.
    It continues as follows.

    \item \textbf{Certified opening of the pad ciphertext.} \label{item:open}
    Let $f_{{\sf pad}}$ be the projection onto the subregister $\widetilde{\calP}$ containing the encrypted pad $(\widehat{x},\widehat{z})$.\footnote{Note that classical projections are linear.}
    The parties run
    \[
        (\widehat{x},\widehat{z})
        \leftarrow
        \mathsf{Open}(I,f_{\mathrm{pad}}).
    \]
    If this invocation outputs $\perp$, then both parties output $\perp$
    and terminate.
    
    Otherwise, $P_\C$ computes
    \[
       x
        =
        \mathsf{Dec}
        \bigl(
            \sk,
            \widehat{x}
        \bigr)~~\mbox{ and }~~ z
=
        \mathsf{Dec}
        \bigl(
            \sk,
            \widehat{z}
        \bigr)
    \]
    If decryption fails, then $P_\C$ sends an abort message, both parties
    output $\perp$, and the protocol terminates. If it succeeds, $P_{\C}$ partitions the pads as
    \[
        \begin{array}{ll}
            x = (x_\calB,\ x_\calC,\ x_\calD) &\: x_0 =(x_{0,\calB},\ x_{0,\calC},\ x_{0,\calD})
            \\
            z = (z_\calB,\ z_\calC,\ z_\calD) &\: z_0 =(z_{0,\calB},\ z_{0,\calC},\ z_{0,\calD})
        \end{array}
    \]
    
    \item \label{item:quantum2PC:out} 
    \textbf{Measured Output.}
    $P_\Q$ and $P_\C$ execute the protocol $\Open(I, f_{\widetilde{\calC}})$ (\Cref{protocol:open}, \Cref{sec:hist-open}) where $I$ is the identity unitary and $f_{\widetilde{\calC}}$ is the function that projects onto the subregister $\widetilde{\calC}$.
    Note that $f_{\widetilde{\calC}}$ is a linear function.
    After running this protocol, $P_\C$ obtains the padded measured output 
    \[
        \widetilde{c} = c \oplus x_\calC \oplus x_{0,\calC}
    \]
    and $P_{\Q}$ holds a commitment opening to state
    \[
        X^{x\oplus x_0} Z^{z\oplus z_0} \ket{\psi_{c}}_{\widetilde{\calB}, \widetilde{\calC}, \widetilde{\calD}}
        \otimes\left(\ket{\widehat{x},\widehat{z}}_{\widetilde{\calP}}\otimes\ket{T}\right)
    \]
    where $\ket{\psi_{c}}$ is the result of $U_G(\ket{\psi}, w)$ post-selected for measured output $c$ and we have omitted the homomorphic computation garbage ancillas.

    \item \textbf{Recovery of the Output Ciphertext.} \label{item:recover}

    Parties $P_\Q$ and $P_\C$ run the partial recovery protocol $\mathsf{Recover}$ to recover $\calR_{\outp}$, containing the full final evaluated ciphertext and no evaluation workspace.
    The opened registers $\widetilde{\calC}$ and $\widetilde{\calP}$ remain part of this recovered ciphertext.
    In the execution of $\Recover$, $P_\Q$ uses input $\calR_\op$ and $P_\C$ uses input $\mathsf{com}$.
    
    If $\mathsf{Recover}$ aborts, then both parties output $\perp$ and
    terminate. Otherwise, $P_\Q$ obtains the state
    \[
        X^{x\oplus x_0} Z^{z\oplus z_0} \ket{\psi_{c}}_{\widetilde{\calB}, \widetilde{\calC}, \widetilde{\calD}} \otimes \ket{\widehat{x},\widehat{z}}_{\widetilde{\calP}}
    \]
    
    \item \textbf{Output.}
    Party $P_\C$ sends $(x_{\calB}\oplus x_{0,\calB},\ z_{\calB}\oplus z_{0,\calB})$ to $P_\Q$.
    Party $P_\Q$ applies the inverse quantum one-time pad $\bigl(X^{x_\calB\oplus x_{0,\calB}} Z^{z_\calB\oplus z_{0,\calB}}\bigr)^\dagger$ to register $\widetilde{\calB}$ and outputs the result.
    Party $P_\C$ outputs $c =\widetilde{c} \oplus x_\calC \oplus x_{0,\calC}$.
 
    \end{enumerate}

\end{protocol}

\subsection{Correctness}

\begin{theorem}
    Protocol~\ref{protocol:quantum} is correct.
\end{theorem}
\begin{proof}
Suppose both parties are honest. As proved in \Cref{sec:certcom}, $\mathsf{CertCom}$ succeeds with overwhelming probability. The clock marginal of a valid history state is uniform over $\{0,1,\ldots,T\}$.
If the measured clock value is $t<T$, recovery returns the time-$t$ state
\[U_
t\cdots U_1(\ket{\psi}\ket{0^a}).
\]
Applying
$U^\dagger_1\cdots U^\dagger_t$
recovers $\ket{\psi}\ket{0^a}$ so the parties can restart without changing the input state.
Note that the probability that for $K=\secpar(T+1)$ attempts, every history-state clock measurement is smaller than $T$ is at most
\[
\left(\frac{T}{T+1}\right)^K=\left(\frac{1}{1+1/T}\right)^K=\left(1+\frac1T\right)^{-K}=\left(1+\frac1T\right)^{-\secpar (T+1)}\leq e^{-\secpar}.
\]
Thus, we can assume that for one of the executions we get $t=T$.
By $\QFHE$ correctness, omitting the homomorphic computation garbage ancillas, the terminal state has the form
\[
    X^{x\oplus x_0}Z^{z\oplus z_0}U_G(\ket{\psi},w)
    \otimes\ket{\widehat{x},\widehat{z}}_{\widetilde{\calP}}\otimes\ket{T},
\]
where $x=\Dec(\sk,\widehat{x})$ and $z=\Dec(\sk,\widehat{z})$.
The opening $f_\mathsf{pad}$ gives $P_\C$ the ciphertexts $(\widehat{x},\widehat{z})$, which it decrypts and partitions as in \Cref{protocol:quantum}.

If the unpadded measured output is $c$, then opening $f_{\widetilde{\calC}}$ gives
\[
    \widetilde{c}=c\oplus x_\calC\oplus x_{0,\calC}.
\]
The $Z$-pad on $\widetilde{\calC}$ contributes only an outcome-dependent global phase.
Conditioned on this outcome, the padded quantum state is
\[
    X^{x\oplus x_0}Z^{z\oplus z_0}
    \ket{\psi_c}_{\widetilde{\calB},\widetilde{\calC},\widetilde{\calD}},
\]
where $\ket{\psi_c}$ is the post-selected state defined in \Cref{protocol:quantum}.
Consequently, $P_\C$ outputs
\[
    \widetilde{c}\oplus x_\calC\oplus x_{0,\calC}=c,
\]
as desired.

Recovery returns a state negligibly far from
\[
    X^{x\oplus x_0}Z^{z\oplus z_0}
    \ket{\psi_c}_{\widetilde{\calB},\widetilde{\calC},\widetilde{\calD}}
    \otimes\ket{\widehat{x},\widehat{z}}_{\widetilde{\calP}}.
\]
After receiving $(x_\calB\oplus x_{0,\calB},z_\calB\oplus z_{0,\calB})$, $P_\Q$ removes the Pauli pad on $\widetilde{\calB}$ and outputs that register.
Its output is therefore negligibly close to the reduced state on $\calB$ of $\ket{\psi_c}$, with $\calC$ and $\calD$ traced out, as prescribed by $G$.
Thus the joint output of the parties is negligibly close to the output in the ideal world.
\end{proof}

\subsection{Security Against Malicious Quantum Parties}

\begin{theorem}
    There exists a uniform sequence of $\QPT$ oracle circuits $\{\Sim_{\secpar, n}\}_{\secpar, n\in \bbN}$ such that for every $\QPT$ adversary $P^*_Q$, every sufficiently large $\secpar \in \bbN$, and every $n = \poly[\secpar]$,
    \[
    \Real_{P^*_\Q}(G,\ket{\psi},w) \overset{1/n}{\approx} \Ideal_\Sim(G,\ket{\psi},w, n)
    \]
\end{theorem}

\begin{proof}

Let $\xi=(\pk,\widehat{w}, \widehat{x}_0,\widehat{z}_0)$ be the first message sent by $P_\C$ in the protocol, and let $C=C_\xi$ the circuit as defined in Step \ref{item:enc} of \Cref{protocol:quantum}.  
The three opening functions used after the certification step (Step \ref{item:hist}) are $f_\mathsf{clock}$, $f_\mathsf{pad}$, and $f_{\widetilde{\calC}}$, all of which are linear.
Here $f_\mathsf{pad}$ projects onto $\widetilde{\calP}$, as in \Cref{protocol:quantum}.
\Cref{protocol:quantum} invokes protocol $\mathsf{CertCom}$, then invokes $\Open$ with respect to the identity unitary and these linear functions, and finally invokes $\Recover$.

\paragraph{The Simulator.} Fix an arbitrary $\QPT$ $P^*_\Q$.
We use the register partition $(\calB,\calC,\calD)$ and the corresponding pad components from \Cref{protocol:quantum}.
The recovered ciphertext includes $(\widetilde{\calB},\widetilde{\calC},\widetilde{\calD},\widetilde{\calP})$, while the ideal $G$-functionality returns only $\calB$ to the simulator.

Let $n = \poly[\secpar]$ and let $\eta = Kn$. Define the following $\QPT$ algorithm $\Sim_{n}$ that on input $(G,\ket{\psi})$ does the following:\footnote{We suppress the dependence on $\secpar$ to reduce notational density.}
 \begin{enumerate}
 
     \item Let $w^*=0^n$ and emulate $P_\C$ in Step \ref{item:enc} of \Cref{protocol:quantum} using $w^*$ in place of the unknown input $w$, and encrypting zero pads:
     \[
         (\widehat{x}_0,\widehat{z}_0)\gets\QEnc(\pk,(0^k,0^k)).
     \]
     Independently sample $x_0,z_0\gets\{0,1\}^k$ for use in the simulated terminal ciphertext and final output.
     Let $C=C_{\pk,\widehat{w^*},\widehat{x}_0,\widehat{z}_0}$, and store $(\sk,x_0,z_0)$.

\item For each attempt $j\in[K]$, emulate
\[
  (\mathsf{CertCom},\Open(I,f_{\mathsf{clock}}),\Recover)
\]
using the stateful simulator
\[
  (\Sim_{\mathsf{Cert}},\Sim_{\Open},\Sim_{\Recover})
\]
with indistinguishability parameter $\eta = Kn$ from \Cref{thm:quantum}, where throughout each attempt the stateful simulator passes the same evolving
pair $(\calV,\tau)$ from one simulated invocation to the next.

If the clock outcome is $t<T$ and recovery succeeds, continue to the next
attempt with the resulting adversarial state. If certification, opening,
or recovery aborts, request abort from the ideal functionality and output
the simulated view.

If all $K$ attempts return a value smaller than $T$, request abort from
the ideal functionality and terminate.
\item
Suppose that for the first time the clock outcome is $t=T$. 
At this point the
clock register has been measured, and the current ideal
register $\calV$ contains a valid terminal state of $C$.
Before either terminal opening, starting with $\Open(I,f_{\mathsf{pad}})$, apply $C^\dagger$ to the current ideal register $\calV$.
Since $\calV$ contains a valid terminal state of $C$, this produces an effective input register $\inputReg$, possibly entangled with the adversary's auxiliary register $\calE$, together with zero ancillas.

Send $\inputReg$ to the ideal functionality.
It applies $U_G(\cdot,w)$, measures $\calC$ to obtain $c$, discards $\calD$, and returns only $\calB$ to $\Sim$.
The classical output $c$ remains hidden inside the ideal functionality pending the final abort decision.

\item Append fresh registers $\calC$ and $\calD$, each initialized to an all-zero state of the appropriate size, to the returned register $\calB$.
Apply $X^{x_0}Z^{z_0}$ to $(\calB,\calC,\calD)$, then apply $\QEnc(\pk,\cdot)$ followed by $\QEval(\pk,I,\cdot)$, retaining the resulting Pauli keys $(x,z)$ in the clear as in \Cref{remark:QFHE} and \Cref{thm:qfhe}.
Use the resulting ciphertext on $(\widetilde{\calB},\widetilde{\calC},\widetilde{\calD},\widetilde{\calP})$ as the terminal ciphertext in the current ideal register $\calV$.

\item Emulate $\Open(I,f_{\mathsf{pad}})$ using $\Sim_{\Open}$.
When the internally emulated classical party would decrypt $(\widehat{x},\widehat{z})$, use the stored values $(x,z)$ instead.
If this opening succeeds, emulate $\Open(I,f_{\widetilde{\calC}})$ using $\Sim_{\Open}$ and ignore the outcome delivered to the internally emulated classical party.
If it also succeeds, emulate partial recovery of $\calR_{\outp}$ using $\Sim_{\Recover}$.
Both measured registers remain in the ciphertext returned by recovery.

\item
If any of the final opening or recovery invocations aborts, send abort
to the ideal functionality. Otherwise, send
\[
  (x_\calB\oplus x_{0,\calB},z_\calB\oplus z_{0,\calB})
\]
to $P_\Q^*$ and send continue to the ideal functionality. The ideal
functionality then releases $c$ to $P_\C$.

 \end{enumerate}

 We prove security by considering the following hybrids:
 \paragraph{Hybrid $H_0$.} This is 
the real execution of \Cref{protocol:quantum} with $P^*_\Q$.
 \paragraph{Hybrid $H_1$: Idealizing $\Pi_{\mathsf{HistCom}}$.}
 Replace each execution of
\[(\mathsf{CertCom},\Open(I,f_\mathsf{clock}), \Recover)
\]
where the clock measures $t<T$,
by its simulated execution from \Cref{thm:quantum}.
Similarly, replace the final execution of 
\[(\mathsf{CertCom},\Open(I,f_\mathsf{clock}), \Open(I,f_\mathsf{pad}), \Open(I,f_{\widetilde{\calC}}), \Recover)
\]
by its simulated  execution from \Cref{thm:quantum} with indistinguishability parameter $\eta = Kn$.
Note that the opening family is fixed, every opening function is linear, and there is at most one recovery invocation per attempt. Since there are at most $K=\poly$ attempts, sequential composition of \Cref{thm:quantum} implies that 
\[
H_0 \overset{1/n}{\approx} H_1.
\]

\paragraph{Hybrid $H_2$: Circuit-private simulation of the final ciphertext.}
Hybrid $H_2$ is identical to $H_1$, except on a non-aborting branch on
which the clock opening returns $t=T$. 

In this case, immediately after the simulated clock opening and before either $\Open(I,f_{\mathsf{pad}})$, $\Open(I,f_{\widetilde{\calC}})$ opening, the history state commitment ideal functionality register contains a terminal state for $C$, which is supported on pure states of the form
\[
    C(\ket{\psi}_{\inputReg} \otimes \ket{\hat{w},\hat{x}_0,\hat{z}_0} \otimes \ket{0^a}_{\calA}),
\]
where $C$ has been applied to an input register $\inputReg$, the encrypted classical inputs and pads, and an ancilla register $\calA$.
Apply $C^\dagger$ to this state to obtain the unprocessed input register $\inputReg$. Then, letting $\widetilde{U}_G$ be the padded unitary from \Cref{protocol:quantum}, apply $\widetilde{U}_G(\cdot, w, x_0,z_0)$ to $\inputReg$ and compute $\Sim_{\QFHE}$ on the result. 
Substitute the simulated ciphertext into the history state commitment ideal functionality register, 
then continue with the simulated executions of $\Open(I,f_{\mathsf{pad}})$, $\Open(I,f_{\widetilde{\calC}})$, and partial recovery of the evaluated ciphertext in register $\calR_{\outp}$.

\begin{claim}
    $H_1\approx_c H_2$
\end{claim}
\begin{proof}
    It is equivalent in $H_1$ to consider applying $CC^\dagger$ to the history state commitment ideal functionality register after measuring the clock opening to get result $t=T$. Additionally, it is equivalent to discard the garbage ancillas from $C$ after applying $CC^\dagger$, since they are not used in future steps (due to being excluded from the partial recovery step).

    Let $\rho_{\inputReg,\calE}$ be the joint state of register $\inputReg$ and all retained auxiliary registers $\calE$, including the adversary and commitment simulator states, the keys, and the external reference right after applying $C^\dagger$. $H_1$ then proceeds by computing
    \[
        \Tr_{\calA'}[C\rho_{\inputReg \calE}C^\dagger] 
        = 
        ((\Eval(\pk, \widetilde{U}_G, \cdot, \hat{w},\hat{x}_0,\hat{z}_0)\circ \Enc_\pk) \otimes I_\calE)(\rho_{\inputReg \calE})
    \]
    where $\calA'$ is the traced out garbage ancilla register. In contrast, $H_2$ first computes the plaintext evaluation
    \[
        \sigma_{\calB\calC\calD\calE}
      :=
      \bigl(\widetilde{U}_G(\cdot,w,x_0,z_0)\otimes I_\calE\bigr)
      \rho_{\inputReg \calE}
      \bigl(\widetilde{U}_G(\cdot,w,x_0,z_0)^\dagger\otimes I_\calE\bigr),
    \]
    then replaces the contents of the commitment ideal functionality register by
    \[
        \bigl(\Sim_{\QFHE}\otimes I_\calE\bigr)(\pk,\sigma_{\calB\calC\calD\calE})
    \]

    After this step, $H_1$ and $H_2$ behave identically. This substitution is indistinguishable by the circuit privacy guarantee for QFHE (\Cref{def:QFHE-circuit-private}) in the presence of external entangement.
\end{proof}

\paragraph{Hybrid $H_3$: Fresh encryption of the simulated plaintext with explicit pads.}

Hybrid $H_3$ is identical to $H_2$, except that the output of the $\QFHE$
circuit-privacy simulator is replaced by a fresh encryption of the same
plaintext, followed by homomorphic evaluation of the identity circuit:
\[
  \bigl(\Sim_{\QFHE}\otimes I_\calE\bigr)(pk,\sigma_{\calB\calC\calD\calE})
  \quad\longrightarrow\quad
  \bigl(\QEval(\pk,I)\otimes I_\calE\bigr)
  \bigl((\QEnc(\pk,\cdot)\otimes I_\calE)(\sigma_{\calB\calC\calD\calE})\bigr).
\]
Moreover, this is done by retaining the pauli keys $x,z$ in the clear. This can be done efficiently by \Cref{remark:QFHE} and \Cref{thm:qfhe}.

Applying circuit privacy to the identity circuit shows that this change
is computationally indistinguishable, jointly with $\sk$, $\calE$, and all
other registers. Crucially, the ancillas used for homomorphic evaluation do not appear in this view as they are held by the commitment ideal functionality. Hence
\[
  H_2\approx_c H_3.
\]

\paragraph{Hybrid $H_4$: Encrypting a dummy classical input and dummy pads.}
Hybrid $H_4$ is identical to $H_3$, except that the first message contains
\[
    \widehat{w^*}\gets\QEnc(\pk,0^n),
    \qquad
    (\widehat{x}_0,\widehat{z}_0)\gets\QEnc(\pk,(0^k,0^k))
\]
in place of encryptions of $w$ and $(x_0,z_0)$.
The independently sampled original pads $x_0,z_0$ are retained privately and are still used to prepare the simulated terminal ciphertext and compute the final $\calB$ sub-pad.
Here $\widehat{x}_0,\widehat{z}_0$ denote the transmitted ciphertexts, which now encrypt zeros.

Write $C_w$ for the public circuit in $H_3$ and $C_0=C_{\pk,\widehat{w^*},\widehat{x}_0,\widehat{z}_0}$ for the circuit defined by the dummy first message in $H_4$.
Every certified-history attempt uses $C_0$.
On a terminal branch, after applying $C_0^\dagger$ to recover the effective input, the hybrid still applies $U_G(\cdot,w)$ and the original outer pads $x_0,z_0$.
Thus the plaintext computation continues to use the honest input $w$ and original pads $x_0,z_0$.

We claim that
\[
    H_3\approx_c H_4.
\]
Replace the encryptions of $w$, $x_0$, and $z_0$ by encryptions of zero in sequence.
For each replacement, a semantic-security reduction samples and retains the original pads, places its challenge ciphertext in the corresponding position of the first message, constructs the resulting public circuit, and runs the remaining experiment.
All certified-history simulations, inverse-circuit operations, and fresh terminal encryptions are efficient public operations.
The reduction uses the retained $w,x_0,z_0$ in the plaintext computation, and knows the fresh inner Pauli keys $(x,z)$ explicitly, so it does not need $\sk$ for the terminal pad opening.
The decryption-failure branch differs from this use of explicit keys only with negligible probability by correctness.
A distinguisher for $H_3$ and $H_4$ would therefore distinguish at least one of these ciphertext replacements, contradicting semantic security.

\paragraph{Hybrid $H_5$: Measuring the plaintext output.}
Hybrid $H_5$ is identical to $H_4$, except that it measures the plaintext register $\calC$ before preparing the fresh terminal ciphertext.
After applying $C_0^\dagger$, let $\rho_{\inputReg\calE}$ denote the recovered effective input and the auxiliary register held by the adversary.
Apply $U_G(\cdot,w)\otimes I_\calE$, measure $\calC$ to obtain $c$, and retain the post-measurement register $\calC$ in state $\ket{c}$.
Let $\rho^c_{\calB\calD\calE}$ denote the corresponding conditional state on the other registers.
In particular, the discarded-output register $\calD$ is still present in this hybrid.
Store $c$ privately as the prospective honest classical output.

Using the pad partition from \Cref{protocol:quantum}, prepare
\[
\begin{aligned}
    \sigma^c_{\calB\calC\calD\calE}
    := {} & (X^{x_0}Z^{z_0}\otimes I_\calE)
    \bigl(|c\rangle\!\langle c|_\calC\otimes\rho^c_{\calB\calD\calE}\bigr)
    (Z^{z_0}X^{x_0}\otimes I_\calE),
\end{aligned}
\]
where the Pauli operators act on the named registers $(\calB,\calC,\calD)$ and leave $\calE$ untouched.
Then prepare the fresh identity-evaluated ciphertext
\[
\begin{aligned}
    \widehat{\sigma}^c_{\widetilde{\calB}\widetilde{\calC}\widetilde{\calD}\widetilde{\calP}\calE}
    := {} & (\QEval(\pk,I)\otimes I_\calE)
    \bigl((\QEnc(\pk,\cdot)\otimes I_\calE)
    (\sigma^c_{\calB\calC\calD\calE})\bigr),
\end{aligned}
\]
retaining the inner pads $x=(x_\calB,x_\calC,x_\calD)$ and $z=(z_\calB,z_\calC,z_\calD)$ explicitly.
The register $\widetilde{\calP}$ contains their classical ciphertexts $(\widehat{x},\widehat{z})$.

Emulate $\Open(I,f_{\mathsf{pad}})$, then $\Open(I,f_{\widetilde{\calC}})$, and then partial recovery of $\calR_{\outp}$, using the retained inner pads in place of decryption.
On successful opening of the measured output, its value is
\[
    \widetilde{c}=c\oplus x_\calC\oplus x_{0,\calC}.
\]
The post-measurement register $\widetilde{\calC}$ is retained, and recovery returns all four ciphertext registers, including $\widetilde{\calC}$ and $\widetilde{\calP}$.
If any invocation aborts, the honest classical party outputs $\bot$.
Otherwise, send $(x_\calB\oplus x_{0,\calB},z_\calB\oplus z_{0,\calB})$ to $P_\Q^*$ and output the stored value $c$ for the honest classical party.
The measured-output opening value is not used in subsequent messages or abort decisions.

\begin{claim}\label{claim:H56}
\[
    H_4\equiv_s H_5.
\]
\end{claim}

\paragraph{Proof of \Cref{claim:H56}.}
In $H_4$, the experiment freshly encrypts the coherent output of $U_G(\cdot,w)$, opens the inner-pad ciphertexts, and then measures $\widetilde{\calC}$.
In $H_5$, it measures the plaintext $\calC$ first and then prepares the fresh encryption of the resulting post-measurement state, retaining $\calC$.
By the explicit Pauli form of the ciphertext and $\QFHE$ correctness, the two measurement orders give the same outcome relation and post-measurement ciphertext, up to negligible error.
The $Z$-pad on $\calC$ contributes only an outcome-dependent global phase, while the $X$-pad changes $c$ to $\widetilde{c}=c\oplus x_\calC\oplus x_{0,\calC}$.
The pad opening acts on $\widetilde{\calP}$, so it commutes with the measurement of $\widetilde{\calC}$.
Ideal binding ensures that the two experiments agree jointly on the opened pad ciphertexts, the post-measurement registers $(\widetilde{\calB},\widetilde{\calC},\widetilde{\calD},\widetilde{\calP})$, and the auxiliary state.
Finally, the classical output in $H_4$ is $\widetilde{c}\oplus x_\calC\oplus x_{0,\calC}=c$, which is the stored output in $H_5$.
\qed

\paragraph{Hybrid $H_6$: Replacing the measured and discarded plaintext registers by zeros.}
Hybrid $H_6$ is identical to $H_5$, except that after obtaining the private outcome $c$, it discards the genuine registers $\calC$ and $\calD$ and replaces them by fresh all-zero registers of the same sizes before preparing the terminal ciphertext.
\[
    H_5 \equiv H_6
\]
by the perfect security of the quantum one-time pad. To see this, observe that the only difference between the two hybrids is whether the masked registers $\widetilde{\calC}$ and $\widetilde{\calD}$ contain the masked original state or masked $\ket{0}$ states, and the sub-pads $(x_{0,\calC}, x_{0,\calD})$ are not used anywhere else in either hybrid experiment.

\paragraph{Hybrid $H_7$: Invoking the ideal functionality.}
Hybrid $H_7$ is identical to $H_6$, except that the plaintext computation, measurement, and discarding are performed by the ideal $G$-functionality.
On the first non-aborting terminal branch, apply $C_0^\dagger$ and send the recovered effective input register $\inputReg$ to the ideal functionality.
It applies $U_G(\cdot,w)$, measures $\calC$ to obtain $c$, discards $\calD$, and returns only $\calB$ to the simulator.
Conditioned on $c$, the joint state of the returned register and the untouched auxiliary register is $\rho^c_{\calB\calE}$.
The functionality retains $c$ pending the simulator's final abort decision.

The simulator appends all-zero registers $\calC$ and $\calD$ of the prescribed sizes and prepares the padded plaintext and its fresh identity-evaluated encryption exactly as in $H_6$.
It retains the inner pads $(x,z)$ explicitly and emulates
\[
    \Open(I,f_{\mathsf{pad}}),\qquad
    \Open(I,f_{\widetilde{\calC}}),\qquad
    \Recover,
\]
where recovery returns $\calR_{\outp}$ and the measured-output opening value is ignored.
If any invocation aborts, the simulator sends an abort request to the ideal functionality, causing the honest classical party to output $\bot$.
Otherwise, it sends
\[
    (x_\calB\oplus x_{0,\calB},z_\calB\oplus z_{0,\calB})
\]
to $P_\Q^*$ and sends a continue request to the ideal functionality, which releases $c$ to the honest classical party.

We have
\[
    H_6\equiv H_7.
\]
Indeed, both hybrids produce the same joint distribution of the private classical outcome $c$ and the residual state $\rho^c_{\calB\calE}$.
After this operation, the simulated ciphertext and all subsequent messages and abort decisions use only the returned $\calB$ register, the dummy registers, and the sampled pads; they do not require knowledge of $c$ or access to the genuine $\calD$.
In $H_6$, the experiment stores $c$ and outputs it on successful completion; in $H_7$, the ideal functionality stores and releases it on the same condition.
By construction, $H_7$ is the ideal execution with the simulator described above:
\[
    H_7=\Ideal_{\Sim}(G,|\psi\rangle,w).
\]
\end{proof}

\subsection{Security Against Semi-Honest Classical Parties}

\begin{theorem}[Security against honest-but-curious $P_\C$]
\label{thm:general-hbc-pc}
    With the commitment and recovery protocols of
    \Cref{constr:com-main,constr:com-recovery-core,constr:com-recovery-recovery},
    \Cref{protocol:quantum} is statistically secure against honest-but-curious classical parties.
\end{theorem}
\begin{proof}
    We use a specific property from these protocols, which is inherited from \Cref{constr:com-main}: a valid opening is an authentication of a string underneath an authentication key $\com$ held by the semi-honest classical party. Thus, given $\com$, it is possible to generate arbitrary openings.

    The simulator sends $w$ to the ideal functionality and obtains its classical output $c$.
    It generates the classical view as follows.
    \begin{enumerate}
        \item Run the honest protocol with classical input $w$ and an all-zero quantum input, stopping immediately after the opening of $(\widehat{x},\widehat{z})$.
        \item Set
        \[
            \widetilde{c}=c\oplus x_\calC' \oplus x_{0,\calC}.
        \]
        Using $\com$, discard the authenticated blocks for $\widetilde{\calC}$ and replace them by an authentication of $\ket{\widetilde{c}}$ under the same key.
        \item Continue the remaining protocol honestly and output the classical view.
        Allow the ideal functionality to release $\calB$.
    \end{enumerate}
    On any abort, the simulator requests abort and returns the view obtained so far.

    \paragraph{Indistinguishability.}
    Using a similar state-preserving argument from the proof of \Cref{lem:linear-2pc-hbc-pc}, 
    any execution that restores every state in its subspace (including its external entanglement) must have a classical view which is independent of that state. We will use this to show that the replacement of the quantum input in the initial history commitment, clock opening, and pad opening steps is indistinguishable.

    First, consider the state immediately after the simulator's initial stopping point.
    If one were to apply the full recovery guaranteed by the history state, decrypt the entire evaluated ciphertext, remove $(x_0,z_0)$, and apply $U_G^\dagger$, then 
    correctness of QFHE implies they would have restored the original quantum input.

    Next, the final opening performs the prescribed projective measurement of $\widetilde{\calC}$ and preserves the logical state within each outcome subspace.
    Following the opening by full recovery therefore returns every state in that subspace.
    The same observation shows that, conditioned on the preceding view and $\widetilde{c}$, the remaining classical view is independent of the residual state and its reference.
    It can therefore be generated from the simulator's modified dummy state, which is correctly authenticated under $\com$.

\end{proof}

\section{2PC with Persistent Memory}
\label{sec:persistent-2pc}
\label{sec:memory-delegation}

Finally, we show how to implement persistent private memory using the single-use 2PC. 
In this, we consider an interactive ideal functionality which works as follows:
\begin{enumerate}
    \item \textbf{Initialization.} The ideal functionality initializes its memory register $\calM$ to $\ket{0^{|\calM|}}$.
    \item \textbf{Evaluation:} Any polynomial number of times, until the adversary aborts:
    \begin{enumerate}
        \item  The classical party $P_C$ specifies a general quantum circuit $G$ and the quantum party $P_Q$ provides a quantum input in register $\calQ$. $G$ also specifies how to split the coherent output register between the ideal functionality and $P_Q$. 
        \item The ideal functionality runs $G$ on registers $\calM$ and $\calQ$.
        \item The ideal functionality sends the classical result to $P_C$ and $P_Q$'s half of the coherent output to $P_Q$. It retains its half of the coherent output in $\calM$.
    \end{enumerate}
    The adversary may abort the execution at any time.
\end{enumerate}
Security is otherwise defined identically to \Cref{sec:def-2PC}, with additional quantification over the multiple rounds of inputs. The established terminology for ideal functionalities which maintain persistent memory across interactions is a ``reactive'' functionality.

\begin{theorem}\label{thm:persistent-memory}
    Assuming an $\epsilon$-secure (respectively, secure) 2PC for general real quantum circuits, there exists an $\epsilon$-secure (respectively, secure) 2PC for polynomial-size general reactive quantum functionalities.
\end{theorem}

\subsection{Construction and Analysis}

The construction enforces persistent memory by having the ideal functionality authenticate its memory register under a key only known to the classical party. Once $\calM$ is protected, the quantum party can hold it until the next evaluation. Such authentication schemes exist unconditionally.

\begin{construction}
    The construction uses a quantum authentication scheme $\Auth.(\KeyGen, \Enc, \Dec, \Ver)$.
    \begin{enumerate}
        \item \textbf{Initialization.}
        \begin{enumerate}
            \item $P_C$ samples an authentication key $k_0\gets \Auth.\KeyGen(1^\secpar, |\calM|)$.
            \item $P_C$ and $P_Q$ engage in a 2PC for the quantum circuit which takes in a key $k$ and $|\calM|$ ancillas initialzed to $\ket{0}$, then outputs $\Auth.\Enc(k_0, \ket{0^{|\calM|}})$ to the quantum party.
        \end{enumerate}
        \item \textbf{Evaluation.} On the $i$'th evaluation, where $P_C$ has input $G_i$ and $P_Q$ has input $\calQ_i$:
        \begin{enumerate}
            \item $P_C$ samples a fresh authentication key $k_i\gets \Auth.\KeyGen(1^\secpar, |\calM|)$.
            \item $P_C$ and $P_Q$ engage in a 2PC for the following circuit. $P_C$ inputs the circuit $G_i$, the old key $k_{i-1}$, and the new key $k_i$. $P_Q$ inputs $\calQ_i$ and the authenticated memory register $\widetilde{\calM}$.
            \begin{enumerate}
                \item The 2PC circuit computes $\calM \gets \Auth.\Dec(k_{i-1}, \widetilde{\calM})$, then computes $G_i((\calM, \calQ_i))$.
                \item It parses the output as $P_C$'s output $y$, $P_Q$'s output $\calQ_i'$, and the updated memory register $\calM'$.
                \item It computes $\widetilde{\calM} \gets \Auth.\Enc(k_i, \calM')$.
                \item $y$ is the classical party's output and $(\calQ'_i, \widetilde{\calM})$ is the quantum party's output.
            \end{enumerate}  
        \end{enumerate}
    \end{enumerate}
\end{construction}

\begin{proof}[Proof of \Cref{thm:persistent-memory}]
    Correctness follows immediately from the correctness of the underlying 2PC for general quantum circuits and the authentication scheme. The addition of the authentication encoding and decoding keeps the circuit as polynomial size.

    Next we show $\epsilon$-security against malicious $P_Q$. Let $T = \poly[\secpar]$ be an upper bound on the number of evaluations. Fix any $\epsilon = 1/\poly[\secpar]$. Consider the hybrid experiment where we simulate each each $\twoPC$ with error $\epsilon/T$. This is computationally indistinguishable from the real execution except with probability $T (\epsilon/T) = \epsilon$. Next, the security of authentication allows replacing $\widetilde{\calM}$ by a dummy authentication of $\ket{0^{|\calM|}}$ and keeping $\calM'$ for direct use in the next execution if $P_Q$ submits a $\widetilde{\calM}$ register which passes $\Auth.\Ver$. At this point, the experiment matches the ideal functionality: $P_Q$ submits its input on each execution and gets its output, and the ideal functionality maintains $\calM$ across executions.

    Security against semi-honest $P_C$ follows immediately from the security of the underlying 2PC and the observation that $\widetilde{\calM}$ is never seen by $P_C$. In particular, simulating each of the underlying 2PCs already matches the ideal functionality: $P_C$ provides its input for each evaluation, then gets its output. 
\end{proof}

\part{Applications}
\section{Highlighted Applications}\label{sec:highlighted-apps}

In this section, we give applications to two open problems: classical verification of quantum sampling and remote state preparation. 
Previously, constructions for both primitives which run in time $T$ incurred a $1/\poly[T]$ security error. In the following subsections, we give a single protocol for each problem which runs in time $T$, but has $\epsilon$ error for \emph{every} $\epsilon = 1/\poly$.
As such, the constructions are suitable for achieving any falsifiable game-based security notion with \emph{negligible} security error. We give explicit example applications in \Cref{sec:other-applications}.
Moreover, our construction of remote state preparation holds for \emph{any efficiently preparable state}. We provide a further discussion of this in \Cref{sec:general-rsp}.

The third highlighted application complements remote state preparation by demonstrating the \emph{receipt} and processing of a quantum message, followed by either storing or discarding it. Specifically, we give a way for a classical verifier to receive, verify, and hold a quantum certificate. We describe it for the specific application of secure key leasing, which allows a lessor to lend out a quantum decryption key to a quantum lessee, who can use it to decrypt ciphertexts. Later, the return of a valid certificate ensures that the quantum sender can no longer decrypt future ciphertexts. We obtain the same guarantees where the revocation process is a classical interaction instead of requiring quantum communication.

\subsection{Classical Verification of Quantum Sampling}\label{sec:CVQS}

The goal of a quantum sampling algorithm is to sample from some uniformly-generated family of distributions $\{\calD_{x}\}_{x\in \{0,1\}^*}$, potentially with some approximation error. More specifically, such an algorithm produces samples from some alternate distribution $\calC_{x}$ and the goal is to minimize the distance between $\calC_{x}$ and $\calD_{x}$.

We show in this section that \emph{every} quantumly-sampleable distribution $\calC_{x}$ can be classically verified with negligible error. 
Thus, if $\|\calC - \calD\|_{\mathsf{TV}} \leq \epsilon$, then a classical party $V$ can verify samples from $\calD_x$ with computational distance $\epsilon$.

\begin{definition}[Quantum-Sampleable Distribution]
    A distribution $\calD$ is \textbf{quantum-sampleable} if there exists a $\QPT$ algorithm $A$ \emph{without quantum advice} such that $A$ outputs samples from $\calD$.
    We say that a family of distributions $\{\calD_x\}_{x\in \{0,1\}^*}$ is (uniformly) quantum-sampleable if there exists a (uniform) sequence of $\QPT$ algorithms $\{A_x\}_x$ such that $A_x$ outputs samples from $\calD_x$.
\end{definition}

\begin{definition}[Classical Verification of Quantum Sampling]
    A classical verification of quantum sampling protocol $\CVQS = \{\CVQS_{x}\}_x$ for a family of distributions $\{\calD_x\}_{x}$ is a uniformly generated set of protocols 
    \[
        \CVQS_x\langle P, V\rangle(1^\secpar) \rightarrow (d, z)
    \]
    between a quantum prover $P$ and a classical verifier $V$. At the end of the protocol, the classical verifier outputs a decision bit $d\in \{\Accept, \Reject\}$ and a sample $z$.
    
    It must satisfy completeness and soundness.
    \begin{itemize}
        \item \textbf{Completeness.} A $\CVQS$ protocol has completeness error $c(\secpar)$ if for all $\secpar \in \bbN$ and all $x \in \{0,1\}^{\poly}$,
        \[
            \Pr[d = \Reject: (d, z) \gets \CVQS_x\langle P, V\rangle(1^\secpar)] \leq c(\secpar)
        \]
        If $c(\secpar)$ is negligible, we simply say that the protocol is complete.

        \item \textbf{Soundness.} 
        For any $\QPT$ cheating prover $P^*$, sufficiently large $\secpar \in \bbN$, and any $x \in \{0,1\}^{\poly}$, consider the following experiment.
        \begin{enumerate}
            \item Sample $(d, z) \gets \CVQS_x\langle P^*, V\rangle(1^\secpar)$.
            \item Define $z_\Ideal$ by
            \[ 
                \begin{cases}
                    z_{\Ideal} = \bot &\text{if } d = \Reject\\
                    z_{\Ideal} \gets \calD_x &\text{if } d = \Accept
                \end{cases}    
            \]
        \end{enumerate}
        $\CVQS$ has computational soundness error $s(\secpar)$ if the real and ideal distributions can be $\QPT$-distinguished with advantage at most $s(\secpar)$:
        \[
            \{(d, z)\} \overset{s(\secpar)}{\approx}_c \{(d, z_\Ideal)\}
        \]
        If $s(\secpar)$ is negligible then we simply say that $\CVQS$ is sound.
    \end{itemize}
\end{definition}

\begin{theorem}
    Assuming the post-quantum hardness of LWE, there is a complete and computationally sound $\CVQS$ protocol for every family of quantum-sampleable distributions $\{\calD_{x}\}_{x}$.
\end{theorem}
\begin{proof}
    Let $\{A_x\}_{x}$ be the sequence of $\QPT$ algorithms without advice which sample from $\{\calD_{x}\}_{x}$. 
    To sample from $\calD_x$, $P$ and $V$ run a state-preserving $2\PC$ (\Cref{sec:general}) for the quantum functionality which computes $A_x$ and outputs its result.
\end{proof}

We remark that since $\{A_x\}_{x}$ is uniformly generated, it is even possible to hide precisely which distribution $\calD_x$ the verifier wishes to obtain samples from, up to the size of the implemented circuit.

As an immediate corollary, it is possible to classically verify samples from $\SampBQP$ with negligible overhead over the approximation error which is baked into the definition of $\SampBQP$.

\begin{definition}[$\SampBQP$]
    A sampling problem $\{D_x\}_{x\in\{0,1\}^*}$ is in $\SampBQP$ if there exists a family of uniformly quantum-sampleable distributions $\{\calC_{x,1^{1/\epsilon}}\}_{x\in \{0,1\}^*, \epsilon \in \bbR^+}$ such that for all $x$ and all $\epsilon$,
    \[
        \|\calD_{x} - \calC_{x,1^{1/\epsilon}}\|_{\mathsf{TV}} \leq \epsilon
    \]
    where $\|\|_{\mathsf{TV}}$ is the total variation distance.
\end{definition}

\subsection{Remote State Preparation}\label{sec:general-rsp}

Remote state preparation allows a classical party to ``send'' a quantum state $\ket{\psi}$ to a quantum party, so long as it knows a classical description of how to efficiently prepare the state. The goal is for the quantum party to receive no more information than if it had directly received $\ket{\psi}$.

There are several limitations to existing constructions of remote state preparation~\cite{FOCS:GheVid19,ICALP:GheMetPor23,ITCS:Zhang25}.
\begin{itemize}
    \item \textbf{$1/\poly$ Error.} Existing protocols are parameterized by a tunable security error $\epsilon = 1/\poly$.\footnote{A protocol set to $\epsilon$ is not guaranteed (and is unlikely) to be, for example, $\epsilon/2$-secure. } Consider a security game which can only be won with $\negl$ probability if directly given $\ket{\psi_x}$. An $\epsilon$ error in remote state preparation allows the adversary's winning probability to rise to $\epsilon + \negl$ if they instead receive $\ket{\psi_x}$ via a remote state preparation protocol.

    \item \textbf{Limited State Families.} Prior works only allow for constructing a limited selection of states $\{\ket{\psi_x}\}_{x}$, such as BB84 states~\cite{FOCS:GheVid19,ICALP:GheMetPor23} or $T^\theta\ket{+}$ states for $\theta \in [8]$~\cite{FOCS:GheVid19}.

    \item \textbf{Randomized States.} Existing approaches produce a random $\ket{\psi_x}$ from the family of possible states (for example, the sender can choose to send a computational basis state, but not which of $\ket{0}$ or $\ket{1}$).
\end{itemize}

In this section, we give a remote state preparation protocol for any state efficiently preparable by a real circuit of the sender's choice, with $\epsilon$-security and negligible-loss falsifiable security. A family of states $\{\ket{\psi_{x}}\}_{x}$ is efficiently preparable if there exists a $\QPT$ algorithm $P$ without advice such that $P(x)$ outputs a state negligibly far from $\ket{\psi_x}$ for all $x$.

\paragraph{Syntax and Correctness.} 
A remote state preparation protocol for a family of states $\{\ket{\psi_x}\}_{x\in\{0,1\}^n}$ is an interactive protocol between a classical sender $S_C$ and a quantum receiver $R_Q$. $S_C$ inputs a description $x\in\{0,1\}^n$. At the end of the protocol, $R_Q$ outputs a state $\ket{\psi'}$. We denote an execution by
\[
    \ket{\psi'} \gets \RSP\langle S_C(x), R_Q\rangle(1^\secpar)
\]
Correctness requires that for all $x\in \{0,1\}^n$ and all $\secpar \in \bbN$, $R_Q$ outputs a state overwhelmingly close to $\ket{\psi_x}$:
\[
    \Pr_{\ket{\psi'} \gets \RSP\langle S_C(x), R_Q\rangle(1^\secpar)}\left[\frac{1}{2}\left\|\ket{\psi'} - \ket{\psi_x}\right\|_{1} \leq \negl \right] \geq 1-\negl
\]

\begin{definition}[General RSP Security]
    A remote state preparation protocol is secure if there exists a $\QPT$ algorithm $\Sim$ such that for all adversarial $\QPT$ receivers $R_Q^*$ and all $x\in\{0,1\}^n$,
    \[
        \left\{
            \calR_{\adv} \gets \RSP\langle S_C(x), R_Q^*\rangle\left(1^\secpar\right)
        \right\}
        \approx_c
        \left\{
            \calR^* \gets \Sim(R_Q^*, \ket{\psi_x})
        \right\}
    \]
    where $\calR^*$ is $R_Q^*$'s output register.

    We say it is $\epsilon$-secure if there is a uniform sequence of $\poly[\secpar, n]$-sized quantum circuits $\{\Sim_{\secpar, n}\}_{n,\secpar\in \bbN}$ such that for all sufficiently large $\secpar$ and all $n = \poly[\secpar]$,
    \[
        \left\{
            \calR_{\adv} \gets \RSP\langle S_C(x), R_Q^*\rangle\left(1^\secpar\right)
        \right\}
        \overset{1/n}{\approx}_c
        \left\{
            \calR^* \gets \Sim(R_Q^*, \ket{\psi_x})
        \right\}
    \]
\end{definition}

There are two main differences from prior works. First, the sender may directly choose the state $\ket{\psi_x}$ to prepare. Second, $\epsilon$ is a parameter of the simulator, \emph{not} the protocol itself. That is, the same protocol must satisfy \emph{every} $\epsilon = 1/\poly[\secpar]$.

Any RSP with $\epsilon$-security may be used to replace one round of quantum communication in any efficiently falsifiable security game with \emph{negligible} security loss. An efficiently falsifiable game $G$ is an interaction between a QPT challenger and a QPT adversary where at the end of the game, the challenger outputs a bit denoting win or loss. Falsifiable security of RSP guarantees that replacing one round of quantum communication in this game changes the adversary's maximum winning advantage negligibly.

\begin{definition}[RSP Falsifiable Security]
    Let $G_Q$ be an efficiently falsifiable game where the challenger sends one quantum message. Let $G_C(\adv)$ be the same game, except the quantum message is replaced by a remote state preparation protocol between the challenger and adversary $\adv$.

    A remote state preparation protocol has falsifiable security if for every efficiently falsifiable game $G_Q$ with one message from the challenger,
    \[
        \max_{\QPT \adv} \Pr[\mathsf{win}\gets G_C] \leq \max_{\QPT \adv} \Pr[\mathsf{win}\gets G_Q] + \negl
    \]
\end{definition}

\begin{theorem}
    Assuming the post-quantum hardness of LWE, there exists a remote state preparation protocol with $\epsilon$-security and (negligible loss) falsifiable security for every family of states $\{\ket{\psi_{x}}\}_{x}$ which can be efficiently prepared by a real circuit.
\end{theorem}
\begin{proof}
    The parties run a state-preserving 2PC (\Cref{sec:general}) for the functionality $P$ which efficiently prepares states from the family. Correctness and security are immediate from the correctness and security of the state-preserving 2PC.

    Falsifiable security follows by observing that the sequence of $\epsilon$-simulators gives a sequence of reductions against every $\epsilon = 1/\poly$. Therefore the adversary's additional advantage in the compiled game can only increase negligibly.
\end{proof}

\subsection{Classical Revocation of Quantum Certificates}\label{sec:classical-revocation}

Many quantum cryptographic primitives involve a quantum certificate that can be
verified by a quantum party.  For example, a user may hold a quantum decryption
key, token, or certificate, and at some later point be required to return it to a
verifier who runs a quantum verification procedure.

Our state-preserving $2\PC$ gives a generic way to make such a verifier classical.
Suppose that a quantum verification procedure $\Ver$ takes as input a classical
verification key $\vk$ and a quantum certificate $\rho$, and outputs a classical
decision bit $b$, and possibly a residual quantum state $\rho'$.  Instead of
sending $\rho$ to a quantum verifier, the certificate holder and a classical
verifier can simply run our $2\PC$ for the functionality
\[
    (b,\rho') \leftarrow \Ver(\vk,\rho),
\]
where the classical verifier inputs $\vk$ and the quantum party inputs $\rho$.
The classical verifier learns only the prescribed verification outcome, while
the residual state is discarded entirely.

Thus, any protocol in which a quantum certificate is checked by an efficient
quantum verification procedure can, in a black-box manner, be transformed into
one in which the verification is performed through classical interaction.
In particular, when successful verification certifies that the quantum
certificate has been revoked, this gives a classical revocation procedure.

\paragraph{Classical Revocation for Secure Key Leasing.}
To make this generic transformation concrete, we instantiate it below for
public-key encryption with secure key leasing (PKE-SKL), where the quantum
certificate is a leased decryption key and revocation consists of verifying its
return.  We emphasize, however, that there is nothing specific to PKE-SKL in
this transformation.  The same approach applies to any primitive in which
revocation is implemented by an efficient quantum verification procedure
applied to a quantum certificate.

In what follows we recall the syntax, correctness, and indistinguishability definitions of PKE-SKL from~\cite[Definitions~3.1 and~3.4 of the \href{https://eprint.iacr.org/2023/264}{full version}]{EC:AKNYY23}.
\newcommand{\sklqfont}[1]{\text{\usefont{OT1}{pzc}{m}{it}#1}}
We use the font $\sklqfont{KG},,\sklqfont{Dec},\sklqfont{Vrfy},\sklqfont{dk}$ for quantum algorithms or quantum states.

\begingroup

\begin{definition}[PKE with Secure Key Leasing]
    A {\rm PKE-SKL} scheme \(\mathsf{SKL}\) is a tuple of four algorithms \((\sklqfont{KG},\Enc,\sklqfont{Dec},\sklqfont{Vrfy})\).
    Below, let \(\calX\) be the message space of \(\mathsf{SKL}\).
    \begin{description}
        \item[\(\sklqfont{KG}(1^\lambda) \to (\mathsf{ek},\sklqfont{dk},\mathsf{vk})\):] The key generation algorithm is a $\QPT$ algorithm that takes as input a security parameter \(1^\lambda\), and outputs an classical encryption key \(\mathsf{ek}\), a quantum decryption key \(\sklqfont{dk}\), and a  classical verification key \(\mathsf{vk}\).

        \item[\(\Enc(\mathsf{ek},\mathsf{m}) \to \ct\):] The encryption algorithm is a $\PPT$ algorithm that takes as input an encryption key \(\mathsf{ek}\) and a message \(\mathsf{m} \in \calX\), and outputs a ciphertext \(\ct\).

        \item[\(\sklqfont{Dec}(\sklqfont{dk},\ct) \to \tilde{\mathsf{m}}\):] The decryption algorithm is a $\QPT$ algorithm that takes as input a quantum decryption key \(\sklqfont{dk}\) and a classical ciphertext \(\ct\), and outputs a classical message \(\tilde{\mathsf{m}}\).

        \item[\(\sklqfont{Vrfy}(\mathsf{vk},\widetilde{\sklqfont{dk}}) \to \top/\bot\):] The verification algorithm is a $\QPT$ algorithm that takes as input a classical verification key \(\mathsf{vk}\) and a (possibly malformed) quantum decryption key \(\widetilde{\sklqfont{dk}}\), and outputs \(\top\) or \(\bot\).

        \item[Decryption correctness:] For every \(\mathsf{m} \in \calX\), 
        \[
            \Pr\left[\sklqfont{Dec}(\sklqfont{dk},\ct) = \mathsf{m} \;\middle|\;
                \begin{array}{l}
                    (\mathsf{ek},\sklqfont{dk},\mathsf{vk}) \gets \sklqfont{KG}(1^\lambda) \\
                    \ct \gets \Enc(\mathsf{ek},\mathsf{m})
                \end{array}
            \right] = 1 - \mathsf{negl}(\lambda).
        \]

        \item[Verification correctness:] 
        \[
            \Pr\left[\sklqfont{Vrfy}(\mathsf{vk},\sklqfont{dk}) = \top \;\middle|\; (\mathsf{ek},\sklqfont{dk},\mathsf{vk}) \gets \sklqfont{KG}(1^\lambda)\right] = 1 - \mathsf{negl}(\lambda).
        \]
    \end{description}
\end{definition}

\begin{definition}[Indistinguishability Against Key Leasing Attacks (\rm{IND-KLA}) Security]
    We say that a \rm{PKE-SKL} scheme \(\mathsf{SKL}\) with the message space \(\calX\) is \rm{IND-KLA} secure, if it satisfies the following requirement, formalized from the experiment \(\Exp^{\mathsf{ind\text{-}kla}}_{\mathsf{SKL},\calA}(1^\lambda,\mathsf{coin})\) between an adversary \(\calA\) and a challenger \(\calC\):
    \begin{enumerate}
        \item \(\calC\) runs \((\mathsf{ek},\sklqfont{dk},\mathsf{vk}) \gets \sklqfont{KG}(1^\lambda)\) and sends \(\mathsf{ek}\) and \(\sklqfont{dk}\) to \(\calA\).

        \item Throughout the experiment, \(\calA\) can access the following (stateful) verification oracle \(O_{\sklqfont{Vrfy}}\) where \(V\) is initialized to be \(\bot\):
        \begin{description}
            \item[\(O_{\sklqfont{Vrfy}}(\widetilde{\sklqfont{dk}})\):] It runs \(d \gets \mathsf{Vrfy}(\mathsf{vk},\widetilde{\sklqfont{dk}})\) and returns \(d\).
            If \(V = \bot\) and \(d = \top\), it updates \(V := \top\).
        \end{description}

        \item \(\calA\) sends \((\mathsf{m}_0^*,\mathsf{m}_1^*) \in \calX^2\) to \(\calC\).
        If \(V = \bot\), \(\calC\) output \(0\) as the final output of this experiment.
        Otherwise, \(\calC\) generates \(\ct^* \gets \Enc(\mathsf{ek},\mathsf{m}_{\mathsf{coin}}^*)\) and sends \(\ct^*\) to \(\calA\).

        \item \(\calA\) outputs a guess \(\mathsf{coin}'\) for \(\mathsf{coin}\).
        \(\calC\) outputs \(\mathsf{coin}'\) as the final output of the experiment.
    \end{enumerate}
    For any QPT \(\calA\), it holds that
    \[
        \begin{aligned}
            \mathsf{Adv}^{\mathsf{ind\text{-}kla}}_{\mathsf{SKL},\calA}(\lambda)
            &:= \left|\Pr\left[\Exp^{\mathsf{ind\text{-}kla}}_{\mathsf{SKL},\calA}(1^\lambda,0) \to 1\right] - \Pr\left[\Exp^{\mathsf{ind\text{-}kla}}_{\mathsf{SKL},\calA}(1^\lambda,1) \to 1\right]\right| \\
            &\leq \mathsf{negl}(\lambda).
        \end{aligned}
    \]
\end{definition}

\paragraph{Classical revocation.}
We now observe that any PKE-SKL scheme with an efficient quantum verification
algorithm can be generically compiled into one where revocation requires only
classical interaction.

Let $\mathsf{SKL}=(\sklqfont{KG},\Enc,\Dec,\sklqfont{Vrfy})$ be a PKE-SKL scheme.  The algorithms
$\sklqfont{KG},\Enc,\Dec$ remain unchanged.  We replace the verification algorithm by an
interactive protocol $\mathsf{Vrfy}_{\sf cl}$ between the holder of the quantum
decryption key and a classical verifier holding $\vk$.  On input a purported
quantum decryption key $\widetilde{\sklqfont{dk}}$ from the quantum party and $\vk$ from
the classical party, the parties run our $2\PC$ from \Cref{sec:general} for the
functionality
\[
    b \leftarrow \sklqfont{Vrfy}(\vk,\widetilde{\sklqfont{dk}}).
\]
The bit $b$ is returned to the classical verifier, and all quantum output
registers are discarded.  Thus, the verification procedure consumes the
submitted quantum key, exactly as in the original PKE-SKL verification
experiment, while requiring only classical communication.

\paragraph{Correctness.} Correctness follows immediately from the correctness of the $2\PC$.  Security
follows from the $\epsilon$-security of the underlying $2\PC$ protocol, together with the fact that IND-KLA security is defined by an efficiently testable game, as follows:

\paragraph{Security.} Security follows from the $\epsilon$-security of the underlying $2\PC$ protocol,
together with the fact that IND-KLA security is defined by an efficiently
testable game, as follows. Suppose towards contradiction that there exists a $\QPT$ adversary that breaks IND-KLA security of the
compiled scheme with advantage at least $1/p(\lambda)$, and let $q(\lambda)$
be a polynomial upper bound on the number of verification queries it makes.
For each verification query, use the $2\PC$ simulator with error
\[
    \epsilon=\frac{1}{4p(\lambda)q(\lambda)}.
\]
For each choice of $\mathsf{coin}$, a sequential hybrid replacing the $q(\lambda)$
executions of the classical verification protocol by their ideal functionalities
changes the output probability by at most
\[
    q(\lambda)\epsilon+\negl.
\]
Therefore, the IND-KLA advantage changes by at most
\[
    2q(\lambda)\epsilon+\negl
    \leq \frac{1}{2p(\lambda)}+\negl.
\]
After these replacements, the experiment is precisely the IND-KLA experiment
for the underlying PKE-SKL scheme, where each ideal verification invocation
corresponds to a query to $\mathcal{O}_{\sklqfont{Vrfy}}$.  Thus, a non-negligible
IND-KLA advantage against the compiled scheme yields a non-negligible
IND-KLA advantage against the underlying scheme, contradicting its security.
Importantly, only the simulator depends on $\epsilon$; the compiled verification
protocol itself is fixed.

Thus, we obtain the following corollary.

\begin{corollary}
Assuming the post-quantum hardness of $\LWE$, every {\rm{IND-KLA}} secure {\rm{PKE-SKL}} scheme
with an efficient quantum verification algorithm can be compiled into an
{\rm{IND-KLA}} secure {\rm{PKE-SKL}} scheme in which revocation is performed by an
interactive protocol between a classical verifier and the quantum key holder,
using only classical communication.
\end{corollary}

\endgroup

\section{Other Applications}
\label{sec:other-applications}

In this section we give some further example applications of our state-preserving 2PC for quantum functionalities. 
Many of these applications are straightforward communication-dequantizations of existing primitives which rely on quantum communication.

We achieve negligible security loss for each application with a falsifiable security property. The reductions all follow along similar lines: fix any $\epsilon = 1/\poly[\secpar]$ and use the quantum-communication messages to $\epsilon$-simulate the 2PC that implements the classical communication. Any adversary with $2\epsilon$ advantage would then violate the security of the underlying primitive. Since this holds for all $\epsilon = 1/\poly[\secpar]$, independently of the protocol description, the adversary's advantage must be negligible in the compiled version.

\subsection{Unclonable Encryption over a Classical Channel}
\label{sec:classical-unclonable-encryption}

Unclonable encryption encodes a classical message in a quantum ciphertext so that an adversary cannot split the ciphertext into two systems that both suffice to recover the message, even after the key is revealed~\cite{TQC:BroLor20}. Our remote state preparation protocol allows the sender of such a ciphertext to be entirely classical.

Let $(\mathsf{Gen},\mathsf{Enc},\mathsf{Dec})$ be an unclonable encryption scheme. $\KeyGen$ and $\Dec$ are identical to the original scheme.
Instead of preparing the ciphertext and sending it over a quantum channel, the sender uses the 2PC from \Cref{sec:general} to remotely prepare
\[
    \rho_{k,m}\leftarrow\mathsf{Enc}(k,m).
\]
Note that the sender does not need to know a classical description of the ciphertext, since the entire encryption process is delegated.

Security follows by replacing the 2PC with its simulation given a single ciphertext, before the adversary splits its state.
Afterwards, the two separated parties run exactly the same algorithms as in the original unclonability game.
Any adversary with, say, $2\epsilon$ cloning advantage in the classical communication game can therefore be converted to an adversary with $\epsilon$ cloning advantage for the original quantum communication game, for any $\epsilon = 1/\poly[\secpar]$. Therefore the adversary's advantage must be negligible.

Combining this compiler with recent advances in unclonable encryption, we obtain the following corollary.

\begin{corollary}
Assuming the post-quantum hardness of LWE, there exists an unclonable encryption scheme with a classical encryptor.
\end{corollary}

Previously, the best known approach allowed $1/\poly[\secpar]$ advantage in the cloning game~\cite{ICALP:GheMetPor23}.

\subsection{Public-Key Quantum Money with a Classical Bank from
Polynomial Hardness}
\label{sec:classical-bank}

A public-key quantum money scheme allows a bank to issue quantum banknotes that can be verified using public information, but cannot be counterfeited. Namely, a QPT adversary that receives $q$ banknotes should not be able to produce $q+1$ valid banknotes. We show how to make the bank classical without strengthening the hardness assumptions used by the underlying money scheme, other than adding the post-quantum hardness of LWE.

To mint a banknote, the classical bank and quantum recipient run the 2PC from~\Cref{sec:general} to implement the minting circuit. The bank provides its secret key as input. The recipient does not input anything, but obtains the banknote and the serial number as a result of the 2PC.\footnote{Note that this approach is not limited to schemes where the bank first samples a classical description of the banknote, then prepares it. It can also implement schemes where the serial number and state are generated as part of an inherently quantum process.}
The verification algorithm is unchanged.

Correctness follows from correctness of the 2PC and the minting procedure. To prove security, we reduce to the security of the original quantum money scheme. Fix any $\epsilon = 1/\poly[\secpar]$. Given a banknote and serial number, $\epsilon/2$-simulate the 2PC, then finish running the cloner. The cloner's success can drop by at most $\epsilon/2$, so any $\epsilon = 1/\poly[\secpar]$ advantage contradicts security of the original quantum money scheme.

In particular, this approach applies to Zhandry's instantiation of subspace-based quantum money using post-quantum indistinguishability obfuscation~\cite{EC:Zhandry19b}. The resulting corollary avoids the subexponential-LWE assumption in the classical-bank construction of~\cite{STOC:Shmueli22}.

\begin{corollary}
Assuming post-quantum indistinguishability obfuscation and the \emph{polynomial} post-quantum hardness of LWE, there exists a public-key quantum money scheme with a classical bank and classical communication during minting.
\end{corollary}

\subsection{Improved Extraction for Classical Verification of QMA}
\label{sec:cvqc-knowledge}

There are two general measures of extraction quality in an argument of knowledge: knowledge error $\kappa$ and extraction error $\delta$. In the case of an argument of quantum knowledge, the number of witnesses extracted also becomes important, which we call the copy error.
\begin{itemize}
    \item \textbf{Knowledge error} $\kappa$ guarantees the extractor's behavior if the prover passes with probability $p \geq \kappa$.
    \item \textbf{Extraction Error} $\delta$ is the loss between the probability that the prover passes $p$ and the probability that the state extracted passes QMA verification with probability $q$. In particular, $\delta$ is a function such that $q = 1 - \delta(p)$. We distinguish between functions $\delta$ and the additive loss $\Delta = p - q$.
    \item \textbf{Copy Error} is the difference between the number of witness copies needed for an honest prover and the number of witness copies the extractor can procure.
\end{itemize}
Previously, \cite{VidickZhang21} achieved constant knowledge error $\kappa$, extraction error $\delta(p) = \poly[p]$ (which translates to additive error $\Delta = p(1 - \poly[p])$, and $0$ copy error. They also present an amplified version which achieves $\kappa = \negl$ and $\delta(p) = p - \negl$ (or $\Delta = \negl$), but at the cost of blowing up the copy error: the honest prover needs $\secpar$ copies of the witness, but the extractor only procures one.
\cite{KKR26} achieved negligible knowledge error $\kappa = \negl$, extraction error $\delta(p) = \poly[p]$, and $0$ copy error.

Our results improve the extraction error while keeping $\kappa = \negl$ and $0$ copy error. In particular, there is a uniform sequence of extractors achieving $\Delta = \epsilon$ in time $\poly[1/\epsilon]$, \emph{all for the same protocol}.
Moreover, the prover may use any witness which the original QMA verifier would accept with probability $\alpha$ to convince the classical verifier with probability $\alpha - \negl$. \cite{KKR26} achieved this for $\alpha = 1-\negl$ and left open the problem for general $\alpha$.

To construct a classical verification of QMA with these properties, simply use the 2PC to implement the original QMA verifier. Correctness of the 2PC ensures that the classical verifier accepts with probability negligibly far from the original QMA verifier. The claimed sequence of extractors is given by $\epsilon$-simulation. $\epsilon$ simulation guarantees that the output distribution of the ideal functionality is at most $\epsilon$ far from the real execution. Therefore the extracted input which the simulator provides to the ideal functionality must be accepted by the original QMA verifier with probability $\geq p - \epsilon$.

\begin{corollary}
    Assuming the post-quantum hardness of LWE, there exists a classical verification for QMA where the prover uses one copy of the witness, with knowledge error $\kappa = \negl$, $0$ copy error, and which simultaneously satisfies additive extraction error $\Delta = \epsilon$ for every $\epsilon = 1/\poly[\secpar]$.
\end{corollary}

\subsection{Proofs of No-Intrusion for Decryption Keys}
\label{sec:poni-skl}\label{sec:PoNI-skl}

A proof of no-intrusion allows a server to convince a client that an intruder has not retained a useful copy of a protected quantum resource, while leaving that resource available to the server.
This was original proposed in the context of protecting ciphertexts, with the protection of other resources left as an open problem~\cite{EC:GoyRai26}.
For decryption keys, the desired guarantee is that the server can continue decrypting, whereas a separated intruder cannot decrypt fresh ciphertexts if the server passes verification.

The construction is very similar to the previously-discussed classical revocation of quantum certificates construction for secure key leasing from \Cref{sec:classical-revocation}. It starts with any secure key leasing scheme which is secure against adversaries with oracle access to the verification algorithm, where the oracle returns the tested state.\footnote{\cite{EC:AKNYY23} considers oracle access to the verification oracle, but their oracle does not return the tested state.}, e.g.~\cite{EC:AKNYY23}. The PoNI is a 2PC for the circuit which runs verification coherently, copies the result to the classical output register, and inverts verification. It also discards the ancillas used for verificaton. The difference from before is that the post-measurement state is retained as output instead of discarded.
We also mention that the initial leasing step can be replaced with a classical interactive protocol.

Correctness and state preservation follows from correctness of the 2PC and correctness of the revocation circuit. The latter uses the gentle measurement principle and the correctness requirement that revocation accepts honest certificates with overwhelming probability.

Security reduces to the security of the original SKL scheme by simulating each of the 2PCs using a verification oracle. Similarly to before, any adversary with advantage $2\epsilon$ translates to an adversary with $\epsilon$ advantage against the original SKL scheme, for any $\epsilon = 1/\poly[\secpar]$. Therefore any adversary's advantage must be negligible.

\begin{corollary}
Assuming the post-quantum hardness of LWE, any key-leasing scheme with security in the presence of a verification oracle that returns the tested state admits proofs of no-intrusion with a classical verifier. Moreover, the key can be leased using classical communication.
\end{corollary}

\subsection{Quantum Obfuscation with Interactive Classical
Communication}
\label{sec:classical-obfuscation}

Finally, our construction allows a classical party to distribute a
quantum obfuscation via an interactive protocol.
Let $\mathsf{Obf}$ be any quantum obfuscator, and let
\[
    \rho_U\leftarrow\mathsf{Obf}(U)
\]
be its output on a classical description of a unitary $U$.
Rather than generating $\rho_U$ locally, the classical party and
the quantum recipient run our 2PC for this preparation procedure,
with $U$ (and optionally the private preparation randomness) supplied by the
classical party. The quantum recipient obtains the prescribed
obfuscation, without learning its preparation description.

Correctness follows from correctness of the 2PC. Security of the
preparation protocol gives a $\epsilon$-simulation using only $\rho_U$, for any $\epsilon = 1/\poly[\secpar]$.
By quantifying over all such $\epsilon$, the classically-sent obfuscated program satisfies the standard notion of indistinguishability obfuscation.

\begin{corollary}
    Any indistinguishability obfuscator for quantum circuits can be generically compiled into an indistinguishability obfuscator for quantum circuits where the obfuscation process is an interactive protocol with classical communication.

    In the classical oracle model, there exists an obfuscation scheme for quantum circuits with an interactive classical obfuscation process satisfying both $\epsilon$-ideal obfuscation and indistinguishability obfuscation.
\end{corollary}

We leave open the question of removing interaction to obtain a \emph{truly classical} obfuscation of quantum circuits.

\newcommand{\etalchar}[1]{$^{#1}$}

\appendix
\part{Appendix}

\section{Proofs of Additional NTCF Properties}
\label{app:NTCF}

\subsection{$\NTCF$ Collapsing with 1 Bit Leakage}

\begin{lemma}[Collapsing with a leaked inner product]
\label{lem:ntcf-collapsing-inner-product:appendix}
Let $(\Setup,\Eval,\Invert,\Chek,\Synd)$ be a $\NTCF$ family.
For a QPT adversary $\mathcal{A}$ and $c\in\{0,1\}$, define the experiment $\CollIP^{\mathcal{A}}_c(1^\lambda)$ as follows:
\begin{enumerate}
    \item The challenger samples $(\pubparams,\td)\gets\Setup(1^\lambda)$ and sends $\pubparams$ to $\mathcal{A}$.

    \item $\mathcal{A}$ sends a classical image $y$ to the challenger. The challenger computes $((0,x_0),(1,x_1))\gets\Invert(\td,y)$.
    If inversion fails, the experiment outputs $0$ and terminates.

    \item The challenger samples $d\gets\{0,1\}^{n+1}$ and defines $\Delta=(0,x_0)\oplus(1,x_1)$. Then, it sends $(d,d\cdot \Delta)$ to $\mathcal{A}$.

    \item $\mathcal{A}$ sends a $(n+1)$-qubit register $\mathcal{X}$ to the challenger, retaining its private registers. The challenger performs the two-outcome projective measurement $\{\Pi_y,I-\Pi_y\}$ on $\mathcal{X}$, where
    \[
        \Pi_y = \sum_{\substack{b\in\{0,1\},\,x\\ \Chek(\pubparams,b,x,y)=1}} \ket{b,x}\bra{b,x}.
    \]
    If the outcome is $I-\Pi_y$, the experiment outputs $0$ and terminates.

    \item If $c=1$, the challenger measures $\mathcal{X}$ in the computational basis and discards the outcome. If $c=0$, it leaves $\mathcal{X}$ unchanged. It then returns $\mathcal{X}$ to $\mathcal{A}$.

    \item $\mathcal{A}$ outputs a bit, which is the output of the experiment.
\end{enumerate}
Then, for every QPT adversary $\mathcal{A}$,
\[
    \left| \Pr\!\left[\CollIP^{\mathcal{A}}_0(1^\lambda)=1\right] - \Pr\!\left[\CollIP^{\mathcal{A}}_1(1^\lambda)=1\right] \right| \leq \negl[\secpar].
\]
\end{lemma}

\begin{proof}\footnote{Proof provided by ChatGPT 6.0 Pro.}
We use the bounded-preimage collapsing-to-collision reduction of Cao--Xue and Zhandry~\cite{C:CaoXue22,C:Zhandry22c}, followed by guessing the leaked bit. We give the two-preimage argument explicitly.

Fix $\mathcal{A}$, and let $\varepsilon$ denote its distinguishing gap in the statement. We construct a QPT claw finder with success probability at least
\[
    \frac{\varepsilon^2}{2}-\negl[\secpar].
\]
We use the branchwise uniqueness and inversion guarantees of the $\NTCF$ family: for a fixed image, there is at most one valid preimage on each branch, and two valid preimages on opposite branches form a claw. Any negligible correctness error is absorbed into the final bound.

\paragraph{Extraction with the correct leakage.}
First consider a procedure that runs $\mathcal{A}$ through the validity check, supplying the correct leakage $(d,d\cdot \Delta)$. If the check rejects, the procedure fails.

To analyze this procedure, fix the key pair and the classical transcript through delivery of $(d,d\cdot \Delta)$.\footnote{$d$ must be fixed at this point. If it can be chosen after the return of $\calX$, the adversarial operations $U$ and $M$ (defined later) will rely on the full $\td$ and not just a single bit of leakage.}
Let $\mathcal{W}$ contain the adversary's private registers and any fresh workspace needed for its final computation, and let $\mathcal{R}$ be an inaccessible reference register. Write $\ket{\psi}_{\mathcal{X}\mathcal{W}\mathcal{R}}$ for a subnormalized purification of the accepting branch of the validity check. Thus $\|\psi\|^2$ equals the probability that the check accepts, conditioned on the fixed key pair and transcript.

For $b\in\{0,1\}$, define
\[
    P_b = \left( \sum_{x:\,\Chek(\pubparams,b,x,y)=1} \ket{b,x}\bra{b,x}_{\mathcal{X}} \right) \otimes I_{\mathcal{W}\mathcal{R}}, \qquad \ket{\psi_b}=P_b\ket{\psi}.
\]
Each sum contains at most one term, and
\[
    \ket{\psi}=\ket{\psi_0}+\ket{\psi_1}, \qquad \langle\psi_0|\psi_1\rangle=0.
\]
In particular, measuring the entire valid preimage register produces the subnormalized state
\[
    \ket{\psi_0}\bra{\psi_0} + \ket{\psi_1}\bra{\psi_1}.
\]

Let $U$ be a coherent implementation of $\mathcal{A}$'s final computation, obtained by deferring its measurements and retaining its workspace. Let $\mathcal{O}$ be its output qubit, and define
\[
    M = U^\dagger \bigl(\ket{1}\bra{1}_{\mathcal{O}}\otimes I\bigr) U, \qquad V = U^\dagger Z_{\mathcal{O}}U = I-2M.
\]
These operators act trivially on $\mathcal{R}$. The circuit $U$, and hence the reflection $V$, is efficiently implementable from the description of $\mathcal{A}$ and the classical transcript. In particular, the final computation makes no further secret-dependent queries to the challenger.

The signed difference between the two experiments' probabilities of outputting $1$, conditioned on the fixed key pair and transcript but not on acceptance of the validity check, is
\begin{align}
    \delta &= \langle\psi|M|\psi\rangle - \sum_{b\in\{0,1\}} \langle\psi_b|M|\psi_b\rangle \notag\\
    &= 2\operatorname{Re}\langle\psi_0|M|\psi_1\rangle \notag\\
    &= -\operatorname{Re}\langle\psi_0|V|\psi_1\rangle.
    \label{eq:coll-ip-cross-term}
\end{align}

The extraction procedure now measures $\mathcal{X}$ in the computational basis and saves the resulting preimage in a separate classical register. It applies $V$ to the adversary's registers, leaving the saved outcome untouched, and measures $\mathcal{X}$ again. If the second outcome is a valid preimage on the opposite branch, it outputs $y$ and the two preimages; otherwise, it fails. Both outcomes can be checked publicly using $\Chek$.

For the fixed key pair and transcript, the probability of finding a claw is
\[
    p_{\mathrm{claw}} = \bigl\|P_1V\ket{\psi_0}\bigr\|^2 + \bigl\|P_0V\ket{\psi_1}\bigr\|^2.
\]
This expression includes the probability of passing the validity check because $\ket{\psi}$ is subnormalized. By~\eqref{eq:coll-ip-cross-term} and Cauchy--Schwarz,
\begin{align*}
    \delta^2 &\leq \bigl|\langle\psi_0|V|\psi_1\rangle\bigr|^2\\
    &= \bigl|\langle\psi_0|P_0V|\psi_1\rangle\bigr|^2\\
    &\leq \|\psi_0\|^2 \bigl\|P_0V\ket{\psi_1}\bigr\|^2\\
    &\leq p_{\mathrm{claw}}.
\end{align*}
For executions on which inversion fails, take $\delta=0$ and let the correct-leakage extraction procedure fail. Averaging over the keys and transcripts therefore gives
\[
    \Pr[\text{correct-leakage extraction finds a claw}] \geq \mathbb{E}[\delta^2] \geq \bigl(\mathbb{E}[\delta]\bigr)^2 = \varepsilon^2,
\]
up to the negligible correctness error.

\paragraph{Guessing the leakage.}
We now construct the actual claw finder $\mathcal{B}$, which is given only $\pubparams$. It runs $\mathcal{A}$ until $\mathcal{A}$ declares $y$, samples
\[
    d\gets\{0,1\}^{n+1}, \qquad \widetilde{\beta}\gets\{0,1\},
\]
and supplies $(d,\widetilde{\beta})$ to $\mathcal{A}$. It then performs the public validity check and the measure--reflect--measure extraction procedure above, using $\widetilde{\beta}$ as the supplied leakage bit.

For every fixed key pair, preceding transcript, and $d$ on which inversion succeeds,
\[
    \Pr[\widetilde{\beta}=d\cdot\Delta]=\frac12.
\]
Conditioned on this event, the adversary's entire execution, including its quantum state, has exactly the distribution of the correct-leakage execution. Since the event has probability $1/2$ for every such fixed classical history, conditioning on it does not bias the distribution of those histories or their correlated quantum states. Hence
\[
    \Pr[\mathcal{B}\text{ finds a claw}] \geq \frac12 \Pr[\text{correct-leakage extraction finds a claw}] \geq \frac{\varepsilon^2}{2}-\negl[\secpar].
\]
Incorrect guesses contribute a nonnegative success probability. Likewise, $\mathcal{B}$ need not determine whether trapdoor inversion would have failed: executions on which the original experiment would abort cannot decrease its claw-finding probability.

The reduction uses only one copy of $\mathcal{A}$'s auxiliary input. That input is included in the adversary's workspace, and the reflection reverses only its final computation; it does not prepare or restore the auxiliary state. The first preimage measurement may disturb that state arbitrarily. The extraction bound already accounts for this disturbance, and no operation acts on the purifying reference register.

Since $\mathcal{B}$ is QPT, claw-freeness implies that its success probability is negligible. Therefore $\varepsilon$ is negligible, as required.
\end{proof}

\subsection{2-Query Adaptive Hardcore Bit Property}

We first recall the 2-query adaptive hardcore bit property.
A $\NTCF$ family is said to satisfy the $2$-query adaptive hardcore bit property if every $\QPT$ adversary $\calA$ wins in the following game with probability $\frac12+\negl$.
\begin{enumerate}
    \item The challenger samples $(\pubparams,\td)\gets\Setup(1^\secpar)$ and sends $\pubparams$ to $\calA$.
    \item $\calA$ sends a classical tuple $(b,\vecx,\vecy)$ to the challenger.  It may retain arbitrary quantum side information.
    \item The challenger checks that $\mathsf{Check}(\pubparams,b,\vecx,\vecy)=1$. If this is not the case then the game is terminated with the default guess $c'=0$.  Otherwise, the challenger does the following:
    \begin{enumerate}
        \item Compute
    $((0,\vecx_0),(1,\vecx_1))=\Invert(\td,\vecy)$.
    \item Sample uniformly at random $c\gets\{0,1\}$ and $\vecd\gets\{0,1\}^{n+1}$.
    \item Compute $p=\vecd\cdot (1,\vecx_0\oplus\vecx_1)\oplus c$.
    \item Send $(\vecd, p)$.
    \end{enumerate}
    \item $\calA$ sends a classical vector $\vece\in\{0,1\}^{n}$.
    \item The challenger checks that $\Synd(b,\vecx_b,\vece)\notin\{\vec0^k,\Synd(b,\vecx_b,\vecd)\}$ for both $b\in\{0,1\}$. If either check fails then the game is terminated with a default guess $c'=0$.  Otherwise, the challenger sends $\beta=\vece\cdot (\vecx_0\oplus \vecx_1)$ to $\calA$.
    \item $\calA$ outputs a bit $c'\in\{0,1\}$.
\end{enumerate}
$\calA$ wins if $c'=c$.

\Cref{lem:2Q-NTCF} states that any $\NTCF$ with linear $\Synd$ satisfies this property.

\paragraph{Proof of \Cref{lem:2Q-NTCF}}
Suppose towards contradiction that there exists a $\QPT$ adversary $\calA$
that wins the 2-query adaptive hardcore bit game with probability
$\frac12+\epsilon$, 
for some non-negligible $\epsilon=\epsilon(\lambda)$.
We construct a $\QPT$ adversary $\calB$ that violates the adaptive hardcore
bit property.

Given $\pubparams$, adversary $\calB$ does the following:
\begin{enumerate}
    \item Run $\calA(\pubparams)$ to obtain a classical tuple $(b,\vecx,\vecy)$.  If
\[
\mathsf{Check}(\pubparams,b,\vecx,\vecy)=0,
\]
then output an honestly generated valid tuple $(b,\vecx,\vecy)$ together with a
uniformly random guess for the hardcore bit.  This succeeds with probability
$\geq \frac12-\negl$.
\item Otherwise, sample uniformly at random $\vecd\gets\{0,1\}^{n+1}$ and $p\gets\{0,1\}$, send $(\vecd,p)$ to $\calA$, and obtain a vector $\vece$.

Notice that this is distributed exactly as in the 2-query game: Denoting by
\[
\Delta:=(1,\vecx_0\oplus\vecx_1),
\]
where $((0,\vecx_0),(1,\vecx_1))=\Invert(\td,\vecy)$,
then in the 2-query game $p=\vecd\cdot\Delta\oplus c$,
and since $c$ is uniform, $p$ is uniform and independent of $\vecd$ and
everything that precedes it.

\item If
\[
\Synd(b,\vecx,\vece)
\in
\{\vec0^k,\Synd(b,\vecx,\vecd)\},
\]
then again output an honestly generated valid adaptive-hardcore-bit
tuple together with a uniformly random guess.

\item If $\Synd(b,\vecx,\vecd)= \vec0^k$ then again $\calB$ outputs an honestly generated valid adaptive-hardcore-bit
tuple together with a uniformly random guess.

Notice that this happens only with negligible probability. 

\item Let $\widetilde{\vece}:=(0,\vece)$
and define
$\widehat{\vecd}:=\vecd\oplus\widetilde{\vece}$.
By linearity of $\Synd$,
\[
\Synd(b,\vecx,\widehat{\vecd})
=
\Synd(b,\vecx,\vecd)
\oplus
\Synd(b,\vecx,\vece)
\neq
\vec0^k.
\]
If the two-query game accepts both branch checks, the same calculation holds for both branches.
Provided $\vecd\in\Good_{0,\vecx_0}\cap\Good_{1,\vecx_1}$, both $\vecd$ and $\widehat{\vecd}$ are then valid hardcore bit vectors.

\item Sample independent uniform bits
$r,i\gets\{0,1\}$,
and send $r$ to $\calA$ in place of the bit $\beta$ in the 2-query game,
and receives a bit $c'$.

\item If $i=0$, then output
$(\vecy,b,\vecx,\vecd,p\oplus c')$.

\item Otherwise, if $i=1$, then $\calB$ output
$(\vecy,b,\vecx,\widehat{\vecd},p\oplus c'\oplus r)$.
\end{enumerate}
We show that $\calB$ predicts an adaptive hardcore bit with non-negligible
advantage.  To this end, fix the transcript up to the point that $\vece$ is chosen, and write
\[
h:=\vecd\cdot\Delta,
\qquad
\beta:=\vece\cdot(\vecx_0\oplus\vecx_1)
      =\widetilde{\vece}\cdot\Delta.
\]
For $r\in\{0,1\}$, let $c'_r$ denote the output of $\calA$ when it is
given $r$, and define
\[
g_r:=p\oplus c'_r.
\]
If $i=0$ then $\calB$ succeeds exactly when
$g_r=h$.  If $i=1$ then
\[
\widehat{\vecd}\cdot\Delta
=
(\vecd\oplus\widetilde{\vece})\cdot\Delta
=
h\oplus\beta,
\]
and hence $\calB$ succeeds exactly when
\[
g_r\oplus r=h\oplus\beta.
\]

Averaging over the uniform choices of $i$ and $r$, we obtain
\[
\Pr[\calB\text{ succeeds}\mid\text{fixed transcript}]
=
\frac14
+
\frac12\cdot
\mathbf{1}[g_\beta=h].
\]
Indeed, if $\beta=0$, the two possible predictions corresponding to
$r=1$ are complementary, whereas the two predictions corresponding to
$r=0$ both succeed exactly when $g_0=h$.  The case $\beta=1$ is symmetric.

Finally, observe that
\[
g_\beta=h
\quad\Longleftrightarrow\quad
p\oplus c'_\beta=h
\quad\Longleftrightarrow\quad
c'_\beta=p\oplus h=c.
\]
Thus, this event is exactly the event that $\calA$ wins the 2-query game.

Let $E$ denote the event that the 2-query game does not terminate before
sending $\beta$.  Since a terminating execution contributes probability
$\frac12$ of guessing $c$, our assumption that $\calA$ wins with probability
$\frac12+\epsilon$ implies
\[
\Pr[E\wedge(c'=c)]
=
\frac12\Pr[E]+\epsilon.
\]
On executions outside $E$, $\calB$ uses a uniformly random guess and
therefore succeeds with probability $\frac12-\negl$.  On
executions in $E$, the calculation above gives
\[
\Pr[\calB\text{ succeeds}\mid E]
=
\frac14
+
\frac12\Pr[c'=c\mid E].
\]
Combining these equations,
\[
\Pr[\calB\text{ succeeds}]
\ge
\frac12+\frac{\epsilon}{2}-\negl.
\]
Since $\epsilon$ is non-negligible, this contradicts the adaptive
hardcore bit property.

Therefore every $\NTCF$ family with linear $\Synd$ satisfies the 2-query
adaptive hardcore bit property.

\qed

\subsection{Phase Injection with 1 Bit Leakage}

Recall the security game from \Cref{lem:phase-injection-leakage}:
\begin{enumerate}
    \item \textbf{Challenger:} Sample $(\pubparams, \td) \gets \Setup(1^\secpar)$ and sample a random bit $p\gets\{0,1\}$. Send $\pubparams$ to $\adv$.
    \item $\adv$ sends back a classical value $y$ and a register $\calR$ (which should store a preimage of $y$).
    \item \textbf{Challenger:} Perform the projective measurement $\{\Pi_y, I - \Pi_y\}$ on $\calR$ where $\Pi_y$ projects onto $(b, r_b)$ such that $\Chek(\pubparams,b,r_b,y)=1$. If the result is $I - \Pi_y$, then $\calA$ needs to guess $p$ (without being given any information about~$p$), and the game is over.
    \item \textbf{Challenger:} Otherwise, if $p=1$ then apply a $Z$ gate to the first qubit of $\calR$.  
    
    Send $\calR$ back to $\adv$.
    \label{step:phase-inj-phase:appendix}
    \item \textbf{Challenger (Leakage):} Compute $((0,r_0),(1,r_1))\gets\Invert(\td,y)$. Set $\Delta = (0, r_0) \oplus (1, r_1)$. Then sample $\vecd\gets\{0,1\}^{n+1}$ and send $(\vecd, \vecd \cdot \Delta)$ to the adversary.
    \item $\adv$ sends a string $\vece$ to the challenger. \label{step:phase-inj-get-d:appendix}
    \item \textbf{Challenger:} If $\Synd(b,r_b,\vece)\notin\{\vec0^k,\Synd(b,r_b,\vecd)\}$ for both $b\in\{0,1\}$, send $(\vece \cdot \Delta) \oplus p$ to $\adv$. Otherwise send a random bit $b'$.

    \label{step:phase-inj-inner-product:appendix}
    \item $\adv$ outputs a bit $p'$. They win if and only if $p'=p$.
\end{enumerate}

\Cref{lem:phase-injection-leakage} states that any QPT adversary wins with probability at most $1/2 + \negl$. The following proof is almost identical to \cite{arxiv:KR26}'s. The only difference is the usage of collapsing and adaptive hardcore bit properties which permit one bit of random inner product leakage.
The latter follows from the standard adaptive hardcore bit property and the 2-query adaptive hardcore bit property, which allows replacing the leakage bit with pure randomness.

\begin{proof}
    We define four hybrid experiments $\Hyb_{ab}$ for $a,b\in \{0,1\}$. $\Hyb_{ab}$ is the same experiment described in the lemma statement, except for two modifications. First, apply $Z^a$ in step \ref{step:phase-inj-phase:appendix}. Second, send $d\cdot (r_0\oplus r_1) \oplus b$ in \ref{step:phase-inj-inner-product:appendix}. Observe that the original game with bit $p$ is exactly $\Hyb_{pp}$.

    Consider the distributions $\Hyb_{a*} = 1/2(\Hyb_{a0} + \Hyb_{a1})$ for $a\in \{0,1\}$.
    These are the distributions resulting from inserting a fixed phase $a$ in step \ref{step:phase-inj-phase:appendix}, but sending a uniformly random bit in step \ref{step:phase-inj-inner-product:appendix} (\emph{regardless of whether both syndrome checks pass}).
    \begin{claim}
         \begin{equation}\label{eq:phase-inj-hyb-phase}
        \left|\Pr_{\Hyb_{0*}}[p' = 1] - \Pr_{\Hyb_{1*}}[p' = 1]\right| = \negl
        \end{equation}
    \end{claim}
    \begin{proof}
        Define the sub-hybrid $\Hyb'_{a*}$ where register $\calR$ is \emph{measured} in the computational basis in step \ref{step:phase-inj-phase:appendix}. Neither $\Hyb_{a*}$ needs the trapdoor since step \ref{step:phase-inj-inner-product:appendix} always reveals a random bit regardless of whether both syndrome checks pass, so $\Hyb_{a*}$ is indistinguishable from $\Hyb'_{a*}$ by the collapsing property of the NTCF with one bit leakage (\Cref{lem:ntcf-collapsing-inner-product:appendix}). Furthermore, $\Hyb'_{0*} = \Hyb'_{1*}$ since the phase is destroyed by measuring in the computational basis.
    \end{proof}
   
    Now consider the distributions $\Hyb_{*b} = 1/2(\Hyb_{0b} + \Hyb_{1b})$ for $b\in \{0,1\}$.
    These are the distributions resulting from inserting a random phase in step \ref{step:phase-inj-phase:appendix}, but sending $d\cdot (r_0 \oplus r_1) \oplus b$ for a fixed $b$ in step $\ref{step:phase-inj-inner-product:appendix}$.
    \begin{claim}
        \begin{equation}\label{eq:phase-inj-hyb-inner-prod}
            \left|\Pr_{\Hyb_{*0}}[p' = 1] - \Pr_{\Hyb_{*1}}[p' = 1]\right| = \negl
        \end{equation}
    \end{claim}
    \begin{proof}
    Conditioned on the $\{\Pi_y, I- \Pi_y\}$ measurement returning $\Pi_y$, the adversary's view after receiving register $\calR$ following step \ref{step:phase-inj-phase:appendix} is an equal mixed state over
    \begin{align*}
        &\alpha \ket{\phi_0} \otimes \ket{0, r_0}_{\calR} + \beta \ket{\phi_1} \otimes \ket{1, r_1}_{\calR}
        \\
        \text{and}\quad &\alpha \ket{\phi_0} \otimes \ket{0, r_0}_{\calR} - \beta \ket{\phi_1} \otimes \ket{1, r_1}_{\calR}
    \end{align*}
    This is precisely the mixed state 
    \[
        |\alpha|^2 \ketbra{\phi_0} \otimes \ketbra{0, r_0}_{\calR} + |\beta|^2 \ketbra{\phi_1} \otimes \ketbra{1, r_1}_{\calR} 
    \]
    Thus, it is equivalent in the adversary's view to \emph{measure} register $\calR$ in the computational basis in step \ref{step:phase-inj-phase:appendix}. By \Cref{lem:2Q-NTCF}, the leaked bit $d\cdot \Delta$ may be indistinguishably replaced by pure randomness.
    
    Now suppose $\Hyb_{*0}$ were distinguishable from $\Hyb_{*1}$, with $\calR$ measurement, with advantage $\epsilon$.
    We construct an adversary for the adaptive hardcore bit property with advantage $\epsilon$. Takes as input public parameters $\pubparams$, then use them to run $b'\gets \Hyb_{*0}$, except report a random bit $b$ instead instead of the inner product $d\cdot(r_0 \oplus r_1)$. As part of this, obtain $x$ from measuring $\calR$ and $d$ from $\calA$ in step \ref{step:phase-inj-get-d:appendix} and obtain $(x, y)$ such that $f_\pubparams(x) = y$ from the measurement of $\calR$. Output $((x,y), (d, b\oplus b'))$.
    
    Observe that if $b = d\cdot (r_0 \oplus r_1)$ and $d$ is good, then this corresponds to $H_{*0}$; if $b \neq d\cdot (r_0 \oplus r_1)$ and $d$ is good then it corresponds to $H_{*,1}$. If $d$ is not good, it is identically distributed to both.
    Since the adversary successfully guesses $b' = b\oplus (d\cdot (r_0 \oplus r_1))$ with $\epsilon$ advantage and their advantage conditioned on invalid $x$ is $0$, the reduction has advantage $\epsilon$ in guessing a valid $x$ and correct inner product $(x, d, d\cdot (r_0 \oplus r_1))$ versus guessing a valid $x$ and incorrect inner product $(x, d, 1\oplus (d\cdot (r_0 \oplus r_1)))$, violating the adaptive hardcore bit property.
    \end{proof}

    Finally, we show that $\adv$ guesses $1$ with almost the same probability in both the $p=0$ and $p=1$ case by combining \Cref{eq:phase-inj-hyb-phase} with \Cref{eq:phase-inj-hyb-inner-prod}.
\ifsubmission
    \begingroup
    \allowdisplaybreaks[1]
    \begin{align*}
        &\left|\Pr_{\Hyb_{00}}[p' = 1] - \Pr_{\Hyb_{11}}[p' = 1]\right|
        \\
        &= \Bigl|1/2\left(\Pr_{\Hyb_{00}}[p' = 1] + \Pr_{\Hyb_{01}}[p' = 1]\right)
        \\
        &\quad+1/2\left(\Pr_{\Hyb_{00}}[p' = 1]) + \Pr_{\Hyb_{10}}[p' = 1]\right)
        \\
        &\quad-1/2\left(\Pr_{\Hyb_{11}}[p' = 1] + \Pr_{\Hyb_{01}}[p' = 1]\right)
        \\
        &\quad-1/2\left(\Pr_{\Hyb_{11}}[p' = 1] + \Pr_{\Hyb_{10}}[p' = 1]\right)\Bigr|
        \\
        &\leq \left|\Pr_{\Hyb_{0*}}[p' = 1]- \Pr_{\Hyb_{1*}}[p' = 1])\right| + \left|\Pr_{\Hyb_{*0}}[p' = 1] - \Pr_{\Hyb_{*1}}[p' = 1]\right| =\negl
    \end{align*}
    \endgroup
\else
    \begin{align*}
        &\left|\Pr_{\Hyb_{00}}[p' = 1] - \Pr_{\Hyb_{11}}[p' = 1]\right|
        \\
        &= \left|\begin{array}{cc}
            &1/2\left(\Pr_{\Hyb_{00}}[p' = 1] + \Pr_{\Hyb_{01}}[p' = 1]\right) + 1/2\left(\Pr_{\Hyb_{00}}[p' = 1]) + \Pr_{\Hyb_{10}}[p' = 1]\right) 
            \\
            - &1/2\left(\Pr_{\Hyb_{11}}[p' = 1] + \Pr_{\Hyb_{01}}[p' = 1]\right) - 1/2\left(\Pr_{\Hyb_{11}}[p' = 1] + \Pr_{\Hyb_{10}}[p' = 1]\right) 
            \end{array}
            \right|
        \\
        &\leq \left|\Pr_{\Hyb_{0*}}[p' = 1]- \Pr_{\Hyb_{1*}}[p' = 1])\right| + \left|\Pr_{\Hyb_{*0}}[p' = 1] - \Pr_{\Hyb_{*1}}[p' = 1]\right|
        \\
       & =\negl
    \end{align*}
\fi
\end{proof}

\end{document}